\documentclass[twocolumn]{aastex62}
\usepackage{amsmath}
\usepackage{longtable}
\usepackage{mathrsfs}
\usepackage[super]{nth}

\usepackage{xcolor}

\usepackage[normalem]{ulem}
\definecolor{mygreen}{RGB}{34, 139, 34}

\newcommand\aastex{AAS\TeX}

\graphicspath{{./}{figures/}}

\shorttitle{VLA-COSMOS 3~GHz: sSFR evolution}
\shortauthors{Leslie et al.}

\begin{document}

\title{The VLA-COSMOS 3~GHz Large Project: Evolution of specific star formation rates out to z$\sim$5}

\correspondingauthor{Sarah Leslie} 
\email{leslie@mpia.de}

\author[0000-0002-4826-8642]{Sarah K. Leslie}
\affil{Max-Planck-Institut f\"{u}r Astronomie, K\"{o}nigstuhl 17, 69117, Heidelberg, Germany}
\affil{Leiden Observatory, Leiden University, PO Box 9513, NL-2300 RA Leiden, the Netherlands}
\altaffiliation{Fellow of the International Max Planck Research School for\\ Astronomy and Cosmic Physics at the University of Heidelberg}

\author{Eva Schinnerer}
\affil{Max-Planck-Institut f\"{u}r Astronomie, K\"{o}nigstuhl 17, 69117, Heidelberg, Germany}

\author{Daizhong Liu}
\affil{Max-Planck-Institut f\"{u}r Astronomie, K\"{o}nigstuhl 17, 69117, Heidelberg, Germany}

\author{Benjamin Magnelli}
\affil{Argelander-Institut f\"{u}r Astronomie, Auf dem H\"{u}gel 71, D-53121 Bonn, Germany}

\author{Hiddo Algera}
\affil{Leiden Observatory, Leiden University, PO Box 9513, NL-2300 RA Leiden, the Netherlands}

\author{Alexander Karim}
\affil{Argelander-Institut f\"{u}r Astronomie, Auf dem H\"{u}gel 71, D-53121 Bonn, Germany}

\author{Iary Davidzon} 
\affil{Cosmic Dawn Center (DAWN), Denmark; Niels Bohr Institute, University of Copenhagen, Lyngbyvej 2, Copenhagen \O{} DK-2100, Denmark}

\author{Ghassem Gozaliasl}
\affil{Department of Physics, University of Helsinki, P. O. Box 64, FI-00014 , Helsinki, Finland.}
\affil{Helsinki Institute of Physics, University of Helsinki, P.O. Box 64, FI-00014, Helsinki, Finland.}

\author{Eric F. Jim\'enez-Andrade}
\affil{Argelander-Institut f\"{u}r Astronomie, Auf dem H\"{u}gel 71, D-53121 Bonn, Germany}

\author{Philipp Lang}
\affil{Max-Planck-Institut f\"{u}r Astronomie, K\"{o}nigstuhl 17, 69117, Heidelberg, Germany}

\author{Mark T. Sargent}
\affil{Astronomy Centre, Department of Physics and Astronomy, University of Sussex, Brighton BN1 9QH, UK}

\author{Mladen Novak}
\affil{Max-Planck-Institut f\"{u}r Astronomie, K\"{o}nigstuhl 17, 69117, Heidelberg, Germany}

\author{Brent Groves}
\affil{Research School of Astronomy and Astrophysics, Australian National University, Canberra, ACT 2611, Australia}

\author{Vernesa Smol\v{c}i\'{c}}
\affil{Faculty of Science University of Zagreb Bijeni\v{c}ka c. 32, 10002 Zagreb, Croatia}

\author{Giovanni Zamorani}
\affil{INAF - Osservatorio di Astrofisica e Scienza dello Spazio di Bologna, via Gobetti 93/3, 40129 Bologna, Italy}

\author{Mattia Vaccari}
\affil{Inter-University Institute for Data Intensive Astronomy University of the Western Cape, Robert Sobukwe Road, 7535 Bellville, Cape Town, South Africa}
\affil{INAF - Istituto di Radioastronomia, via Gobetti 101, 40129 Bologna, Italy
}

\author{Andrew Battisti}
\affil{Research School of Astronomy and Astrophysics, Australian National University, Canberra, ACT 2611, Australia}

\author{Eleni Vardoulaki}
\affil{Argelander-Institut f\"{u}r Astronomie, Auf dem H\"{u}gel 71, D-53121 Bonn, Germany}

\author{Yingjie Peng}
\affil{Kavli Institute for Astronomy and Astrophysics, Peking University, 5 Yiheyuan Road, Beijing 100871, China}

\author{Jeyhan Kartaltepe}
\affil{School of Physics and Astronomy, Rochester Institute of Technology, Rochester, NY 14623, USA}



\begin{abstract}
We provide a coherent, uniform measurement of the evolution of the logarithmic star formation rate (SFR) -- stellar mass ($M_*$) relation, called the main sequence of star-forming galaxies (MS), for star-forming and all galaxies out to $z\sim5$. We measure the MS using mean stacks of 3\,GHz radio continuum images to derive average SFRs for $\sim$\,200,000 mass-selected galaxies at $z>0.3$ in the COSMOS field. We describe the MS relation adopting a new model that incorporates a linear relation at low stellar mass (log($M_*$/M$_\odot$)$<$10) and a flattening at high stellar mass that becomes more prominent at low redshift ($z<1.5$).
We find that the SFR density peaks at $1.5<z<2$ and at each epoch there is a characteristic stellar mass ($M_* = 1 - 4 \times 10^{10}\mathrm{M}_\odot$) that contributes the most to the overall SFR density. This characteristic mass increases with redshift, at least to $z\sim2.5$.
We find no significant evidence for variations in the MS relation for galaxies in different environments traced by the galaxy number density at $0.3<z<3$, nor for galaxies in X-ray groups at $z\sim0.75$. We confirm that massive bulge-dominated galaxies have lower SFRs than disk-dominated galaxies at a fixed stellar mass at $z<1.2$. As a consequence, the increase in bulge-dominated galaxies in the local star-forming population leads to a flattening of the MS at high stellar masses.
This indicates that ``mass-quenching'' is linked with changes in the morphological composition of galaxies at a fixed stellar mass.

\end{abstract}



\section{Introduction} \label{sec:intro}
Observations in the last century concluded that the global star formation rate (SFR) in a co-moving volume, the SFR density (SFRD) of the Universe is a rapidly evolving quantity (e.g., \citealt{Tinsley1980, Gallego1995, Cowie1996, Lilly1996, Connolly1997, Madau1998, Pascarelle1998, Tresse1998, Cowie1999, Flores1999, Steidel1999, Blain1999}).
Reviewed by \cite{Madau2014}, the cosmic SFRD has undergone a rapid decline over the last $\sim8$ billion years after
having peaked at redshift $\sim2$. The SFRD at early cosmic times ($z>4$), as inferred from ultraviolet luminosity functions, declines steeply to higher redshifts (out to $z\sim10$; \citealt{Oesch2013, Bouwens2014a, Bouwens2015, Oesch2018} \citealt{Livermore2017}). In the presence of interstellar dust, UV light is heavily obscured, and attenuation corrections have to be applied. The amount of dust-obscured star formation at these redshifts is highly uncertain and could be between 0 to $\sim10$ times that currently observed \citep{Casey2018}. Uncertain dust correction factors required for early cosmic times highlight the critical need for dust-unbiased measurements at these high redshifts ($z>4$, e.g. \citealt{Bowler2018}). 


Measuring the star formation rates (SFRs) of galaxies using the long-wavelength radio continuum can provide a dust-unbiased view of the cosmic SFR history of the Universe.
Radio continuum SFRs have been successfully applied at low redshifts ($z\lesssim 0.1$; e.g., \citealt{Condon2002, Hopkins2003, Tabatabaei2017}), and at higher redshifts ($0<z<3$ e.g., \citealt{Haarsma2000},\citealt{Pannella2009,Karim2011,Zwart2014,Pannella2015, Novak2017}, \citealt{Upjohn2019}). Radio continuum observations of the high-redshift Universe are not affected by source confusion that limits current deep, wide-area, infrared (IR) data from, e.g., \textit{Spitzer} and \textit{Herschel}, primarily due to the higher angular resolution observations that can be achieved with radio interferometers (e.g. $\sim1''$ at 3\,GHz with the VLA compared to $>10''$ at $>160$\,$\mu$m with \textit{Herschel}). 

Most radio SFR calibrations rely on the tight correlation (scatter $\sim0.2-0.3$\,dex; \citealt{Molnar2020, Yun2001}) between galaxy IR and radio emission in the local Universe.
 At frequencies $<20$\,GHz, the radio continuum emission of star-forming galaxies is typically composed of a dominant synchrotron component ($\sim90\%$ at $\nu=1.4$\,GHz, e.g., \citealt{Condon1992, Tabatabaei2017}) from supernovae and their remnants, and a thermal flat-spectrum element from warm HII regions; these two physical processes render radio continuum emission a robust SFR tracer on timescales of 0-100 Myr \citep{Murphy2011, Kennicutt2012}. 
The radio-infrared correlation has recently been reported to evolve mildly with redshift and thus it can be used to constrain an empirical radio-SFR calibration (e.g. \citealt{Delhaize2017, Magnelli2015}).
A widespread concern about using the radio as a SFR tracer is the fact that an AGN can be hidden at all other wavelengths but contribute to, or dominate (in the case of radio-AGN), the radio continuum emission (e.g. \citealt{Wong2016})\footnote{We would like to caution that measuring the SFR of AGN-host galaxies remains a fundamental challenge for all tracers including SED models.}. However, studies such as \citet{Novak2018} report that faint radio sources observed at 3\,GHz are overwhelmingly star-forming sources (e.g. 90-95\% at fluxes between 0.1 and 10\,$\mu$Jy).

Despite our growing knowledge about the cosmic SFRD, details about the key factors driving its evolution remain unclear.
In the SFR-stellar-mass plane, observations indicate that galaxies reside in two populations. One population consists of star-forming galaxies whose SFR is positively correlated with stellar mass out to redshifts of at least 4 (e.g. \citealt{Brinchmann2004, Elbaz2007, Daddi2007, Pannella2009, Magdis2010, Karim2011, Rodighiero2011, Wuyts2011, Whitaker2012, Sargent2012, Whitaker2014, Rodighiero2014, Speagle2014, Lee2015, Tomczak2016, Pearson2018}). Since first reported, this tight relationship, referred to as the main sequence of star-forming galaxies (MS; \citealt{Noeske2007}), has been used widely by the astronomical community as a tool for understanding galaxy evolution, from sample selections to constraints or validation tests for simulations. The second population consists of quiescent galaxies that are not actively forming stars. Quiescent galaxies are those that fall $\sim1$ dex below the MS (e.g. \citealt{Renzini2015}) and typically reside at the high-mass end (i.e. they have lower specific star formation rates; sSFR = SFR/$M_*$).
However, the sharp bimodality in sSFR seen in optical or UV-selected samples in the local Universe disappears, if far-infrared SFRs are used as revealed by \textit{Herschel} Surveys \citep{Eales2018a, Eales2018}, where a single galaxy sequence is found. 

There is no established consensus in the literature on the proper form of the MS; whether it is linear across all redshifts (e.g., \citealt{Wuyts2011, Speagle2014, Pearson2018}), or has a flattening or turn-over at stellar masses $\log(M_*/\mathrm{M}_\odot)> 10.5$ (e.g., \citealt{Whitaker2014, Lee2015, Tomczak2016, Schreiber2015}) remains a matter of debate. This discrepancy seems to be driven by selection effects; for instance, studies sampling more active or bluer star-forming galaxies generally report a linear MS relation (e.g., \citealt{Johnston2015}), similar to studies that include only the SFR and stellar mass residing in a disk component \citep{Abramson2014} or if only disk galaxies are selected \citep{Whitaker2015}. Furthermore, the normalization of the MS relation depends on the SFR tracer and calibrations used (e.g., \citealt{Speagle2014, Bisigello2018}).

The position of galaxies on the SFR-stellar mass plane is intimately related to both galaxy color and structural morphology, with galaxies lying on the local MS being predominantly blue and disk-like and most galaxies below the MS are red and bulge-dominated or spheroidal \citep{Kauffmann2003, Wuyts2011, McPartland2019}. In the past, this was discussed in terms of galaxy bimodality \citep{Strateva2001, Madgwick2003, Baldry2004, Bell2004, Hogg2004}. Various evolutionary channels from the blue cloud to the red sequence have been proposed; in this context, pathways that involve the rapid shutdown of star formation are referred to as quenching \citep{Bundy2006, Brown2007, Cattaneo2006, Dekel2006, Faber2007, Arnouts2007, Ilbert2010, Brammer2011}.
\citet{Peng2010} popularised the idea of a mass-dependent quenching process, ``mass-quenching'', which is dominant above stellar masses of $\sim10^{10.2}$\,M$_\odot$. 
In contrast, at low stellar masses, $M_* < 10^{10}$\,M$_\odot$, environmental quenching effects like satellite quenching and merging are believed to be the dominate source of quenching \citep{Hashimoto1998, Baldry2006, Peng2012}.

In previous work, to push to lower radio luminosities, \citet{Karim2011} performed stacking on the 1.4~GHz data in the COSMOS field (VLA-COSMOS Large, Deep and Joint projects; \citealt{Schinnerer2004, Schinnerer2007, Schinnerer2010}), drawing on a deep 3.6\,$\mu$m-selected parent sample of $>10^5$ galaxies. \citet{Karim2011} found that the sSFRs of star-forming and quiescent galaxies demonstrate a linear relationship with stellar mass and that both populations show a strong mass-independent increase in their sSFR with redshift out to $z\sim3$.

The COSMOS field with a coverage of 2 deg$^2$ is the largest cosmological deep field to-date with \textit{Hubble Space Telescope (HST)} coverage \citep{Scoville2007} and contains a rich set of panchromatic and ancillary data products, making it an ideal choice for a consistent study of galaxy properties over a sufficient range of stellar masses, redshifts, and environments. 
Now, with updated multi-wavelength photometry \citep{Laigle2016} and stellar mass functions \citep{Davidzon2017} available in the COSMOS field, it is timely to revisit the analysis of \citet{Karim2011} and better constrain the MS, the evolution of sSFR, and the SFRD of the Universe out to $z\sim5$. With the latest COSMOS2015 catalog \citep{Laigle2016} containing new K$_s$-band photometry 0.7 mag deeper than previously available (e.g. \citealt{Ilbert2013}), we are able to select a 90\% mass-complete sample down to 10$^{10}$ M$_\odot$ at $z=4$. Recent radio data at 3\,GHz \citep{Smolcic2017} is also deeper (assuming a standard spectral index of $S_\nu \propto \nu^{-0.7}$) than the previous 1.4~GHz imaging maps available \citep{Schinnerer2010}, allowing for better constraints on the average radio flux. The higher angular resolution of the radio images also allows for better matching to multi-wavelength counterparts.

We introduce the relevant datasets in Section \ref{sec:datasets} and determine the average SFRs of galaxies, employing a stacking analysis outlined in Section \ref{sec:methods}. We present our results for the MS relation for star-forming and all galaxies in Section \ref{sec:MS}. In Section \ref{sec:charmass} we combine our SFR measurements with the stellar mass functions of \citet{Davidzon2017} to explore the cosmic SFR activity in the Universe as a function of mass and redshift. In Section \ref{sec:csfrd}, we compare our cosmic SFRD measurements with the literature and discuss systematic uncertainties in these measurements. We report SFR -- $M_*$ measurements for galaxies with different morphological classifications in Section \ref{sec:morph} and for galaxies in different bins of local density, including a comparison between X-ray group members and field galaxies in Section \ref{sec:env}. We discuss the implications of our most important results in Section \ref{sec:discuss} and provide a summary and outlook in Section \ref{sec:conc}. In the Appendices we give further details on tests we have performed (e.g., with mock galaxy samples) to verify our results, and compare different MS relations using different selections, functional forms, and different literature studies.
We use a \citet{Chabrier2003} initial mass function (IMF), AB magnitudes, and the cosmological parameters ($\Omega_M ,\Omega_\Lambda, h$)=(0.30,0.70,0.70).

\section{Datasets and sample selection}\label{sec:datasets}

\subsection{Radio data}
\begin{figure}
\includegraphics[width =\linewidth]{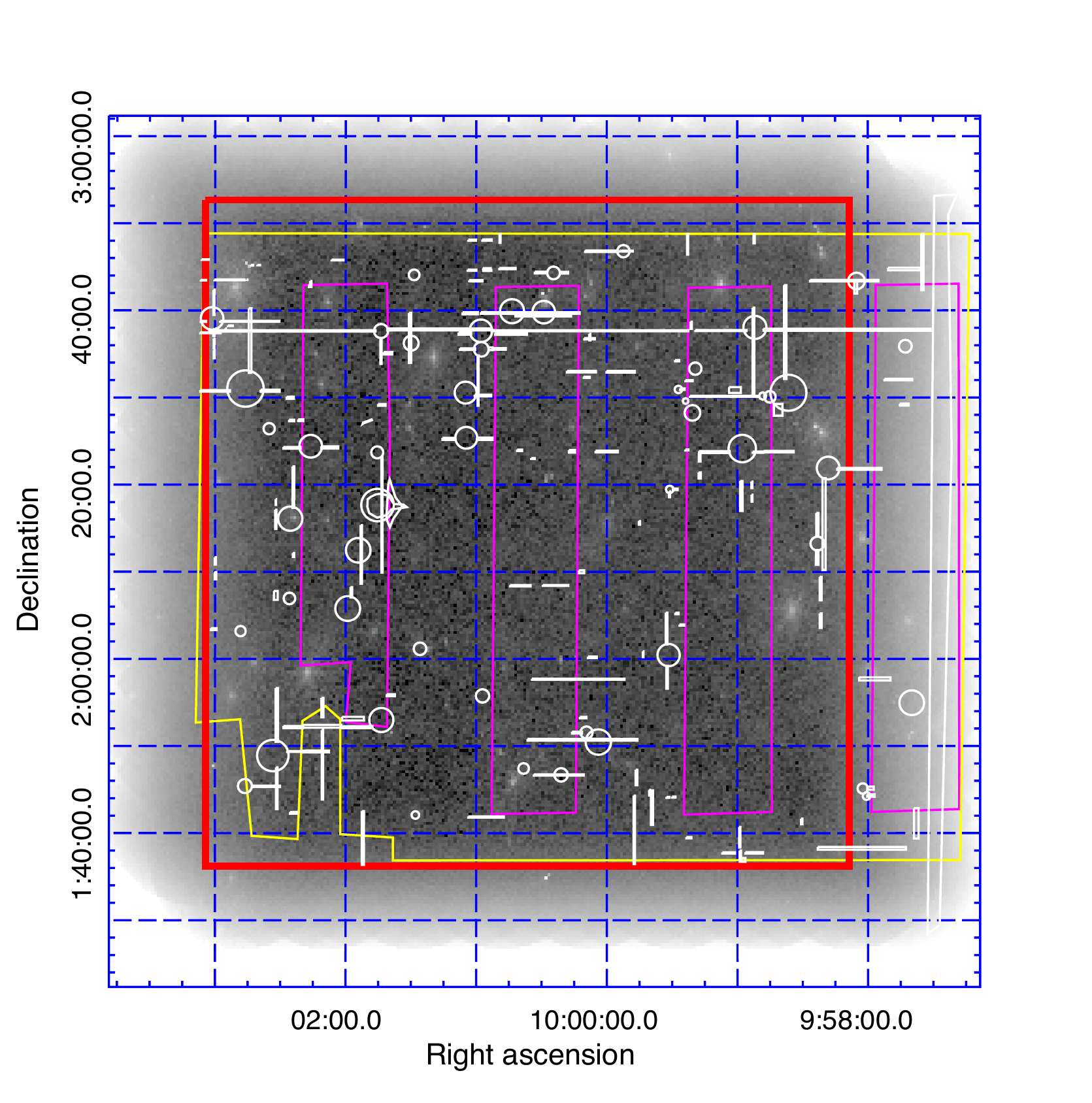}
\caption{Schematic of the COSMOS field showing the regions of data used. The background image is the VLA-3\,GHz LP RMS image \citep{Smolcic2017}. The region used for stacking (where the RMS noise is $<3$\,$\mu$Jy/beam is drawn in red. 
The yellow area is the region covered by the UltraVISTA-DR2 observations \citep{Laigle2016, McCracken2012}. The ultra-deep stripes in the UltraVISTA-DR2 observations are indicated in magenta. White regions indicate masked areas in the optical images, mostly due to bright stars. 
}\label{rms}
\end{figure}

The VLA-COSMOS 3\,GHz Large Project (hereafter VLA-3\,GHz LP), described in \citet{Smolcic2017}, observed the COSMOS field for 384 hours using the VLA S-band centered at 3\,GHz with a 2048\,MHz bandwidth. Imaging was performed with a multiscale, multifrequency synthesis algorithm for each pointing separately, tapering each with a Gaussian to achieve a circular beam before creating a final mosaic in the image plane. Across the entire 2.6 square degrees surveyed, 10,830 sources were detected above $5\sigma$ using the Blobcat software of \citet{Hales2012}. 
After visual inspection, we identified 67 sources composed of two or more detached components \citep{Smolcic2017, Vardoulaki2019}: these sources were removed from our analysis. 20\% of 3\,GHz sources lie outside the UltraVISTA regions (see Figure \ref{rms}). The astrometry is estimated to be accurate to 0.01$''$ based on a comparison to the Very Long Baseline Array - COSMOS survey \citep{HerreraRuiz2018}.
 
The observing layout was designed to achieve a uniform rms (median 2.3~$\mu$Jy at 0.75$''$ resolution) over the inner 2 square degrees with 192 pointings. Because the outermost regions of the map do not contain overlapping pointings, the noise increases rapidly towards the edge. We define a region where the rms of the 3\,GHz map is $<$ 3 $\mu$Jy/beam to use for the stacking experiment (red box in Figure \ref{rms}). 
This area can be defined by $149.53<\mathrm{RA (J2000)}<150.76$ deg and $1.615<\mathrm{DEC (J2000)}<2.88$ deg.

\subsection{COSMOS2015 photometry for a stellar mass-selected sample}
The COSMOS2015 catalog published in \citet{Laigle2016} provides optical and NIR photometry (in 31 bands) for over 1 million sources detected in $z^{++}$ (SuprimeCam) $YJHK_s$ (UltraVISTA-DR2; \citealt{McCracken2012}) stacked detection images.  
The catalog covers a square of 2 deg$^{2}$ and uses the UltraVISTA-DR2 ``deep'' and ``ultra-deep'' stripes, resulting in a depth and completeness that is not uniform across the field. The increased exposure time in the ``ultra-deep'' region, covering an area of 0.62 deg$^{2}$, doubles the number of sources compared to the previous version of the catalog \citep{Ilbert2013}. 
Regions saturated by stars or bright sources are masked out in the optical to NIR bands, resulting in a total coverage of 1.77 square degrees. 

Photometric redshifts ($z_{p}$) were computed using LePhare (\citealt{Arnouts1999, Ilbert2009}) and have been calibrated using $\sim20,000$ spectroscopic targets from the literature. The spectral energy distribution (SED) fitting library is the same as \cite{Ilbert2013}. 

Stellar masses are derived as described in \citet{Ilbert2015}, using a grid of synthetic spectra created using the stellar population synthesis models of \citet{Bruzual2003} with a \citet{Chabrier2003} IMF for two metallicities, $\tau$ (i.e. declining and delayed) star formation histories, two different dust attenuation curves, and emission-line prescriptions. 

We not only require accurate photometric redshifts and stellar masses for selecting our parent samples and sub-samples for stacking and SFR calculations, but accurate source positions are also required in order to stack the radio images at the locations of the galaxies of interest. 
A positional matching of the VLA-3\,GHz LP sources with the COSMOS2015 catalog performed using a search radius of 0.8$''$ by \citet{Smolcic2017b} found small systematic astrometric offsets that vary across the field\footnote{These offsets arise due to the astrometry of the COSMOS2015 being tied to Megacam $i-band$ data \citep{McCracken2010,McCracken2012} and will not be an issue in future releases of Ultra-VISTA data which are to be tied to Gaia astrometry.}. Therefore, before using the optical positions reported in COSMOS2015 as inputs in our stacking routine, we correct for the systematic offset using the best fitting linear relations reported in \citet{Smolcic2017b}: 
\begin{eqnarray}
\text{RA} = \text{RA}_{L16}+(-0.041\text{RA}_{L16}+6.1)/3600\\
\text{Dec} = \text{Dec}_{L16}+(0.058\text{Dec}_{L16}-0.147)/3600
\end{eqnarray}

\subsubsection{High-z optimised parameters from \citet{Davidzon2017}}

To optimize the SED fitting for high redshift galaxies ($z>2.5$) \citet{Davidzon2017} expanded the grid of allowed redshifts out to $z=8$ 
 and re-fit the COSMOS2015 photometry with LePhare using additional high-z templates of extremely active galaxies with a rising star formation history (SFH) as well as allowing highly attenuated galaxies. 
The \citet{Davidzon2017} catalog also provides improved removal of stellar contaminants. 
Overall, there is good agreement between the \citet{Laigle2016} and \citet{Davidzon2017} redshifts, however, there is a subsample of objects that moved from $z_{L16}<1$ to $z_{D17}\sim3$ because they are now classified as dusty galaxies at high redshift. Other groups of galaxies that changed redshift between the catalogs had no statistical impact on the analysis of \citet{Davidzon2017}. 
In Sections \ref{sec:charmass} and \ref{sec:csfrd}, we use the stellar mass functions calculated from \citet{Davidzon2017}. 

Due to the range of templates used, the \citet{Davidzon2017} catalog is not as robust for low-z galaxy properties as the \citet{Laigle2016}, increasing the photometric redshift errors. To combine the low- and high-z optimised catalogs of \citet{Laigle2016} and \citet{Davidzon2017} we adopt the \citet{Davidzon2017} values (such as stellar mass, photometric redshift, and rest-frame colors) for all galaxies with $z_{D17}>2.5$, and the \citet{Laigle2016} values for all galaxies with $z_{D17}<2.5$. In this way, we do not double count any galaxies in our sample, however, we note that 18,902 galaxies with $z_{D17}<2.5$ still have $z_{L16}>2.5$. Therefore, we are biasing our results towards higher redshifts by adopting the $z_{L16}$ in these cases.

\subsection{Sample selection: a stellar-mass complete sample.}

The $K_s$ band traces the stellar mass of galaxies out to $z\sim4$. At $z>4$, the $K_s$ band lies blueward of the Balmer break and therefore the 3.6$\mu$m band must be used. The latter is also a better stellar-mass proxy than the $K_s$ band between $2.5<z<4$ (see \citealt{Davidzon2017}). 
The limiting magnitude of both the $K_s$ and 3.6\,$\mu$m catalogs is $\mathcal{M}_{\mathrm{lim}}=24$ across the full field.
For our highest redshift-bin only, we exclusively select galaxies from the ultra-deep UltraVISTA regions where the limiting magnitude is fainter than in the deep regions ($\mathcal{M}_{\mathrm{lim}}=25$ at 3.6\,$\mu$m, for a 70\% completeness; \citealt{Davidzon2017}). 
To maximize the number of galaxies in each stack, we use the deep UltraVISTA limits ($K_s<24$) across the entire UltraVISTA COSMOS region out to $z<2.5$. 

To select a stellar mass-based sample, we apply the following selection criteria:
\begin{itemize} 
\item $K_s<24.0$ for $0.2<z_{p}<2.5$,
\item $3.6\mu$m $< 23.9$ for $2.5<z_{p}<4$, and
\item $3.6\mu$m $< 25.0$ and FLAG\_DEEP = 1 for $z_{p}>4$.
\end{itemize}
where the FLAG\_DEEP=1 requirement in the COSMOS2015 catalog means that the galaxies are in the ultra-deep UltraVISTA regions. 206,674 galaxies meet these criteria.

\begin{figure*}
\includegraphics[width=\linewidth]{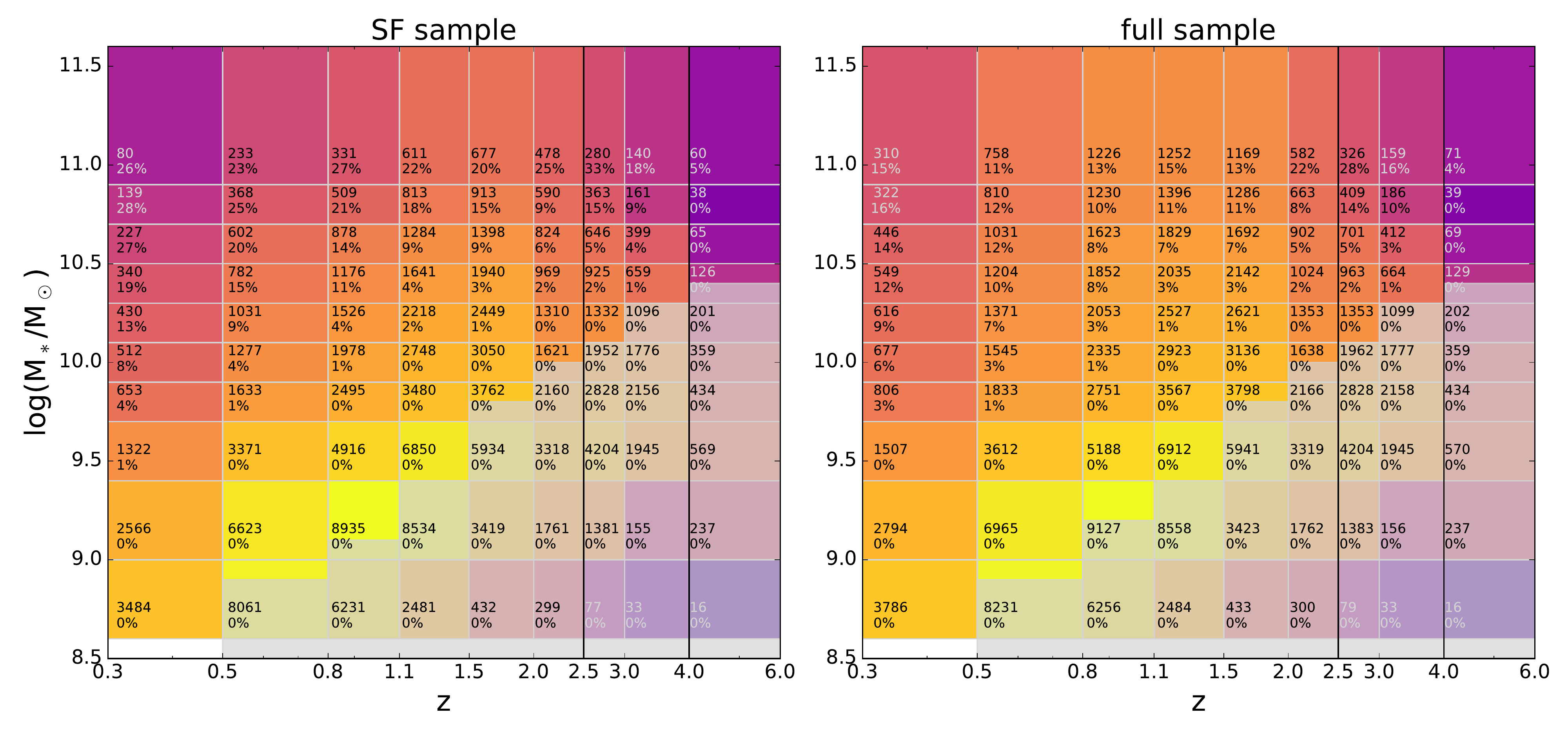}
\caption{
Binning scheme in stellar mass and redshift used for our primary stacking analysis for galaxies that meet our magnitude and area cuts. The color represents the number of galaxies in the bin (also indicated by the top number in the box). The left panel shows galaxies classified as star-forming according to their $NUVrJ$ colors and the right panel shows our scheme for all galaxies. The vertical black lines at $z=2.5$ and $z=4$ show where we transition from using the \citet{Laigle2016} catalog to the \citet{Davidzon2017} catalog and from selecting galaxies across the full COSMOS field to selecting galaxies only from the UltraVISTA Ultra-Deep regions, respectively. The bottom number in each bin shows the percentage of galaxies detected in the $5\sigma$ VLA-3\,GHz catalog; ranging from $\sim$30\% in the most massive bins to 0\% at stellar masses $<10^{9.6}$ M$_\odot$. Regions below the stellar mass-completeness limits have grey shading overlaid.}\label{fig:binning}
\end{figure*}

Stellar mass completeness is calculated empirically across our subsamples following \citet{Pozzetti2010} (see also \citealt{Moustakas2013}, \citealt{Laigle2016}, and \citealt{Pearson2018}). For each galaxy with a redshift and $K_s$ (or 3.6\,$\mu$m) band detection, we calculate an estimate of the stellar mass it would need in order to be observed at the magnitude limit (in $K_s$ band for galaxies $z<2.5$ and at 3.6\,$\mu$m for galaxies $z>2.5$):
$\log(M_{*\text{,lim}}) = \log(M_*) - 0.4 (\mathcal{M_*}_{K_s,\text{lim}}-\mathcal{M}_{K_s})$. 
In each redshift bin, the faintest 20\% of objects were selected and the $M_{*\text{lim}}$, above which 90\% of these faint galaxies lie, is adopted as our stellar mass completeness limit (e.g., \citealt{Pozzetti2010}). 
Using the 3$\sigma$ limiting magnitudes, we find limiting masses, shown in Figure \ref{fig:binning}, that are consistent with \citet{Laigle2016} and \cite{Davidzon2017}. 
For Sections \ref{sec:morph} and \ref{sec:env}, where different selections and binning schemes are used, mass completeness limits are calculated as stated above. We only show data above this mass-complete limit unless stated otherwise. 

\subsection{Galaxy type and AGN classifications}\label{sec:classify}
The galaxy population is often quoted as being bi-modal, with star-forming galaxies and quiescent galaxies having different sSFR distributions\footnote{However, we note that measurements of SFR for quiescent galaxies tend to be highly uncertain, e.g., \cite{Salim2016}.}.
However, without a measure for SFR to begin with (obtaining SFRs is the aim of our radio-stacking), 
we instead rely on galaxy colors to select star-forming (SF) galaxies. Another concern to be addressed in this section is the uncertain contribution of AGN emission to the radio fluxes used to measure star-formation activity. 

\citet{Speagle2014} showed that studies selecting bluer galaxies, such as those using a Lyman Break or $BzK$ selection, find steeper (linear) MS slopes of 0.75--1 than ``mixed'' studies, such as those using rest-frame color-color cuts, which tend to find slopes of $\sim0.6$, and that studies making no pre-selection for star-forming galaxies find slopes $\leq0.4$. 
In this work, we adopt the color-color selection of \citet{Ilbert2013}; quiescent galaxies are those with rest-frame $M_{NUV}-M_r>3(M_r - M_J)+1$ and $M_{NUV}-M_r>3.1$. 
This method has the advantage of separating dusty star-forming galaxies and quiescent galaxies (e.g. \citet{Laigle2016}, \citet{Ilbert2017} and \citet{Davidzon2017}).
We want to emphasize that measuring the MS depends critically on the sample selection, so we discuss it in further detail in Appendix \ref{sec:select}, and show our results for the commonly used $UVJ$ selection as well as our $NUVrJ$ selection. 
Out of the 206,674 galaxies that meet our selection criteria, 183,987 are classified as star-forming according to our $NUVrJ$ selection. 

Besides sample selection effects, one might expect AGN contamination to be a problem for radio emission since there are radio-loud AGN which cannot be easily identified as AGN at other wavelengths. However, these particular objects tend to live in systems with redder colors (e.g. \citealt{Smolcic2009, Moric2010, Brown2001}) and should have been removed from our star-forming galaxy sample at low redshift. 
For our main analysis we remove galaxies classified as multi-component radio sources\footnote{Multi-component radio sources are sources whose radio emission breaks into multiple components (e.g. a radio jet comprised of a core and two radio lobes). Including the 9/57 multi-component galaxies of \citet{Vardoulaki2019} classified as star-forming in our stacks does not change our measurements of median SFRs.} in \citet{Vardoulaki2019}, and galaxies classified as AGN according to their X-ray luminosity (Using [0.5-2]\,keV \textit{Chandra} data, if the luminosity was $L_X>10^{42}$~erg s$^{-1}$), or observed MIR colors \citep{Donley2012}. 
 For sources in the 3\,GHz multiwavelength counterpart catalog of \citet{Smolcic2017b}, we have flagged and removed sources whose SED is best fit with an AGN + galaxy template and radio excess sources ($\log(L_{1.4}/\text{W Hz}^{-1})>\log(\text{SFR}_{\text{IR}}/\text{M}_\odot \text{yr}^{-1})+ 21.984\times(1+z)^{0.013}$) from \citet{Delvecchio2017}. 
 
In Appendix \ref{sec:agn}, we show how the results would be affected by including AGN in our SF galaxy sample: 
 the median SFR is not significantly altered from before, however, the mean SFR is, particularly for the most massive galaxies where it can be up to a factor of 3 higher with AGN. 
When considering all galaxies, we include galaxies of red colors, which are more likely to host a radio AGN, especially at the massive end (e.g. \citealt{Auriemma1977}, \citealt{Sadler1982}, \citealt{Fabbiano1989}, \citealt{Smolcic2009}, \citealt{Brown2011}). In the local Universe, all galaxies with stellar mass $M_*>10^{11}$ show radio-AGN activity with $L_\mathrm{150 MHz}>10^{21}$\,W Hz$^{-1}$ \citep{Sabater2019}. The most massive bins ($10.9<\log(M/\mathrm{M}_{\odot}<11.6$) are the most affected by our AGN removal. For these most massive galaxies, the 3\,GHz flux over-estimates the SFR when considering galaxies with red $NUVrJ$ colors due to flux contribution from the AGN, and the difference is strongest at $z<2$. We also note that this discrepancy at high mass is stronger ($>4\sigma$) when comparing mean rather than median SFRs for the two different AGN de-selection methods. 
Our final sample after AGN removal consists of 204,903 galaxies, including 182,730 SF galaxies.

\subsection{Morphological parameters}
Morphological measurements were carried out on HST/Advanced Camera for Surveys (ACS) F814W ($I$-band) images with a resolution of $\sim$ 0.15$''$ \citep{Koekemoer2007} to create the Zurich Structure and Morphology Catalog (ZSMC\footnote{available on IRSA https://irsa.ipac.caltech.edu/data/COSMOS/ tables/morphology/}). 
We have matched the ZSMC, complete down to $I\le$24 mag, with the COSMOS2015 photometric catalog of \citet{Laigle2016}. 
Galaxies in ZSMC were classified as ``early-type'', ``late-type'' or ``irregular/peculiar'' according to the Zurich Estimator of Structural Types algorithm (ZEST; \citealt{Scarlata2007}). The ZEST algorithm performs a principal component analysis on five non-parametric structural estimators: asymmetry, concentration, Gini coefficient, $M_{20}$, and ellipticity. 
The bulginess classification of late-type galaxies was additionally based on S\'{e}rsic indices derived from single-component GIM2D fits \citep{Sargent2007}, available for the brightest galaxies $I\le22.5$. 

To study the sSFR of galaxies in different morphology classes, we separate galaxies into four classes, ZEST type = 1 (early-type; ET), ZEST type = 2.0 or 2.1 (bulge-dominated late type), ZEST type = 2.2 or 2.3 (disk-dominated late type), and ZEST type = 3 (irregular). Because we only have HST imaging in one band, we are unable to account for morphological k-corrections (due to color gradients in galaxies), and, as such, we limit our analysis to
 $z<1.5$, where the $I$-band traces the stellar light (see \citealt{Scarlata2007}).

\subsection{Environmental parameters}

\citet{Scoville2013} probed the large-scale structure of the COSMOS field by measuring galaxy surface density in 127 redshift slices between $0.15<z<3.0$ using 155,954 $K_s$ band-selected galaxies from the photometric redshift catalog of \citet{Ilbert2013}. 
We use the Voronoi tessellation results which provide a density estimate for all galaxies because it yields a 2D surface density in each redshift slice. 
The local environment is described by $\delta$, the density per comoving Mpc${^2}$ at the location of each source with the mean density of the redshift slice subtracted.

We also investigate the average radio-based SFR properties of galaxies lying inside X-ray groups in the COSMOS field. The primary X-ray galaxy groups catalogs were presented by \citet{Finoguenov2007, George2011} and used available X-ray data of Chandra and XMM-Newton with photometric datasets and identified groups with secure redshift out to $z=1.0$. 
The sample of X-ray galaxy groups used in this study is selected from two recent catalogs of 247 and 73 groups presented by \citet{Gozaliasl2019} and Gozaliasl et al. (in preparation). 
The X-ray emission peak/center of groups is determined with an accuracy of $\sim 5\arcsec$, using the smaller scale emission detected by high-resolution \textit{Chandra} imaging. The new X-ray groups have a mass range of $M_{200c}=$ 8$\times 10^{12}$ to $3\times10^{14}$ M$_\odot$ with a secure redshift range of $0.08<z<1.53$. $M_{200c}$ is the total mass of groups which is determined using the scaling relation $L_{X}-M_{200c}$ calibrated by weak lensing \citep{Leauthaud2010}. For the full details of group identification and their properties, we refer readers to \citet{Gozaliasl2019}.

 For our study of the MS relation in X-ray groups, we focus our analysis on one particular redshift bin, $0.64<z<0.88$, which includes 73 groups with $M_{200c}=2\times10^{13}$ to $2.5\times10^{14}$ M$_\odot$. The X-ray group members are selected within $R_{200c}$ from the group X-ray centers using the COSMOS2015 photometric redshift catalog. 
 We note that $\sim 31\%$ of the group galaxies have spectroscopic redshifts.

\section{Methods: radio stacking, flux measurement, and SFR calculation}\label{sec:methods}
We summarize our stacking workflow here and will justify the choices made in Appendix \ref{sec:appendixmethod}. In principle, stacking is a straightforward process, but in practice, there are many subtleties which we discuss here and in Appendix \ref{sec:appendixmethod}.

\begin{enumerate}
\item First, an input list of coordinates is created for the $N_\mathrm{objs}$ galaxies to be stacked, taking into consideration the selection criteria and positional offsets described in the previous section. 
\item Using the stacking routine developed by \citet{Karim2011}\footnote{
We have verified the performance of the routine by inserting artificial Gaussian sources into the 3\,GHz map and recovering the stacked images. }, we create $40''\times40''$ (200$\times$200 pixels) cutouts of the 3\,GHz image, centered on the input position of each galaxy. The cutouts are saved together as a data-cube of size $N_\mathrm{objs} \times 200\times 200$. 
The stacking routine also calculates the median image of the cutouts and saves the central pixel value (which should be the peak flux) and the rms.
\item To calculate the total flux, we fit a 2D elliptical Gaussian function to the mean image\footnote{We discuss the advantages and disadvantages of a median and mean stacking in Appendix \ref{sec:meanvsmed}}, restricted to the central $8''\times8''$ (40$\times$40 pixels). We input initial conditions for the fit using the peak flux from the stacking routine and the beam size.
\item To estimate the uncertainty on the total flux, we perform a bootstrap analysis. This involves randomly drawing, with replacement, $N_\mathrm{objs}$ galaxy cutouts (from our initial list of galaxies) and creating a new mean stack. We then record the total flux for each resulting stack and repeat the process 100 times. We report the 5th, 16th, 50th, 84th, and 95th percentiles of the total fluxes. Tables 4 and 5 show the 5th and 95th percentile as the lower and upper errors. In this way, our errors on the total flux are indicative of the sample variance.
\end{enumerate}
\paragraph{A note on total fluxes}
Point sources can be described entirely by their peak cleaned flux, which, if the optical and radio centers are aligned, should correspond to the central pixel in each cutout. Galaxies in the 3\,GHz image are not just point sources, and so we have to take into account the source extent. Indeed, \citet{Bondi2018} found 77\% of star-forming galaxies are resolved in the VLA-3\,GHz LP. \citet{JimenezAndrade2019} shows that the physical effective radius of the 3\,GHz component is 1-2~kpc in star-forming galaxies and is relatively constant with stellar mass and evolves shallowly with redshift out to $z\sim2.25$.
Astrometric uncertainty also plays a role in the determination of total flux and is a second key reason why the peak flux from the stack cannot be used to represent the total flux. 
Jittering of sources due to astrometric offsets between the optical catalog and the true 3\,GHz source position causes an effective blurring of the stacked image. For sources in the $5\sigma$ VLA catalog, the difference between the 3\,GHz and COSMOS2015 positions have a spread of $\sigma=0.1''$ (after correcting for the systematic offset; \citealt{Smolcic2017b}), which is half the size of a pixel.
We discuss more details about the tests we have performed to verify our methods in Appendix \ref{sec:appendixmethod}.

\subsection{Radio SFR calibration}\label{sec:sfrcalib}
The most commonly used radio SFR calibrations are bootstrapped from the empirical IR-radio correlation. For our results we have adopted a SFR calibration based on the IR-radio correlation determined by \citet{Molnar2020}, and a radio spectral index\footnote{$S_\nu \propto \nu^\alpha$} of $\alpha=-0.7$ unless stated otherwise. The following explains how we convert observed flux at 3~GHz to a SFR. 

Observed 3~GHz fluxes ($S_{3\text{GHz}}$; W Hz$^{-1}$ m$^{-2}$) are converted to rest-frame 1.4 GHz luminosities ($L_{1.4 \text{GHz}}$; W Hz$^{-1}$) where the infrared-radio correlation is traditionally calibrated, via:
\begin{equation}
L_{1.4\text{GHz}} = \frac{4\pi D_L^2}{(1+z)^{\alpha+1}}\left(\frac{1.4\mathrm{GHz}}{3\mathrm{GHz}}\right)^\alpha S_{3 \text{GHz}},
\label{eq:Lradio}
\end{equation}
where $D_L$ is the luminosity distance to the galaxy and $\alpha$ is the spectral index.
It is standard in the literature to assume a single spectral index for the radio spectral energy distribution (usually taken to be $\alpha=- 0.7$ or $\alpha = - 0.8$). The spread in spectral indices is observed to be $\sigma=0.35$ (e.g., \citealt{Smolcic2017}) and the uncertainty of the spectral index can induce significant errors in the derived radio luminosity for a single object. However, on a statistical basis, the symmetry of the spread is expected to cancel out the variations, typically yielding a valid average luminosity for the given population (see \citealt{Novak2017, Delhaize2017, Smolcic2017b} for more specific discussions on this topic). 

The relation between the radio and FIR luminosities of star-forming galaxies is assumed to arise because both emissions depend on the recent massive star formation (e.g., \citealt{Condon1992, Yun2001, Bell2003, Lacki2010}).
\citet{Helou1985} first introduced the commonly used parameter $q_\mathrm{TIR}$, the ratio of infrared-to-radio luminosity:
\begin{equation} 
\label{eq:q}
q_\mathrm{TIR} = \log\left(\frac{L_{\text{TIR}}}{3.75\times 10^{12} \text{Hz}}\right) - \log\left(\frac{L_{1.4\text{GHz}}}{\text{W Hz}^{-1}}\right).
\end{equation}
The infrared-to-radio ratio $q_\mathrm{TIR}$ has been observed to decrease with redshift, (e.g.  \citealt{Seymour2009},\citealt{Ivison2010a}, \citealt{Ivison2010b},    \citealt{Basu2015}, \citealt{Magnelli2015}, \citealt{CalistroRivera2017}, \citealt{Delhaize2017}, and \citealt{Ocran2019}), but we note some studies have found no significant evolution (e.g. \citealt{Garrett2002, Appleton2004, Sargent2010}). The reason for this observed evolution is currently unclear, however, evolution in the SFR surface density, selection effects, and the presence of radio AGN have been suggested as possible explanations \citep{Magnelli2015, Molnar2018, Molnar2020}. 

\cite{Molnar2020} found, for a flux-matched sample of star-forming galaxies at $z<0.2$, that $q_\mathrm{TIR}$ depends on radio luminosity;
\begin{equation}
\label{eq::l_vs_q}
 q_{\mathrm{TIR, SF}} = (-0.153\pm0.008) \cdot \,\log({L_{\rm 1.4GHz}}) + (5.9\pm0.2),
\end{equation}
We combine Equations \ref{eq:Lradio}, \ref{eq:q}, and \ref{eq::l_vs_q} to compute $L_\mathrm{TIR}$ for a given 3GHz flux and redshift.
Finally, we derive SFRs using the total infrared calibration in \citet{Kennicutt2012}, for a Chabrier IMF. 


%

\begin{figure}
\includegraphics[width=\linewidth]{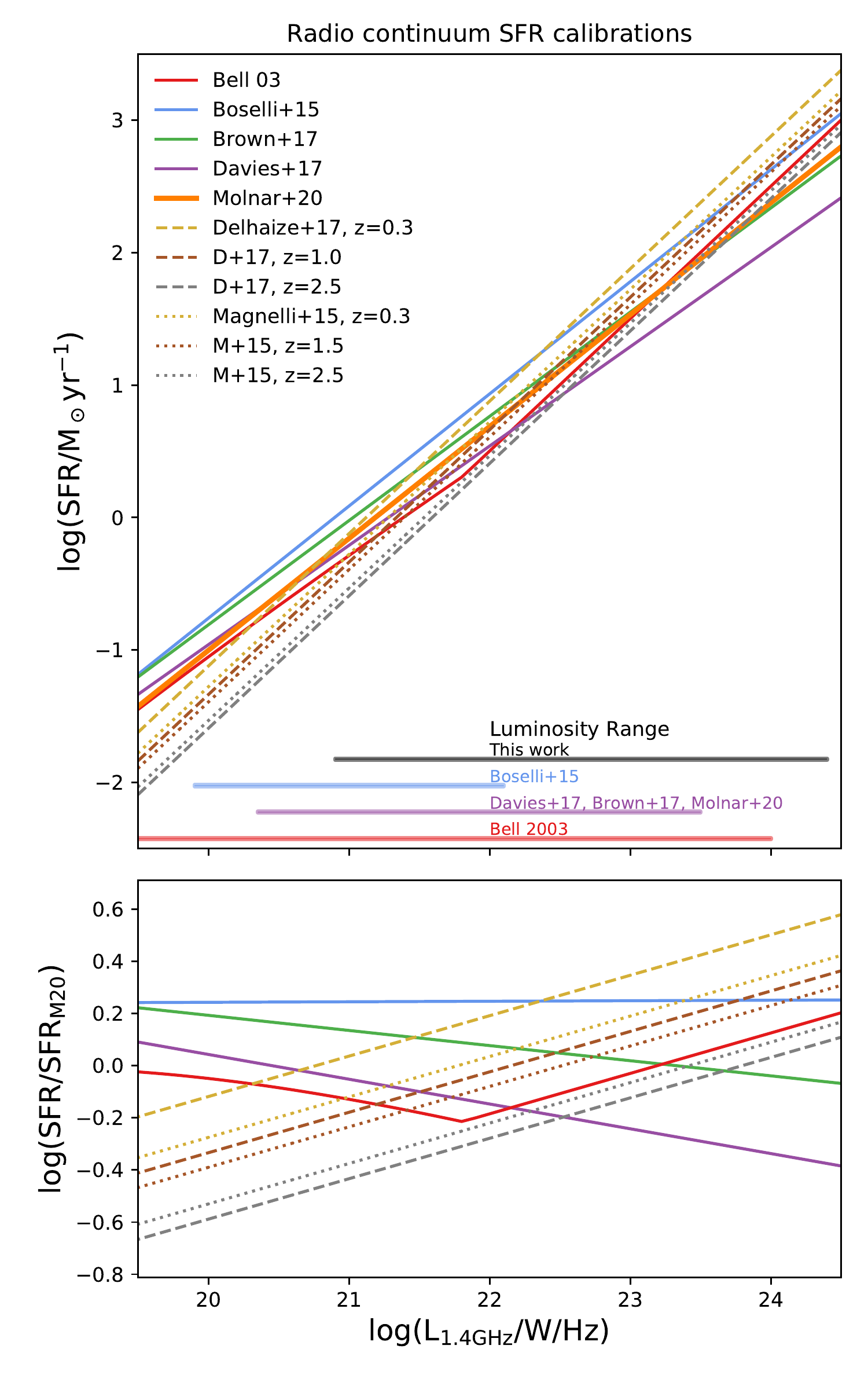}
\caption{Comparison of 1.4~GHz radio continuum luminosity SFR calibrations from the literature. For the redshift dependent ``$q_\mathrm{TIR}$" from \cite{Delhaize2017} and \cite{Magnelli2015}, we show the calibration at $z=0.3$ (gold), $z=1.5$ (plum), and $z=2.5$ (gray), as solid, dashed, and dotted lines, respectively. Horizontal bars at the bottom of the top panel illustrate the radio luminosity range over which different SFR calibrations were derived. Red shows the range from \cite{Bell2003} and purple shows the range from \cite{Davies2017}, which is comparable to other studies that also rely on the FIRST survey (e.g. \cite{Brown2007}, \cite{Molnar2020}). The rest-frame 1.4\,GHz luminosity range probed by our 3GHz COSMOS stacks is shown in black and extends to higher luminosities than probed by the local studies. The bottom panel shows the difference between calibrations normalized to the \cite{Molnar2020} calibration adopted in this work.}\label{fig:radiocalib}
\end{figure}

In Figure \ref{fig:radiocalib} we compare some literature SFR relations over the radio luminosity probed by our 3\,GHz data. We compare recipes that directly calibrate L$_{1.4\mathrm{~GHz}}$ to SFR, where the SFR comes from another tracer; namely multiband UV+IR \citep{Davies2017}, H$\alpha$ \citep{Brown2017}, and 
\cite{Boselli2015}, and FUV+IR \citep{Bell2003}. We also include two calibrations for $q_\mathrm{TIR}$ as a function of redshift determined by \cite{Delhaize2017} and \cite{Magnelli2015}.
\cite{Magnelli2015} also performed stacking with a mass-selected sample and confirmed that $q_\mathrm{TIR}$ does not depend significantly on the offset from the MS, nor does the radio-spectral index, at $M_*>10^{10}M_\odot$ and to $z<2.3$. 

The relevant radio powers used to derive the calibrations are highlighted as horizontal bars in Figure \ref{fig:3GHzcalib}.
For the local results (shown as solid lines), the calibrations are more consistent with each other at lower luminosities (with the exception of \cite{Bell2003}).
However, even in the range of radio luminosities covered by all samples $21<\log(L_{1.4\mathrm{~GHz}})<23$ there are significant discrepancies of up to 0.4 dex.
The largest discrepancy between calibrations occurs for radio luminosities above those probed in the local Universe and used as calibration samples ($L_{1.4\mathrm{\,GHz}}>10^{24}$\,W/Hz). At $z=0.3$, switching from a \cite{Delhaize2017} to \cite{Bell2003} SFR calibration at lowest radio luminosities would decrease the SFR by 0.2 dex, whereas at $z=2.5$ the difference would be 0.6 dex. At the highest luminosities, the \cite{Bell2003} SFRs are 0.4 dex higher than \cite{Delhaize2017} SFRs at $z=0.3$ and $\sim$0.1 dex lower at $z=2.5$.
We also note the above SFR calibrations have not been tested on quiescent galaxies, therefore our SFR measurements for ``all'' galaxies must be taken with caution.

\section{Results}
\subsection{Galaxy SFR--$M_*$ relation}\label{sec:MS}
\begin{figure*}
\centering
\includegraphics[height=11cm, trim={0 0 1.5cm 0},clip]{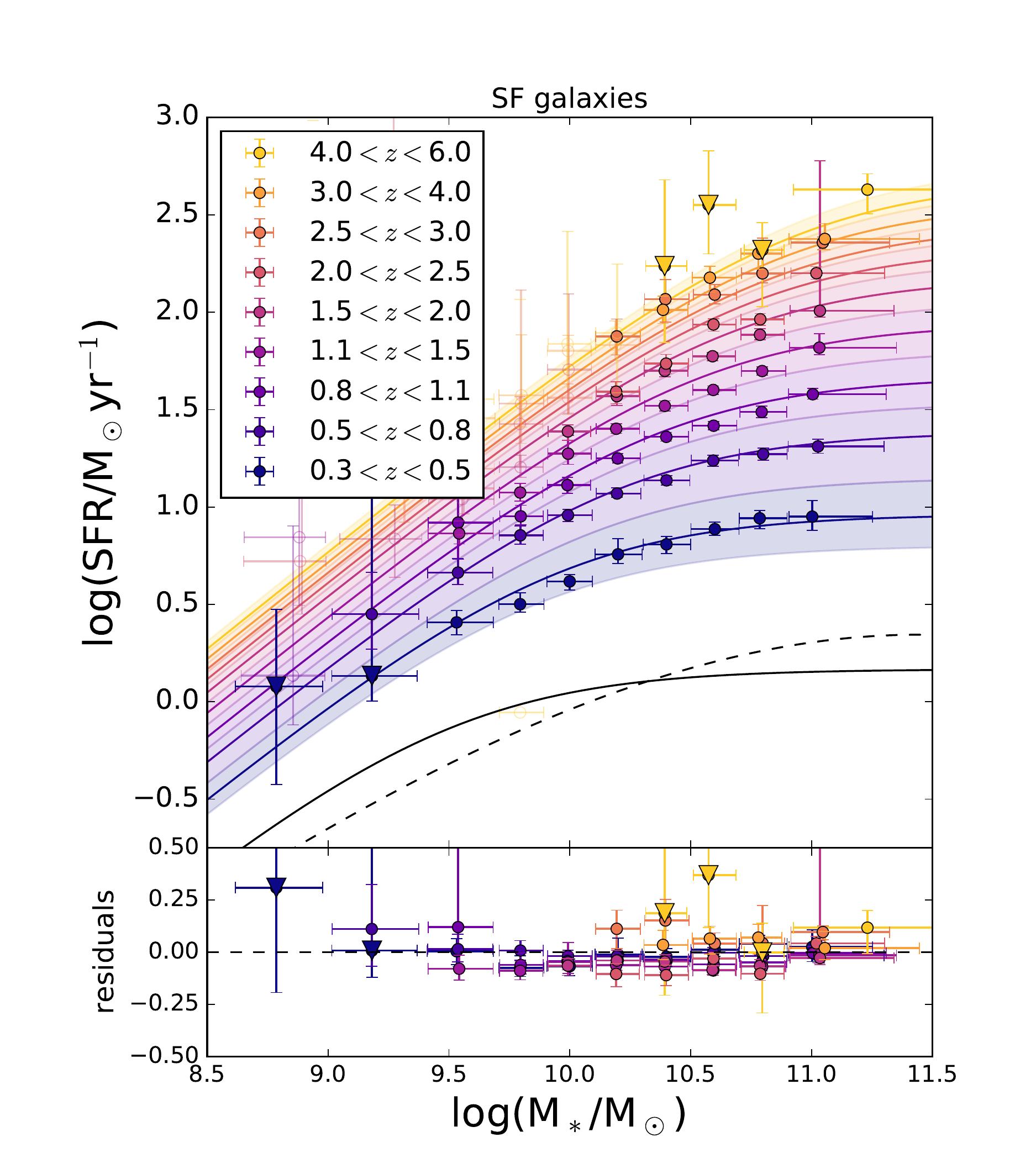}
\includegraphics[height=11cm, trim={1cm 0 0 0},clip]{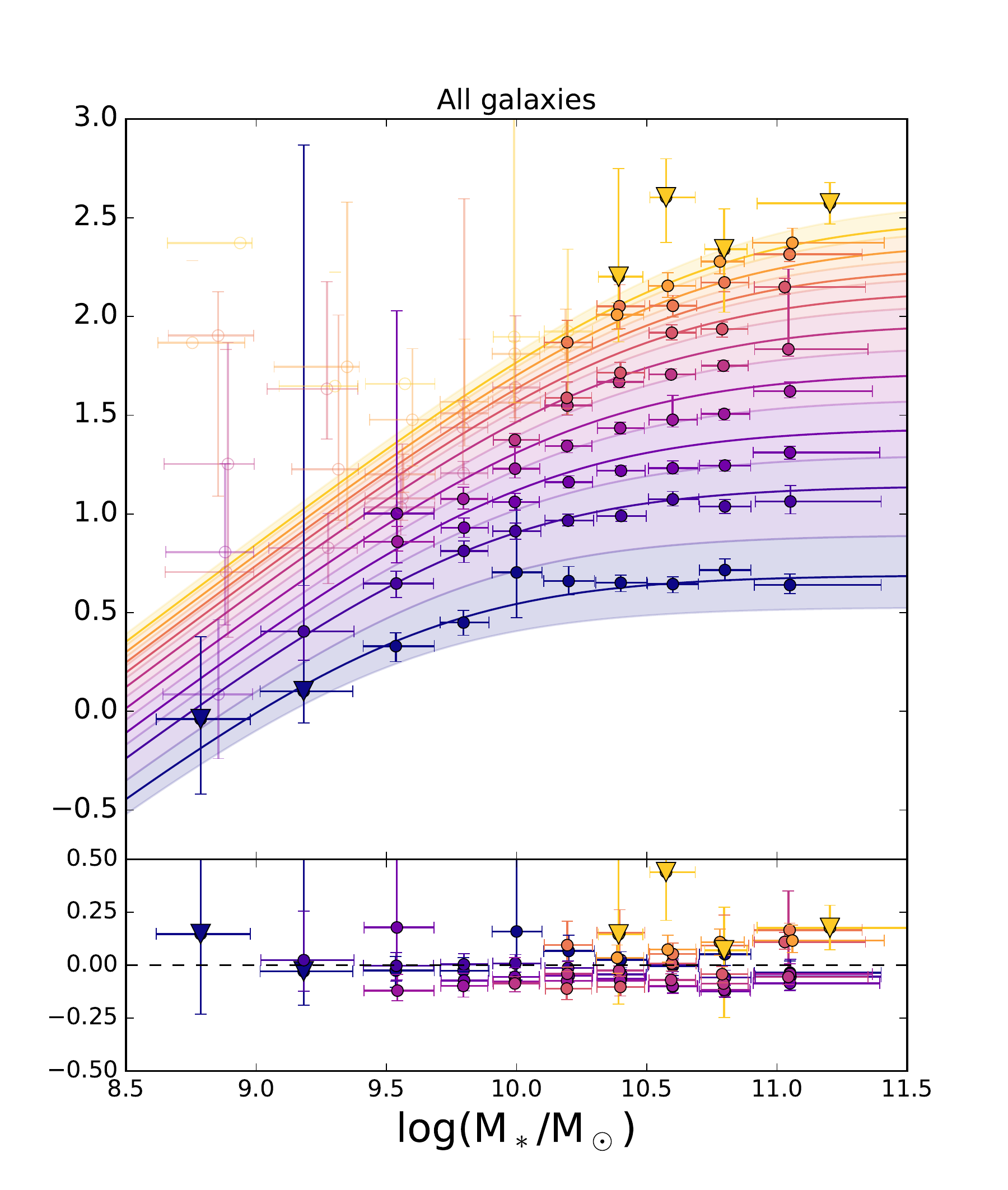}
\caption{Galaxy radio SFR--stellar mass relation for SF galaxies (left panels) and all galaxies (right panels). We show errors (2$\sigma$) on the median SFR from our bootstrapping analysis of the stacked fluxes. The error bars in the $x$-direction show the 2$\sigma$ variation of stellar masses within each redshift bin. Only stellar-mass-complete bins (shown as opaque circles) with peak SNR $>10$ are used for the fitting. Bins suffering from mass incompleteness are shown as transparent symbols, and radio stacks with SNR $<10$ are shown as down-facing triangles. We parametrize the relation according to Eq. \ref{eq:myform}, allowing for a flattening at high stellar masses. The solid line shows the relation at the median galaxy redshift across each redshift bin, and the shaded areas show the evolution of the MS over the redshift bin. The dashed line in the left panel is the MS for SDSS galaxies from \cite{Saintonge2016}. The solid black line shows an extrapolation of our MS to $z=0.035$. Bottom panels show residuals measured as log(SFR) - best fit model log(SFR).}\label{fig:ms}
\end{figure*}

In this section, we present our measurement of the MS relation and discuss its functional form. Measurements for median redshift, peak 3\,GHz flux, total flux, and SFR are given at the end of the paper for mean stacks of star-forming and all galaxies. The MS derived from these data is shown in Figure \ref{fig:ms}. 

Star-forming galaxies show a positive correlation between SFR and stellar mass at all redshifts.
As expected, the SFRs of ``all" galaxies agree well with ``SF galaxies" at lower masses where most galaxies are classified as SF but disagree at higher masses, showing a ``turnover''. Including quiescent galaxies in the sample of all galaxies results in a flatter MS relation at $z<2$ for $\log(M_*/\mathrm{M}_\odot)>9.5$. A less-extreme flattening is also present in the star-forming galaxy sample at low-z for high masses. 


How should we define the MS? Is there a definition that gives a natural insight into the physical processes involved, or one that makes it the most straight-forward to conduct inter-sample studies? 
\citet{Renzini2015} suggested the MS should be defined as the ridge-line of the star-forming peak in a 3D logarithmic SFR -- $M_*$--number plane, and similar definitions have been adopted by, e.g., \citet{Hahn2019} and \citet{Magnelli2014}. 
Naturally, this method requires robust SFR and $M_*$ measurements for large numbers of galaxies.
Many studies find that star-forming galaxies follow a log-normal SFR distribution at a fixed stellar mass \citep{Popesso2019}. For a log-normal distribution, the mode and the median of the (linear) SFR are related to each other by the width of the distribution, and we note that the mean SFR will be higher than the median.

In our analysis, we do not distinguish starburst galaxies lying above the log-normal MS for SF galaxies; they are all included in the stack. Results do not yet agree about how the fraction of starburst galaxies varies with stellar mass and redshift (e.g. \citealt{Sargent2012}, \citealt{Schreiber2015}, and \citealt{Bisigello2018}). However, the number of starburst galaxies that lie above the MS is small (5-20\% of SF galaxies), and should not drastically affect our results.


The appropriate form of the MS depends on the stellar mass range under consideration. Using the deep GOODS fields in addition to the COSMOS field allowed \citet{Whitaker2014} to constrain the steeper lower-mass end (log$(M_*$/M$_\odot)<10.0$) of the MS. To combine the steep low and shallow high mass relations, a turn-over is required (see also \citealt{Bisigello2018}). 
At the high-mass end in the local Universe, $M_*>10^{11}$\,M$_\odot$ there are very few star-forming galaxies and their SFRs are highly uncertain, and depend sensitively on the SFR calibration used \citep{Popesso2019}; these galaxies are critical for determining the flattening of the MS. 


Figure \ref{fig:ms} shows that below $z<1$, a flattening in the MS is observed in our sample, therefore we would like to fit a functional form that allows this\footnote{We show in Appendix \ref{sec:msform} that a flattening of the form proposed by \citet{Lee2015} is strongly preferred over a linear model at $z<1.5$.}.
We cannot constrain the low-mass slope at high redshifts due to the mass completeness limitations imposed by the current COSMOS data, nor at low redshifts, due to insufficient signal-to-noise in our stacks (see e.g., Section \ref{sec:appendixsim}). Recent simulations predict a constant log(sSFR) at low stellar masses ($M_*<10^{9.5}$\,M$_\odot$), out to $z\sim5$ (e.g., \citet{Torrey2018,Matthee2019}, see also \citet{Iyer2018} for low-mass slope constraints based on star-formation histories reconstructed from observations), supporting our decision to fix the low-mass power-law slope to 1 (see also e.g. \citealt{Schreiber2015}). 
In Appendix \ref{sec:msform}, we have tested the performance of the linear form of \citet{Speagle2014} and the nonlinear MS forms of \citet{Schreiber2015} and \citet{Tomczak2016} at describing our data.
Here, we introduce a new form in Equation \ref{eq:myform}, inspired by \citet{Lee2015}:
\begin{equation}\begin{aligned}
\log(\langle\mathrm{SFR}\rangle) &= S_o - a_1 t - \log\Bigg(1+\left(\frac{10^{M_t'}}{10^{M}}\right)\Bigg),\\
M_t'& = M_0-a_2 t,\\
\end{aligned}\label{eq:myform}
\end{equation}
where $M$ is $\langle\log(M_*$/M$_\odot)\rangle$ and $t$ is the age of the Universe in Gyr. 
We found that the choice of redshift or time evolution, e.g. $z$, $\log(1+z)$ or $t$, plays an important role in the number of parameters required for a good fit. 
Our new parametrization takes into account the evolution of the normalization ($S_0+a_1r$), turn-over mass ($M_t'$), and assumes a low-mass slope of 1.

\begin{figure*}
\includegraphics[width=0.5\linewidth]{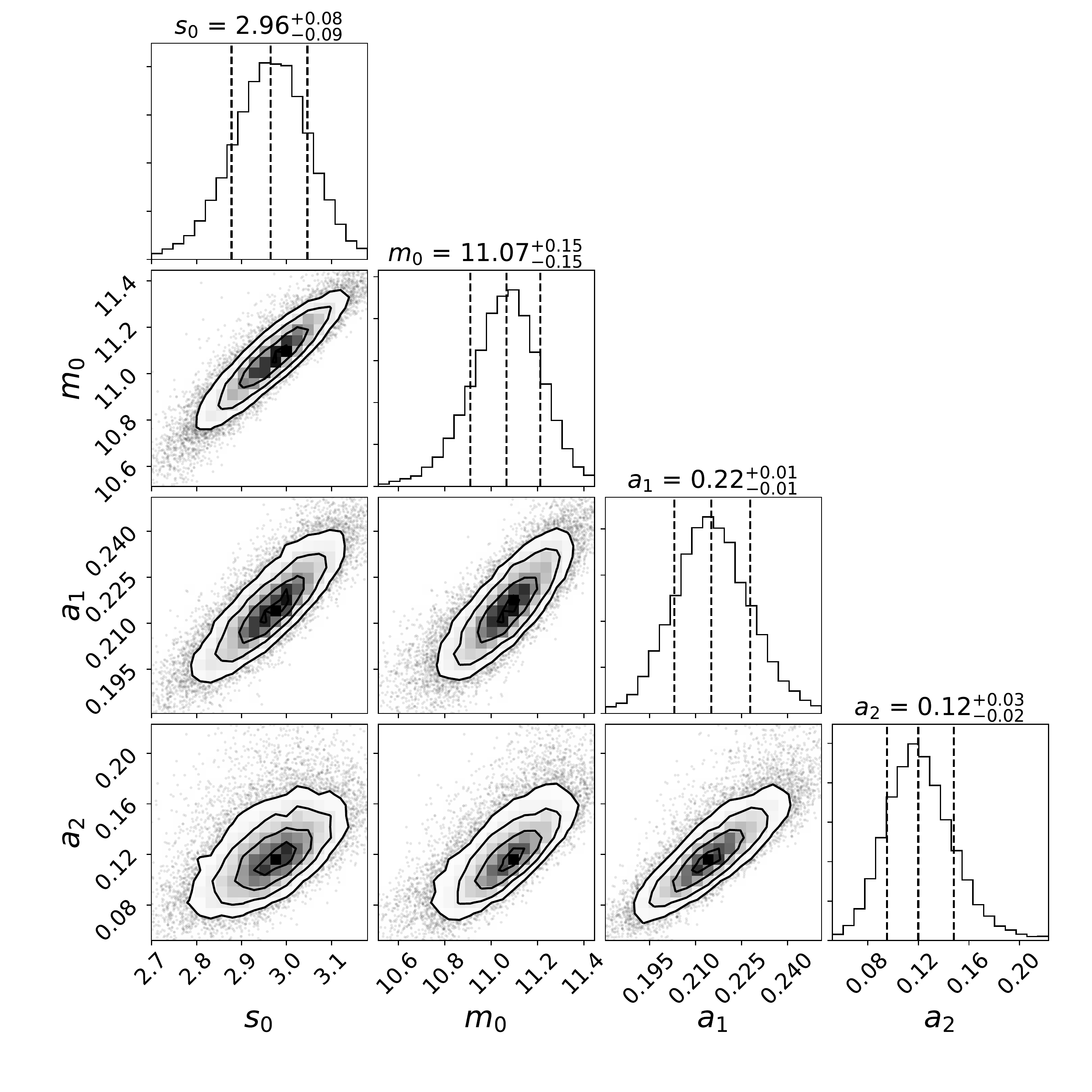}
\includegraphics[width=0.5\linewidth]{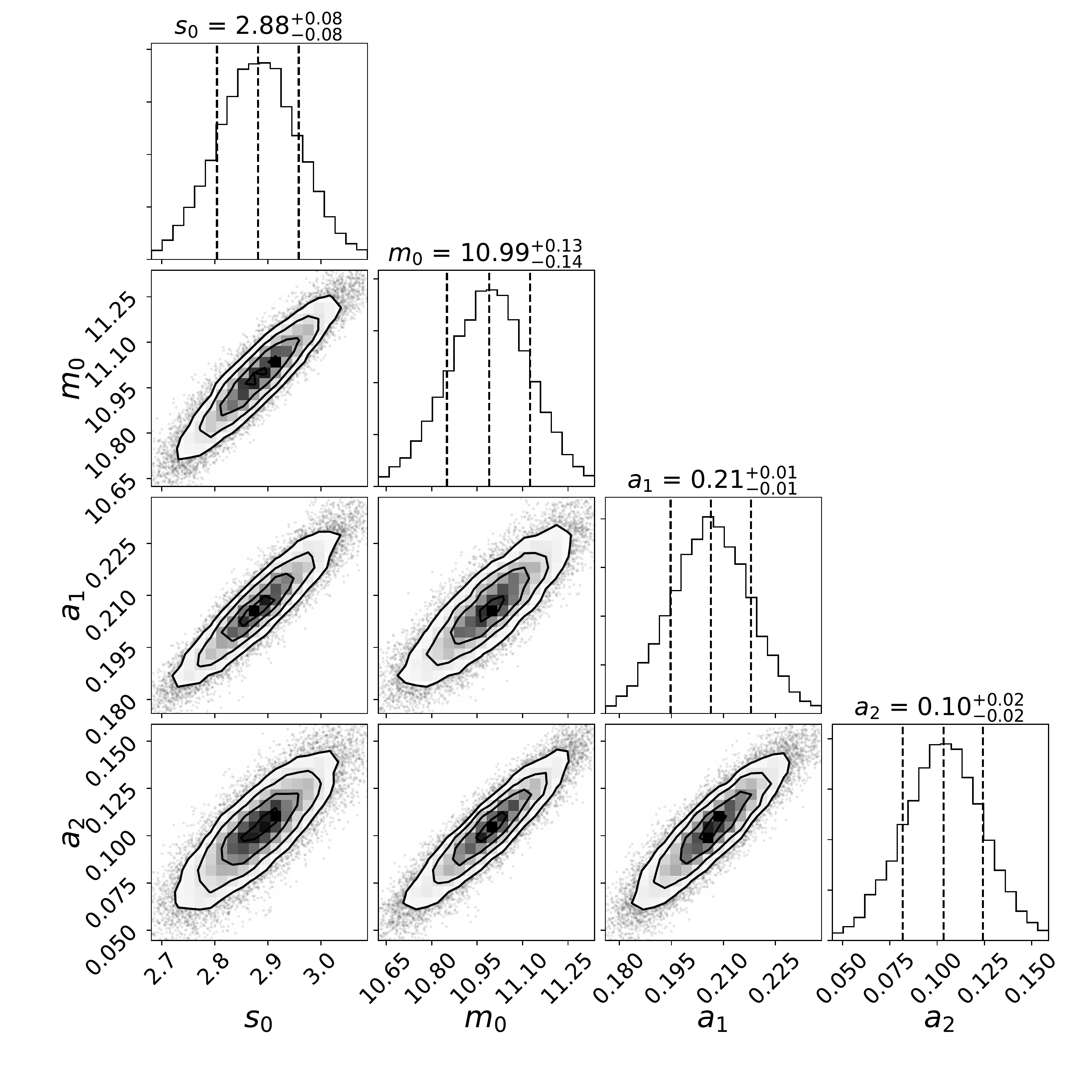}
\caption{The one and two dimensional projections of the margianalized posterior probability distributions of parameters in our model. The left panels show results from fitting the model from Eq. \ref{eq:myform} to star-forming galaxies. Right-hand panels show results for fitting our model to all galaxies. Medians and widths of the marginalised distribution for each parameter are given above the histograms of each parameter and are also reported in Table \ref{tab:mybestfits}.}\label{fig:cornerplot}
\end{figure*}

We fit our model to the data using \textsc{scipy}'s \textsc{optimize.curve\_fit}, that uses the Levenberg-Marquardt least-squares algorithm, only considering mass complete data (opaque points in Figure \ref{fig:ms}). Based on our simulations (see Appendix \ref{sec:appendixsim}), we find that stacks with $S_p/\mathrm{rms}<10$ can bias the recovered mean fluxes by more than 10-20\%, therefore we only use $S_p/\mathrm{rms}<10$ data for our fitting. 
To estimate the parameter uncertainties we resample the data and vary the stellar mass, SFR, and redshift values of each stack by a random amount that follows a Gaussian distribution with $\sigma$ given by our upper and lower errors. The SFR error comes from the flux error (of our bootstrapping analysis) and for this, we have assumed that the errors follow a normal distribution in flux. The $\sigma$ for stellar mass and redshift values come from the standard deviations of stellar masses and redshifts of galaxies within the bin. We have resampled the data with replacement 50,000 times\footnote{However, generally, less than half of these runs converge.} and recorded the best-fit for each new sample. The median and 16th and 84th percentile from all our converged runs are reported for the model parameters in Table \ref{tab:mybestfits}.
The best-fit parameter distributions are also shown in Figure \ref{fig:cornerplot}. For the remainder of this work, we use our MS results from Equation \ref{eq:myform} unless stated otherwise. 
In the left panel of Figure \ref{fig:ms}, we show the extrapolation of our MS model to low redshift, $z=0.035$, in comparison to the MS found by \cite{Saintonge2016} for SDSS galaxies between $8.5< \log(M_*/\textrm{M}_\odot)<11.5$. 
We find a turn-over mass that increases with redshift. Studies such as \citet{Tomczak2016} and \citet{Lee2018} have also found that the turnover mass increases with redshift. The model of \cite{Peng2010}, expects mass-quenching processes to be the dominant cause of quiescent galaxies at high stellar mass $M_*>10^{10.2}$\,M$_\odot$. The high-mass MS flattening may be a result of including galaxies that are in the process of ``mass-quenching'' in our sample.


\begin{table}[]
  \centering
  \begin{tabular}{|c|cccc|}
  \hline
    Sample & $S_0$ & $M_0$ & $a_1$ & $a_2$ \\
    \hline
     SF &$2.97^{+0.08}_{-0.09}$ & $11.06^{+0.15}_{-0.16}$ & $0.22^{+0.01}_{-0.01}$ & $0.12^{+0.03}_{-0.02}$ \\
     All& $2.80^{+0.08}_{-0.09}$ & $10.8^{+0.15}_{-0.17}$ & $0.23^{+0.01}_{-0.01}$ & $0.13^{+0.03}_{-0.02}$\\
     \hline
  \end{tabular}
  \caption{Best fit parameters (and 1$-\sigma$ errors) of the MS (Equation \ref{eq:myform}) fit to SFRs determined from mean 3GHz fluxes for galaxies as a function of stellar-mass and redshift. These fits are demonstrated in Figure \ref{fig:ms}.
  }
  \label{tab:mybestfits}
\end{table}

\begin{figure} 
\includegraphics[width = \linewidth]{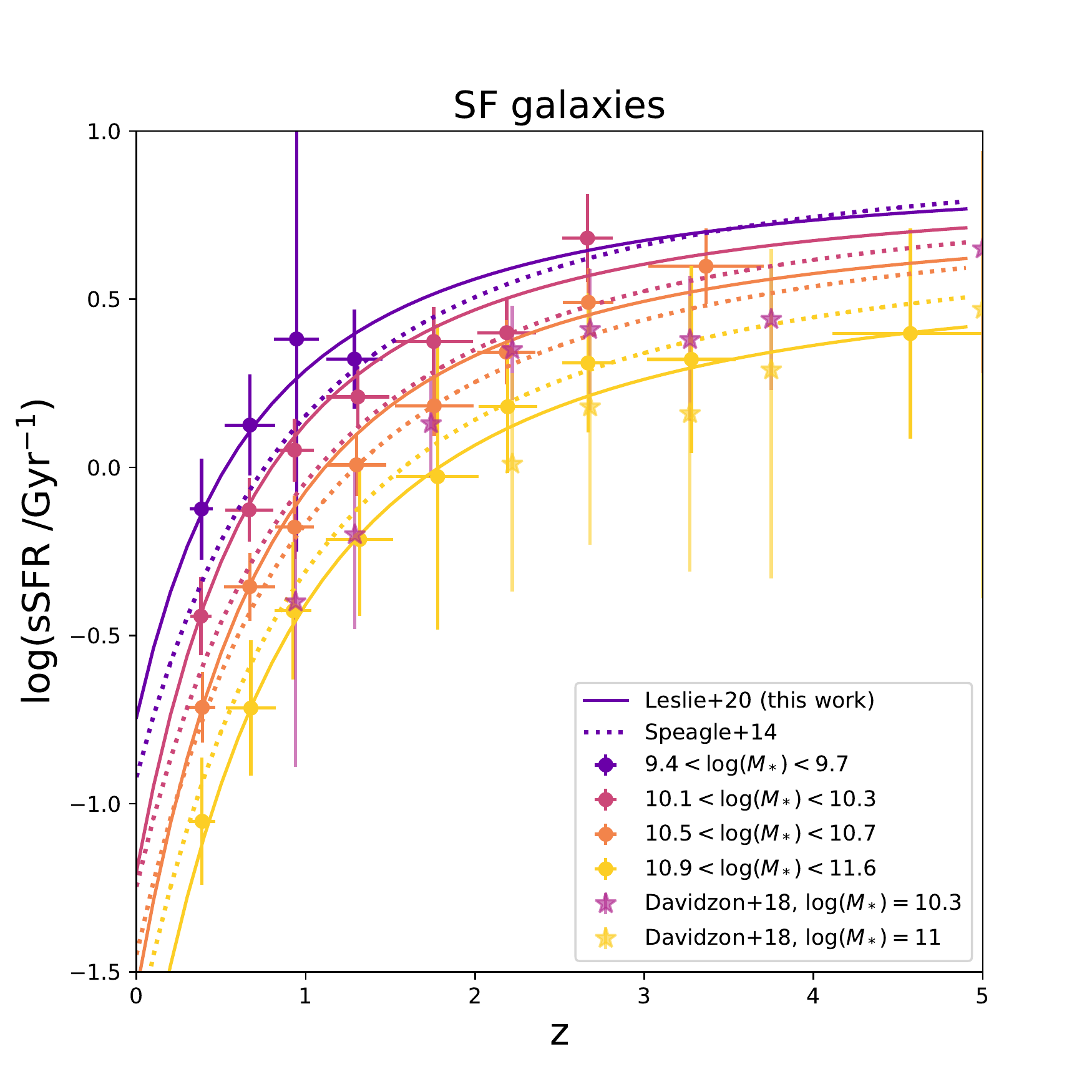}
\caption{Evolution of sSFR for galaxies at four representative stellar masses. Solid lines show the best fit of Equation \ref{eq:myform}. The \citet{Speagle2014} MS, converted to our \citet{Chabrier2003} IMF, is shown in dotted lines for comparison. Only four stellar mass bins are shown for clarity, and data-points indicate redshifts where the COSMOS sample is complete. Our data and fitted MS is particularly different from the \citet{Speagle2014} MS at the low- and high-mass ends. Stars show sSFRs in two stellar mass bins inferred from the differential evolution of the stellar mass function in COSMOS \citep{Davidzon2018}.}\label{fig:ssfrz}
\end{figure}

Figure \ref{fig:ssfrz} shows the sSFR of SF galaxies as a function of redshift. At all epochs, the most massive galaxies have the lowest sSFR. For galaxies following the functional form proposed in this work, the sSFRs span a smaller range at high redshift, and become more divergent towards the present time, with massive galaxies evolving faster, decreasing their sSFR at earlier epochs. 
Figure \ref{fig:ssfrz} shows how our form compares with the canonical \citet{Speagle2014} MS at four different mass bins. 
The \citet{Speagle2014} model shows a smaller range of sSFR values, under-predicting our measured sSFRs at low mass and over-predicting them at high stellar mass. 

\citet{Davidzon2018} used the differential evolution of the galaxy stellar mass function to infer the sSFR evolution of galaxies. Using the stellar mass functions from \citet{Davidzon2017} and \citet{Grazian2015} (for $z>4$), they report a shallow evolution of the sSFR $\sim(1+z)^{1.1}$ at $z>2$, consistent with e.g., \citet{Tasca2015}. We also show the results from \citet{Davidzon2018} in Figure \ref{fig:ssfrz} for their lowest and highest stellar mass bins, $\log(M_*$/M$_\odot$) = 10.3, and 11 respectively. The evolution with redshift matches well the evolution seen in our data, and our results are consistent within the large error bars. However, our mean-stacked sSFRs are systematically higher than those inferred by \cite{Davidzon2018}.

\subsection{Characteristic Stellar Mass}\label{sec:charmass}
\citet{Karim2011} found that the majority of new stars formed since $z\sim3$ were formed in galaxies with a mass of $M_* = 10^{10.6\pm0.4}$ M$_\odot$.
Stellar mass functions reported in \citet{Davidzon2017} provide information on the number of galaxies that exist per co-moving cubic Mpc as a function of stellar mass and at each redshift. 
In this section, we combine our 3\,GHz MS results with the stellar mass functions reported for star-forming galaxies (``active galaxies'') in \citet{Davidzon2017} to determine the cosmic SFRD.
At $0.2<z<3.0$, a double \citet{Schechter1976} stellar mass function function was fit to the data:
\begin{equation}
\Phi(M)\mathrm{d}M = e^{-\frac{M_*}{M^\star}}\left[ \Phi_1^\star \left(\frac{M_*}{M^\star}\right)^{\alpha_1} + \Phi_2^\star\ \left(\frac{M_*}{M^\star}\right)^{\alpha_2}\right] \frac{\mathrm{d}M_*}{M^\star},
\end{equation}
whereas at higher redshift, a single Schechter function was used \citep{Davidzon2017}. The low-mass slope shows a progressive steepening moving towards higher redshifts, decreasing from $-1.29\pm0.03$ at $z\sim0.35$ to $-2.12\pm0.05$ at $z\sim5$ for star forming galaxies \citep{Davidzon2017}.

\begin{figure*}
\includegraphics[width = \linewidth]{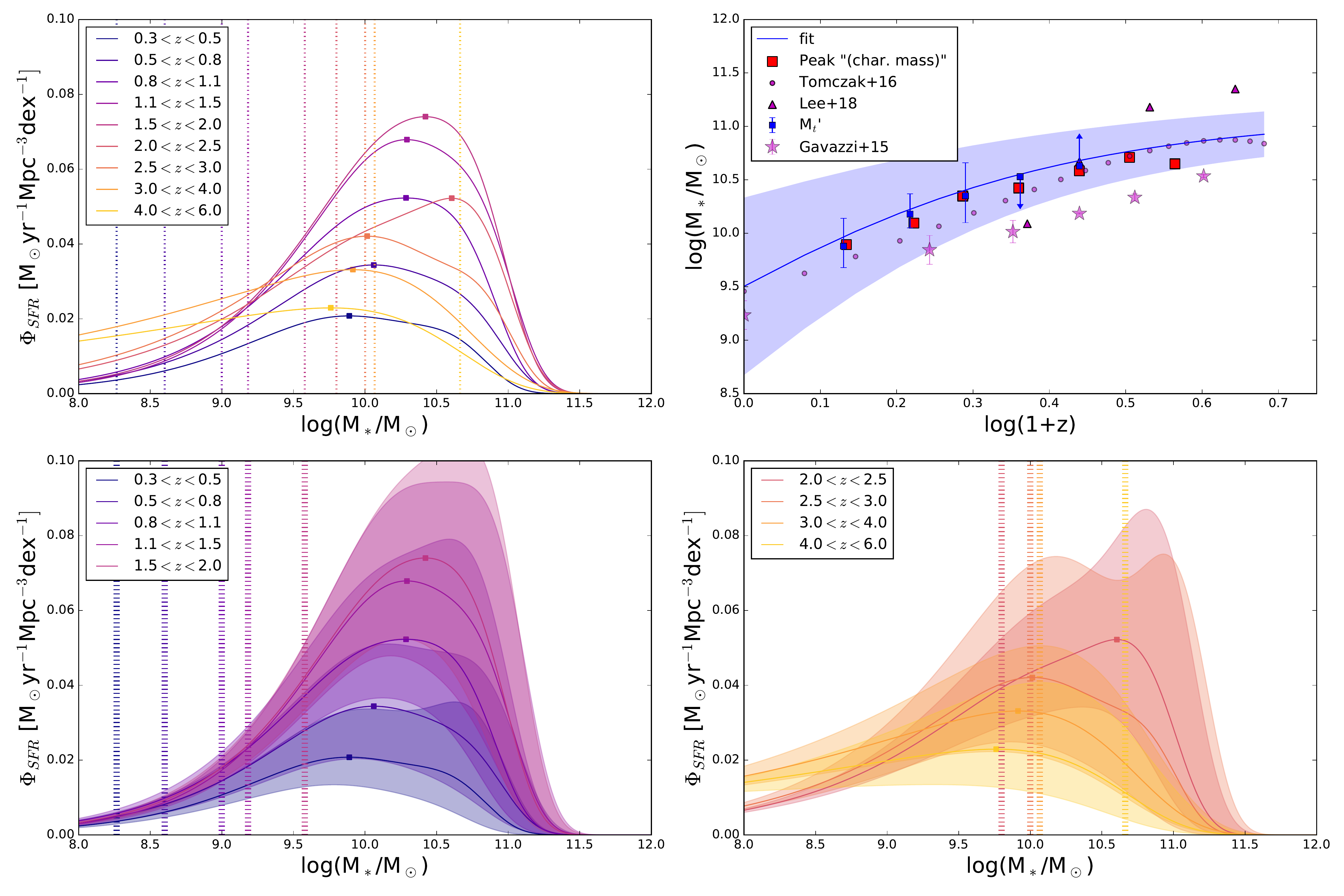}
\caption{The distribution of the SFR density as a function of stellar mass at different epochs out to $z\sim4$. We plot the function that results from multiplying the best-fit radio derived SFR-sequence at a given epoch with the corresponding stellar-mass function for star-forming galaxies from \citet{Davidzon2017}. The top left panel shows a summary of all redshift bins together. The bottom left panel shows data at $z<2$, and the bottom-right panel shows $z>2$. The uncertainty range was obtained by combining the uncertainties on the stellar mass function's Schechter parameters with the MS at the minimum and maximum redshift of our bins. At each epoch, our COSMOS data is not mass-complete below the stellar mass limits indicated by vertical dotted lines. The stellar mass corresponding to the peak SFR density is indicated as a square. The top right panel shows how this ``characteristic'' mass evolves with redshift (out to $z=3$; red squares). The evolution in characteristic mass is driven by the evolving turn-over mass in our MS. We fit the \citet{Lee2015} MS form for each $z$ bin individually in Appendix \ref{sec:msform}, which gives $M_0$ (blue squares; see Table \ref{tab:bestfitz}). Our best-fit functional form $M_{t} = M_0+a_2 t$ in Eq. \ref{eq:myform} is shown as a blue line (with $1\sigma$ confidence level shaded).}\label{fig:charmass}
\end{figure*}

Firstly, the SFR density as a function of stellar mass is estimated by multiplying the stellar mass function with the mean SFR derived for each mass and redshift bin from our mean stacking. The results are shown in Figure \ref{fig:charmass}. 
Vertical lines in the figure show the stellar mass above which the COSMOS sample is 90\% complete. 
The top left panel, showing all redshifts together, reveals that the SFR density peaks at $1.5<z<2$. The SFR density per stellar mass bin is peaked, showing a characteristic stellar mass that contributes most to the SFR density at a given redshift out to $z=2.5$. We find that the SFR density becomes more and more tightly peaked out to $z=2.5$ and that most new stars are formed in galaxies with a characteristic stellar mass that increases with redshift (see upper right panel of Fig. \ref{fig:charmass}). Contrary to the finding of \citet{Karim2011} (whose MS was well fit with a power-law relation), with our deeper parent catalogs, we find an evolving characteristic mass that shows signs of cosmic downsizing, in which the most massive galaxies formed first \citep{Rodighiero2010, Thomas2010, Dave2016, Siudek2017}. 

The evolving characteristic mass is closely connected to the turn-over mass of the MS. In Fig. \ref{fig:charmass} (top right panel), we show the evolution in the characteristic mass corresponding to the peak SFR density and the evolution of the turn-over mass of the MS from both this work and from \citet{Gavazzi2015}, \citet{Tomczak2016}, and \citet{Lee2018}. Our MS turnover mass (shown in blue) matches well with \citet{Tomczak2016} at $z<2$ and matches with \citet{Lee2018} for $z>2$. 
At $z>3$ our COSMOS data is no-longer complete past the characteristic mass, so it is unconstrained in the highest $z$-bin shown ($3.5<z<4$). 


\subsection{Cosmic SFR density of the Universe}\label{sec:csfrd}
By integrating the curves shown in Figure \ref{fig:charmass} over the stellar mass range $M=10^8-5\times10^{12}$\,M$_\odot$, we report the contribution of star-forming galaxies to the SFR density of the Universe as a function of redshift in Figure \ref{fig:csfh}. The MS is parametrized in terms of log(SFR), however, when we multiply with galaxy stellar mass functions, we need to convert to a linear SFR scale; log(mean(SFR)). Assuming a log-normal SFR distribution, these means have an offset of $\mathrm{mean}(\log(\mathrm{SFR})) - \log(\mathrm{mean}(\mathrm{SFR})) = -0.5\mathrm{ln}(10)(\sigma/\mathrm{dex})^2$ \citep{Padoan2002}. Assuming a scatter of 0.29\,dex \citep{Popesso2019b}, we have therefore scaled our SFRD down by 0.1\,dex.

In addition to our results, we show the \citet{Madau2014} literature compilation curve and recent radio (purple symbols) or IR (red symbols) studies. Our SFRs (blue squares) lie above the classic \citet{Madau2014} results at $z<2$.

We also calculate lower limits on the cosmic SFRD, free from assumptions about the MS form and stellar mass function, by multiplying the average radio flux in each stack by the number of galaxies in the stack and normalize by the co-moving volume probed by the redshift range considered. The area used for our stacking experiment is 1.55\, deg$^2$ (Section \ref{sec:datasets}), and the cosmological volume was calculated for each redshift slice using the \textsc{celestial} package by A. Robotham. The resulting (lower limit) SFRD values are shown as stars in Figure \ref{fig:systematics}. We only sum over mass-complete bins and do not apply a completeness correction because that would require assumptions about the stellar mass function or luminosity function (e.g., \citealt{Liu2018}). This is why our cosmic SFRD limits calculated in this way drop to low values at high redshift in Figure \ref{fig:systematics} where we are no longer probing past the knee of the stellar mass function.

\subsubsection{Systematic Uncertainties}

In Figure \ref{fig:systematics}, we show the systematic variations that different choices of SFR calibration can have on the cosmic SFRD. 
The purple triangles in Figure \ref{fig:systematics} show the cosmic SFRD inferred from integrating the \cite{Novak2017} luminosity functions over the luminosity range actually covered by the 5$
\sigma$ detections. Large extrapolations are required to account for the unobserved population of galaxies. 

\paragraph{SFR calibrations}
To illustrate the differences of SFR calibrations we show in Figure \ref{fig:systematics} lower-limits of the cosmic SFRD from summing our average SFR multiplied by the number of galaxies in the stack divided by the volume probed in the stack (series labeled ``sum'').

When we re-run our MS analysis using the \citet{Delhaize2017} SFR calibration adopted by \cite{Novak2017}, we find a worse agreement (SFRD values too high) at low redshifts between the \citet{Madau2014}. We report the best-fitting MS parameters found for this calibration, and the others mentioned below, in Table \ref{tab:systematics}.

\begin{figure}
\includegraphics[width=\linewidth]{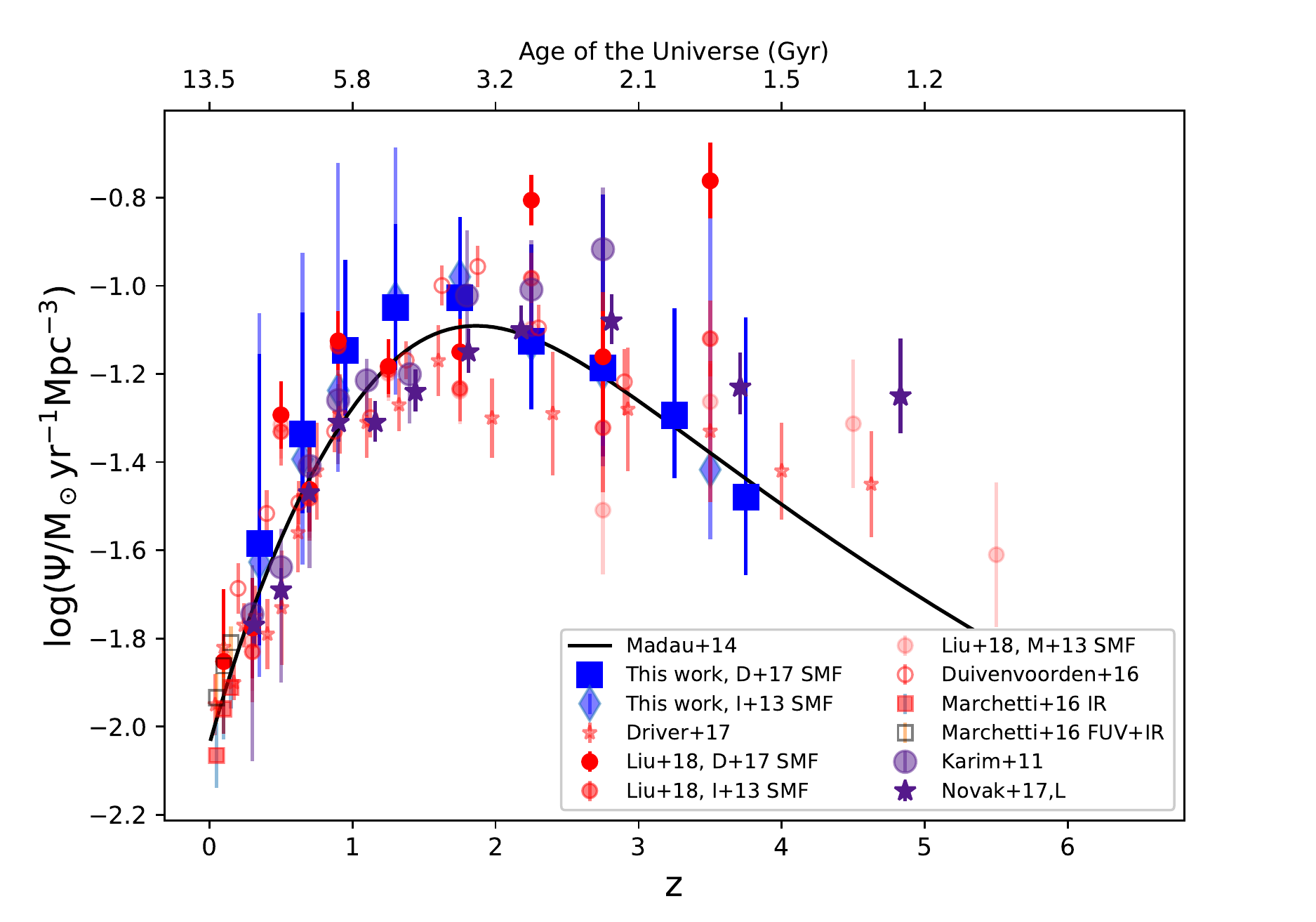}
\caption{SFR density resulting from integrating the SFR density as a function of stellar mass shown in Figure \ref{fig:charmass} between $M_*=10^8$ and $5\times10^{12}$ M$_\odot$. Results from this work are shown in blue, results from radio-based observations are shown in purple, and results from IR-based observations are indicated in red.}\label{fig:csfh}
\end{figure}

\begin{figure}
\includegraphics[width =\linewidth]{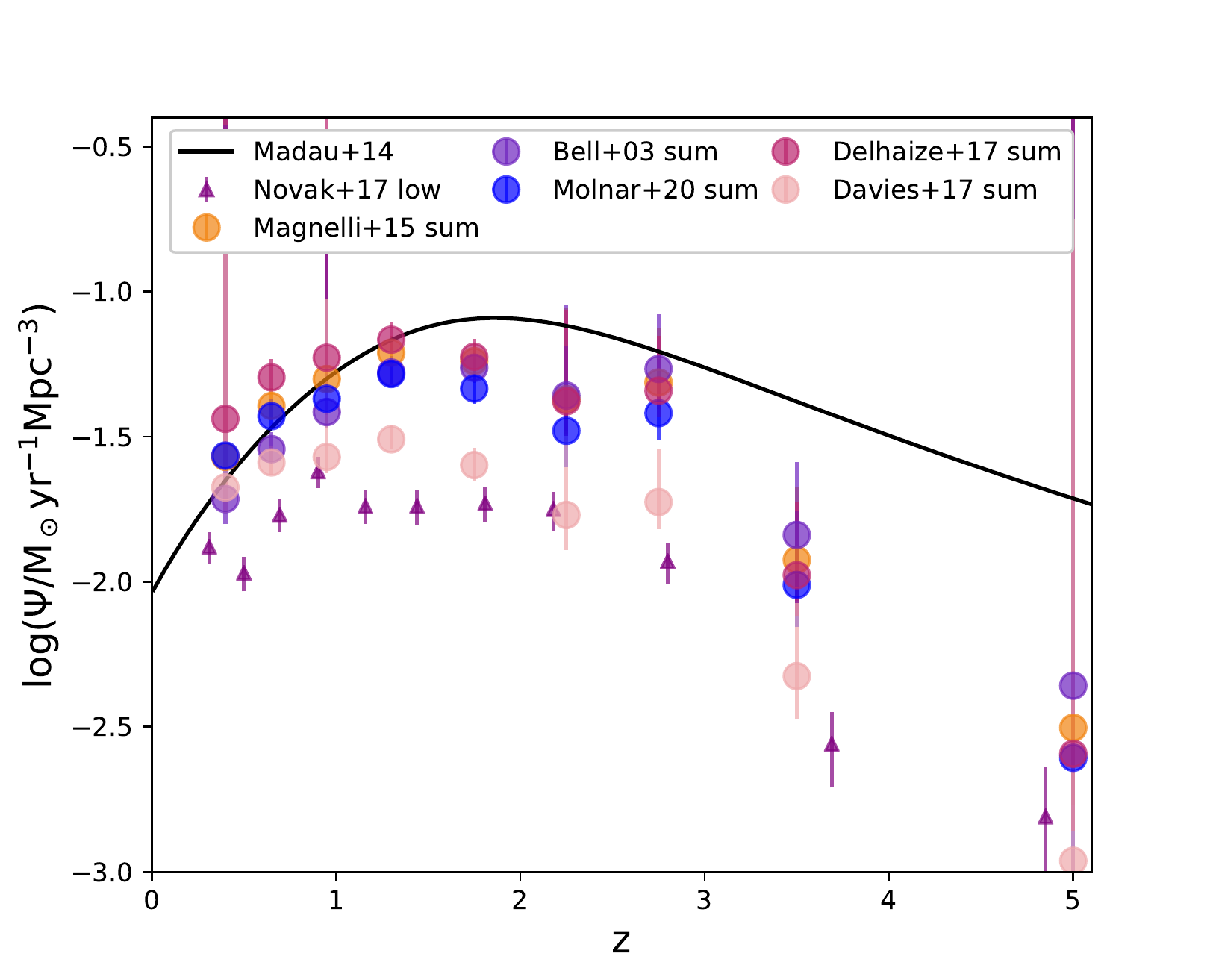}
\caption{Effect in the cosmic star-formation rate density for our galaxy sample caused by assuming different SFR-calibration. 
Circles represent the total cosmic SFRD inferred by summing the fluxes measured in the mass-complete stacks directly and are not extrapolated to lower mass bins. Purple triangles show the cosmic SFRD inferred from integrating the \cite{Novak2017} luminosity functions over the range covered by the detected sources.
}\label{fig:systematics}
\end{figure}

We have also tested the power-law 1.4\,GHz SFR calibration from \citet{Davies2017}. Although this non-linear SFR calibration works well with our 3\,GHz data (light pink stars in Figure \ref{fig:systematics}) for reproducing the local literature cosmic SFRD values at $z<0.7$, the SFRDs implied at $z>2$ are much lower than canonical values (even with completeness corrections applied).
\citet{Karim2011} adopted the \citet{Bell2003} 1.4\,GHz radio SFR calibration (purple circles in Figure \ref{fig:csfh}). This SFR calibration applied to our 3\,GHz stacks, gives lower SFRs (purple) than our \citet{Molnar2020} prescription (blue) at $z<2$, more consistent with the \citet{Madau2014} relation (once completeness corrections are applied), but slightly higher SFRs at $z>3$. Nevertheless there is relatively good agreement between the cosmic SFRDs derived from the \cite{Bell2003}, \cite{Magnelli2015}, and \cite{Molnar2020} calibrations. 
One particular avenue which we will explore in the future is whether the evolving $q_\mathrm{TIR}$ prescriptions give an accurate conversion from radio luminosity to SFR. For example, whether or not the total infrared luminosity of a galaxy gives a reliable SFR should depend on stellar mass according to studies of the IR/UV ratios in galaxies out to $z<5$ (e.g., \citealt{Bouwens2016, Whitaker2017, Fudamoto2017}).

The radio spectral index plays a role in k-corrections, correcting observed 3GHz to rest-frame 1.4GHz. Studies such as \cite{Tisanic2019} found that starburst galaxies have spectral steepening at high frequencies ($\nu>1.4$\,GHz). The evolution of the radio-IR relation has been calibrated on bright detected galaxies in our field, however, it has not yet been tested for low-mass galaxies. Local studies show that dwarf galaxies are often fainter than expected in radio continuum emission (e.g., \citealt{Bell2003,Leroy2005, Paladino2006, Filho2019}), with studies such as \cite{Hindson2018} suggesting that the non-thermal synchrotron emission is suppressed. Measuring the radio-SED for a range of representative galaxies will be essential to resolve these issues. 

\paragraph{Uncertainties involved with fitting functions} 
Calculating the cosmic SFRD involves extrapolation beyond the data, and so the shape of the low-mass MS makes a significant difference. 
For example, repeating our main analysis assuming a fixed low-mass slope of 0.6 and 1.2 (minimal and maximal values reported in the literature), results in a 0.001\,dex difference in the cosmic SFRD at $z=0.35$, but a $>0.2$\,dex difference at $z>2$, where our sample is not complete below the turn-over mass.

\paragraph{Choice of stellar mass function}
The size of this effect depends on the low-mass slope of the stellar mass function.
To investigate the effect of the stellar mass functions used, we have repeated our main analysis using the \citet{Ilbert2013} stellar mass function, which generally results in slightly lower SFR densities (transparent blue diamonds in Figure 
\ref{fig:csfh}). The difference in cosmic SFRD is at most $0.09$\,dex at $z\sim0.95$.

We also show the cosmic SFRD results from \citet{Liu2018}, who constrained the contribution from optically undetected dusty star-forming galaxies at $z<6$ in the GOODS-North field. They reported the SFRD obtained using either the \citet{Davidzon2017}, \citet{Ilbert2013}, or \citet{Muzzin2013} (dark red circles to transparent red circles respectively in Figure \ref{fig:csfh}) stellar mass function to correct for incompleteness. The choice of stellar mass function has a large effect on the measured SFRD at $z>2$ where current IR and radio data are incomplete.

\paragraph{Integration ranges} Typically, studies integrate luminosity functions to obtain the cosmic SFR. In our case, we integrate over the cosmic SFRD as a function of mass.
Our study is more sensitive to the choice of the low-mass limit of integration rather than the high-mass limit. However, the effect is small, $<0.1$\,dex at both high and low redshifts when changing the low mass limit of integration from $10^7$ to $10^9$ M$_\odot$.

\paragraph{Scatter of the MS}
We correct our cosmic SFRD when calculated by combining the SMF and MS, by a factor that depends on the log-normal dispersion of the MS. Some studies report this scatter to evolve with redshift, finding $\sigma$ to decrease at high-z. 
\cite{Pearson2018} reported a scatter of only $\sim
0.15$\,dex at $z\sim 4.3$. This would have an impact on our correction (e.g. move $z\sim4$ SFRD values 0.07\,dex relative to the $z\sim0.4$ value). Observationally, the measured scatter should depend on the SFR indicator chosen. Instantaneous SFR tracers such as nebular emission lines return higher $\sigma$ than tracers that are sensitive to longer timescales because they are sensitive to rapid variations in the star-formation histories \citep{Davies2016,Davies2019,Caplar2019}.
There is also some evidence for a stellar-mass dependent scatter, for example, \citet{Guo2013,Ilbert2015,Popesso2019,Donnari2019} (however, see also \citealt{Whitaker2012, Speagle2014, Tomczak2016, Pearson2018}).

\paragraph{Other systematic differences} When comparing results with other studies, systematic differences also come from cosmic variance. Some regions in the COSMOS field contain over- or under-densities (e.g. \citealt{Darvish2015}) which will affect the cosmic SFRD due to the decreased or increased number of galaxies compared to the cosmic average (e.g. \citealt{Madau2014, Driver2018}). 
For a survey the size of COSMOS, \cite{Moster2011} finds the relative cosmic variance is $\sim6$\% for galaxies with stellar masses $\sim10^{10}$\,M$_\odot$ in a redshift slice of $\Delta z= 0.5$. This increases to $\sim21$\% for galaxies more massive than $10^{11}$\,M$_\odot$.

In this section, we have discussed some of the assumptions and systematic uncertainties that go into calculating a cosmic SFRD. These issues need to be resolved in order to use upcoming radio continuum (and other wavelength) surveys for studying both dust-obscured and unobscured star-formation across cosmic time. 

\subsection{Morphology trends}\label{sec:morph}

In this section, we study the MS as a function of galaxy morphology out to $z\sim$1.5 using HST morphological measurements in the ZEST catalog \citep{Scarlata2007} probing rest-frame optical wavelengths (ACS $I$ band). 
It has long been known that the sSFR of a galaxy is anti-correlated with central mass concentration or the presence of a bulge (e.g. \citealt{Kauffmann2003b, Franx2008, Wuyts2011, Bell2012, Lang2014, Bluck2014, Whitaker2015, Parkash2018, McPartland2019}), but here we show the trends still hold using dust-unbiased radio-continuum sSFRs.

\begin{figure*}
\includegraphics[width = \linewidth]{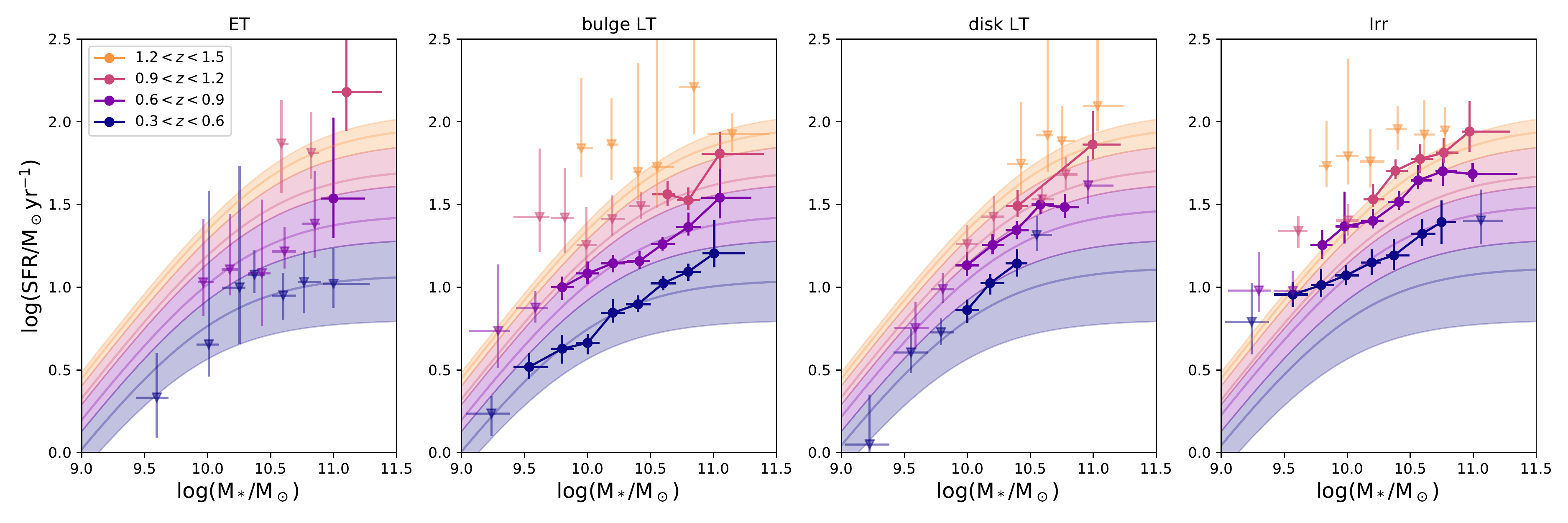}
\includegraphics[width = \linewidth]{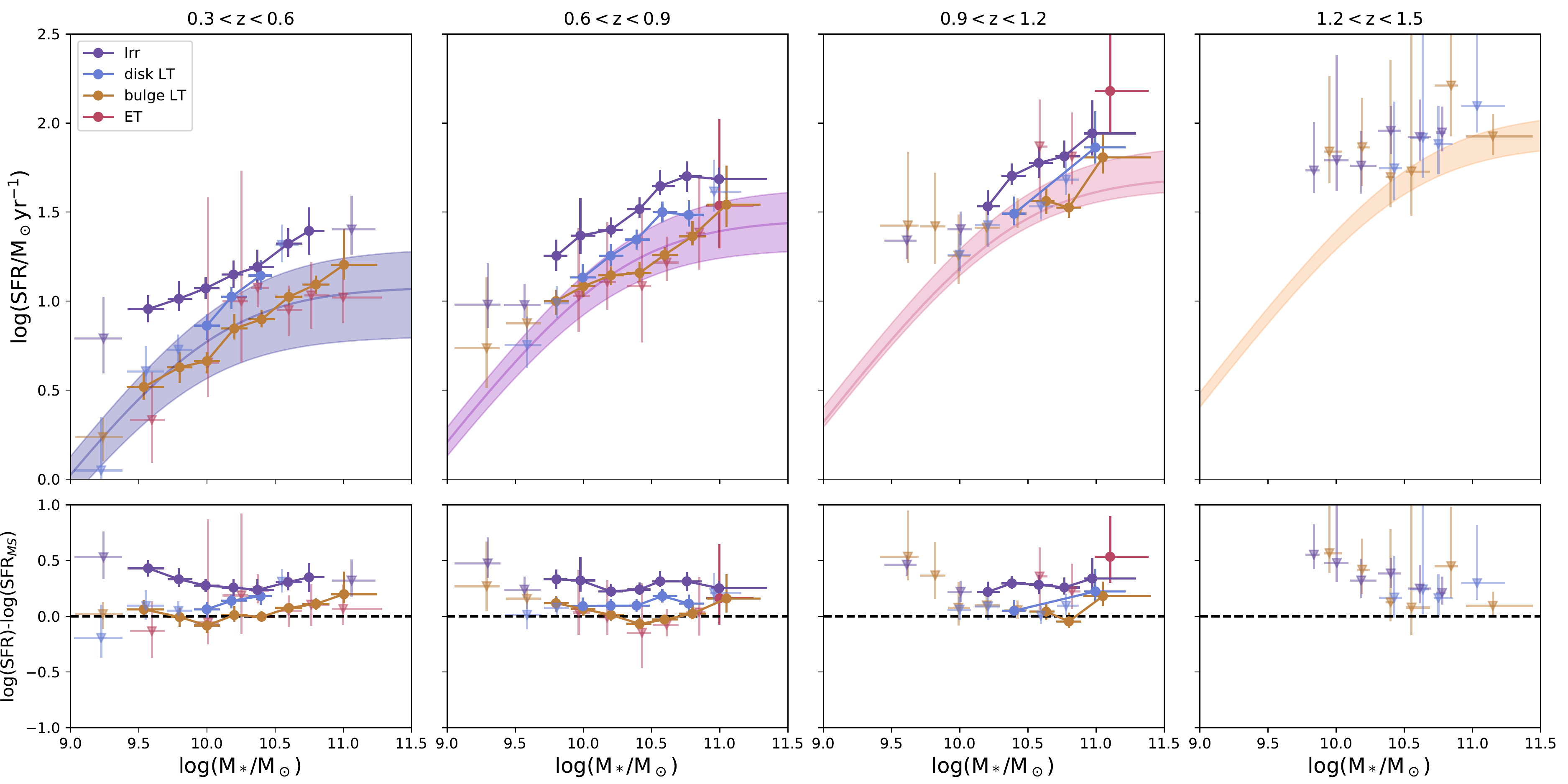}
\caption{ Star-forming galaxy MS for different morphological types. 
The top panels show, from left to right, the MS from a stacking analysis for star-forming galaxies classified as ZEST types 1 (early-type), 2.0-2.1 (late-type with bulges), 2.2-2.3 (disk-dominated late-type), and type 3 (irregular). The MS relation determined in Section \ref{sec:MS} for all star-forming galaxies is shown in the background for comparison. Stacks with SNR $<$ 10 are shown as downward-facing triangles because their fluxes are likely overestimated. 
The middle panels show the four redshift bins separately, allowing for an easy comparison of the SFRs at a fixed stellar mass for the different morphological classes (in particular, see the bottom ``residuals'' panels where the average MS from Eq. \ref{eq:myform} has been subtracted). Disk galaxies (type disk LT) follow a steeper SFR-stellar mass relation than other types, showing no clear evidence for a flattening at high stellar masses. Irregular galaxies have systematically higher SFRs than the other morphological classes.}\label{fig:morph}
\end{figure*}

Taken separately, the SFR -- $M_*$ relations for the different morphological classes appear to follow linear relations but have different slopes and normalizations (Figure \ref{fig:morph}). 
We find that massive galaxies dominated by a bulge tend to have lower SFRs at fixed stellar mass and redshift. 
The lower panels of Figure \ref{fig:morph} show that the difference between SFRs of different morphology types becomes prominent at low-$z$; the sSFR of high-mass bulge-dominated and spheroidal galaxies drops faster than the sSFR of disk-dominated galaxies.
Disk-dominated star-forming galaxies follow a steeper SFR -- $M_*$ relation than other types, particularly at $M_*>10^{10}$\,M$_\odot$ (with slopes of $\sim0.75$ compared to $\sim0.5$). When the disk-dominated galaxies are combined with the ET and bulge-dominated galaxies, a flattening in the average MS becomes apparent at higher stellar masses where the ET and bulge-dominated galaxies dominate. Irregular galaxies have SFRs above the MS at all redshifts, and are more star-forming than pure disks, at least at $z<1.2$, and $M_* <10^{11}$\,M$_\odot$.

\begin{figure*}
\includegraphics[width = \linewidth]{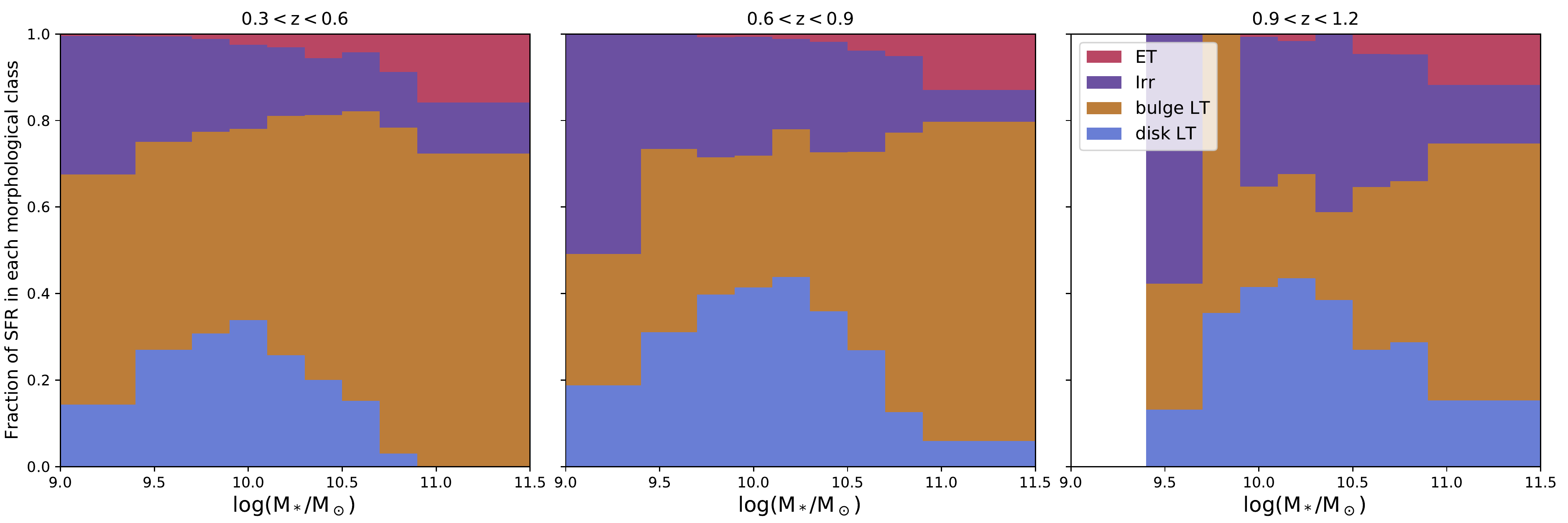}
\caption{Contributions from different morphological types to the total SFR as a function of mass in three redshift bins. The total SFR is given by multiplying the average from the stack with the number of galaxies in said stack. Only bins mass-complete in all four ZEST types are shown. Results for $1.2<z<1.5$ are not shown because all the fluxes were measured with SNR$<10$ (see Figure \ref{fig:morph}). The fractional contribution from disk-dominated galaxies declines as the contribution from bulge-dominated galaxies increases towards lower redshifts.}\label{fig:morphfrac}
\end{figure*}

Figure \ref{fig:morphfrac} shows the fraction of the total SFR occurring in galaxies in each morphological class as a function of stellar mass. 
The fraction of SFR occurring in disk galaxies decreases with redshift as the number of pure-disk galaxies decreases. By $z<0.6$, there are no pure disks in the most massive bin of star-forming galaxies\footnote{Based on the volume probed in this redshift bin and the way pure disk galaxies has been defined, this is not inconsistent with SDSS studies that do find massive $M_*>10^{11}$\, M$_\odot$ pure disk galaxies at $z\sim0.1$ \citep{Thanjavur2016,Ogle2016}}. The fraction of star formation in irregular galaxies is roughly constant at around $20\%$ in the three redshift bins shown but is slightly decreased in favor of galaxies classified as spheroidal for masses $10^{11}$\,M$_\odot$.
\cite{Grossi2018} found that, integrated over all luminosities, pure disk galaxies contribute significantly more to the cosmic SFRD at $z<1$ and that the decline in sSFR with redshift is faster for bulge-dominated systems than for pure disks. 

Interestingly, the SFR contribution from disk-dominated galaxies in Figure \ref{fig:morphfrac} is peaked around log($M_*$/M$_\odot)=10$ at each redshift. To explain this constant stellar mass, as disk-dominated galaxies have a single power-law MS relation, disk-dominated galaxies should, therefore, have an approximately redshift-independent stellar mass function shape to $z<1$. Indeed, \citet{Pannella2009b} also found that the stellar mass function of disk-dominated galaxies in the COSMOS field is consistent with being constant with redshift out to $z\approx1.2$. However, we note that only the shape needs to be constant for the SFR contribution from disk galaxies to be peaked at the same stellar mass, not the normalization. 


In this section, we have used the $q_\mathrm{IR}$ from \citet{Molnar2020} to calculate SFR from radio luminosity. More tests (e.g., including UV emission) need to be done to ensure we can recover the correct SFR in galaxies of different morphological types given that the infrared -- radio correlation \citep{Molnar2018} and perhaps radio spectral index \citep{Gurkan2018} vary between them.

To summarise, we find the flattening of the high-mass slope can be, in part, explained by the inclusion of massive bulge-dominated galaxies which follow a shallower SFR -- $M_*$ relation than disk-dominated galaxies. As bulges grow more prominent in the low-redshift galaxy population (especially at large stellar masses), the flattening of the main sequence becomes more significant. 
We discuss these results further in Section \ref{sec:discussionmorph}.

\subsection{Environmental trends}\label{sec:env}
\begin{figure*}
\includegraphics[width = \linewidth]{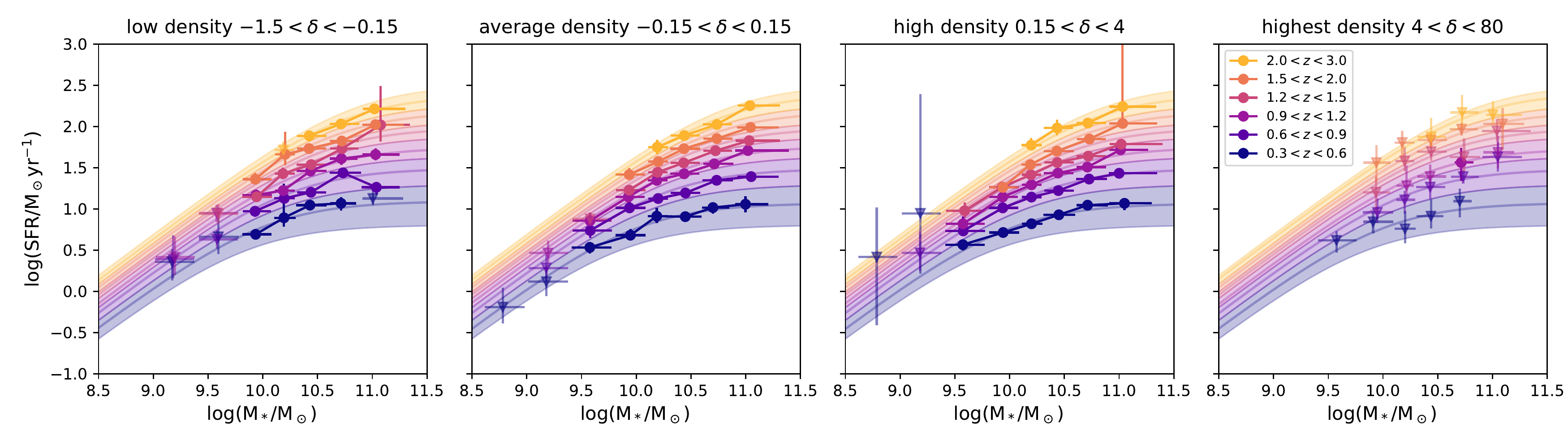}
\includegraphics[width = \linewidth]{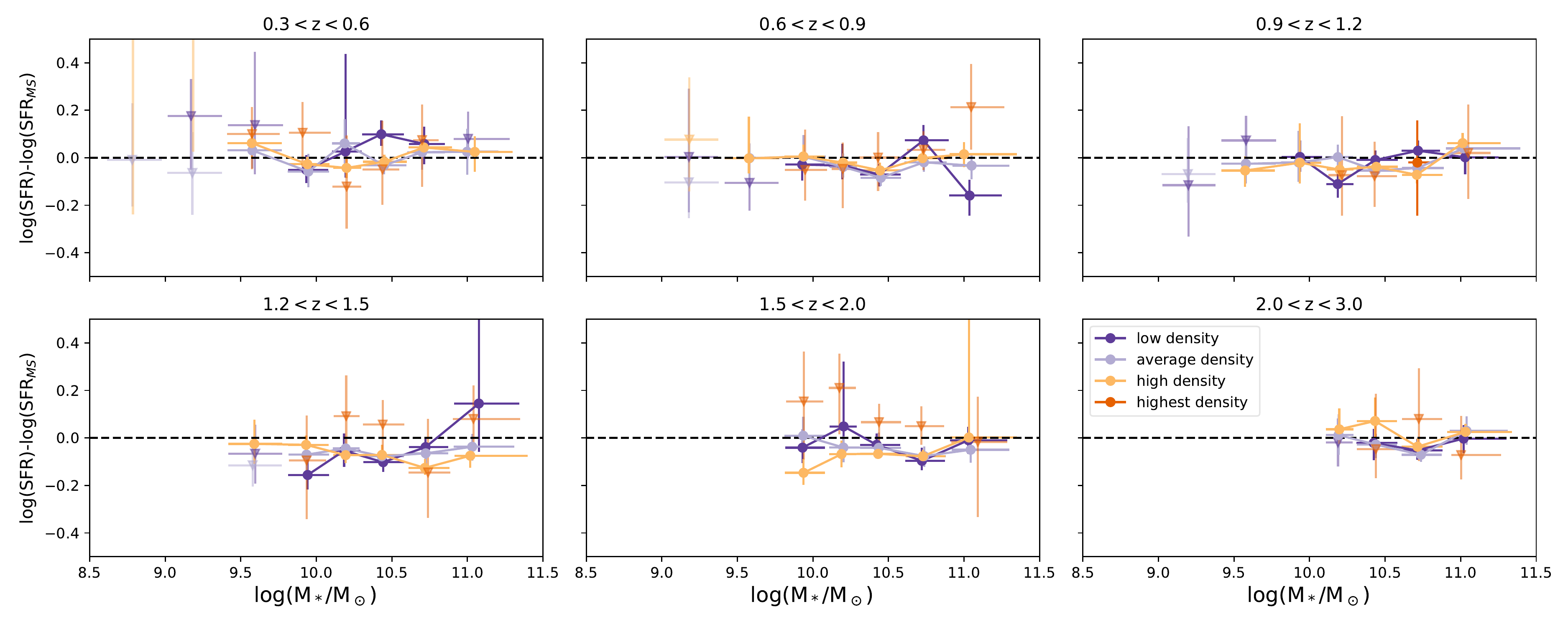}
\caption{Effect of local number density on the MS. The top panels show SFR as a function of stellar mass for star-forming galaxies in different local densities (as defined by \citealt{Scoville2013}) out to $z\sim3$. The relationship between sSFR and $M_*$ remains constant over these different local densities (explained in Section \ref{sec:env}), from galaxies in under-dense regions (left) to galaxies in over-dense regions (right). Our best fit relation for all SF galaxies is shown in the background, and no panel shows significant deviations from this relation. Data with SNR $<$ 10 are shown as downward-facing triangles as their fluxes are likely overestimated.
The last two rows show the difference between the SFRs of star-forming galaxies compared to the average MS in different local environments for different redshifts. }\label{fig:env1}
\end{figure*}

\begin{figure}
\includegraphics[width = 0.9\linewidth]{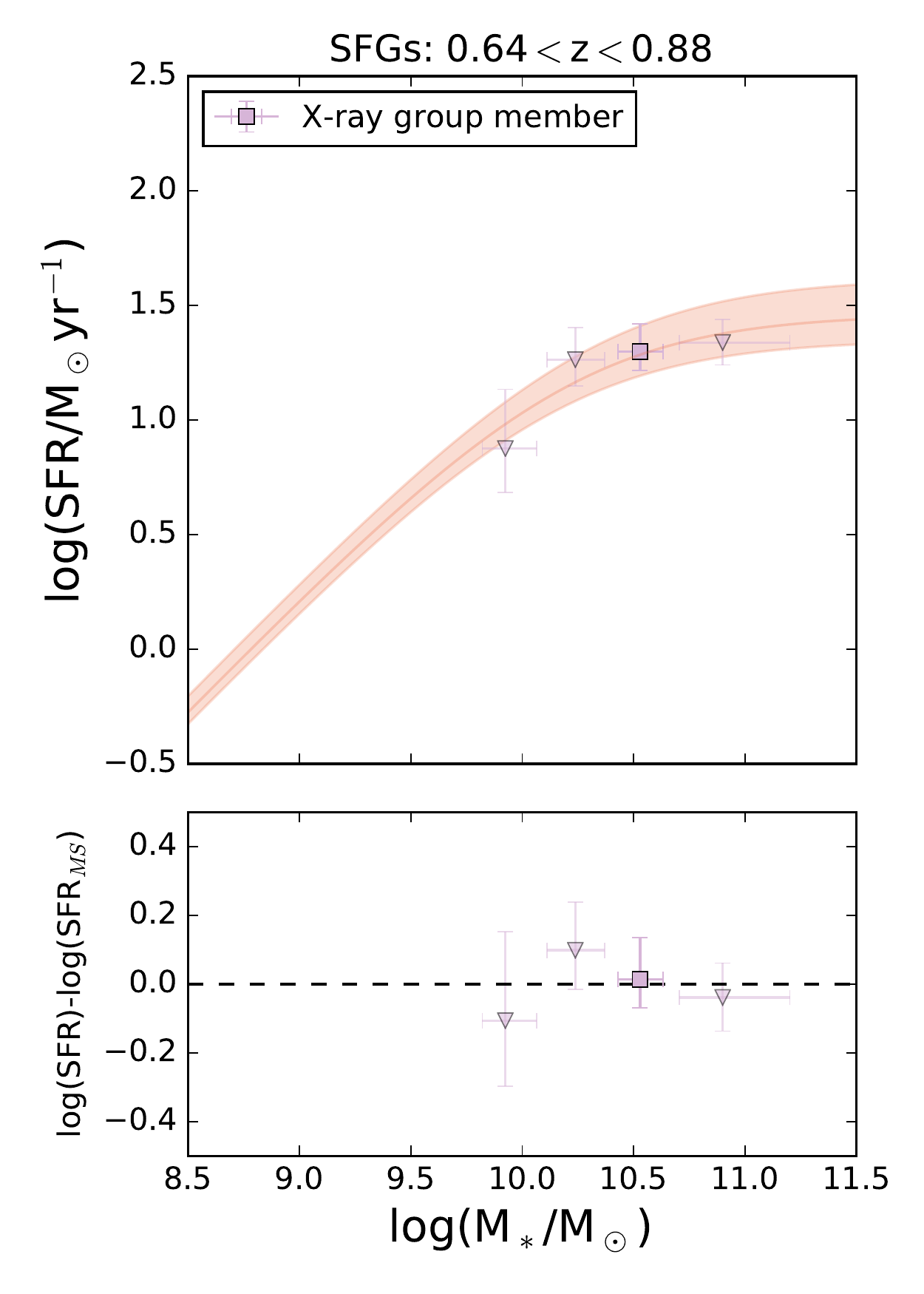}
\caption{The SFR-stellar mass relation for galaxies residing in X-ray groups and clusters at $z\sim0.76$ (top panel). 
The background curve shows the corresponding SFR -- $M_*$ relation (Eq. \ref{eq:myform}). The bottom panel shows the difference in the SFRs of galaxies in X-ray groups and the model MS. Stacks with SNR $<$ 10 are shown as downward-facing triangles because their fluxes are likely overestimated. }\label{fig:env2}
\end{figure}

\citet{Scoville2013} found that the median SFR of galaxies in the COSMOS field is not dependent on their environment at $z>0.8$, while at lower redshift the median SFR in the highest densities is almost 1 dex lower than the SFR in low-density environments. 
Figures \ref{fig:env1} and \ref{fig:env2} show the MS for SF galaxies in different environments. 
Using the local density as probed by the number of galaxies per Mpc$^{3}$ normalized by the mean number in each redshift slice \citep{Scoville2013}, we look for variations in sSFR within four local density bins:$ -1.5<\delta<-0.15$, $-0.15<\delta<0.15$, $0.15<\delta<4$, $4<\delta<80$, (where -1.5~Mpc$^{-2}$ and 80~Mpc$^{-2}$ are the minimum and maximum values in the catalog). These very roughly trace void, field, filament and group, and cluster environments \citep{Paulino-Afonso2019}.
 Figure \ref{fig:env1} shows no statistically significant difference between the MS relation for galaxies in different local densities. The same is true when considering ``all'' galaxies, shown in the Appendix.
In the bottom panel of Fig \ref{fig:env1}, the difference between the SFRs (normalized to the MS for all star-forming galaxies) is shown in each redshift bin. 
\citet{Duivenvoorden2016} found that the (IR) SFR decreases (by $<0.2$ dex) with local density out to $z<2$ in the COSMOS field. However, both our results have large uncertainties and the decreasing trend reported by \cite{Duivenvoorden2016} was only significant at $1.5<z<2$. The completeness corrected total cosmic SFRD from \citet{Duivenvoorden2016} is shown in Figure \ref{fig:csfh}, and is consistent with our SFRD at $z\sim2$.

Figure \ref{fig:env2} shows the MS for SF galaxies in X-ray groups at $0.64<z<0.88$ where the statistics are robust. The best-fitting MS determined in Section \ref{sec:MS} is shaded in the background. 
We only show SFRs derived for stellar-mass complete bins. When the stacked 3\,GHz image has $S_p$/rms$>$10, the SFR is shown as a solid square; the SFR is shown as a downward-facing triangle otherwise. 

We find no significant difference in the SFRs of star-forming galaxies in X-ray groups (halo mass $12.5<\log(M_{200c}/$M$_\odot)<14.5$) compared to star-forming galaxies in the field, consistent with our findings above which used local galaxy number density as a probe for the environment. However, the scatter is very large, and larger samples will be required to confirm this result. When we focus on the three most massive X-ray groups (which we refer to as clusters, with halo masses $14<\log(M_{200c}/$M$_\odot)<14.5$), there is an enhancement of the SFR of intermediate-mass $10.0<\log(M_*/$M$_\odot)<10.5$ members compared to the field. However, there are only 4-7 galaxies per bin when considering only the clusters at $0.64<z<0.78$, so this result is not reliable. That being said, at higher redshifts, starburst galaxies could preferentially lie in denser regions, consistent with the observed clustering of bright \textit{Herschel} sources \citep{Cooray2010, Amblard2011, Viero2013} and luminous blue galaxies at $z\sim1$ \citep{Cooper2006}.


Using ancillary data from the COSMOS field, we find that galaxy environment probed by X-ray-groups, clusters, and local galaxy number density has little effect on the shape of the radio-derived MS at $z>0.3$. This could indicate that environmental quenching (e.g., gas strangulation or stripping) is not effective at these redshifts, or that the colors redden before the SFR is suppressed, resulting in the affected galaxy being classified as quiescent and not being included in the MS analysis. However, we also find no significant trends when including all galaxies, which points to the former scenario being most likely. 

\section{Discussion}\label{sec:discuss}

\subsection{Comparisons with literature}\label{discussionscatter}

\begin{figure}
\centering
\includegraphics[width = \linewidth]{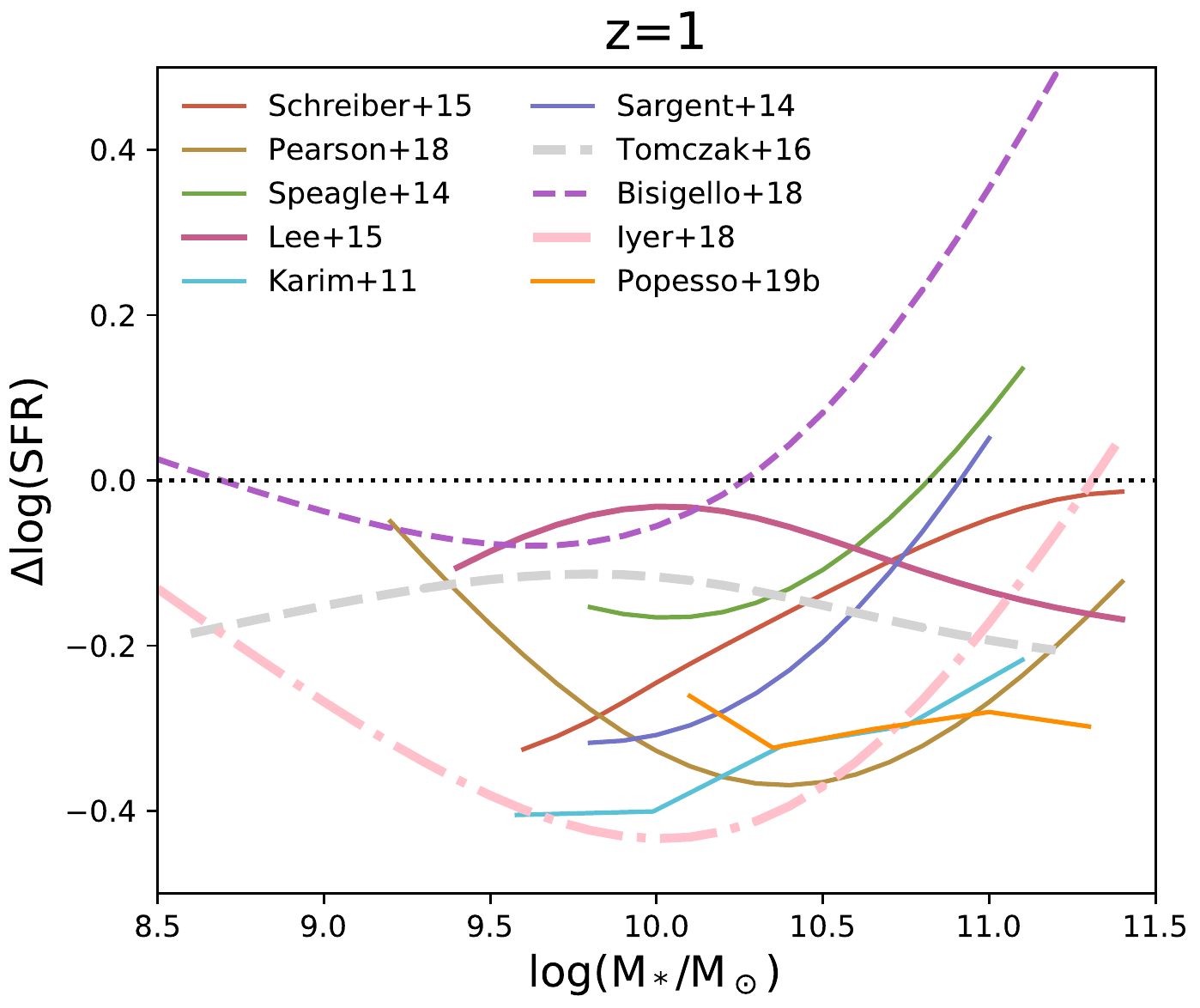}
\caption{Comparison with literature MS relations at $z=1$. $\Delta$log(sSFR) is the ratio of the MS SFR from the literature relative to the MS SFR from this work (Eq. \ref{eq:myform}). Dotted line at $\Delta$log(SFR) = 0 indicates where the relations agree. For more redshift bins, see Figure \ref{fig:comp}.}\label{fig:comp0p5}
\end{figure}

In Figure \ref{fig:comp0p5}, we show the difference between a selection of literature determinations of the MS for star-forming galaxies and the 3~GHz-determined MS we found in Sect. \ref{sec:MS}. Variations of up to $\pm0.4$ dex occur, and results are particularly divergent at low and high stellar masses, where the effects of completeness and star-forming sample selection become most important, respectively. We give more information about the studies included, and show the differences in literature studies at 6 redshift bins, out to $z=4.5$, in Appendix \ref{sec:comp}. Our stacking analysis probes stellar masses up to 0.5\,dex below that of \cite{Schreiber2015}.
In general, our SFRs tend to be higher than measured by other studies, but given the large sample-to-sample variation present, and the fact that different SFRs are sensitive to different stellar populations, we do not consider this to be an important shortcoming of our analysis\footnote{We note that our MS derived from median SFRs is in slightly better agreement with the literature at $z\sim1$ because the median SFR can be $\sim0.04$\,dex below the mean. However, our MS derived from median fluxes still lies at the extrema of the literature}. But, as shown by simulations in Appendix \ref{sec:appendixsim}, we are not able to well-reproduce sample averages using median stacking.. 
In addition to observations showing large discrepancies in the MS, when compared at $z=0$, different simulations also have a MS whose amplitude varies by up to 0.7\,dex between studies, and the power-law slopes range from 0.7-1.2 \citep{Hahn2019}.
Observationally, the different indicators used either to define the MS location or to estimate the SFR, combined with galaxy sample selection effects can cause the large variations between studies \citep{Popesso2019}, and will be discussed in this Section.

The scatter of the MS increases towards higher masses \citep{Guo2013, Davies2019, Popesso2019,Popesso2019b} (although see, e.g., \citealt{Rodighiero2010, Schreiber2015}), which can complicate the MS determination at the massive-end $(\log(M_*/\mathrm{M}_\odot) \sim 11)$. Studies such as \cite{Whitaker2014} find strong evolution in the slope for more massive galaxies, evolving from 0.6 at $z=2$ to 0.3 at $z=1$. But other studies such as \cite{Popesso2019b}, and \cite{Karim2011}, indicate no slope evolution. Discrepancies between SFR tracers also tend to be largest at $\log(M_*/\mathrm{M}_\odot) > 10.5$ (e.g. \citealt{Katsianis2020}).
Here, where the scatter is large, and not normally distributed, defining the location of the MS as the mean, median, or mode of the SFR distribution can have an important impact. Also, the decision to model the whole galaxy population or only SF galaxies and whether SF galaxies are selected by color or a sSFR threshold has an important impact on the determined MS at high masses. To summarise, the MS is not well defined at $\log(M_*/M)\odot)\sim 11$.

Most discrepancies in the MS can be explained by the different SFR tracers used by different studies \citep{Katsianis2020}.
Studies that use IR SFRs often find a turn-over in the MS at $z<2$, whereas studies using optical and SED tracers tend to find a linear MS (see Appendix \ref{sec:comp}). Recently, \citet{Pearson2018} used spectral energy distribution (SED) (UV - 160\,$\mu$m) priors to de-blend the Herschel SPIRE maps and overcome the limitations of confusion in measuring SFRs in the COSMOS field. \citet{Pearson2018} report no evidence for a high-mass MS flattening. The SED SFRs obtained with CIGALE (Code Investigating GALaxy Emission; \citealt{Burgarella2005}) are $\sim0.4$ dex lower than the SFRs obtained by combining UV and IR SFRs, the latter of which has been a standard method since introduced in the MS study by \citet{Wuyts2011}. \citet{Schreiber2015} used UV + IR SFRs in a stacking analysis in the COSMOS and GOODS fields to push to deeper IR luminosities, and reported a high-mass turnover that vanishes by $z\sim2$ (confirmed by \cite{Schreiber2017} who found a simple linear MS relation at $z\sim4$ from ALMA-derived SFRs).
Both \citet{Pearson2018} and \citet{Schreiber2015} (as well as \citealt{Lee2015}), used the same \textit{Herschel} data in the same fields, and yet these studies resulted in different interpretations.
In general, for more robust results, uncertainties specific to each dataset and from the sample selection could be better understood using simulated observations, and systematic uncertainties from stellar mass and SFR derivations should be analyzed.
Such analysis has begun, for example, \citeauthor{Davies2019} (2019; observations) and \citeauthor{Katsianis2020} (2020; simulations). Using the SKIRT full 3D radiative transfer post-processing of the EAGLE simulations by \cite{Camps2018}, \cite{Katsianis2020} found that methods relying on IR wavelengths have SFRs that systematically exceed the intrinsic relation by 0.2 to 0.5\,dex. They suggest this may be due to contamination of the IR luminosity from dust that is heated by an evolved stellar population rather than by newly born stars, particularly in the outskirts of EAGLE+SKIRT galaxies. This effect is also found in observations of local galaxies (e.g. \citealt{Helou2000}, \cite{Bendo2015}). Because the 3\,GHz SFR calibration used in our work was anchored to local infrared data, this could explain why our SFRs are higher than some of the literature, and by extension, why our cosmic SFRD lies above the \cite{Madau2014} curve.

\subsection{Cosmic SFRD evolution and characteristic stellar mass}\label{discussioncharmass}
The overall normalization of our SFR density, seen in Figure \ref{fig:csfh}, lies $\lesssim$ 0.15\,dex above what is expected from the literature (e.g. \citealt{Madau2014}) at $z<3$. 

Our SFR density -- stellar mass relationship (Fig \ref{fig:charmass}) shows a peak at a characteristic stellar mass of $M_\star\sim10^{9.9}$\,M$_\odot$ at $z\sim0.35$ and a characteristic mass that increases with redshift out to $z\sim 2.5$ ($M_\star\sim10^{10.6}$\,M$_\odot$). 
This trend was also predicted by \citet{Bethermin2013} who showed that the majority (90\%) of the star formation activity is hosted in halo masses between $11.5<\log(M_h$/M$_\odot)<13$ regardless of redshift\footnote{Indeed, many studies have found that halos more massive than $\sim10^{12}$\,M$_\odot$ can be quenched by the shock heating of the cold inflowing gas, e.g., \citet{Dekel2006}, \cite{Birnboim2007}, \cite{Dekel2008}, \cite{FaucherGiguere2011}}. 
However, halos that have a mass of $10^{11.5}<M_h<10^{13.5}$\,M$_\odot$ at $z>4.7$ will become massive clusters with halo masses $M_h>10^{14.5}$ by $z=0$. Similarly, halos in the characteristic halo mass range at $0.5<z<3.7$ will become small clusters or groups by $z=0$, while at $z<0.5$, the SFR density peaks in Milky Way-like halos \citep{Bethermin2013}. In the empirical model of \cite{Bethermin2013}, the bulk of cosmic star formation initially occurs in the progenitors of the most massive halos before becoming less efficient and propagating to less massive halos; a trend referred to as ``downsizing''(\citealt{Cowie1996}; see also e.g., \citealt{Tinsley1968}). 

Similarly, \citet{Legrand2018} studied the stellar-to-halo mass relationship in the COSMOS field using a parametric abundance matching technique and find that the ratio peaks at a characteristic halo mass of $\sim10^{12}$\,M$_\odot$ at $z=0.2$, and that this characteristic mass increases with redshift to $z\sim2.3$ and then remains flat until $z=4$ (see also e.g., \citealt{Leauthaud2012}, and \citealt{Moster2010}). They argue that this is consistent with the picture that star formation is quenched in more massive halos first. Below the characteristic halo mass, the stellar mass build-up (i.e. SFR) ``keeps up'' with the dark matter accretion rate, whereas above the characteristic mass the galaxy is more likely to be quenched. Interestingly, \cite{Mowla2019} found that the galaxy size-- stellar mass relation has a broken power-law shape, whose slope changes at a pivot mass M$p$ that also increases with redshift in the same manner as the peak of the stellar-to-halo mass relation from \cite{Leauthaud2012}. 
We interpret the evolution in turn-over mass or characteristic mass towards $z=0$ to support the idea that quenching starts in the most massive galaxies at high redshift, and then proceeds to lower-mass galaxies with time. 


\subsection{Morphology trends}\label{sec:discussionmorph}

We found in Section \ref{sec:morph} that for massive galaxies ($\log(M_*/$M$_\odot)>10.0$), galaxy morphology is correlated to the position on the MS: at a fixed stellar mass, early-type and bulge-dominated galaxies have lower SFRs than disk-dominated galaxies, which in turn have lower SFRs than irregular galaxies.

Galaxy interactions could account for irregular morphologies and also cause a burst of star-formation resulting in irregular type galaxies lying above the MS of disk-dominated galaxies. \citet{Cibinel2019} reported that the starburst population across redshifts $0.2<z<2$ predominately consists of late-stage mergers. We note that the slope of the SFR-stellar mass relation for these irregular galaxies is slightly shallower than for the disks, inconsistent with results from \citet{Bisigello2018} for starburst galaxies. 
To further investigate the composition of our ZEST-classified irregular (Type 3) galaxies, we cross-matched our parent sample using the \citet{Capak2007} position coordinates with the 70\,$\mu$m selected catalog of 1,503 sources from \citet{Kartaltepe2010a, Kartaltepe2010b}, finding 1,142 overlapping sources. Out of the $\sim$\,2,300 irregular ZEST sources, we found 216 with visual morphological classifications in \citet{Kartaltepe2010b}. The authors 
classified source morphology (e.g. spiral, elliptical) and noted whether each source was undergoing a minor or major merger. We found that for galaxies in both the ZEST and 70$\mu$m catalogs, 23.6\% are minor mergers and 28.8\% are major mergers (total 52.5\% mergers). For irregular ZEST galaxies the fraction of mergers is much higher; 80\% (30.1\% minor mergers and 50\% major mergers). 
This supports the idea that our ZEST Type 3 irregular galaxies have higher star formation rates (and more irregular morphologies) than the disk-dominated sub-sample because they are undergoing mergers that trigger gas inflows resulting in an enhanced SFR \citep{Barnes1996}. 


One key result from our work is that bulge-dominated galaxies follow a shallower MS relation than disk-dominated galaxies.
Trends between sSFR and morphology have also been found in spiral galaxies, which are presumably not merger driven. For example, \cite{Parkash2018} found that earlier type spiral galaxies (Sa) have lower sSFRs than later-type spirals (Sc).
\citet{Schreiber2016} performed bulge-disk decompositions of galaxies with $\log(M_*/$M$_\odot)>10.2$ in CANDELS, and found that the disk component of galaxies at $z=1$ follows a similar flattening of the MS as the total (stacked) population, even for disk-dominated galaxies, in disagreement with our results. They also reported that the lower sSFRs of massive galaxies corresponds to a lower star formation efficiency (not present at $z=2$) rather than a lower gas fraction. However, the study used \textit{Herschel} data to also infer the gas content which should be confirmed with independent observations. 
Although \citet{Schreiber2016} found that disk-dominated galaxies exhibit the same flattening as bulge-dominated galaxies, other studies such as \citet{Erfanianfar2016} found that the disk galaxies have a more linear MS relation than the bulge-dominated galaxies, consistent with our findings. The number of pure disks at high stellar mass is very low in the low redshift Universe \citep{Ogle2016}; our lowest redshift bin $0.3<z<0.6$ contains only 4 disk-dominated galaxies above $10^{10.7}$\,M$_\odot$.
Larger samples at low redshifts, such as those provided by SDSS, have shown that the disk component of galaxies follows a linear MS \citep{Abramson2014}. 

\begin{figure*}
\includegraphics[width = 0.5\linewidth]{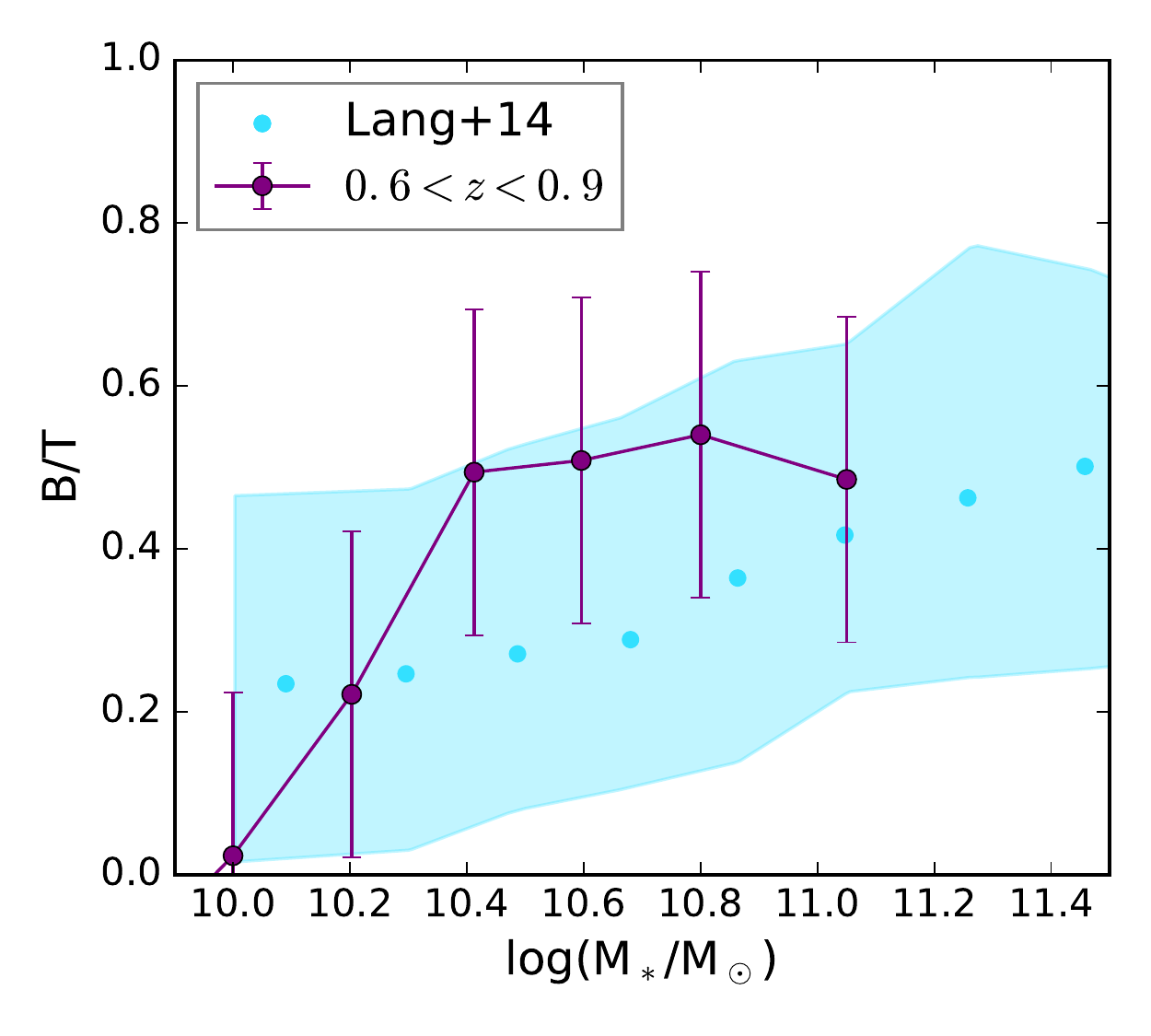}
\includegraphics[width = 0.5\linewidth]{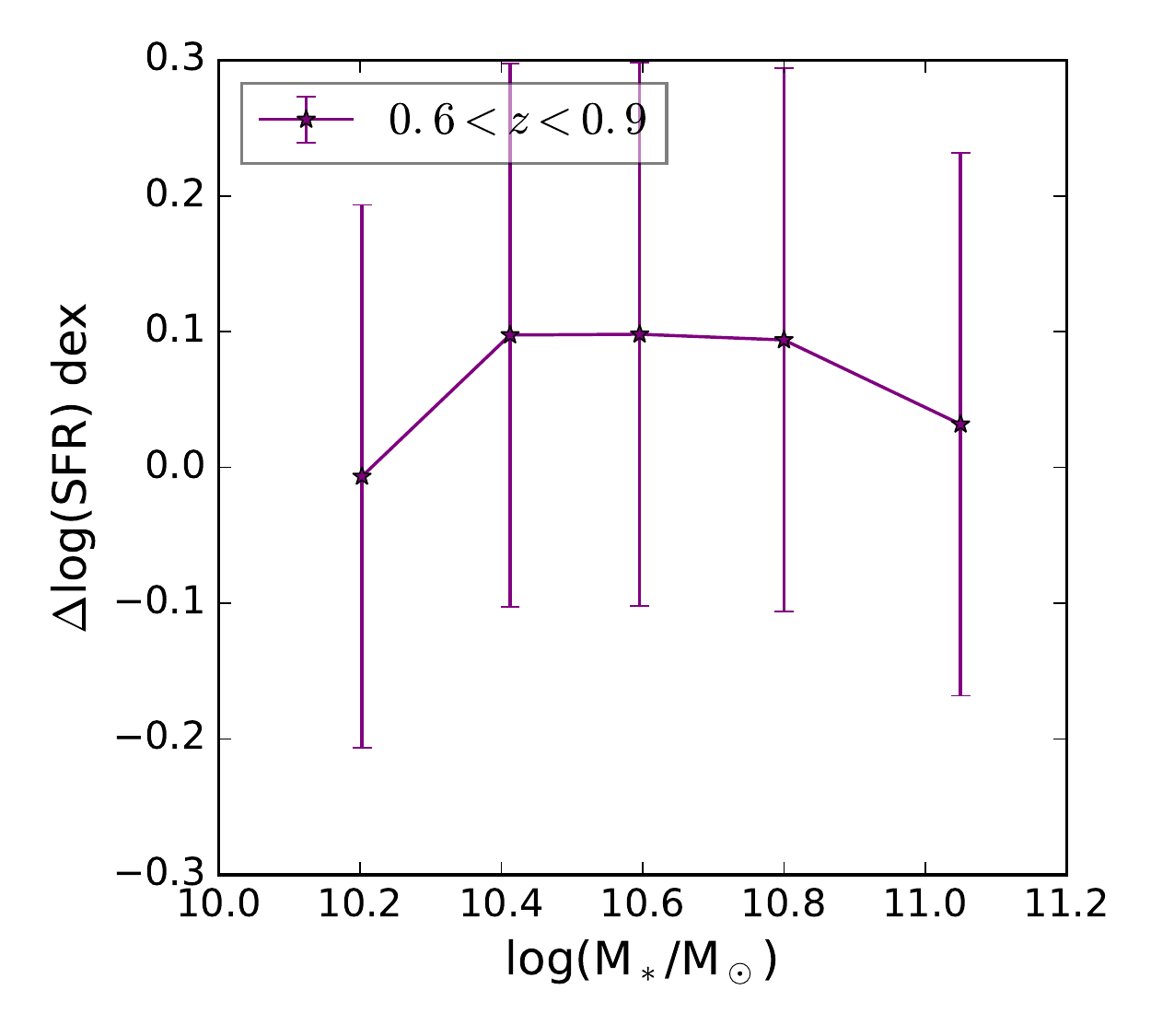}
\caption{Do bulges suppress SFR? The left panel shows the typical bulge-to-total ratio of our bulge-dominated late-type galaxies required to explain their measured sSFR, assuming they consist of a disk component that follows the same SFR -- $M_*$ relation as the disk-dominated galaxies, and a bulge that has no star formation (providing a lower limit to the B/T). We fix the error on our B/T ratio to 0.2, which comes from the range of the B/T ratios obtained repeating this method for disk-dominated galaxies (which should have returned zeros).
The B/T ratios inferred in this way are consistent with those measured by decompositions of the stellar mass by \citet{Lang2014} at $0.5<z<2.5$ (cyan dots indicate median measured B/T and the shaded region shows the range of values).
The right panel shows the amount of star-formation activity that needs to be suppressed in the disk in order to recover the SFR -- $M_*$ relation for the bulge-dominated galaxies if the bulge-dominated galaxies have B/T ratios observed by \citet{Lang2014} and bulges that have no star formation. Our results set an upper limit for the morphological SFR suppression in bulge dominated galaxies, if any, to be less than 0.2-0.3\,dex.}\label{fig:bt}
\end{figure*}

We explore how the steeper MS of disk-dominated galaxies can be related to the shallower MS for bulge-dominated galaxies at a fixed redshift in Figure \ref{fig:bt}. We infer how much stellar mass (assumed to be in a quiescent bulge) must be added to the disk-dominated galaxies to shift the SFR -- $M_*$ relation to be consistent with that of the bulge-dominated late-type galaxies. 
First, we fit a linear relation to the disk-dominated galaxies and compute the disk mass required to provide the SFR observed for the bulge-dominated galaxies. We then infer the bulge to disk ratio assuming that the bulge-dominated galaxies consist of a disk (following the MS of disk-dominated galaxies) plus a bulge with zero star-formation activity. This assumption will give a lower-limit to the B/T ratio, as bulges can conceivably contribute a small amount of star formation activity. These resulting B/T ratios are shown in Figure \ref{fig:bt} for $0.6<z<0.9$ where we have the most robust results.
The resulting SFR-derived B/T (bulge mass to total mass) ratios show a clear increase with stellar mass and values consistent with the result from \citet{Lang2014} for star-forming (sSFR selected) galaxies in the CANDELS/3D-HST fields.

If we adopt the B/T ratios given by the \citet{Lang2014} measurements, then the amount of SFR that needs to be suppressed in the disk is shown in the right-hand panel of Figure \ref{fig:bt} as a function of total galaxy mass. In other words, we calculate the difference between the expected SFR of the disk components based on the linear fits to the disk-galaxy MS and the measured average SFR of the bulge-dominated galaxies.
The amount of inferred SFR suppression is consistent with zero, implying that the disks of bulge-dominated systems with a non-starforming bulge have similar sSFR as pure disks. 
The same result was also recovered when we compared stacks of ZEST Type 2.3 (pure-disks) and ZEST Type 2.1 (2.0 is more likely contaminated by early-type galaxies).

We note that if the bulge component also contains any star-formation activity, then a trend towards SFR-suppression in the disk component could be stronger and that the $\Delta\log(\mathrm{SFR})$ values in Figure \ref{fig:bt} are lower limits. There is also the possibility that our SFR calibrations based on the infrared-radio correlation should be different for different galaxy types \citep{Molnar2018}. If we use a 3\,GHz-SFR calibration for our disk-dominated galaxies based on the $q_\mathrm{TIR}(z)$ relation reported by \cite{Molnar2018} for disk-dominated galaxies and use their spheroidal-galaxy $q_\mathrm{TIR}(z)$ values for the bulge-dominated galaxy SFR, then our inferred suppression $\Delta\log(\mathrm{SFR})$ would increase by $\sim 0.1$\,dex. Again, our result (calculated using the same \cite{Molnar2020} calibration for all galaxies) can be considered as a lower limit for the effect of bulges.

Studies such as \citet{Popesso2019} and \citet{Schreiber2016} have reported that the presence of bulges alone is not sufficient to explain the flattening of the MS at very high masses; in addition to the increased mass given by a bulge component, there is also an indication for a decrease of star-formation activity of the disks along the MS.
Theoretical arguments suggested by \citet{Meidt2018}, explain that the gravitational potential of a centrally concentrated bulge can help stabilize the gas against fragmenting (see also simulations from \citealt{Su2018} and \citealt{Gensior2020}). Bulges can also decrease the strength of bars, thereby decreasing the efficiency that gas is funneled into the galaxy center \citep{Barazza2008,Fragkoudi2016}. \citet{Popesso2019} speculated that while the quiescence of the bulge component could be explained by super-massive black hole feedback, the lower SFR of the disk at high masses could be due to gas starvation induced by gravitational heating in massive halos. However, we find that a quiescent bulge is sufficient to explain the decrease of SFR along the MS for SF galaxies. 
Nevertheless, larger sample sizes and a better understanding of SFR calibrations will be required to get the uncertainties on B/T ratios and sSFR suppression to within $\sim 0.2$\,dex, the amount of influence that morphological suppression due to the presence of a bulge is expected to have on the sSFR.


\subsection{Environmental trends}\label{sec:discussionenv}

Results tend to agree that higher density environments increase the fraction of quiescent (red) galaxies and we do see this in the COSMOS field (see top right panel of Figure \ref{fig:envall}; \citealt{Peng2010, McGee2011, Scoville2013, Darvish2014, Darvish2016}).
Whether or not the normalization of the MS (defined for star-forming galaxies) depends on galaxy environment is still under debate, but many studies identify no variations of MS with environments such as clusters and voids, e.g., \citet{Tyler2013}, \citet{Koyama2014}, \citet{Ricciardelli2014}, \citet{Tyler2014}, \cite{Grossi2018}, \citet{Paulino-Afonso2019}, \citet{Pharo2020} (although see \citealt{Duivenvoorden2016} who found a difference at $1.5<z<2$ in COSMOS). 
However, at low redshift ($z<0.3$), studies have reported a clear dependence of the MS on galaxy environment (e.g., \citealt{vonderLinden2010}, \citealt{Haines2013}, \citealt{Gu2018}, \citealt{Paccagnella2016}).
Studies such as \cite{Balogh1998}, \cite{Couch2001}, \cite{Balogh2002}, and \cite{Gomez2003} have found that SFRs of cluster galaxies were lower relative to field galaxies of similar bulge-to-disk ratio and luminosity, which suggests the decrease in star-formation may not be fully explained by the density--morphology relation \citep{Dressler1980} alone.
Environmental effects appear important at low redshift, where spatially resolved studies show that environmental quenching works on galaxies outside-in (e.g. \citealt{Schaefer2019}).

Recently, \cite{Old2020}, showed that environmental suppression of SFR, relative to the MS, indeed becomes more important towards low redshift $(z<1)$ and lower stellar masses. However, the amount of suppression was only a factor of 1.4 (3.3\,$\sigma$ significance). 
It is not surprising that our COSMOS sample does not show clear evidence for an environmental dependence of the sSFR when probing local galaxy number density at $z>0.3$, or in X-ray groups at $0.64<z<0.88$. 
In simulations, \citet{Matthee2019} found that star-forming satellite galaxies only account for $\approx0.04$ dex of the scatter in the MS indicating satellites-specific processes are either weak or strong and rapid (such that the satellites quickly drop out of the sample of star-forming galaxies), consistent with our results. In the local universe $z<0.3$, there is evidence for both slow ($\sim$2-4\, Gyr; e.g. strangulation) and rapid (e.g. stripping of cold gas) quenching processes associated with galaxy environment \citep{Paccagnella2016, Maier2019}.


Unlike our results, recent work by \citet{RodriguezMunoz2019} found that not only is the fraction of star-forming galaxies lower\footnote{We find that the fraction of SF galaxies is lower in X-ray environments by a factor that depends on stellar mass, but is generally $<2$, see Figure \ref{fig:envall}} in X-ray group environments by a factor of $\sim2$, but also the star-forming galaxies in groups have an average sSFR $\sim0.3$ dex lower than the field across all redshift ranges probed ($0.1<z<0.9$; they have a redshift bin $0.60<z<0.89$, similar to ours which consists of 2 clusters). 
One might also expect the resolution of the photometry and SFR tracer to have a significant impact on environmental studies because source blending can be an issue if not properly taken into account. 
Larger cosmological volumes should be probed to understand the importance of the environment on the average sSFR of galaxies.

\citet{Erfanianfar2016} separated group and field environments for two redshift bins ($0.15<z<0.5, 0.5<z<1.1$) and found at $z<0.5$, that group and cluster galaxies have a reduction of SFR compared to their counterparts in the field (with a larger fraction of disks being redder). They also report that the flattening of the MS for field galaxies is due to an increased fraction of bulge-dominated galaxies at high masses and that the associated quenching process must already be in place before $z\sim1$. 
\cite{Tomczak2019} and \cite{Koyama2013} found that SF galaxies show a difference of, at most, 0.2 dex in sSFR in dense environments as compared to the field from $z=0.4$ to $z=2$. 

Once a group or cluster is in gravitational equilibrium, the increased local density should decrease the timescales involved in evolution compared to the relevant timescales in the field because of the larger gravitational potential. Indeed \citet{Popesso2015b} shows that the cosmic SFR activity declines faster (towards $z=0$) in group size haloes than in the field. This faster evolution increases the scatter around the MS and could decrease the average SFR when including all galaxies. To summarize, the local density affects the speed of the star-formation evolution more than the mode (altering the MS) of evolution (at least from $0.3<z<3$).


\section{Conclusion}\label{sec:conc}
Radio continuum emission is a useful dust-unbiased star formation tracer and stacking is an effective way to push current stellar mass limits achievable through direct detections.
We have mean stacked 3\,GHz images for galaxies in the latest COSMOS photometric catalogs \citep{Laigle2016,Davidzon2017} to measure the average dust-unbiased SFR for galaxies as a function of stellar mass, redshift, environment, and morphology. Our findings can be summarized as follows: 
\begin{itemize}
\item We find that star-forming galaxies follow a SFR -- $M_*$ relation that is steeper at low masses than at high-masses (i.e. a flattening is present). We model the MS relation for star-forming galaxies (and all galaxies) with a new function that allows (if needed) for the flattening to occur at increasing stellar mass and for the increase in overall normalization with redshift.
\item The cosmic star formation rate density peaked at $1.5<z<2$, but the measurements still contain large systematics such as SFR calibrations and the necessity of assuming a stellar mass function to account for incompleteness. 
\item Our results support the downsizing scenario, that massive galaxies are formed and quenched first. At higher redshift, more massive galaxies contributed more to the cosmic SFR density than at lower redshift, consistent with literature findings. This result is driven by the increasing turn-over mass with redshift in our MS.
\item We find no significant difference in the MS for star-forming galaxies located in different environments probed by the local number density, or by X-ray group membership. 
\item Early-type galaxies have the lowest SFRs at a fixed stellar mass, followed by bulge-dominated galaxies, disk-dominated galaxies, and irregulars. We find this to be the case at $z<1.5$. 
\item Massive bulge-dominated galaxies follow a shallower SFR -- $M_*$ relation than disk-dominated galaxies. As the number of bulge-dominated galaxies increases towards low redshift, the contribution of disk-dominated galaxies to the total SFR occurring declines, particularly at high stellar mass. This increase in bulge-dominated galaxies could be related to the mechanisms responsible for downsizing or mass quenching.
\item Combining samples of bulge-dominated galaxies and disk-dominated galaxies can explain, in part, the flattening of the MS observed to occur at low redshifts and high stellar masses. 
\end{itemize}
The decrease in the cosmic SFRD from $z\sim2$ to $z\sim0$ is caused by a combination of the decrease in the sSFR of star-forming galaxies and the quenching of star-forming galaxies (e.g., \citealt{Renzini2016}). 
The decrease in the sSFR of star-forming galaxies since $z\sim2$ is likely due to the decrease in cold gas accretion. Quenching transforms star-forming galaxies into quiescent galaxies and is likely driven by feedback (stellar and AGN), environmental effects, and/or morphological quenching.
With the latest COSMOS data, we see a mass-dependence in this decrease of the MS (a flattening at high stellar mass that shifts to lower masses at low-z). We also found that the stellar mass in which most new stars are formed (a characteristic mass) increases with redshift. The increasing characteristic mass with redshift and the evolving flattening in the MS corresponds to bulge formation and a decline in SFR in massive galaxies. The mechanism that reduces the SFR proceeds to be effective at lower stellar masses towards the present epoch. 
Our results on the sSFR of galaxies with different morphological types could imply that the presence of a bulge affects the star-formation process. However, future observations are required to confirm these conclusions and to determine whether the gas in bulge-dominated galaxies is stabilized against fragmentation, or is not present to form stars (either it has been removed by feedback or cold gas is no longer able to be accreted onto the galaxy disk). 






\vspace{1cm}
\acknowledgments
We would like to thank the anonymous referee for their careful comments which lead to a highly improved paper. We would like to thank D. Molnar and K. Cooke for helpful discussions and feedback. SL acknowledges funding from Deutsche Forschungsgemeinschaft (DFG) Grant BE 1837 / 13-1 r. ES, DL, and PL acknowledge funding from the European Research Council (ERC) under the European Union's Horizon 2020 research and innovation programme (grant agreement No. 694343). 
VS and MN acknowledge the European Union's Seventh Framework programme under grant agreement 337595 (CoSMass). BG acknowledges the support of the Australian Research Council as the recipient of a Future Fellowship (FT140101202). YP acknowledges the National Key R\&D Program of China, Grant 2016YFA0400702 and NSFC Grant No. 11773001, 11721303, 11991052.
This publication has received funding from the European Union's Horizon 2020 research and innovation programme under grant agreement No 730562 [RadioNet]
SL would like to thank L. Kewley and ASTRO3D for providing the excellent conditions at Bateman's Bay writing retreat during which part of this manuscript was prepared. 



{\it Software:}
Figure \ref{fig:cornerplot} made use of the \textsc{corner} python package \citep{ForemanMackey2016}.
This research has made use of NASA's Astrophysics Data System. This research has made use of the NASA/IPAC Infrared Science Archive, which is operated by the Jet Propulsion Laboratory, California Institute of Technology, under contract with the National Aeronautics and Space Administration. This research made use of NumPy \citep{van2011numpy} This research made use of matplotlib, a Python library for publication quality graphics \citep{Hunter:2007}. This research made use of ds9, a tool for data visualization supported by the Chandra X-ray Science Center (CXC) and the High Energy Astrophysics Science Archive Center (HEASARC) with support from the JWST Mission office at the Space Telescope Science Institute for 3D visualization. This research made use of Astropy, a community-developed core Python package for Astronomy \citep{2013A&A...558A..33A}. This research made use of TOPCAT, an interactive graphical viewer and editor for tabular data \citep{2005ASPC..347...29T}. 
\facility{
Based on data products from observations made with ESO Telescopes at the La
Silla Paranal Observatory under ESO programme ID 179.A-2005 and on
data products produced by TERAPIX and the Cambridge Astronomy Survey
Unit on behalf of the UltraVISTA consortium. Based on observations obtained with XMM-Newton, an ESA science mission with instruments and contributions directly funded by ESA Member States and NASA. The National Radio Astronomy Observatory is a facility of the National Science Foundation operated under cooperative agreement by Associated Universities, Inc. }



\appendix

\section{Stacking analysis: Details and tests}\label{sec:appendixmethod}
In this section, we discuss some of the issues concerning stacking radio images, in particular, the choice of method to measure total flux, and the choice of whether to represent the average as the mean or the median. 

Radio images with bright sources must be cleaned to have any hope at detecting sources amongst the side-lobes of the bright source. 
For this reason, most stacking analysis to date has been performed on cleaned radio images.
Components of sources that lie above the clean threshold will have a point spread function (PSF) corresponding to the Gaussian-shaped clean beam, whereas fainter sources with no cleaned components will have a PSF corresponding to the dirty beam. 
Stacking a population that includes sources with different PSFs is not an intuitive situation. For the VLA 3~GHz LP, the entire map was cleaned down to 5$\sigma$ and further down to 1.5$\sigma$ using tight masks around 5$\sigma$ sources \citep{Smolcic2017}. 
The noise level in each pointing was around 4-5\,$\mu$Jy beam$^{-1}$.
In addition, each pointing was tapered with a Gaussian to achieve a circular clean beam with 0.75$''$ FWHM. However, the sizes vary across the COSMOS field due to different uv coverage (with a difference between the major and minor axis being most 3\%), so even for ``clean sources'' there is no ``single PSF.''
We have tested further cleaning our stacks (after averaging in the image plane) with the mean beam shape, but this tended to remove flux from the already cleaned sources, rather than adding back flux from the side-lobes. One way to minimize this could be to stack sources below 1.5$\sigma$ and then clean the resulting stack, but this would have to be done separately for the 192 different VLA pointings due to variations in the dirty beam across the field, resulting in a loss of signal-to-noise. We note that every source will have some residual dirty flux, for example, a source detected at 5$\sigma$, if it was included in the clean masks, will have 30\% of its flux ``dirty'' (below 1.5$\sigma$). This percentage will go down for higher S/N sources, but be 100\% for non-cataloged sources. 

Studies such as \citet{Mao2011} with similar numbers of detections and non-detections in each bin, combine detected and non-detected sources separately, weighing the flux from each detected source by 1/N and the stack of undetected sources as n/N, where N is the total number of sources in the bin and n is the number of sources in the stack.  
There is a significant variation in the fraction of sources detected at $>5\sigma$ between stellar mass and redshift bins, ranging from 0 to 30\% from low to high stellar masses. We decided to treat all sources homogeneously by stacking detected and non-detected sources together, thereby not introducing an arbitrary flux limit above which we consider a source detected. This uniform treatment also has the advantage of allowing errors to be readily calculated by bootstrapping methods. 

In the region selected for this analysis, the rms variation is minimal (e.g., rms variations on scales of $\sim1'$ are less than 2\%, \citealt{Smolcic2017}) and we found no difference in our mean or median stacked image fluxes by weighting each cutout by its inverse variance. 

\subsection{Flux Measurement: Comparison with 1.4\,GHz and IR stacks.}\label{sec:fluxes}
In Section \ref{sec:methods}, we mentioned that astrometric uncertainties, and intrinsic source extent require us to measure a total flux rather than the flux contained within one beam around the central object. 
For our source fluxes, we adopt a simple elliptical 2D Gaussian fitting. We input initial conditions for the fit using the peak flux from the stacking routine and the beam size and, find best-fit parameters using \textsc{optimize.curve\_fit}.
The fitting procedure was tested allowing for non-zero background values on cataloged and simulated sources. We found background fluxes were always $<$1\,$\mu$Jy/beam, so have adopted the fits with no background subtraction for our analysis.



\begin{figure}
\centering
\includegraphics[width=0.8\linewidth]{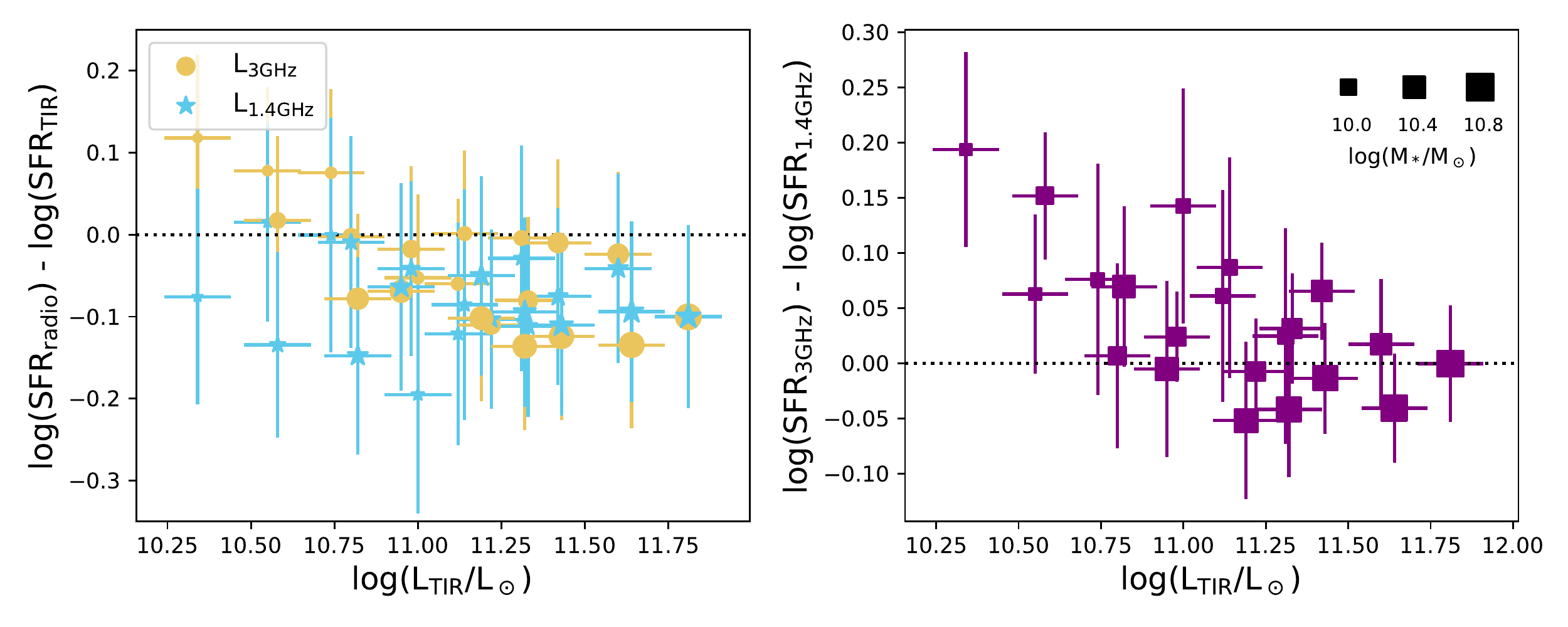}
\caption{Comparison of 1.4\,GHz, 3\,GHz, and TIR-derived SFRs from median stacked images of star-forming galaxies at $z<2$. The left panel shows radio SFRs compared to IR SFR. 1.4\,GHZ fluxes are from \cite{Karim2011}. The right panel shows the difference between 1.4\,GHz and 3\,GHz SFRs as a function of IR luminosity. In both panels, the symbol size represents the average mass of stacked galaxies. }\label{fig:3GHzcalib}
\end{figure}

We have measured total fluxes of 3~GHz stacks that were created using the input SF galaxy sample of \citet{Karim2011}, allowing us to compare our results with their published median stacked 1.4~GHz fluxes. To be consistent with \cite{Karim2011}, we show results from median stacking in our comparison (Figure \ref{fig:3GHzcalib}).
We have also stacked these sources at infrared wavelengths following \citet{Magnelli2015} and derived a (median) FIR luminosity for the stacks. We chose to focus on $z<2$ where all samples are reliable, and stellar masses $>10^{10.0}$ M$_\odot$ for completeness. 
For our comparison, we have converted radio fluxes to infrared luminosities assuming a spectral index of $-0.7$, and the radio-infrared correlation of \cite{Molnar2020}.
Figure \ref{fig:3GHzcalib} shows the difference between the 3~GHz, 1.4\,GHz, and IR SFRs. 
All stacks shown have $S_p$/rms$>10$ at 3\,GHz.
The 1.4~GHz and 3~GHz SFRs measured from Gaussian fits to determine the total fluxes agree within 0.2\,dex, as illustrated in Figure \ref{fig:3GHzcalib}. For the low-mass galaxies (log($M_*$/M$_\odot)\approx 10.0$), the scatter is larger, in part due to the lower SNR in the 1.4\,GHz stack. However, we also cannot rule out a stellar mass-dependent radio spectral index from causing the 3\,GHz derived SFRs to be higher than the 1.4\,GHz SFRs at low IR luminosities.

\subsection{Mean or median}\label{sec:meanvsmed}
While the mean of a flux distribution is natural to interpret mathematically, it can be sensitive to outliers. Because the median is robust to the presence of outliers, it has the advantage that all data can be used. However, the median value recovered from a stacked image depends not only on the underlying distribution but also on the noise level. \citet{White2007} showed that in the limit where the individual sources are well below the rms level, the median values trace the population means. These arguments only strictly apply to point sources.
The underlying distribution of source fluxes is also not likely a simple Gaussian\footnote{The distribution of galaxies in SFR at a fixed $M_*$ and z is thought to consist of two or three components \citep{Bisigello2018, Hahn2019}, that could be log-normal (or not; \citealt{Eales2018}).}, thus leaving us with a difficult problem that deserves more attention.


Figure \ref{fig:meanvsmedian} shows the ratio of total fluxes derived from mean and median stacking analysis. In all cases, the mean reported is larger than the median; which is expected for most SFR distributions. In particular, for a Gaussian distribution in log(SFR), with a given $\sigma$, the ratio between the mean SFR and the SFR corresponding to the median log(SFR) (i.e. the peak of the Gaussian) is $e^{2.652\sigma^2}$. 
We find that the ratio of $S_{\rm{mean}}/S_{\rm{med}}$ increases with stellar mass; this is mostly due to the presence of AGN outliers affecting the mean fluxes.
The bottom panels show the results when AGN are excluded, as done for our main analysis. Here, when considering all galaxies, there seems to be a residual trend of increasing ratio (scatter) with stellar mass even once AGN are removed. On one hand, radio-AGN removal in passive galaxies is complicated, but on the other hand, there are likely more quenched galaxies with low SFRs at high-masses included in the full sample which would increase the measured scatter.

\begin{figure}
  \centering
  \includegraphics[width = \linewidth]{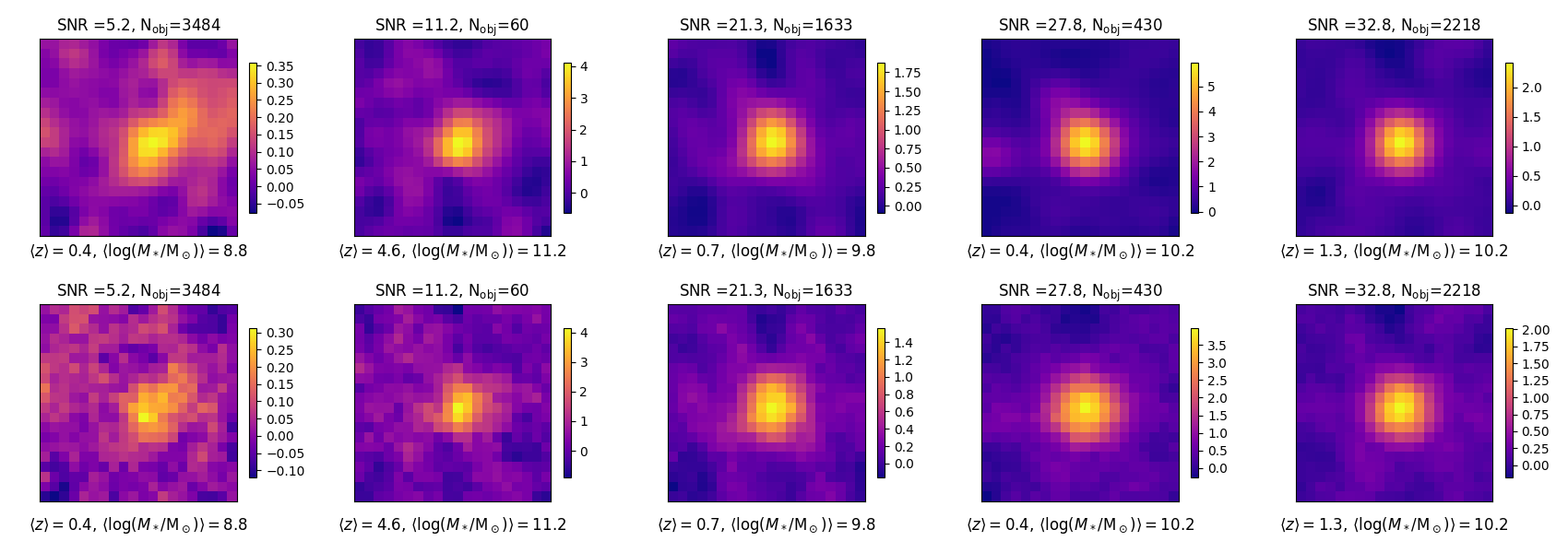}
  \caption{Some example 3\,GHz stacks. The top row shows the mean stack, and the bottom row shows the median stack for the same image. The peak flux / rms ratio for the median stacks is given as the SNR and increases from left to right. For our analysis, the left-most stack (low-mass) would be excluded because the fluxes tend to be over-estimated for low SNR sources (see Figure \ref{fig:simulation}).}
  \label{fig:examplestacks}
\end{figure}

\begin{figure}
  \centering
  \includegraphics[width = 0.49\linewidth]{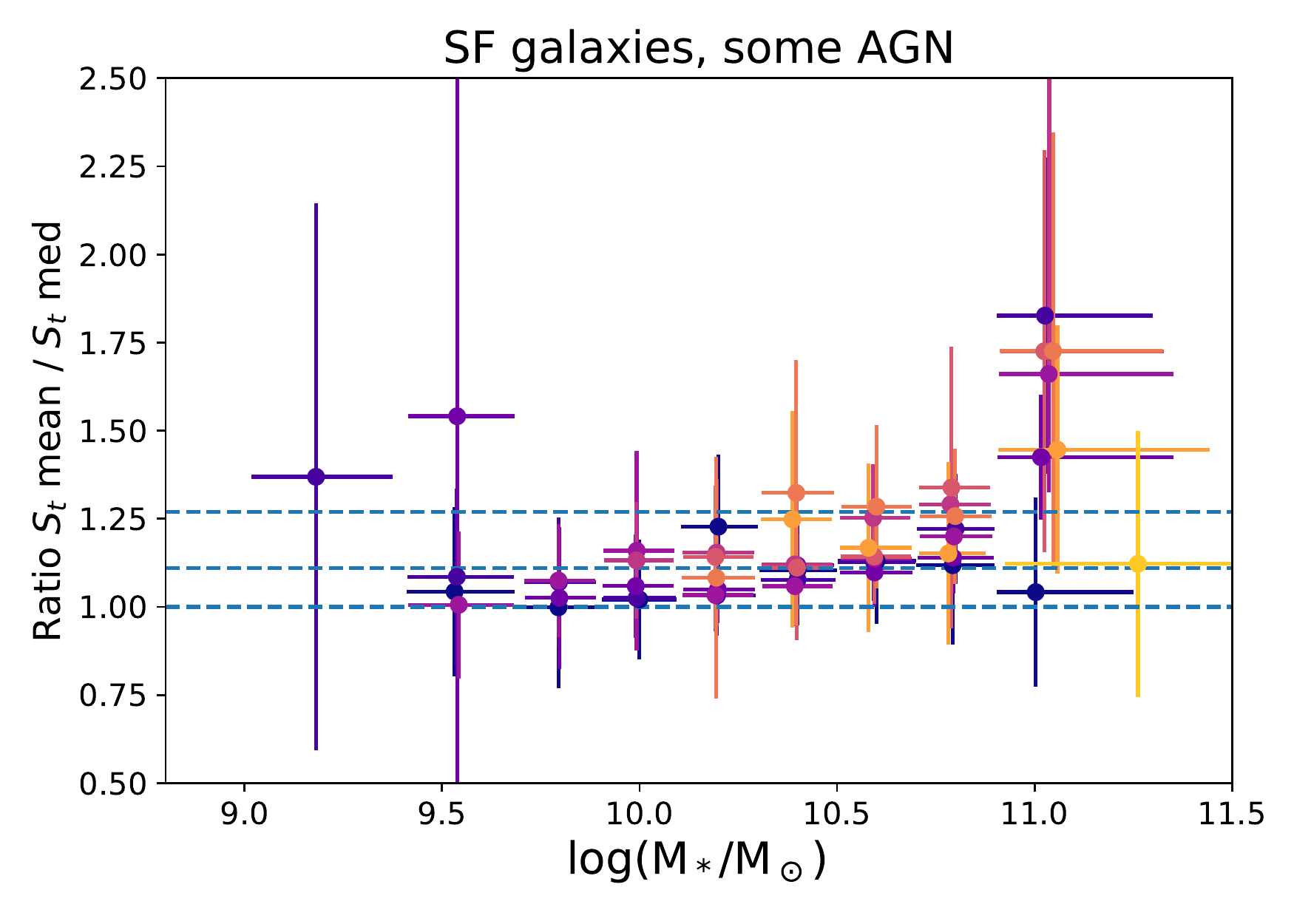}
  \includegraphics[width = 0.49\linewidth]{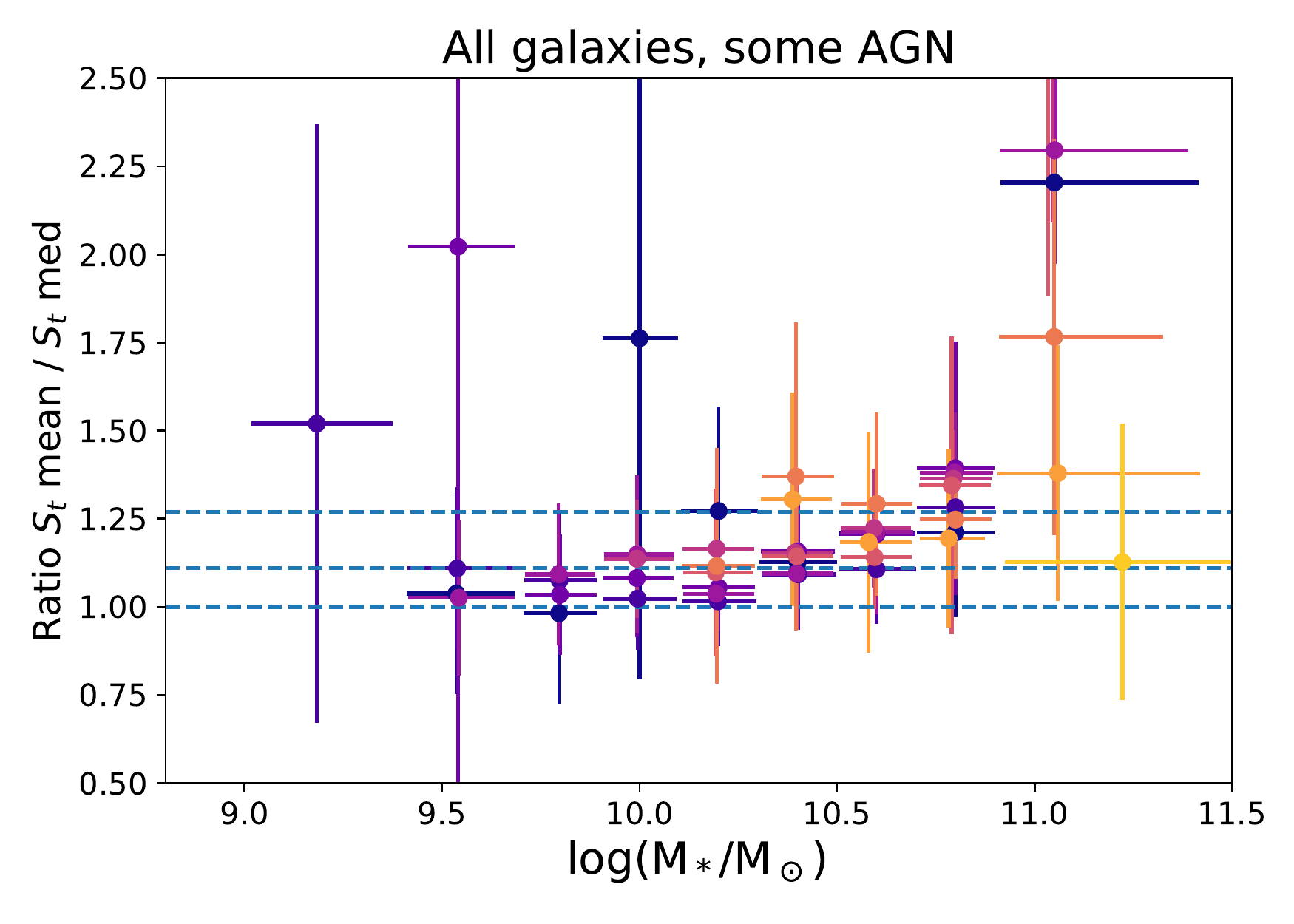}
  \includegraphics[width = 0.49\linewidth]{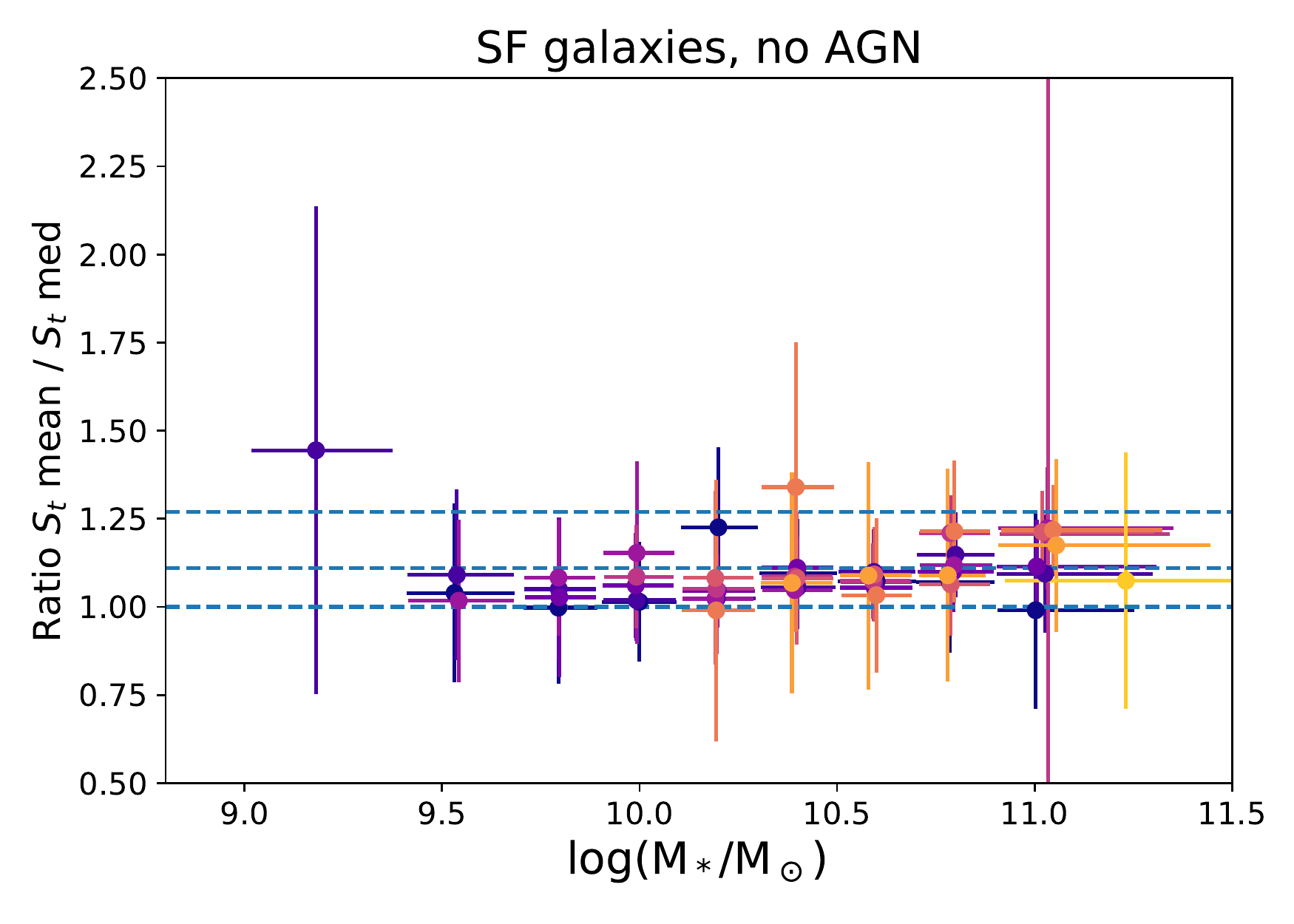}
  \includegraphics[width = 0.49\linewidth]{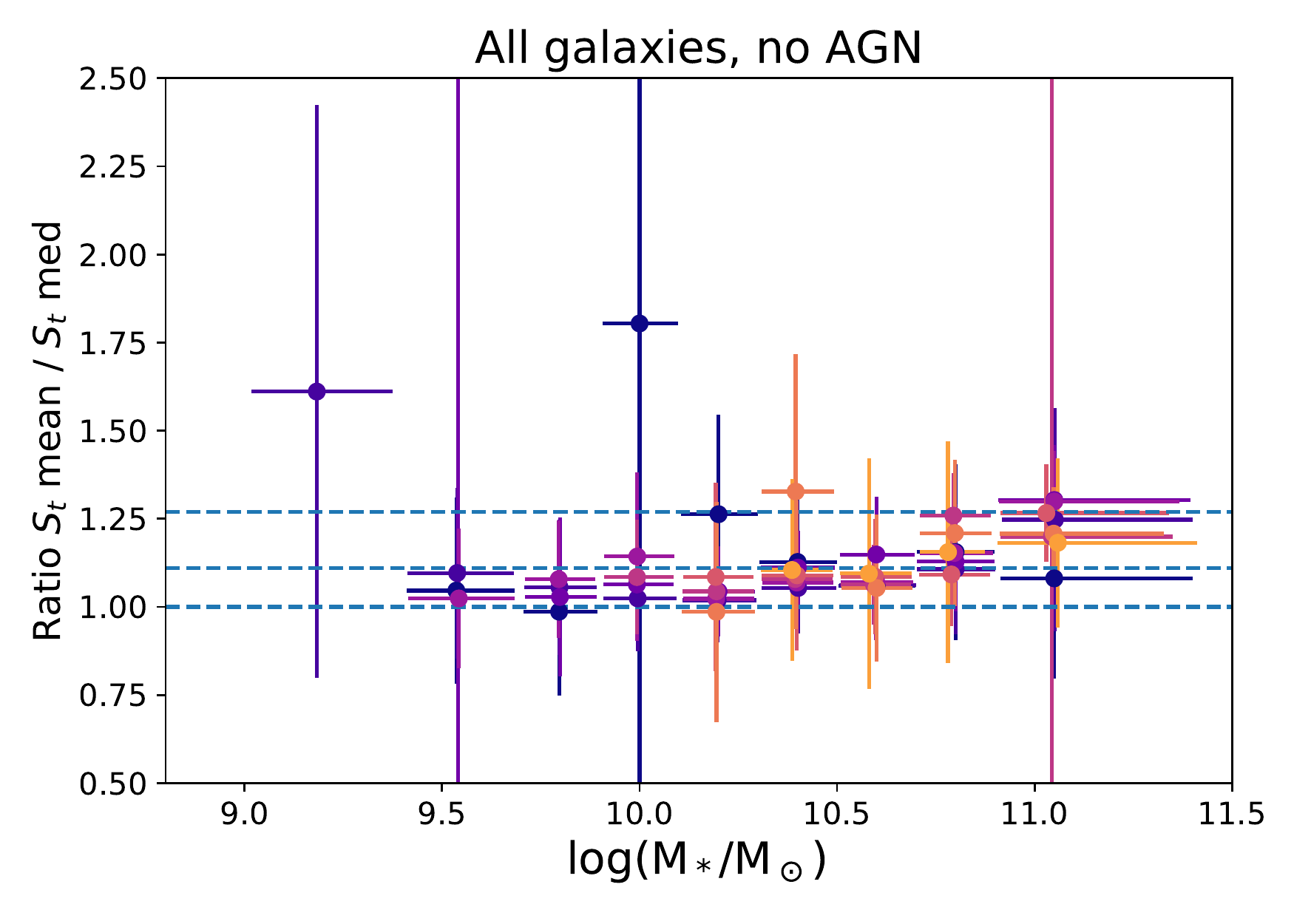}
  \caption{Ratio of total fluxes derived from mean and median stacked images as a function of the median stellar mass. Left panels show SF galaxies, selected by $NUVrJ$ colors. Right panels show all galaxies. Top panels show results with only multi-component radio AGN removed. Bottom panels show the results using our sample with multiwavelength AGN removal applied. Color represents redshift (purple to yellow is low to high redshift using the same scale as e.g Figures \ref{fig:ms}, and \ref{fig:uvj}).
  Error-bars represent the 5-95\,\% range. Horizontal lines are drawn at S$_\mathrm{t,mean}$/S$_\mathrm{t,median}$=1.0, 1.11, 1.27, which correspond to the values expected if SFR is symmetrically distributed, or if log(SFR) is normally distributed with a $\sigma=0.2$, $\sigma=0.3$ dex, respectively (and if flux is linearly related to SFR). Only stacks with $S_p$/rms$>$10 are shown. 
Radio emission of AGN host galaxies is dominated by the AGN emission, making them outliers affecting the mean-to-median ratio.
The MS dispersion is consistent with 0.2 dex.}
  \label{fig:meanvsmedian}
\end{figure}

The uncertainty of the median fluxes from our bootstrapping analysis are smaller than the uncertainties on the mean because the median is more robust to the presence of outliers or interloping sources. In Figure \ref{fig:examplestacks}, we show five example images resulting from our stacking analysis, with mean stacks on the top panels and median stacks shown at the bottom. 

Mean SFRs follow a less-smooth MS relation than the median SFRs, with higher SFRs at low masses (log($M_*$/M$_\odot)<9.5$) and high masses (log($M_*$/M$_\odot)>11$).
However, our simulations indicate that our fluxes are more robust for mean stacks than median stacks (see next section).


\subsection{Realistic galaxy simulations}\label{sec:appendixsim}

We adopted a realistic mock galaxy Monte Carlo simulation to test the reliability of our stacking method. Following the method of \cite{Liu2019}, we generate mock galaxies within the VLA-COSMOS survey area ($\sim1.55\,\mathrm{deg}^2$) which obey the observed galaxy stellar mass functions (down to $10^{8}\,\mathrm{M_{\odot}}$) and a MS correlation at each redshift from 0 to 10 (31 redshift bins). The simulation has three steps: (1) generating mock galaxy catalog with redshift, stellar mass, SFR, and coordinate properties; (2) assigning galaxy sizes and radio 3~GHz fluxes; and (3) making a simulated radio 3~GHz image. We then repeated our analysis procedure using the generated mock galaxy catalog and simulated 3\,GHz image; here we focus on our results from stacking and measuring fluxes.

In the first step, we loop over redshift bins and construct a star-forming galaxy stellar mass function (as shown in Fig.~23 of \cite{Liu2019b}) to compute the number of star-forming galaxies in each stellar mass bin. Then we do a Monte Carlo procedure to generate the redshifts and stellar masses of the corresponding number of galaxies in each stellar mass and redshift bin. By adopting a MS, we then generate log-normal distributions of SFRs for galaxies in each stellar mass bin\footnote{For our test, no starburst or quiescent galaxies were included. The input MS prescription and scatter of the log-normal was kept hidden until after the analysis was complete, to minimize researcher bias.}. The spatial coordinates of the mock galaxies are assigned in the same way as \cite{Liu2019}, i.e., inserting into the simulation area but avoiding being too close (0.7$''$) to real galaxies (\citealt{Laigle2016}). 

In the second step, the assigned SFR for each galaxy was converted to an IR luminosity and then to a radio luminosity following \cite{Molnar2020}: 
$q_{\mathrm{TIR}} = -0.155 \times \log L_{\mathrm{1.4GHz}} + 5.96$. 
Rest-frame 1.4GHz luminosities were converted to observed 3GHz luminosities using a spectral index of -0.7.

The galaxies were modeled as Gaussian sources with sizes depending on their redshifts and stellar masses following (extrapolating from) \cite{vanderWel2014}.
We also took into account that dust sizes and radio sizes are observed to be a factor of about 2 smaller than the optical size (e.g. \citealt{Fudamoto2017, Bondi2018, JimenezAndrade2019}).
A random minor/major axis ratio from 0.2 to 1 was assigned to each source.

These model galaxies are then convolved with a Gaussian with a FWHM 0.75 arcseconds, to emulate the observations with the VLA LP beam. 
Our mock catalog tests have been restricted to using Gaussian sources and thus effects related to differences between the CLEAN and dirty beam are neglected. This effect is described at the beginning of this section, however, quantifying this effect is beyond the scope of this paper. 
In the last step, we first create a noise image by taking the RMS values from \cite{Smolcic2017} 
and generating normal distribution pixel values across the field. Then we insert galaxies as Gaussian shapes into the noise image.

\begin{figure}
  \centering
  \includegraphics[width = \linewidth]{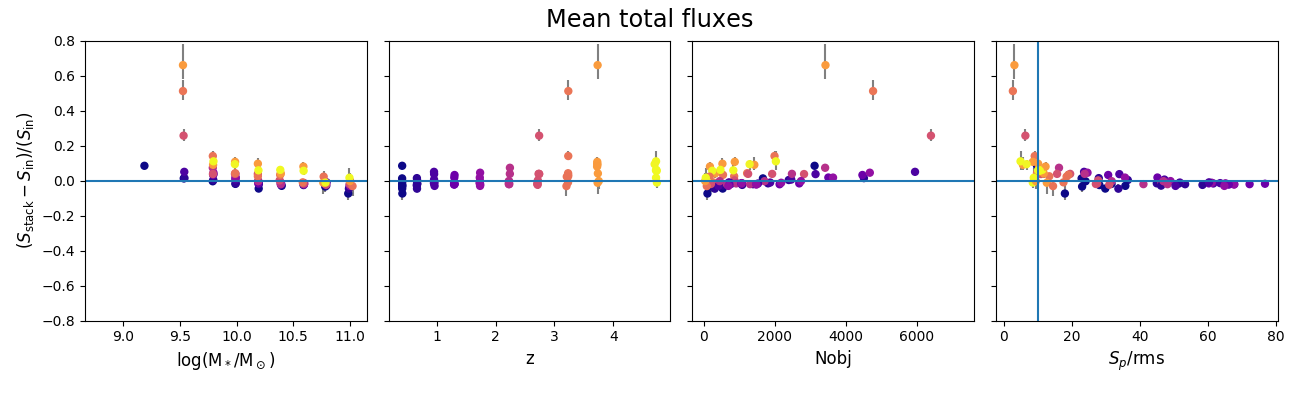}
  \includegraphics[width = \linewidth]{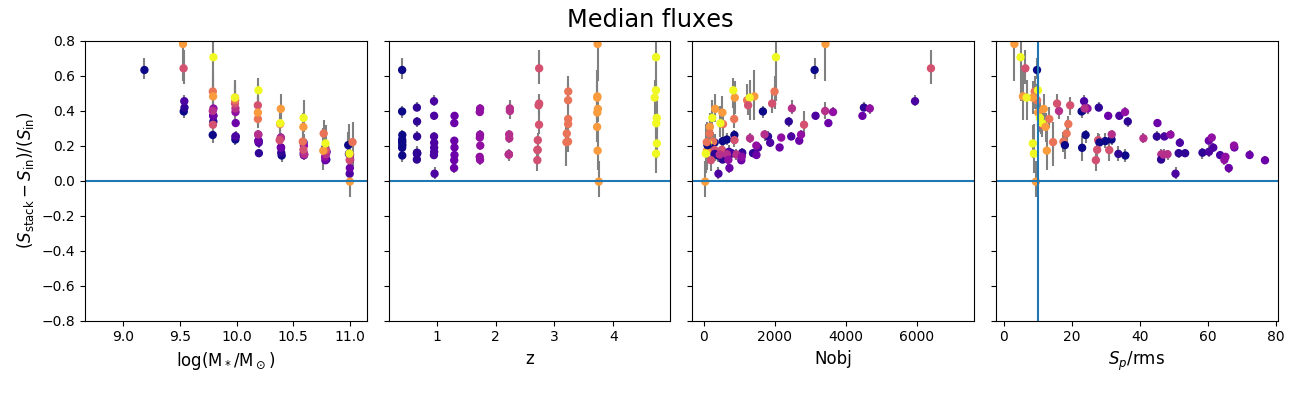}
  \caption{Results from stacking simulated galaxies. From left to right, fractional error on the total flux recovered from mean stacking as a function of stellar mass, redshift, number of sources, and SNR ($S_p$/rms). In bins of stellar mass and redshift, we compare in the top panels the mean of the input galaxy 3GHz total fluxes with the total flux measured from the stacked cutout of those sources. Bottom panels show the comparison of the median fluxes. The y-axis represents the fractional error on the fluxes. A blue vertical line in the last panel is drawn at $SNR=10$. Data are colored by redshift (see the second panel).}
  \label{fig:simulation}
\end{figure}

We tested our flux measurement method on individual galaxies and the stacks of simulated galaxies. According to our simulations, the fluxes for stacks with a low $S_p$/rms$>10$ (SNR) are over-estimated, as shown in Figure \ref{fig:simulation}. For our analysis, we use a cut of $S_p$/rms$>10$, where we can recover mean fluxes to within $\sim$10\%.
In the bottom panels of Figure \ref{fig:simulation}, we compare the median stacked fluxes with the median and mean fluxes from the input catalog. The resulting fractional errors are higher than for the mean fluxes, with the median flux from the stack being biased high. Because the bias in the median fluxes is a complicated function of both SNR and source size, there was no simple cut we could make to be confident in the fluxes from the median stacks (Figure \ref{fig:simulation}). Therefore, we have adopted the mean fluxes for our main analysis. We do, however, include MS fitting results using SFRs derived from median fluxes (which are systematically lower than our mean-SFRs) in Table \ref{tab:bestfits}.

\subsection{Comparison with Novak et al. 2017}
As a consistency check, we test whether our model of the MS given in Equation \ref{eq:myform}, with best-fit parameters for star-forming galaxies, can reproduce the luminosity functions of \cite{Novak2017}. 
This test involves generating a mock galaxy catalog using the stellar mass functions of SF galaxies from \cite{Peng2010} and \cite{Davidzon2017}. We assign SFRs following a log-normal distribution with the mean from our MS model and scatter $=0.29$\,dex (\citealt{Popesso2019}) to mock sources in bins of redshift and stellar mass bin as described in Section \ref{sec:appendixsim}. 

\begin{figure}
  \centering
  \includegraphics[width=\linewidth]{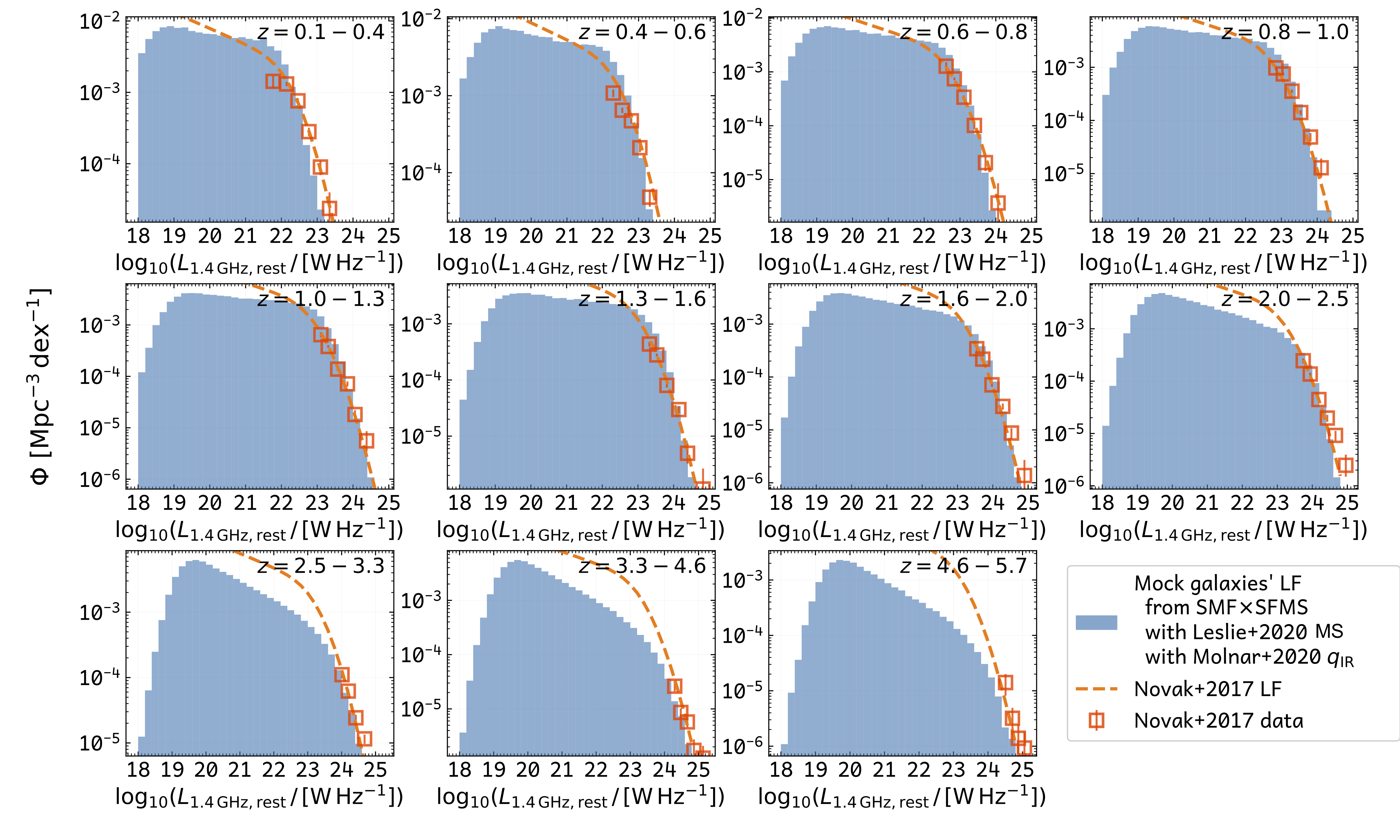}
  \caption{Luminosity functions generated from our MS model, compared with radio luminosity functions measured using the 3GHz LP detections by \cite{Novak2017}. Our mock sources, derived by combining the MS from Eq. \ref{eq:myform}, and the SMF in the COSMOS field, agree well with the luminosity functions from \cite{Novak2017}, where they are constrained by detected sources (orange squares).}
  \label{fig:LF}
\end{figure}

We calculate radio luminosity functions from our mock sources and, in Figure \ref{fig:LF}, compare them to the 1.4\,GHz luminosity functions of \cite{Novak2017} derived using the $\sim6,000$ radio sources without radio excess (SF galaxies) in the 3\,GHz LP. \cite{Novak2017} fixed the faint and bright end shape of the radio luminosity function to the local values. The redshift bins used by \cite{Novak2017} are displayed in Figure \ref{fig:LF}.

The luminosity functions generated by our mock sources agree well with the \cite{Novak2017} luminosity functions, where both our MS and the 
\cite{Novak2017} luminosity functions are constrained (i.e. the orange squares match the blue histograms at $0.3<z<4.7$).

\begin{figure}
  \centering
  \includegraphics[width = \linewidth]{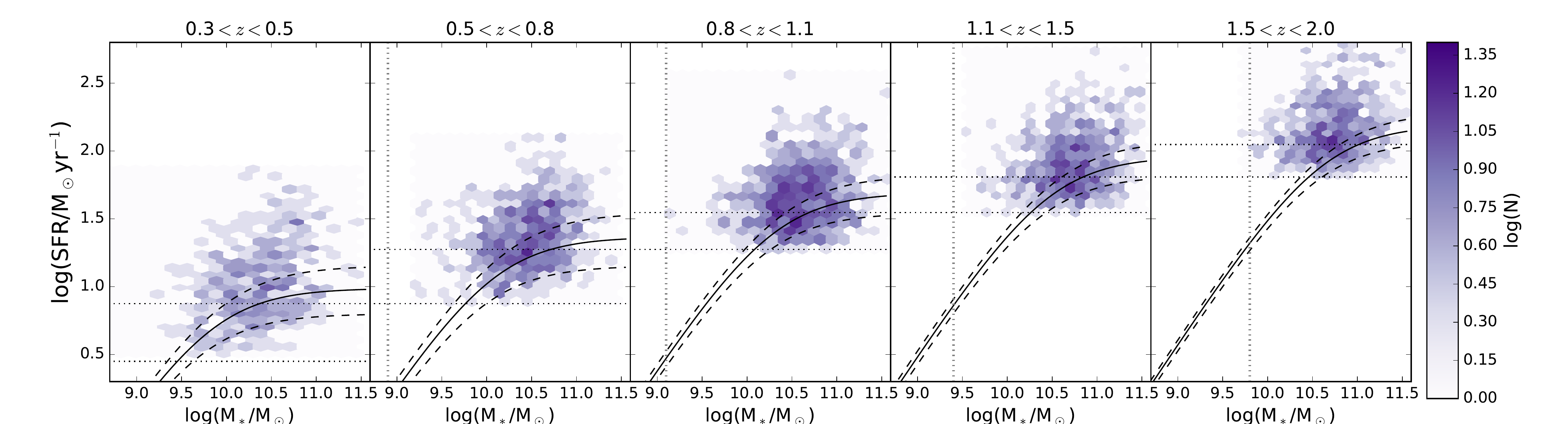}
  \caption{MS for COSMOS2015 galaxies detected in the 5$\sigma$ VLA 3GHz catalog and classified as pure SF galaxies. The solid line shows the best fit MS at mid-point of the redshift bin displayed, with the dashed lines showing the MS evolution between the bounds of the redshift bin. The color shows the number of galaxies in each hexbin. Vertical dotted lines show the mass completeness limit for this redshift. Horizontal dotted lines show the 5$\sigma$ 3GHz flux limit converted to a SFR at the lowest and highest redshift in the bin. This flux cut limits the direct detections to galaxies that lie above the MS.} 
  \label{fig:msdetect}
\end{figure}

In Figure \ref{fig:msdetect} we show the SFR--M$_*$ relation for SF galaxies listed in the \cite{Smolcic2017} catalog. The horizontal dashed lines show the 5$\sigma$ flux limit converted to a SFR at the limits of the redshift range shown. From this, it is clear that to constrain the MS from the 3\,GHz LP, we must include sources below the detection threshold, for example, via a stacking analysis.

\section{Selection effects}

\subsection{selecting star-forming galaxies}\label{sec:select}
\begin{figure}
\includegraphics[width = 0.49\linewidth]{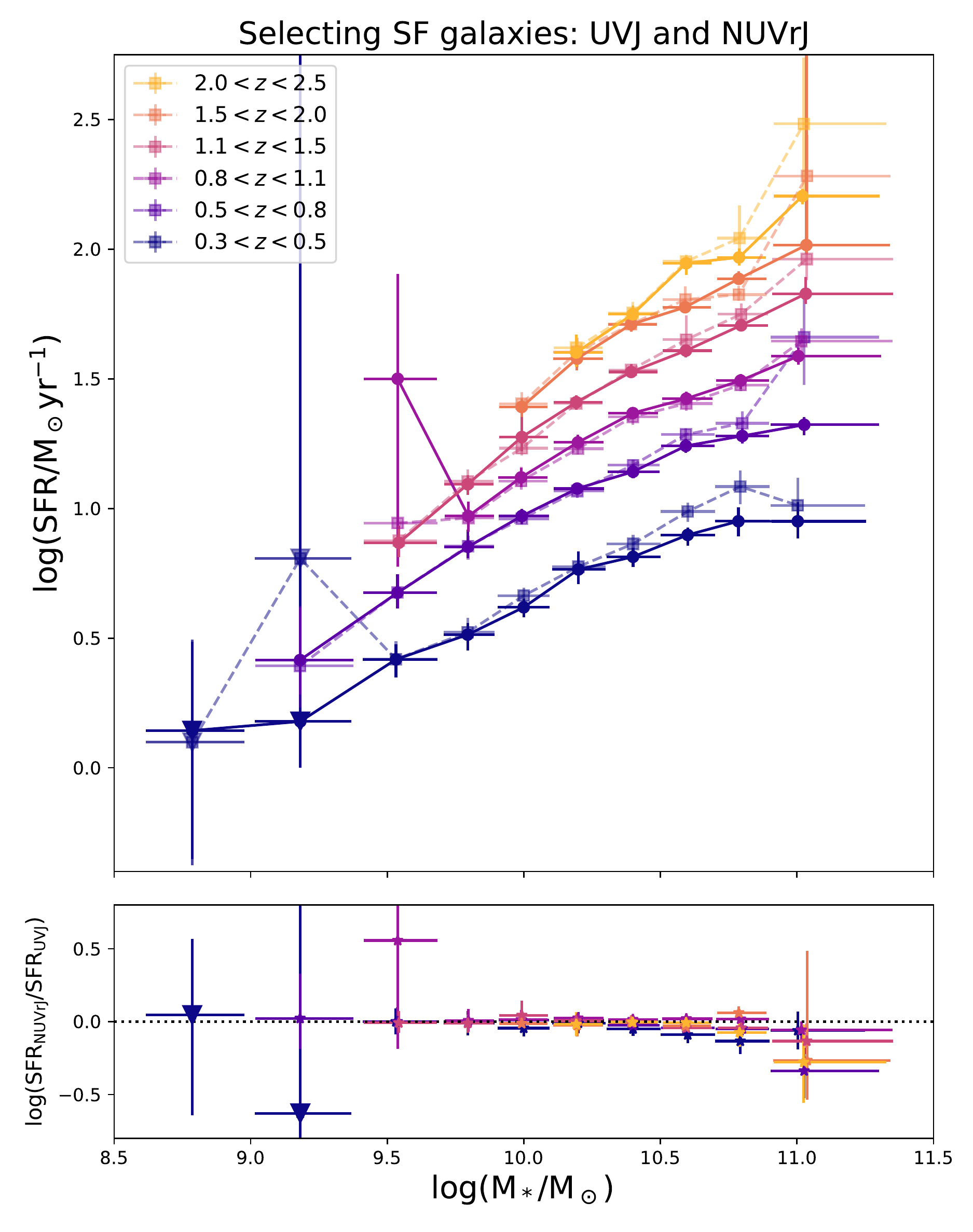}
\includegraphics[width = 0.49\linewidth]{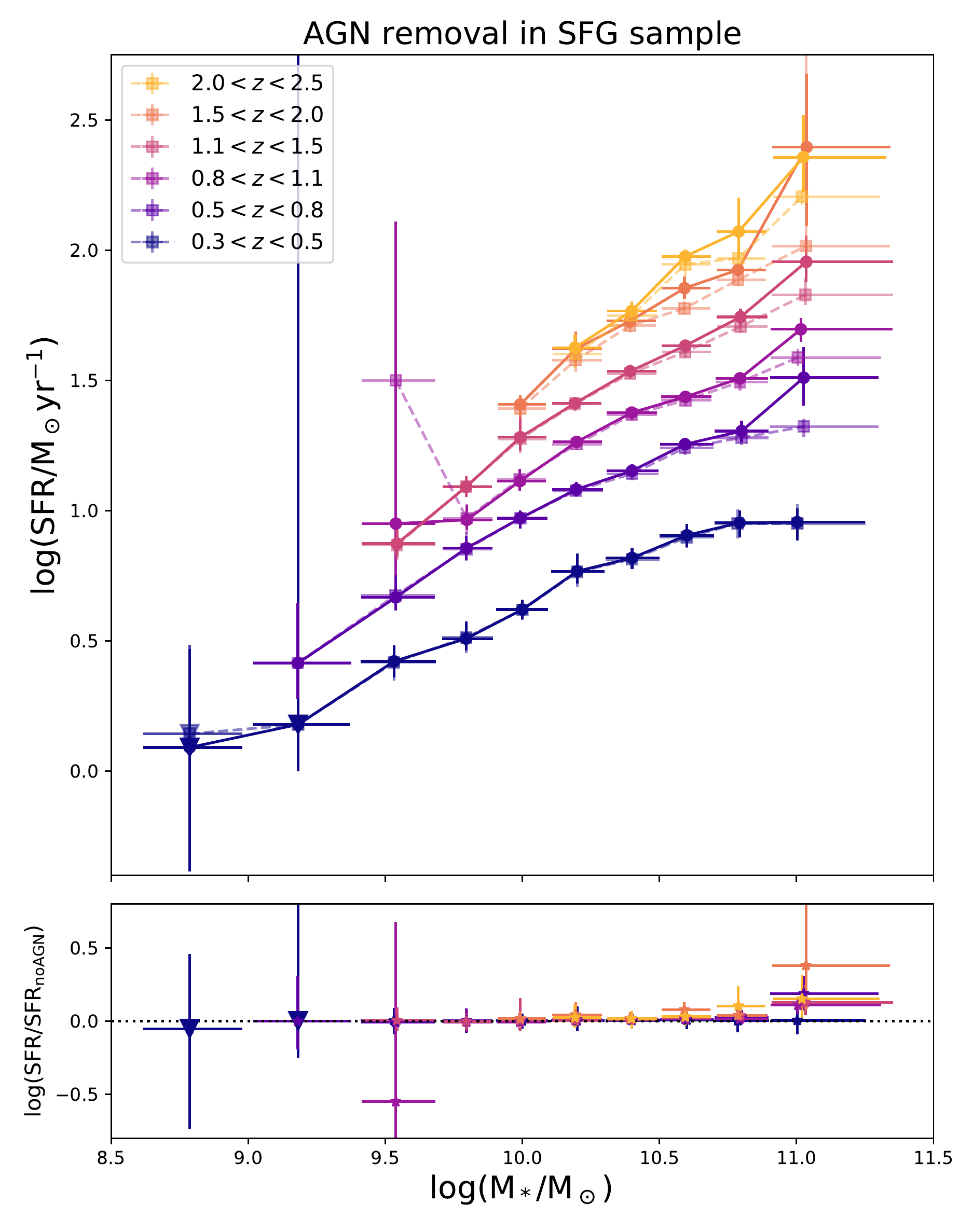}
\caption{Different methods for selecting star-forming galaxies result in different SFR-stellar mass relations. Left: We show results using the \citet{Laigle2016} catalog photo-$z$ and stellar mass determinations and from rest-frame $UVJ$ or $NUVrJ$ colors for selecting star-forming galaxies. The $NUVrJ$ selection used for this work is shown as solid circles with a solid line. We show $UVJ$ selected data as transparent squares with dashed lines between data points. The bottom panel shows the difference between log(SFR) of galaxies selected using the two criteria.
Right: AGN removal. Circles are no AGN (as used for this work) transparent squares are with AGN.}\label{fig:uvj}
\end{figure}

The selections used in constructing the parent sample affect the resulting MS. Studies without a cut to separate star-forming and quiescent galaxies (e.g. \citealt{Sobral2014}) show reduced MS slopes because the quiescent galaxies ``contaminate'' the highest mass bins across a wide range of redshift.
Using stricter SF selections results in a straighter or steeper MS (e.g. \citealt{Karim2011, Johnston2015}). 
In Figure \ref{fig:uvj}, we compare the difference between two commonly used color-color cuts from the literature, the $UVJ$, and $NUVrJ$ selections. 
 Color-color cuts are empirical criteria based on the number density of galaxies in these 2D planes. The $UVJ$ selection uses the redshift-dependent rest-frame U-V, V-J color-color cuts described in \citet{Whitaker2011}. The number density of galaxies in the COSMOS field on the $NUVrJ$ diagrams can be seen in \citet{Laigle2016}.
 
 The resulting SFR -- M$_*$ relations are consistent between the two selections within the uncertainties except for our lowest redshift bin ($0.3<z<0.5$), and at the highest masses, where the $UVJ$ color cut selects galaxies with a higher mean SFR as compared to the $NUVrJ$ color cut. 
We note that, for the COSMOS sample, at $z<0.3$, the results are even more discrepant because our NUV-r color requirement misses the most star-forming sources at $z<0.3$ that are included the U-V criteria because of the different sensitivities of the NUV and U band observations at this low redshift, and the small volume probed. Thus, we have limited our analysis to $z>0.3$.

\subsection{more on AGN removal}\label{sec:agn}
AGN activity and the SFR activity of their hosts are likely connected, possibly driven by a common fueling mechanism \citep{Vito2014}. The cosmic star formation history and black hole accretion history follow similar volume averaged evolution \citep{Madau2014}. 
Identifying AGN is a challenging task because the SED of an AGN and host galaxy can vary widely depending on the geometry and accretion properties in addition to the AGN energy output varying on timescales that are dependent on the wavelength (see e.g, \citealt{Elitzur2014, Schawinski2015, Buisson2017, Noda2018}).
Disentangling the radio emission from the AGN from that related to star-formation is beyond the scope of this paper. 
We combine multiwavelength AGN diagnostics to reach an inclusive AGN sample as described in Section \ref{sec:classify}.

\begin{figure}
\includegraphics[width = \linewidth]{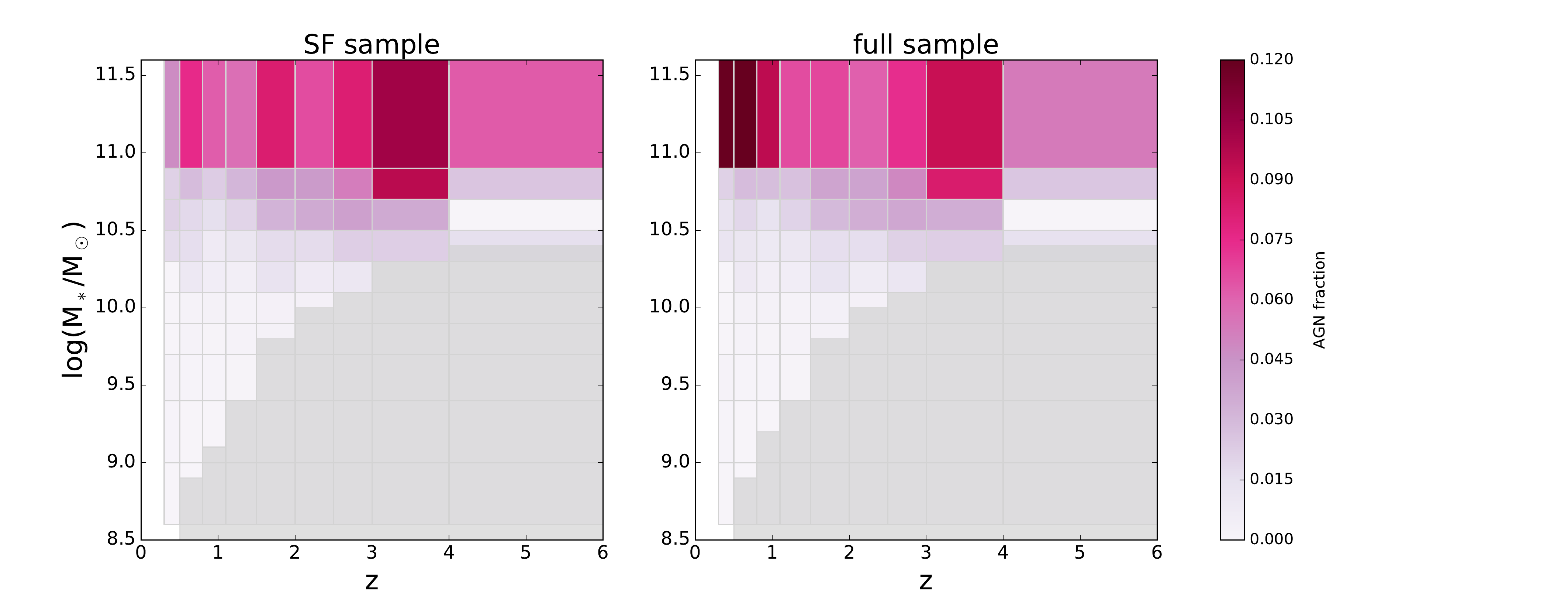}
\caption{Fraction of galaxies hosting an AGN selected by multiwavelength criteria (see Section \ref{sec:select}) in bins of stellar mass and redshift. Grey shading indicates where the COSMOS2015 catalog is not mass complete.}
\label{fig:agnfrac}
\end{figure}

Figure \ref{fig:agnfrac} shows the fraction of galaxies hosting an AGN identified by at least one of our criteria. 
At all redshifts, a higher fraction of massive galaxies host detectable AGN than for lower mass galaxies. 
At higher redshift, a larger fraction of massive galaxies with blue colors host AGN compared to at low redshift, with the fraction of massive ($\log(M_*/M_\odot)>10.7$) SF galaxies hosting an AGN peaking at 10\% at $z\sim3$. On the other hand, when all galaxies including passive galaxies are considered, the full sample shows a peak of AGN at low redshift (12\%). We note that AGN fractions are highly dependent on selection criteria, for example, the threshold fraction of the host galaxy luminosity coming from AGN emission required for a system to be classified as an AGN at a given wavelength. Our AGN fractions of $<12\%$ for massive galaxies are broadly consistent with the literature, for example, \citealt{Assef2013} (2-8\%\,;\,$z<3$) \citealt{Haggard2010} (0.15-4\%\,;\,$z<0.7$), \citealt{Mishra2020} (0.3 to 8\%\,;\,$z<0.7$). It is not surprising that our fractions are, in some cases, higher than these literature values, due to our inclusive multi-wavelength AGN criteria. 

Figure \ref{fig:uvj} shows the difference in average SFR inferred between no AGN-deselection and our strict AGN de-selection. The mean fluxes are more affected than the median fluxes by the presence of AGN because the mean is more sensitive to outliers (shown in Figure \ref{fig:meanvsmedian}). 
Radio-loud AGN are often found in host galaxies with red colors and so are excluded using our passive criteria (e.g. jet-mode AGN; \citealt{Heckman2014}). In radiative-mode AGN, (e.g. Seyfert galaxies), the radio emission comes from a combination of star formation and AGN activity (see, e.g. \citealt{Heckman2014} for a review). Mid-IR criteria are often able to select these, often quite powerful, AGN (e.g. \citealt{Jarrett2011, Mateos2012, Williams2018}).

\section{The functional form of the MS across redshift}\label{sec:msform}

In this section, we are concerned with how to best represent the SFR--$M_*$ relation given by our data. 
First, we fit the MS relation at each redshift bin comparing a power-law and broken power-law model, and then we discuss different functional forms that take both stellar mass and redshift as independent variables. 
For readability, we will omit units in the equations of this section, but all SFRs have units of $\text{M}_\odot\text{yr}^{-1}$, and stellar mass is denoted $M_*$ with units of M$_\odot$.

As described in Section \ref{sec:MS}, we fit models to our mass complete data using a simple least-squares minimization. To estimate the parameter uncertainties we resample the data 50,000 times (or 10,000 times in our first experiment where we fit a function at each redshift) varying the stellar mass, SFR and redshift (if included as an independent variable) of each stack by a random amount that follows a Gaussian distribution with $\sigma$ given by our upper and lower errors. We then report the median, \nth{16}, and \nth{84} percentile from all our runs for the best-fitted parameter distributions. 

For the best fits, we report reduced $\chi^2$, which is the weighted sum of the squared deviations in log(SFR) per degree of freedom (number of observations minus the number of fitted parameters). A $\chi_r^2>1$ indicates that the fit has not fully captured the data or that the error variance has been underestimated and a $\chi_r^2<1$ indicates the model is over-fitting the data or the error variance has been overestimated. 

To compare the relative quality of different models we have also computed the Akike Information Criterion (AIC):\begin{equation} 
\mathrm{ AIC} = 2k -2\ln\left(\sum_n\frac{(\log(\mathrm{SFR})-\widehat{\log(\mathrm{SFR})})^2}{n}\right) ,\label{eq:AIC}
\end{equation} 
where $\widehat{\log(SFR)}$ is the predicted value, $n$ is the number of observations, and $k$ is the number of parameters estimated by the model plus the constant variance of the errors. The preferred model is the one with the minimum AIC. 
We also report a second definition of the AIC, with a correction for small sample sizes that addresses the tendency for overfitting (selecting a model with too many parameters) that often occurs for small sample sizes (e.g. \citealt{Hurvich1989,Cavanaugh1997}), 
\begin{equation} 
\mathrm{ AICc} = 2k -2\ln\left(\sum_n\frac{(\log(\mathrm{SFR})-\widehat{\log(\mathrm{SFR})})^2}{n}\right) + \frac{2k^2 +2k}{n-k-1},\label{eq:AICc}
\end{equation} 
The AIC does not give information on the absolute quality of the model, but rather only the quality of a model relative to other models. Only differences in AIC are meaningful; we interpret $\Delta$AIC$>$2 to be evidence against the model with the higher AIC, $>6$ to be strong evidence and $>10$ to be very strong evidence against the model with the higher AIC (or strong evidence for preferring the model with the lower AIC).

Table \ref{tab:bestfitz} gives the results of fitting the log(SFR)--log($M_*$) relation for individual redshift slices (out to $z=2$, where we have enough data in each bin to constrain a 3 parameter model) using a linear form: 
\begin{equation}
\log(SFR) = a_1\log(M_*/10^9) + a_2
\label{eq:linform}
\end{equation}

and the \citet{Lee2015} form: 
\begin{equation}\label{eq:leeform}
\log(\mathrm{SFR}) = S_0 - \log\left(1+\left(\frac{10^M}{10^{M_0}}\right)^{-\gamma}\right),
\end{equation} 
where $M=\log(M_*/\mathrm{M}_\odot)$, $\gamma$ is the power-law slope at lower stellar masses, and $S_o$ is the maximum value of $\log$(SFR) approached at higher stellar masses. 

In general, the bent MS produces more favorable reduced $\chi^2$ (closest to 1), and there is very strong evidence that a bent MS relation is preferred to a linear relation at $z<1.1$, and strong evidence at $1.1<z<1.5$ as well. However, using the AIC criteria based on Eq. \ref{eq:AIC} (rather than Eq. \ref{eq:AICc}) implies that the linear form is preferred at $z>1.1$. This is likely due to the smaller number of mass-complete measurements available for low-mass constraints at higher redshift. These results are consistent with 
\citet{Lee2018}, who found that the MS can be described by a simple linear relation in log($M_*$) and log(SFR) at $z>2$, whereas below $z\sim1.5$, there is a statistically significant flattening at high stellar masses.

Looking at the fitted parameters of the \citet{Lee2015} form in Table \ref{tab:bestfitz}, it is clear that $S_0$ and $M_0$ increase with redshift ($M_0$ evolution is illustrated in Figure \ref{fig:charmass}). 
We considered this when developing the functional form of Eq. \ref{eq:myform}.

It is most convenient to provide a functional form that takes stellar mass and redshift as independent parameters and returns a SFR. Next, we will fit our data with functional forms adopted in the literature that allow this, including the new parameterization proposed in Eq. \ref{eq:myform}. 

A simple functional form adopted in the literature is that of a linear relationship between log(SFR) and log($M_*$). \citet{Speagle2014} compiled 25 studies from the literature and fitted the following linear form to the MS that allows for evolution of both the slope and normalisation\footnote{Although most studies agree that the slope remains relatively constant over time, e.g., \citet{Wuyts2011}.}:
\begin{equation}\label{eq:speagle}
\log(\mathrm{SFR}) = ( a_1 - a_2 t)\log(M_*) - (a_3 - a_4 t),
\end{equation}
where $t$ is the age of the Universe in Gyr. We have fit this functional form to our data and report the best fitting parameters $\Theta=(a_1,a_2,a_3,a_4)$ in Table \ref{tab:bestfits}. This form performs poorly for our low redshift ($z<1$) data, but matches our high-z data well.

The following functional forms allow for a flattening at high-masses in the MS.
We have investigated fitting the functional form used by \citet{Schreiber2015}, namely: 
\begin{equation}
\begin{split}
\log(\mathrm{SFR})= m -m_0 +a_0 r - a_1[\max(0,m-m_1 -a_2 r)]^2
\end{split}\label{eq:schreiber}
\end{equation}
where $r = \log(1+z)$ and $m=\log(M_*/10^9)$. This relation constrains the low-mass slope by assuming a constant sSFR (i.e. a low-mass slope of 1). The posterior distribution of these parameters is not smoothly distributed. To produce reasonable constraints on our parameters, we added the following bounds: $a_0>0$, $a_1>0$, and $a_2>0$ and $-3<m_1<12$. For this case, the
``Trust Region Reflective'' algorithm, was used for optimization in \textsc{scipy.optimize}.

\citet{Tomczak2016} fit the \citet{Lee2015} function with a quadratic evolution of parameters with redshift: 
\begin{equation}\begin{aligned}
\log(\mathrm{SFR}) &= S_0' - \log\left(1+\left(\frac{M}{10^{M_0'}}\right)^{-\gamma}\right), \\
S_0' &=s_0 + a_1 z - a_2 z^2\\
M_0' &= m_0 + a_3 z - a_4 z^2\\
\gamma &= 1.091.
\end{aligned}\label{eq:tomczak}
\end{equation}
Again, all best fits are reported in Table \ref{tab:bestfits}.

\section{Comparison with Literature}\label{sec:comp}
\begin{figure*}
\includegraphics[height =5.5cm]{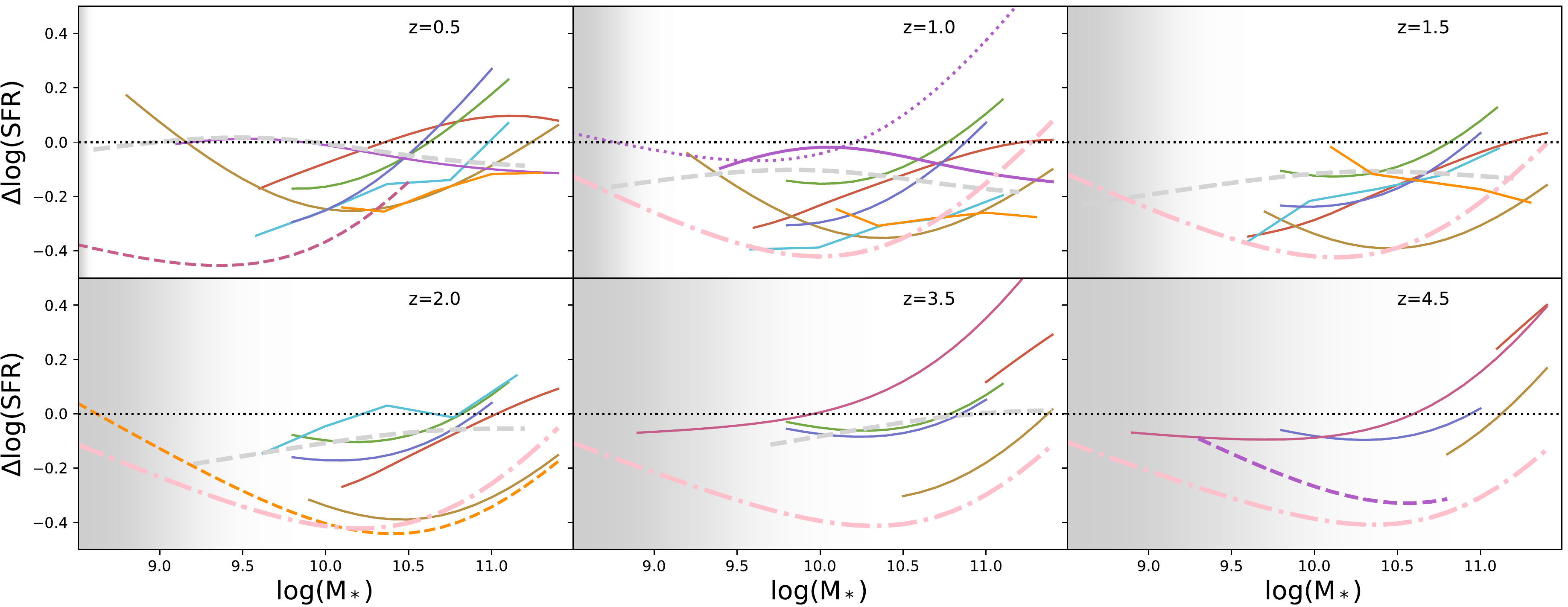}
\includegraphics[trim = {0cm 0 1cm 0},height=5.5cm]{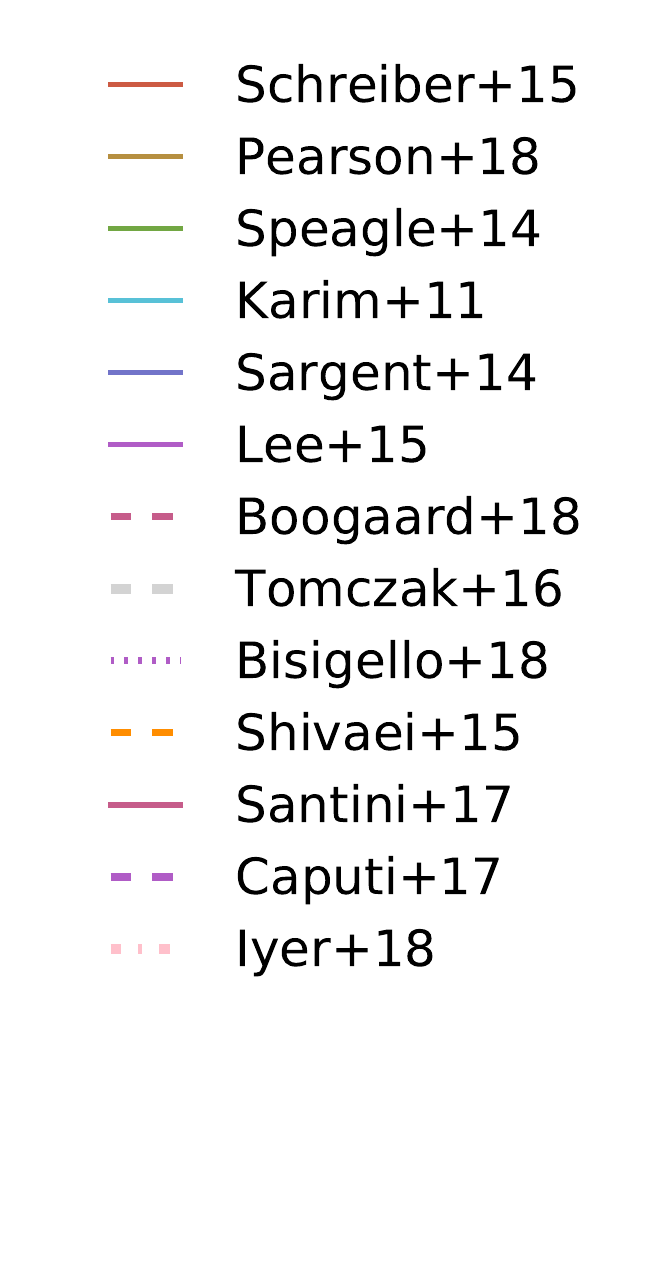}
\caption{The difference between SFRs resulting from the literature MS relations and the MS relation in this work (Equation \ref{eq:myform}). Grey shaded area illustrates where this work is incomplete in stellar mass. We note that our SFRs lie above the literature at $M_*=10^{10}$\,M$_\odot$, which is the mass that contributes the most to the cosmic SFRD. This offset might lead to the discrepancies seen in Figure \ref{fig:csfh} between our result and the \cite{Madau2014} compilation curve.}\label{fig:comp}
\end{figure*}

The diversity of sSFRs from different studies as a function of stellar mass and redshift, normalized by the relation given in Equation \ref{eq:myform} is shown in Figure \ref{fig:comp}. All literature relations were converted to the same Chabrier IMF used in this study when necessary (as mentioned below). We do not correct for the small effect that different cosmologies introduce into the relations. The mass range shown in Figure \ref{fig:comp} illustrates where the literature studies are complete. 
There is a spread of $>0.2$\,dex between different MS relations. 
Below we discuss the pertinent details on how MS relations shown in Figure \ref{fig:comp} were derived.

\begin{itemize}
\item \citet{Bisigello2018} fitted linear SFR -- $M_*$ relations for three populations of galaxies in the CANDELS/GOODS-S field; a starburst, main-sequence, and quiescent population, assuming each population follows a log-normal sSFR distribution. They fit galaxies in three bins between $0.5<z<3$, and stellar masses down to log($M_*$/M$_\odot)>7.5$. \citet{Bisigello2018} adopted a ($\Omega_M, \Omega_\Lambda, H_0$) = (0.27, 0.73, 70) cosmology and Salpeter IMF. In Figure \ref{fig:comp} we include the $1<z<2$ relation whose SFRs were calculated from UV plus IR luminosities (where IR luminosity is from \citet{Chary2001} template fitting).

\item \citet{Boogaard2018} selected star-forming galaxies at $0.11<z<0.91$, with two Hydrogen Balmer emission lines detected with SNR $>$ 3 from which they calculated SFR. Star-forming galaxies were further required to have $D_n(4000)<1.5$, $EW_{H\alpha} >2$\,\AA~, $EW_{H\beta}>2$\,\AA. \citet{Boogaard2018} found a linear relation between log(SFR) and $\log(M_*/\mathrm{M}_\odot)$ over the mass range probed ($7<\log(M_*/\mathrm{M}_\odot)<10.5$). 

\item \citet{Caputi2017} obtained H$\alpha$-derived SFRs from H$\alpha$+[NII]+[SII] equivalent width measurements of galaxies in the \textit{Spitzer} Matching Survey of the UltraVISTA Ultra-Deep Stripes over the range $3.9<z<4.9$. 
They report a bi-modal distribution in sSFR interpreted as a main-sequence and starburst population. For the main-sequence population, they report a linear relation, constrained across a mass range $9.2<\log(M_*/\mathrm{M}_\odot)<10.8$. Above this stellar mass, they postulate AGN contamination influences the SFR measurements. 

\item \citet{Iyer2018} used the $UVJ$ colors to select $\sim18,000$ galaxies in the CANDELS GOODS-S field at $0.5<z<6$. The galaxy SED-fits provide stellar mass and SFR measurements. In addition, the star formation histories were reconstructed, allowing for more data-points on the MS relation at high redshift down to a representative mass of $\log(M_*/\mathrm{M}_\odot)\sim 7$ (90\% representative at $z\sim7.5$). 

\item \citet{Karim2011} used 1.4\,GHz data and the \citet{Bell2003} SFR calibration (assuming $\alpha=-0.8$) to calculate SFRs for COSMOS galaxies across $0.2<z<3$. SF galaxies were selected using NUV-r rest-frame color cuts. We show individual stacked mass complete data points (rather than the fit-function) in Figure \ref{fig:comp}.

\item \citet{Lee2015} studied star-forming galaxies selected by a $NUVrJ$ color-color cut \citep{Ilbert2013}. The SFRs were calculated from IR and UV data assuming \citet{Hinshaw2013} cosmology ($\Omega_M, \Omega_\Lambda, H_0$) = (0.28, 0.72,70.4 km s$^{-1}$ Mpc$^{-2}$). The mass limit used for fitting was given in their Table 1. \citet{Lee2015} fit the MS for galaxies at $0.25<z<1.3$ with the functional form \ref{eq:leeform}.

\item \citet{Pearson2018}, fit a linear MS relation over a redshift range $0.2<z<6$ in the COSMOS field, calculating SFRs and stellar masses via SED fitting. SF galaxies were selected using a redshift-dependent $UVJ$ color-color cut and a slightly different cosmology \citep{Larson2011} was assumed ($\Omega_M, \Omega_\Lambda, H_0) = (0.273,0.727,70.4$\,km\,s$^{-1}$\,Mpc$^{-2})$. 
Figure \ref{fig:comp} compares our result to the \citet{Pearson2018} MS showing the 90\% mass completeness limits calculated for the COSMOS UDeep field $K_s$ band magnitude limit. 

\item \citet{Popesso2015b} constrained the shape of the MS to a local value (MS$_0$; \citealt{Popesso2019}), and fit an evolution of the overall normalization out to $z<2.5$; $\mathrm{SFR}(z,M_*) = \mathrm{MS}_0\times 1.01(1+z)^{3.21}$, for galaxies more massive than $\log(M_*/\mathrm{M}_\odot)>10$.
Data in the GOODS+CANDELS and COSMOS-PEP (with the latter being limited to $z<0.7$ or $\log(M_*/\mathrm{M}_\odot)>10.8$) fields containing \textit{Herschel} and \textit{Spitzer} coverage were used to calculate SFRs by a combining FIR and UV luminosities. The location of the MS in their work is defined as the location of the peak of the log-normal SFR distribution, as in \cite{Renzini2015}, which is the median SFR of the distribution. 

\item \citet{Santini2017} studied the MS in the HST frontier fields at $1.3<z<6$. They were able to probe $\log(M_*/\mathrm{M}_\odot)>7.5$ at $z<4$ and $\log(M_*/\mathrm{M}_\odot)>8$ at $z>4$. They fit the MS separately for different redshift bins with a linear relation, selecting main-sequence galaxies by a sigma clipping. Figure \ref{fig:comp} only shows their fits for $3<z<4$ and $4<z<5$ because these match the z slices at which we are demonstrating. \citet{Santini2017} adopted a Salpeter IMF and calculate SFRs using rest-frame UV data \citep{Castellano2012}. 

\item \citet{Sargent2014} compiled data from 16 different studies spanning from $0<z<7$ and fit a power-law fit to the sSFR-z relation. Two mass scales were considered, $M_*\sim5\times10^9$\,M$_\odot$ and $ M_*\sim5\times10^{10}$\,M$_\odot$. The samples used by \citet{Sargent2014} include a mix of selection methods and SFR tracers, but all were converted to WMAP-7 cosmology \citep{Larson2011}.

\item \citet{Schreiber2015} stacked \textit{Herschel} data in the extragalactic GOODS-N, GOODS-S, UDS, and COSMOS fields to derive SFRs (they also added unobscured UV SFRs). A Salpeter IMF was used. In Figure \ref{fig:comp} we show their function (of form \ref{eq:schreiber}) only over a mass range where a stacked signal was detected. 

\item \citet{Shivaei2015} used H$\alpha$ and H$\beta$ spectroscopy to derive dust-corrected instantaneous SFRs for 261 galaxies spanning $9.5<\log(M_*/\mathrm{M}_\odot)<11.5$, $1.37 < z <2.61$ in the MOSFIRE Deep Evolution Field (MOSDEF) survey. 

\item \citet{Speagle2014} compiled literature data from 25 studies and we show their best fit for studies that have a``mixed'' SF selection. \citet{Speagle2014} report a $1\sigma$ inter-sample scatter of $<0.1$~dex. They fit a linear form (of form Eq. \ref{eq:speagle}) over a stellar mass range $9.7<\log(M_*/\mathrm{M}_\odot)<11.1$, and a redshift range of $0.16< z< 3.15$ (data from the first and last 2~Gyr were removed). They adopted a \citet{Kroupa2001} IMF. 

\item \citet{Tomczak2016} report a MS with a turn-over (of the form \ref{eq:tomczak}) over a redshift range $0.5 < z < 4$ and mass range 8.5$<\log(M)<11.0$. SF galaxies were selected using a $UVJ$ selection. Their SFRs come from UV+IR data from the FourStar Galaxy Evolution Survey (ZFOURGE) in combination with far-IR imaging from the \textit{Spitzer} and \textit{Herschel} observatories. 

\end{itemize}

\section{Extra Figures} 
\begin{figure}
\includegraphics[width = \linewidth]{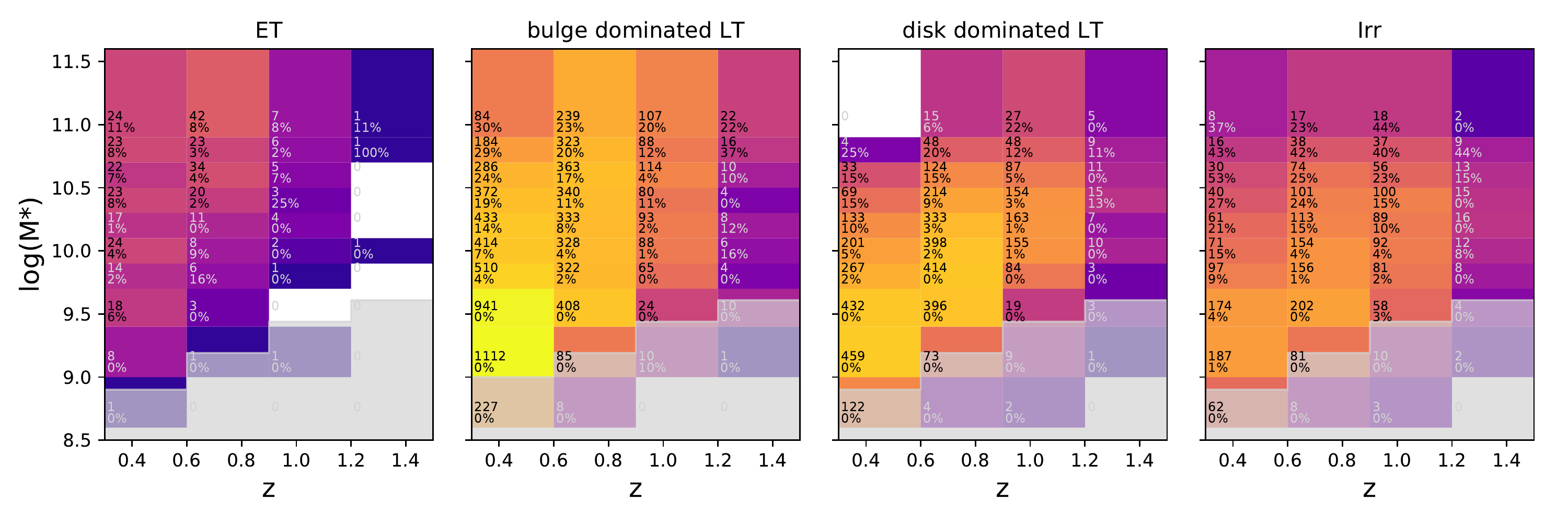}
\caption{Binning scheme for our morphology analysis using morphological classes based on the Zurich Structure and Morphology Catalog \citep{Scarlata2007}. Star-forming galaxies are selected using $NUVrJ$ color-color cut. Numbers for each bin are provided analogous to Fig. \ref{fig:binning}}
\label{fig:binningmorph}
\end{figure}

\begin{figure}
\includegraphics[width = \linewidth]{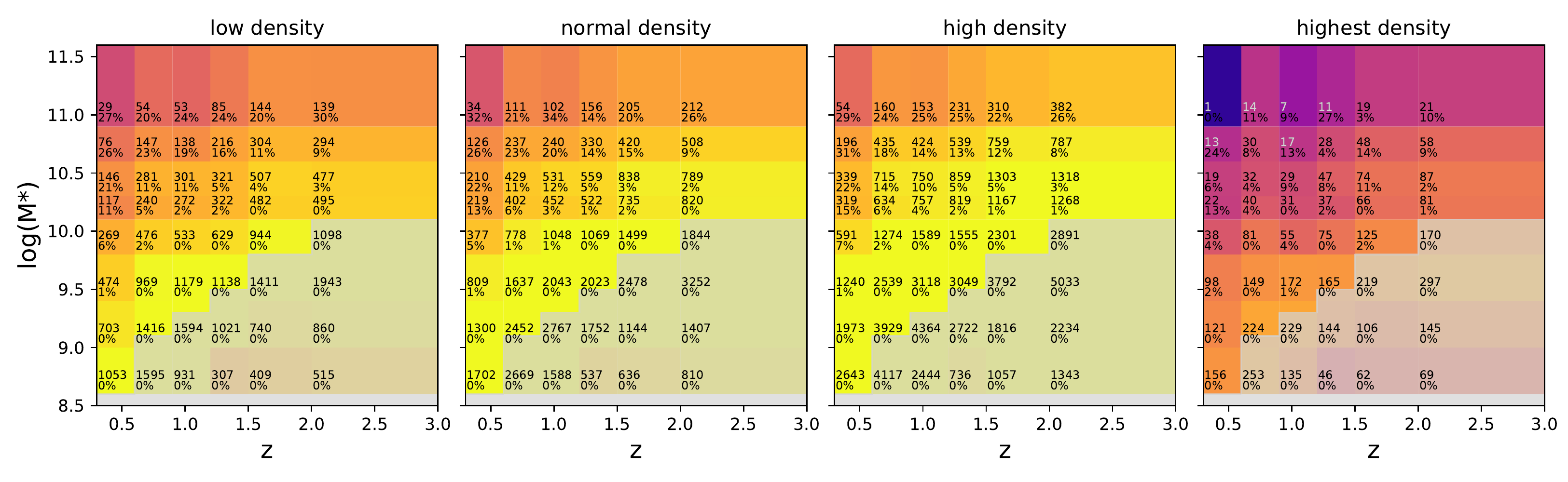}
\includegraphics[width = \linewidth]{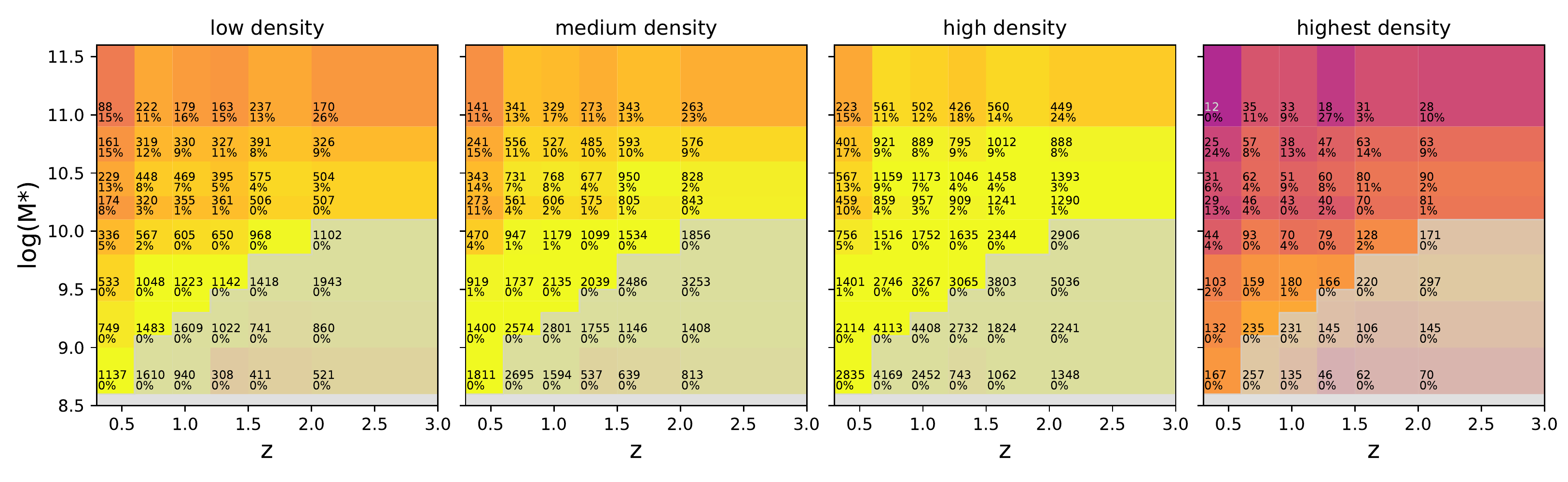}
\caption{Binning scheme for our environment analysis using local density estimates from \citet{Scoville2013}. Star-forming galaxies (top panels) are selected using $NUVrJ$ color-color cut. Numbers for each bin are provided analogous to Fig. \ref{fig:binning}}.
\label{fig:binningenv}
\end{figure}

\begin{figure}[h!]
\centering
\includegraphics[width=\linewidth]{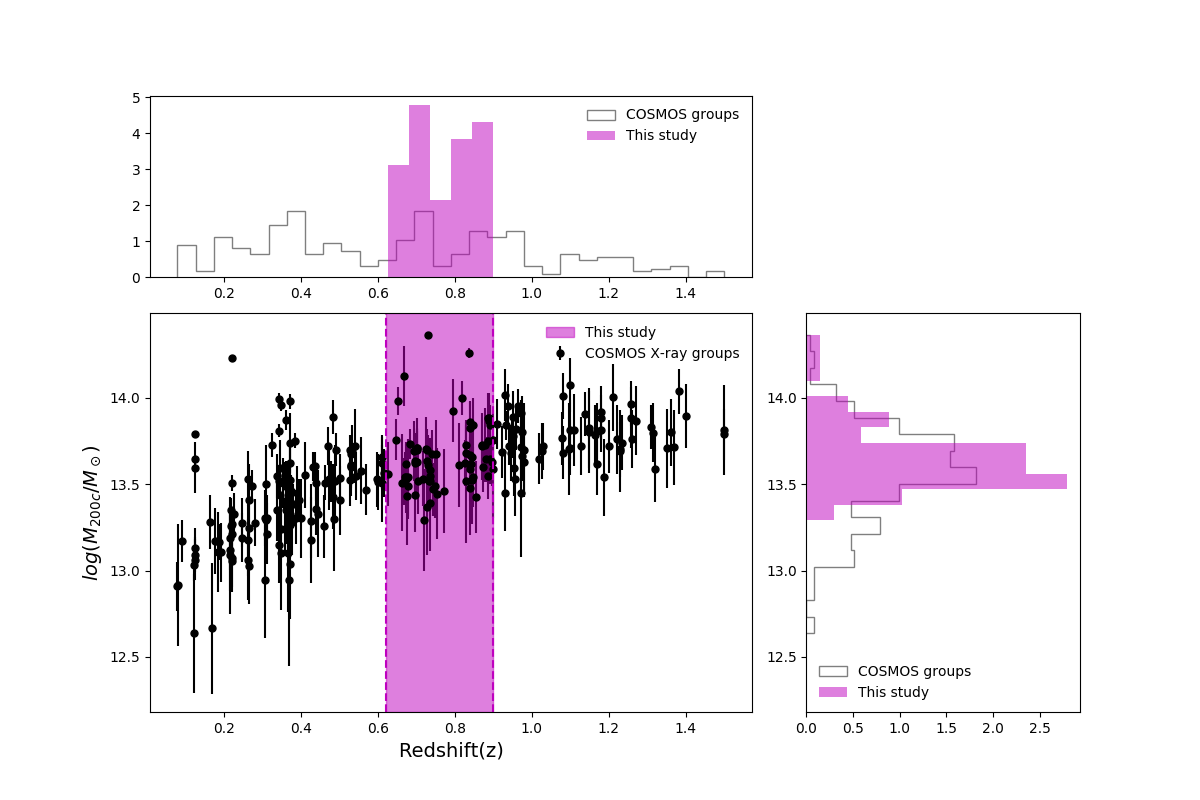}
\caption{The halo mass of X-ray selected groups ($M_{200c}$) in the COSMOS field as a function of their redshift (filled black circles; \citealt{Gozaliasl2019}). The highlighted magenta region represents 73 groups within a redshift range of $0.62<z<0.88$ used in this study. The redshift and halo mass distributions of all X-ray groups and those used in this study are presented in the upper left and lower right panel. }
\label{fig:m200-z}
\end{figure}

Binning schemes for our galaxy morphology analysis (Section \ref{sec:morph}), and local environment analysis (Section \ref{sec:env}) are shown in Figures \ref{fig:binningmorph} and \ref{fig:binningenv}, respectively. Figure \ref{fig:m200-z} shows halo mass--redshift distribution of X-ray groups in the COSMOS field.

\begin{figure}
  \centering
  \includegraphics[width=0.32\linewidth]{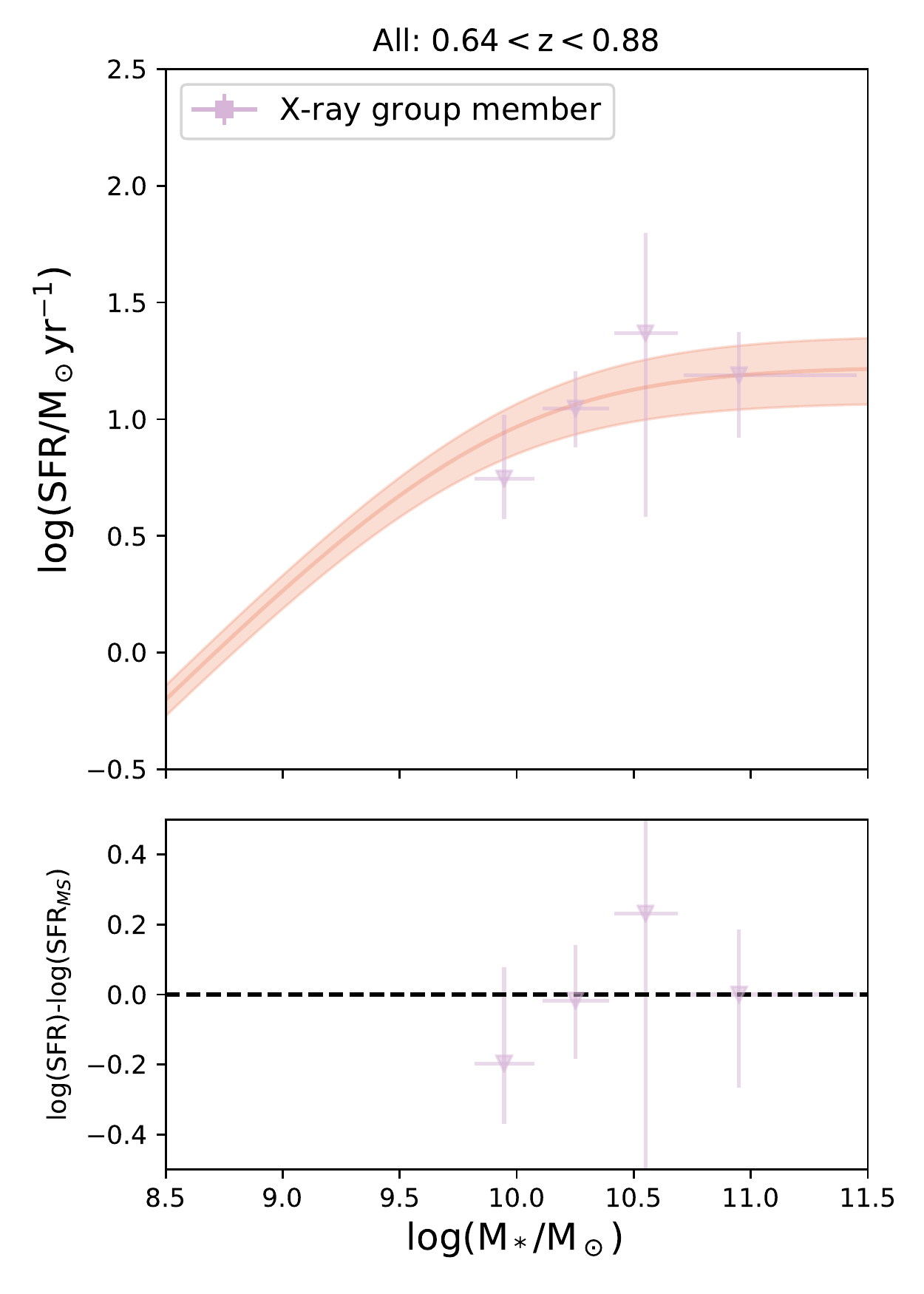}
  \includegraphics[width=0.67\linewidth]{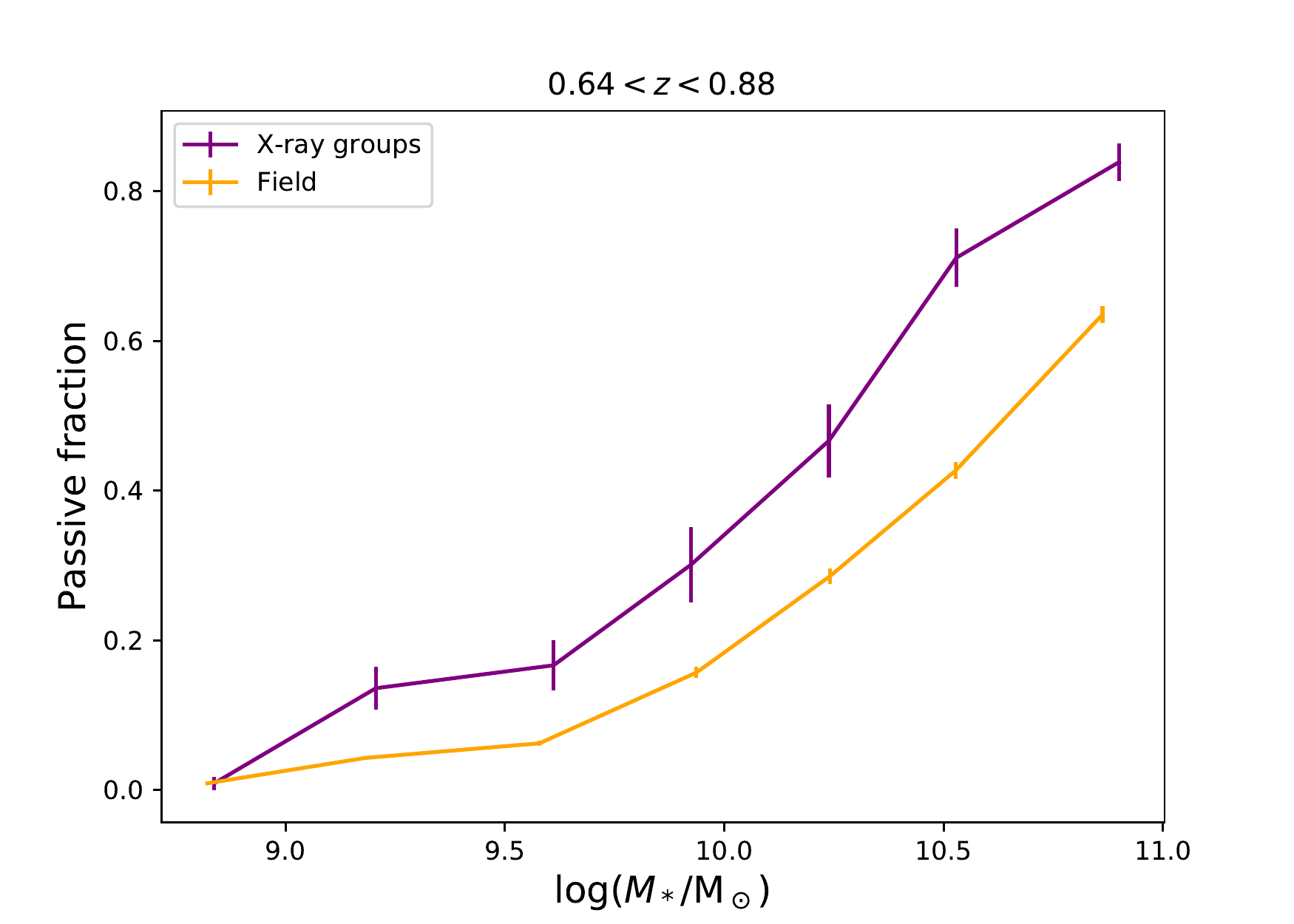}
  \includegraphics[width=\linewidth]{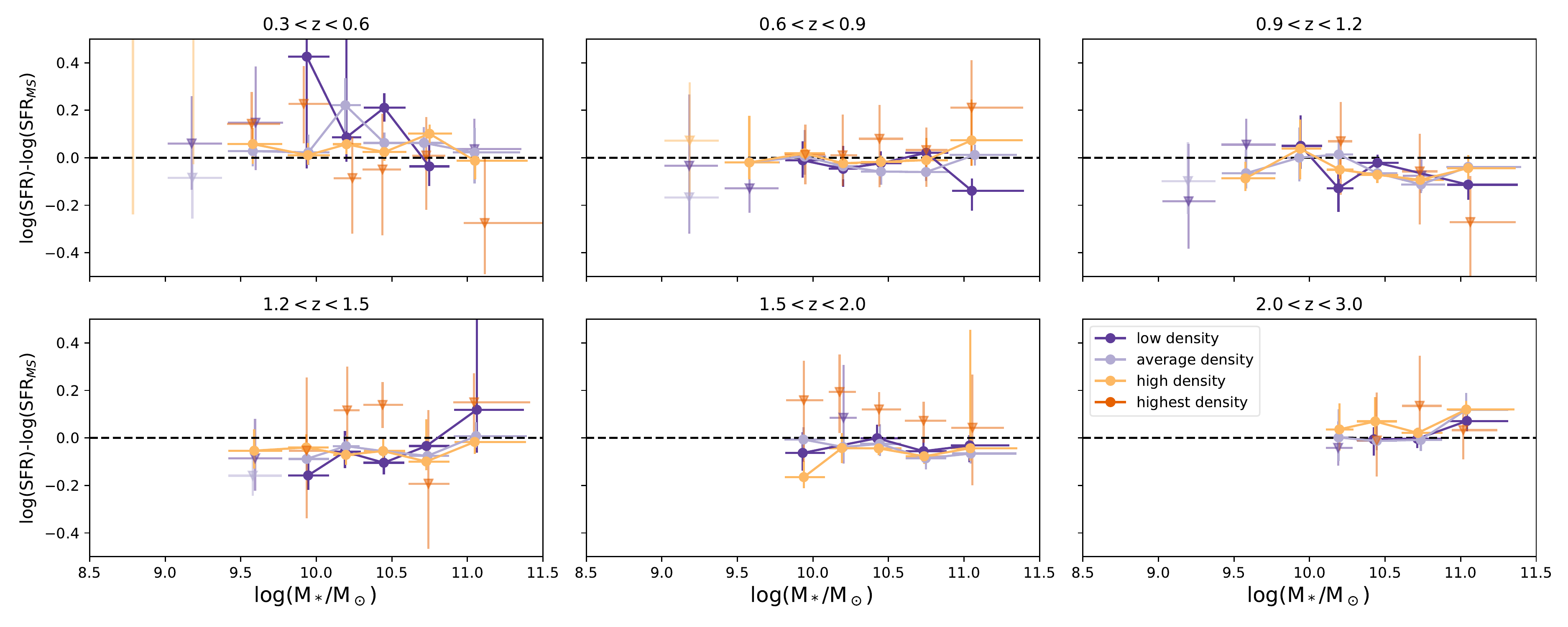}
  \caption{Environmental trends for all galaxies (including passive). Top left: shows the MS relation of galaxies in X-ray groups compared to all galaxies regardless of environment. Top right: fraction of $NUVrJ$-selected passive galaxies as a function of stellar mass for X-ray groups, compared to galaxies regardless of environment. Bottom panels show the difference between the SFR from the MS (calculated for all galaxies) and the SFR of galaxies in different groups. In all cases there is no clear difference between the different environments, as traced by the density fields of \cite{Scoville2013}.}
  \label{fig:envall}
\end{figure}

We show in Figure \ref{fig:envall}, that the MS derived relation for all galaxies shows no statistically significant difference between low and high-density local environments, nor in X-ray groups at $z\sim0.76$.

\newpage

\begin{deluxetable*}{c|cc|cc|cc}
\tablewidth{0pt}
\tablecaption{Linear (Eq. \ref{eq:linform}) and non-linear (Eq. \ref{eq:leeform};\citealt{Lee2015}) MS model fits to mass-complete data at $z<2$ with SNR$>10$. $\Delta \text{AIC}$($c$) is AIC$_\text{linear}$-AIC$_\text{Lee}$ calculated using Equation \ref{eq:AIC} (Eq. \ref{eq:AICc}). The more positive $\Delta \text{AIC(c)}$, the more the Lee et al. form is preferred. The more negative $\Delta$AIC, the more the linear form is preferred. Note the large uncertainties for the $0.8<z<1.1$ results. \label{tab:bestfitz}.}
\tablehead{\colhead{Redshift} & \colhead{Linear form $(a_1,a_2)$} & \colhead{
$\chi^2_r$} & \colhead{Lee form ($S_0, M_0, \gamma$)} & \colhead{$\chi^2_r$} & \colhead{$\Delta \mathrm{AIC}$}&\colhead{$\Delta$AICc}} 
\startdata
$0.2<z<0.5$ & ($0.40^{+0.05}_{-0.05}, 0.22^{+ 0.06}_{-0.06}$)& 1.5 &($1.09^{+0.28}_{-0.10},10.33^{+0.86}_{-0.27}, -0.77^{+0.22}_{-0.31}$)& 0.7 & 7.2 & 0.2\\
$0.5<z<0.8$ & ($0.46^{+0.07}_{-0.06},0.45^{+0.09}_{-0.08}$)& 8.0& ($1.42^{+0.13}_{-0.06},10.3^{+0.39}_{-0.17},-0.83^{+0.18}_{-0.17}$)& 0.57 & 22.8 & 16.9\\
$0.8<z<1.1$ & ($0.46^{+0.08}_{-0.3}, 0.68^{+0.45}_{-0.13}$)& 4.1 & ($1.83^{+7.1}_{-0.20}, 10.7^{+33}_{-0.44},-0.69^{+0.54}_{-0.35}$)& 25.0 & 14.9 & 21.9 \\
$1.1<z<1.5$ & ($0.63^{+0.05}_{-0.06}, 0.59^{+0.09}_{-0.06}$)& 5.8& ($1.93^{+0.18}_{-0.19},10.62^{+0.56}_{-0.3},-0.93^{+0.23}_{-0.16}$)& 1.3 & 9.1 & 2.1\\
$1.5<z<2.0$ & ($0.58^{+0.05}_{-0.06},0.86^{+0.11}_{-0.07}$)& 2.8&($2.22^{+6.3}_{-0.27}, 10.95^{+13.7}_{-0.5}, -0.82^{+0.30}_{-0.35}$)& 10.77& -8.4 & -36.3\\
\hline
\enddata
\end{deluxetable*}

\begin{deluxetable*}{cccc}
\tablewidth{0pt}
\tablecaption{MS model fits to all mass-complete data with SNR$>10$, adopting the \cite{Molnar2020} SFR calibration.\label{tab:bestfits}}
\tablehead{\colhead{MS form ($\mathscr{M}$)} & \colhead{parameters ($\Theta$)} &\colhead{best fit values} &\colhead{$\chi^2_r$}}
\startdata
SF, \citet{Speagle2014}, Eq. \ref{eq:speagle} & ($a_1,a_2,a_3,a_4$) & ($0.76^{+0.08}_{-0.10},0.041^{+0.012}_{-0.013},5.6 ^{+0.9}_{-1.0}, 0.26^{+0.13}_{-0.13}$) & 4.7 \\
SF, \citet{Schreiber2015}, Eq. \ref{eq:schreiber} & ($m_0, a_0, a_1, m_1, a_2$) & ($-0.60^{+0.49}_{-0.15}, 0.11^{+0.81}_{-0.11}, 0.10 ^{+0.06}_{-0.01}, -2.8^{+1.5}_{-1.5}, 4.7^{+0.6}_{-1.5}$)&30\\
SF, \citet{Tomczak2016}, Eq. \ref{eq:tomczak} & ($s_0, m_0, a_1, a_2, a_3, a_4)$ & (0.52$^{+0.12}_{-0.14}$, 9.39$^{+0.30}_{-0.44}$, 1.22$^{+0.22}_{-0.18}$, 0.16$^{+0.07}_{-0.04}$, 0.96$^{+0.48}_{-0.36}$, 0.15$^{+0.13}_{-0.08}$) & 7.3\\
SF, This work, Eq. \ref{eq:myform} & ($s_0, m_0, a_1, a_2$) & ($2.965^{+0.081}_{-0.088}, 11.07^{+0.15}_{-0.16}, 0.215^{+0.012}_{-0.012}, 0.120^{+0.027}_{-0.024}$) & 6.5\\
\hline
All, \citet{Speagle2014} Eq. \ref{eq:speagle} & ($a_1,a_2,a_3,a_4$) & ($0.75^{+0.11}_{-0.13},0.063^{+0.019}_{-0.018}, 5.5^{+1.2}_{-1.4}, 0.48^{+0.20}_{-0.18}$)& 6.9\\
All, \citet{Schreiber2015}, Eq. \ref{eq:schreiber} & ($m_0, a_0, a_1, m_1, a_2$)& ($-0.233^{+0.32}_{-0.60}, 0.76^{+0.50}_{-0.76}, 0.19^{+0.09}_{-0.07}, -1.40^{+0.78}_{-1.41}, 3.43^{+0.92}_{-0.87})$ & 2.4 \\
All, \citet{Tomczak2016}, Eq. \ref{eq:tomczak} & ($s_0, m_0, a_1, a_2, a_3, a_4)$ & ($0.23^{+0.11}_{-0.11}. 8.9^{+0.35}_{-0.64}, 1.33^{+0.16}_{-0.17}, 0.199^{+0.049}_{-0.050}, 1.26^{+0.67}_{-0.44}, 0.226^{+0.186} _{-0.116}$) & 2.9\\
All, This work, Eq. \ref{eq:myform} & ($s_0, m_0, a_1, a_2$) & ($2.80^{+0.080}_{-0.085}, 10.82^{+0.15}_{-0.16}, 0.226^{+0.011}_{-0.011}, 0.131^{+0.028}_{-0.024}$) & 9.0\\
\enddata
\end{deluxetable*}

\begin{table*}
\centering
\caption{Fitted parameters of our MS model (Eq. \ref{eq:myform}) to all mass-complete data for SF galaxies under different assumptions. See \ref{sec:sfrcalib} for more information about the different SFR calibrations.}\label{tab:systematics}
\begin{tabular}{cccc}
\hline
SFR calibration & best fit values ($s_0, m_0, a_1, a_2$) &$\chi^2_r$\\ 
\hline
\citet{Molnar2020} & ($2.965^{+0.081}_{-0.088}, 11.07^{+0.15}_{-0.16}, 0.215^{+0.012}_{-0.012}, 0.120^{+0.027}_{-0.024}$) & 6.5 \\
\citet{Magnelli2015} & ($3.20^{0.10}_{-0.10}, 11.28^{+0.17}_{-0.17},0.228^{+0.015}_{-0.014}, 0.125^{+0.029}_{-0.025}$) & 4.4 \\
\citet{Delhaize2017} & ($2.96^{0.10}_{-0.11}, 11.29^{+0.16}_{-0.17},0.205^{+0.015}_{-0.014}, 0.127^{+0.030}_{-0.026}$) & 3.6\\
\citet{Davies2017} & ($2.54^{0.07}_{-0.07}, 10.94^{+0.14}_{-0.15},0.190^{+0.011}_{-0.010}, 0.120^{+0.028}_{-0.024}$) &7.2 \\
\citet{Bell2003} & ($3.33^{0.11}_{-0.12}, 11.28^{+0.17}_{-0.18},0.264^{+0.015}_{-0.016}, 0.123^{+0.027}_{-0.025}$) & 6.4\\
\hline
\end{tabular}
\end{table*}

\bibliographystyle{mnras}
\bibliography{stacking_bib}

\begin{thebibliography}{}
\makeatletter
\relax
\def\mn@urlcharsother{\let\do\@makeother \do\$\do\&\do\#\do\^\do\_\do\%\do\~}
\def\mn@doi{\begingroup\mn@urlcharsother \@ifnextchar [ {\mn@doi@}
  {\mn@doi@[]}}
\def\mn@doi@[#1]#2{\def\@tempa{#1}\ifx\@tempa\@empty \href
  {http://dx.doi.org/#2} {doi:#2}\else \href {http://dx.doi.org/#2} {#1}\fi
  \endgroup}
\def\mn@eprint#1#2{\mn@eprint@#1:#2::\@nil}
\def\mn@eprint@arXiv#1{\href {http://arxiv.org/abs/#1} {{\tt arXiv:#1}}}
\def\mn@eprint@dblp#1{\href {http://dblp.uni-trier.de/rec/bibtex/#1.xml}
  {dblp:#1}}
\def\mn@eprint@#1:#2:#3:#4\@nil{\def\@tempa {#1}\def\@tempb {#2}\def\@tempc
  {#3}\ifx \@tempc \@empty \let \@tempc \@tempb \let \@tempb \@tempa \fi \ifx
  \@tempb \@empty \def\@tempb {arXiv}\fi \@ifundefined
  {mn@eprint@\@tempb}{\@tempb:\@tempc}{\expandafter \expandafter \csname
  mn@eprint@\@tempb\endcsname \expandafter{\@tempc}}}

\bibitem[\protect\citeauthoryear{{Abramson}, {Kelson}, {Dressler}, {Poggianti},
  {Gladders}, {Oemler}  \& {Vulcani}}{{Abramson} et~al.}{2014}]{Abramson2014}
{Abramson} L.~E.,  {Kelson} D.~D.,  {Dressler} A.,  {Poggianti} B.,  {Gladders}
  M.~D.,  {Oemler} Jr. A.,   {Vulcani} B.,  2014, \mn@doi [\apjl]
  {10.1088/2041-8205/785/2/L36}, \href
  {http://adsabs.harvard.edu/abs/2014ApJ...785L..36A} {785, L36}

\bibitem[\protect\citeauthoryear{{Amblard} et~al.,}{{Amblard}
  et~al.}{2011}]{Amblard2011}
{Amblard} A.,  et~al., 2011, \mn@doi [\nat] {10.1038/nature09771}, \href
  {https://ui.adsabs.harvard.edu/abs/2011Natur.470..510A} {470, 510}

\bibitem[\protect\citeauthoryear{{Appleton} et~al.,}{{Appleton}
  et~al.}{2004}]{Appleton2004}
{Appleton} P.~N.,  et~al., 2004, \mn@doi [The Astrophysical Journal Supplement
  Series] {10.1086/422425}, \href
  {https://ui.adsabs.harvard.edu/\#abs/2004ApJS..154..147A} {154, 147}

\bibitem[\protect\citeauthoryear{{Arnouts}, {Cristiani}, {Moscardini},
  {Matarrese}, {Lucchin}, {Fontana}  \& {Giallongo}}{{Arnouts}
  et~al.}{1999}]{Arnouts1999}
{Arnouts} S.,  {Cristiani} S.,  {Moscardini} L.,  {Matarrese} S.,  {Lucchin}
  F.,  {Fontana} A.,   {Giallongo} E.,  1999, \mn@doi [\mnras]
  {10.1046/j.1365-8711.1999.02978.x}, \href
  {https://ui.adsabs.harvard.edu/abs/1999MNRAS.310..540A} {310, 540}

\bibitem[\protect\citeauthoryear{{Arnouts} et~al.,}{{Arnouts}
  et~al.}{2007}]{Arnouts2007}
{Arnouts} S.,  et~al., 2007, \mn@doi [\aap] {10.1051/0004-6361:20077632}, \href
  {http://adsabs.harvard.edu/abs/2007A%26A...476..137A} {476, 137}

\bibitem[\protect\citeauthoryear{{Assef} et~al.,}{{Assef}
  et~al.}{2013}]{Assef2013}
{Assef} R.~J.,  et~al., 2013, \mn@doi [\apj] {10.1088/0004-637X/772/1/26},
  \href {https://ui.adsabs.harvard.edu/abs/2013ApJ...772...26A} {772, 26}

\bibitem[\protect\citeauthoryear{{Astropy Collaboration} et~al.,}{{Astropy
  Collaboration} et~al.}{2013}]{2013A&A...558A..33A}
{Astropy Collaboration} et~al., 2013, \mn@doi [\aap]
  {10.1051/0004-6361/201322068}, \href
  {http://adsabs.harvard.edu/abs/2013A%26A...558A..33A} {558, A33}

\bibitem[\protect\citeauthoryear{{Auriemma}, {Perola}, {Ekers}, {Fanti},
  {Lari}, {Jaffe}  \& {Ulrich}}{{Auriemma} et~al.}{1977}]{Auriemma1977}
{Auriemma} C.,  {Perola} G.~C.,  {Ekers} R.~D.,  {Fanti} R.,  {Lari} C.,
  {Jaffe} W.~J.,   {Ulrich} M.~H.,  1977, \aap, \href
  {https://ui.adsabs.harvard.edu/abs/1977A&A....57...41A} {57, 41}

\bibitem[\protect\citeauthoryear{{Baldry}, {Glazebrook}, {Brinkmann},
  {Ivezi{\'c}}, {Lupton}, {Nichol}  \& {Szalay}}{{Baldry}
  et~al.}{2004}]{Baldry2004}
{Baldry} I.~K.,  {Glazebrook} K.,  {Brinkmann} J.,  {Ivezi{\'c}} {\v{Z}}.,
  {Lupton} R.~H.,  {Nichol} R.~C.,   {Szalay} A.~S.,  2004, \mn@doi [\apj]
  {10.1086/380092}, \href
  {https://ui.adsabs.harvard.edu/abs/2004ApJ...600..681B} {600, 681}

\bibitem[\protect\citeauthoryear{{Baldry}, {Balogh}, {Bower}, {Glazebrook},
  {Nichol}, {Bamford}  \& {Budavari}}{{Baldry} et~al.}{2006}]{Baldry2006}
{Baldry} I.~K.,  {Balogh} M.~L.,  {Bower} R.~G.,  {Glazebrook} K.,  {Nichol}
  R.~C.,  {Bamford} S.~P.,   {Budavari} T.,  2006, \mn@doi [\mnras]
  {10.1111/j.1365-2966.2006.11081.x}, \href
  {https://ui.adsabs.harvard.edu/abs/2006MNRAS.373..469B} {373, 469}

\bibitem[\protect\citeauthoryear{{Balogh}, {Schade}, {Morris}, {Yee},
  {Carlberg}  \& {Ellingson}}{{Balogh} et~al.}{1998}]{Balogh1998}
{Balogh} M.~L.,  {Schade} D.,  {Morris} S.~L.,  {Yee} H.~K.~C.,  {Carlberg}
  R.~G.,   {Ellingson} E.,  1998, \mn@doi [\apjl] {10.1086/311576}, \href
  {https://ui.adsabs.harvard.edu/abs/1998ApJ...504L..75B} {504, L75}

\bibitem[\protect\citeauthoryear{{Balogh} et~al.,}{{Balogh}
  et~al.}{2002}]{Balogh2002}
{Balogh} M.~L.,  et~al., 2002, \mn@doi [\apj] {10.1086/338056}, \href
  {https://ui.adsabs.harvard.edu/abs/2002ApJ...566..123B} {566, 123}

\bibitem[\protect\citeauthoryear{{Barazza}, {Jogee}  \& {Marinova}}{{Barazza}
  et~al.}{2008}]{Barazza2008}
{Barazza} F.~D.,  {Jogee} S.,   {Marinova} I.,  2008, \mn@doi [\apj]
  {10.1086/526510}, \href
  {https://ui.adsabs.harvard.edu/abs/2008ApJ...675.1194B} {675, 1194}

\bibitem[\protect\citeauthoryear{{Barnes} \& {Hernquist}}{{Barnes} \&
  {Hernquist}}{1996}]{Barnes1996}
{Barnes} J.~E.,  {Hernquist} L.,  1996, \mn@doi [\apj] {10.1086/177957}, \href
  {https://ui.adsabs.harvard.edu/abs/1996ApJ...471..115B} {471, 115}

\bibitem[\protect\citeauthoryear{{Basu}, {Wadadekar}, {Beelen}, {Singh},
  {Archana}, {Sirothia}  \& {Ishwara-Chandra}}{{Basu} et~al.}{2015}]{Basu2015}
{Basu} A.,  {Wadadekar} Y.,  {Beelen} A.,  {Singh} V.,  {Archana} K.~N.,
  {Sirothia} S.,   {Ishwara-Chandra} C.~H.,  2015, \mn@doi [\apj]
  {10.1088/0004-637X/803/2/51}, \href
  {https://ui.adsabs.harvard.edu/\#abs/2015ApJ...803...51B} {803, 51}

\bibitem[\protect\citeauthoryear{{Bell}}{{Bell}}{2003}]{Bell2003}
{Bell} E.~F.,  2003, \mn@doi [\apj] {10.1086/367829}, \href
  {http://adsabs.harvard.edu/abs/2003ApJ...586..794B} {586, 794}

\bibitem[\protect\citeauthoryear{{Bell} et~al.,}{{Bell}
  et~al.}{2004}]{Bell2004}
{Bell} E.~F.,  et~al., 2004, \mn@doi [\apj] {10.1086/420778}, \href
  {https://ui.adsabs.harvard.edu/abs/2004ApJ...608..752B} {608, 752}

\bibitem[\protect\citeauthoryear{{Bell} et~al.,}{{Bell}
  et~al.}{2012}]{Bell2012}
{Bell} E.~F.,  et~al., 2012, \mn@doi [\apj] {10.1088/0004-637X/753/2/167},
  \href {https://ui.adsabs.harvard.edu/abs/2012ApJ...753..167B} {753, 167}

\bibitem[\protect\citeauthoryear{{Bendo} et~al.,}{{Bendo}
  et~al.}{2015}]{Bendo2015}
{Bendo} G.~J.,  et~al., 2015, \mn@doi [\mnras] {10.1093/mnras/stu1841}, \href
  {https://ui.adsabs.harvard.edu/abs/2015MNRAS.448..135B} {448, 135}

\bibitem[\protect\citeauthoryear{{B{\'e}thermin}, {Wang}, {Dor{\'e}},
  {Lagache}, {Sargent}, {Daddi}, {Cousin}  \& {Aussel}}{{B{\'e}thermin}
  et~al.}{2013}]{Bethermin2013}
{B{\'e}thermin} M.,  {Wang} L.,  {Dor{\'e}} O.,  {Lagache} G.,  {Sargent} M.,
  {Daddi} E.,  {Cousin} M.,   {Aussel} H.,  2013, \mn@doi [\aap]
  {10.1051/0004-6361/201321688}, \href
  {https://ui.adsabs.harvard.edu/abs/2013A&A...557A..66B} {557, A66}

\bibitem[\protect\citeauthoryear{{Birnboim}, {Dekel}  \& {Neistein}}{{Birnboim}
  et~al.}{2007}]{Birnboim2007}
{Birnboim} Y.,  {Dekel} A.,   {Neistein} E.,  2007, \mn@doi [\mnras]
  {10.1111/j.1365-2966.2007.12074.x}, \href
  {https://ui.adsabs.harvard.edu/abs/2007MNRAS.380..339B} {380, 339}

\bibitem[\protect\citeauthoryear{{Bisigello}, {Caputi}, {Grogin}  \&
  {Koekemoer}}{{Bisigello} et~al.}{2018}]{Bisigello2018}
{Bisigello} L.,  {Caputi} K.~I.,  {Grogin} N.,   {Koekemoer} A.,  2018, \mn@doi
  [\aap] {10.1051/0004-6361/201731399}, \href
  {http://adsabs.harvard.edu/abs/2018A%26A...609A..82B} {609, A82}

\bibitem[\protect\citeauthoryear{{Blain}, {Smail}, {Ivison}  \&
  {Kneib}}{{Blain} et~al.}{1999}]{Blain1999}
{Blain} A.~W.,  {Smail} I.,  {Ivison} R.~J.,   {Kneib} J.~P.,  1999, \mn@doi
  [\mnras] {10.1046/j.1365-8711.1999.02178.x}, \href
  {https://ui.adsabs.harvard.edu/abs/1999MNRAS.302..632B} {302, 632}

\bibitem[\protect\citeauthoryear{{Bluck}, {Mendel}, {Ellison}, {Moreno},
  {Simard}, {Patton}  \& {Starkenburg}}{{Bluck} et~al.}{2014}]{Bluck2014}
{Bluck} A. F.~L.,  {Mendel} J.~T.,  {Ellison} S.~L.,  {Moreno} J.,  {Simard}
  L.,  {Patton} D.~R.,   {Starkenburg} E.,  2014, \mn@doi [\mnras]
  {10.1093/mnras/stu594}, \href
  {https://ui.adsabs.harvard.edu/abs/2014MNRAS.441..599B} {441, 599}

\bibitem[\protect\citeauthoryear{{Bondi} et~al.,}{{Bondi}
  et~al.}{2018}]{Bondi2018}
{Bondi} M.,  et~al., 2018, \mn@doi [\aap] {10.1051/0004-6361/201834243}, \href
  {https://ui.adsabs.harvard.edu/\#abs/2018A&A...618L...8B} {618, L8}

\bibitem[\protect\citeauthoryear{{Boogaard} et~al.,}{{Boogaard}
  et~al.}{2018}]{Boogaard2018}
{Boogaard} L.~A.,  et~al., 2018, \mn@doi [\aap] {10.1051/0004-6361/201833136},
  \href {https://ui.adsabs.harvard.edu/\#abs/2018A&A...619A..27B} {619, A27}

\bibitem[\protect\citeauthoryear{{Boselli}, {Fossati}, {Gavazzi}, {Ciesla},
  {Buat}, {Boissier}  \& {Hughes}}{{Boselli} et~al.}{2015}]{Boselli2015}
{Boselli} A.,  {Fossati} M.,  {Gavazzi} G.,  {Ciesla} L.,  {Buat} V.,
  {Boissier} S.,   {Hughes} T.~M.,  2015, \mn@doi [\aap]
  {10.1051/0004-6361/201525712}, \href
  {https://ui.adsabs.harvard.edu/abs/2015A&A...579A.102B} {579, A102}

\bibitem[\protect\citeauthoryear{{Bouwens} et~al.,}{{Bouwens}
  et~al.}{2014}]{Bouwens2014a}
{Bouwens} R.~J.,  et~al., 2014, \mn@doi [\apj] {10.1088/0004-637X/795/2/126},
  \href {http://adsabs.harvard.edu/abs/2014ApJ...795..126B} {795, 126}

\bibitem[\protect\citeauthoryear{{Bouwens} et~al.,}{{Bouwens}
  et~al.}{2015}]{Bouwens2015}
{Bouwens} R.~J.,  et~al., 2015, \mn@doi [\apj] {10.1088/0004-637X/803/1/34},
  \href {http://adsabs.harvard.edu/abs/2015ApJ...803...34B} {803, 34}

\bibitem[\protect\citeauthoryear{{Bouwens} et~al.,}{{Bouwens}
  et~al.}{2016}]{Bouwens2016}
{Bouwens} R.~J.,  et~al., 2016, \mn@doi [\apj] {10.3847/1538-4357/833/1/72},
  \href {https://ui.adsabs.harvard.edu/abs/2016ApJ...833...72B} {833, 72}

\bibitem[\protect\citeauthoryear{{Bowler}, {Bourne}, {Dunlop}, {McLure}  \&
  {McLeod}}{{Bowler} et~al.}{2018}]{Bowler2018}
{Bowler} R.~A.~A.,  {Bourne} N.,  {Dunlop} J.~S.,  {McLure} R.~J.,   {McLeod}
  D.~J.,  2018, \mn@doi [\mnras] {10.1093/mnras/sty2368}, \href
  {https://ui.adsabs.harvard.edu/abs/2018MNRAS.481.1631B} {481, 1631}

\bibitem[\protect\citeauthoryear{{Brammer} et~al.,}{{Brammer}
  et~al.}{2011}]{Brammer2011}
{Brammer} G.~B.,  et~al., 2011, \mn@doi [\apj] {10.1088/0004-637X/739/1/24},
  \href {http://adsabs.harvard.edu/abs/2011ApJ...739...24B} {739, 24}

\bibitem[\protect\citeauthoryear{{Brinchmann}, {Charlot}, {White}, {Tremonti},
  {Kauffmann}, {Heckman}  \& {Brinkmann}}{{Brinchmann}
  et~al.}{2004}]{Brinchmann2004}
{Brinchmann} J.,  {Charlot} S.,  {White} S.~D.~M.,  {Tremonti} C.,  {Kauffmann}
  G.,  {Heckman} T.,   {Brinkmann} J.,  2004, \mn@doi [\mnras]
  {10.1111/j.1365-2966.2004.07881.x}, \href
  {http://adsabs.harvard.edu/abs/2004MNRAS.351.1151B} {351, 1151}

\bibitem[\protect\citeauthoryear{Brown, Webster  \& Boyle}{Brown
  et~al.}{2001}]{Brown2001}
Brown M. J.~I.,  Webster R.~L.,   Boyle B.~J.,  2001, \mn@doi [The Astronomical
  Journal] {10.1086/320410}, 121, 2381

\bibitem[\protect\citeauthoryear{{Brown}, {Dey}, {Jannuzi}, {Brand }, {Benson},
  {Brodwin}, {Croton}  \& {Eisenhardt}}{{Brown} et~al.}{2007}]{Brown2007}
{Brown} M. J.~I.,  {Dey} A.,  {Jannuzi} B.~T.,  {Brand } K.,  {Benson} A.~J.,
  {Brodwin} M.,  {Croton} D.~J.,   {Eisenhardt} P.~R.,  2007, \mn@doi [\apj]
  {10.1086/509652}, \href
  {https://ui.adsabs.harvard.edu/abs/2007ApJ...654..858B} {654, 858}

\bibitem[\protect\citeauthoryear{{Brown}, {Jannuzi}, {Floyd}  \&
  {Mould}}{{Brown} et~al.}{2011}]{Brown2011}
{Brown} M. J.~I.,  {Jannuzi} B.~T.,  {Floyd} D. J.~E.,   {Mould} J.~R.,  2011,
  \mn@doi [\apjl] {10.1088/2041-8205/731/2/L41}, \href
  {https://ui.adsabs.harvard.edu/abs/2011ApJ...731L..41B} {731, L41}

\bibitem[\protect\citeauthoryear{{Brown} et~al.,}{{Brown}
  et~al.}{2017}]{Brown2017}
{Brown} M. J.~I.,  et~al., 2017, \mn@doi [\apj] {10.3847/1538-4357/aa8ad2},
  \href {https://ui.adsabs.harvard.edu/abs/2017ApJ...847..136B} {847, 136}

\bibitem[\protect\citeauthoryear{{Bruzual} \& {Charlot}}{{Bruzual} \&
  {Charlot}}{2003}]{Bruzual2003}
{Bruzual} G.,  {Charlot} S.,  2003, \mn@doi [\mnras]
  {10.1046/j.1365-8711.2003.06897.x}, \href
  {http://adsabs.harvard.edu/abs/2003MNRAS.344.1000B} {344, 1000}

\bibitem[\protect\citeauthoryear{{Buisson}, {Lohfink}, {Alston}  \&
  {Fabian}}{{Buisson} et~al.}{2017}]{Buisson2017}
{Buisson} D.~J.~K.,  {Lohfink} A.~M.,  {Alston} W.~N.,   {Fabian} A.~C.,  2017,
  \mn@doi [\mnras] {10.1093/mnras/stw2486}, \href
  {https://ui.adsabs.harvard.edu/abs/2017MNRAS.464.3194B} {464, 3194}

\bibitem[\protect\citeauthoryear{{Bundy} et~al.,}{{Bundy}
  et~al.}{2006}]{Bundy2006}
{Bundy} K.,  et~al., 2006, \mn@doi [\apj] {10.1086/507456}, \href
  {https://ui.adsabs.harvard.edu/abs/2006ApJ...651..120B} {651, 120}

\bibitem[\protect\citeauthoryear{{Burgarella}, {Buat}  \&
  {Iglesias-P{\'a}ramo}}{{Burgarella} et~al.}{2005}]{Burgarella2005}
{Burgarella} D.,  {Buat} V.,   {Iglesias-P{\'a}ramo} J.,  2005, \mn@doi
  [\mnras] {10.1111/j.1365-2966.2005.09131.x}, \href
  {https://ui.adsabs.harvard.edu/\#abs/2005MNRAS.360.1413B} {360, 1413}

\bibitem[\protect\citeauthoryear{{Calistro Rivera} et~al.,}{{Calistro Rivera}
  et~al.}{2017}]{CalistroRivera2017}
{Calistro Rivera} G.,  et~al., 2017, \mn@doi [\mnras] {10.1093/mnras/stx1040},
  \href {https://ui.adsabs.harvard.edu/\#abs/2017MNRAS.469.3468C} {469, 3468}

\bibitem[\protect\citeauthoryear{{Camps} et~al.,}{{Camps}
  et~al.}{2018}]{Camps2018}
{Camps} P.,  et~al., 2018, \mn@doi [\apjs] {10.3847/1538-4365/aaa24c}, \href
  {https://ui.adsabs.harvard.edu/abs/2018ApJS..234...20C} {234, 20}

\bibitem[\protect\citeauthoryear{{Capak} et~al.,}{{Capak}
  et~al.}{2007}]{Capak2007}
{Capak} P.,  et~al., 2007, \mn@doi [The Astrophysical Journal Supplement
  Series] {10.1086/519081}, \href
  {https://ui.adsabs.harvard.edu/abs/2007ApJS..172...99C} {172, 99}

\bibitem[\protect\citeauthoryear{{Caplar} \& {Tacchella}}{{Caplar} \&
  {Tacchella}}{2019}]{Caplar2019}
{Caplar} N.,  {Tacchella} S.,  2019, arXiv e-prints, \href
  {https://ui.adsabs.harvard.edu/abs/2019arXiv190107556C} {p. arXiv:1901.07556}

\bibitem[\protect\citeauthoryear{{Caputi} et~al.,}{{Caputi}
  et~al.}{2017}]{Caputi2017}
{Caputi} K.~I.,  et~al., 2017, \mn@doi [\apj] {10.3847/1538-4357/aa901e}, \href
  {https://ui.adsabs.harvard.edu/\#abs/2017ApJ...849...45C} {849, 45}

\bibitem[\protect\citeauthoryear{{Casey} et~al.,}{{Casey}
  et~al.}{2018}]{Casey2018}
{Casey} C.~M.,  et~al., 2018, \mn@doi [\apj] {10.3847/1538-4357/aac82d}, \href
  {https://ui.adsabs.harvard.edu/\#abs/2018ApJ...862...77C} {862, 77}

\bibitem[\protect\citeauthoryear{{Castellano} et~al.,}{{Castellano}
  et~al.}{2012}]{Castellano2012}
{Castellano} M.,  et~al., 2012, \mn@doi [\aap] {10.1051/0004-6361/201118050},
  \href {https://ui.adsabs.harvard.edu/abs/2012A&A...540A..39C} {540, A39}

\bibitem[\protect\citeauthoryear{{Cattaneo}, {Dekel}, {Devriendt}, {Guiderdoni}
   \& {Blaizot}}{{Cattaneo} et~al.}{2006}]{Cattaneo2006}
{Cattaneo} A.,  {Dekel} A.,  {Devriendt} J.,  {Guiderdoni} B.,   {Blaizot} J.,
  2006, \mn@doi [\mnras] {10.1111/j.1365-2966.2006.10608.x}, \href
  {https://ui.adsabs.harvard.edu/abs/2006MNRAS.370.1651C} {370, 1651}

\bibitem[\protect\citeauthoryear{{Cavanaugh} \& {Shumway}}{{Cavanaugh} \&
  {Shumway}}{1997}]{Cavanaugh1997}
{Cavanaugh} J.~E.,  {Shumway} R.~H.,  1997, Statistica Sinica, 7, 473

\bibitem[\protect\citeauthoryear{{Chabrier}}{{Chabrier}}{2003}]{Chabrier2003}
{Chabrier} G.,  2003, \mn@doi [\pasp] {10.1086/376392}, \href
  {http://adsabs.harvard.edu/abs/2003PASP..115..763C} {115, 763}

\bibitem[\protect\citeauthoryear{{Chary} \& {Elbaz}}{{Chary} \&
  {Elbaz}}{2001}]{Chary2001}
{Chary} R.,  {Elbaz} D.,  2001, \mn@doi [\apj] {10.1086/321609}, \href
  {https://ui.adsabs.harvard.edu/abs/2001ApJ...556..562C} {556, 562}

\bibitem[\protect\citeauthoryear{{Cibinel} et~al.,}{{Cibinel}
  et~al.}{2019}]{Cibinel2019}
{Cibinel} A.,  et~al., 2019, \mn@doi [\mnras] {10.1093/mnras/stz690}, \href
  {https://ui.adsabs.harvard.edu/abs/2019MNRAS.485.5631C} {485, 5631}

\bibitem[\protect\citeauthoryear{{Condon}}{{Condon}}{1992}]{Condon1992}
{Condon} J.~J.,  1992, \mn@doi [\araa] {10.1146/annurev.aa.30.090192.003043},
  \href {https://ui.adsabs.harvard.edu/abs/1992ARA&A..30..575C} {30, 575}

\bibitem[\protect\citeauthoryear{{Condon}, {Cotton}  \& {Broderick}}{{Condon}
  et~al.}{2002}]{Condon2002}
{Condon} J.~J.,  {Cotton} W.~D.,   {Broderick} J.~J.,  2002, \mn@doi [\aj]
  {10.1086/341650}, \href
  {https://ui.adsabs.harvard.edu/abs/2002AJ....124..675C} {124, 675}

\bibitem[\protect\citeauthoryear{{Connolly}, {Szalay}, {Dickinson}, {SubbaRao}
  \& {Brunner}}{{Connolly} et~al.}{1997}]{Connolly1997}
{Connolly} A.~J.,  {Szalay} A.~S.,  {Dickinson} M.,  {SubbaRao} M.~U.,
  {Brunner} R.~J.,  1997, \mn@doi [\apjl] {10.1086/310829}, \href
  {https://ui.adsabs.harvard.edu/abs/1997ApJ...486L..11C} {486, L11}

\bibitem[\protect\citeauthoryear{{Cooper} et~al.,}{{Cooper}
  et~al.}{2006}]{Cooper2006}
{Cooper} M.~C.,  et~al., 2006, \mn@doi [\mnras]
  {10.1111/j.1365-2966.2006.10485.x}, \href
  {https://ui.adsabs.harvard.edu/abs/2006MNRAS.370..198C} {370, 198}

\bibitem[\protect\citeauthoryear{{Cooray} et~al.,}{{Cooray}
  et~al.}{2010}]{Cooray2010}
{Cooray} A.,  et~al., 2010, \mn@doi [\aap] {10.1051/0004-6361/201014597}, \href
  {https://ui.adsabs.harvard.edu/abs/2010A&A...518L..22C} {518, L22}

\bibitem[\protect\citeauthoryear{{Couch}, {Balogh}, {Bower}, {Smail},
  {Glazebrook}  \& {Taylor}}{{Couch} et~al.}{2001}]{Couch2001}
{Couch} W.~J.,  {Balogh} M.~L.,  {Bower} R.~G.,  {Smail} I.,  {Glazebrook} K.,
   {Taylor} M.,  2001, \mn@doi [\apj] {10.1086/319459}, \href
  {https://ui.adsabs.harvard.edu/abs/2001ApJ...549..820C} {549, 820}

\bibitem[\protect\citeauthoryear{{Cowie}, {Songaila}, {Hu}  \& {Cohen}}{{Cowie}
  et~al.}{1996}]{Cowie1996}
{Cowie} L.~L.,  {Songaila} A.,  {Hu} E.~M.,   {Cohen} J.~G.,  1996, \mn@doi
  [\aj] {10.1086/118058}, \href
  {https://ui.adsabs.harvard.edu/abs/1996AJ....112..839C} {112, 839}

\bibitem[\protect\citeauthoryear{{Cowie}, {Songaila}  \& {Barger}}{{Cowie}
  et~al.}{1999}]{Cowie1999}
{Cowie} L.~L.,  {Songaila} A.,   {Barger} A.~J.,  1999, \mn@doi [\aj]
  {10.1086/300959}, \href
  {https://ui.adsabs.harvard.edu/abs/1999AJ....118..603C} {118, 603}

\bibitem[\protect\citeauthoryear{{Daddi} et~al.,}{{Daddi}
  et~al.}{2007}]{Daddi2007}
{Daddi} E.,  et~al., 2007, \mn@doi [\apj] {10.1086/521818}, \href
  {https://ui.adsabs.harvard.edu/abs/2007ApJ...670..156D} {670, 156}

\bibitem[\protect\citeauthoryear{{Darvish}, {Sobral}, {Mobasher}, {Scoville},
  {Best}, {Sales}  \& {Smail}}{{Darvish} et~al.}{2014}]{Darvish2014}
{Darvish} B.,  {Sobral} D.,  {Mobasher} B.,  {Scoville} N.~Z.,  {Best} P.,
  {Sales} L.~V.,   {Smail} I.,  2014, \mn@doi [\apj]
  {10.1088/0004-637X/796/1/51}, \href
  {https://ui.adsabs.harvard.edu/\#abs/2014ApJ...796...51D} {796, 51}

\bibitem[\protect\citeauthoryear{{Darvish}, {Mobasher}, {Sobral}, {Scoville}
  \& {Aragon-Calvo}}{{Darvish} et~al.}{2015}]{Darvish2015}
{Darvish} B.,  {Mobasher} B.,  {Sobral} D.,  {Scoville} N.,   {Aragon-Calvo}
  M.,  2015, \mn@doi [\apj] {10.1088/0004-637X/805/2/121}, \href
  {https://ui.adsabs.harvard.edu/abs/2015ApJ...805..121D} {805, 121}

\bibitem[\protect\citeauthoryear{{Darvish}, {Mobasher}, {Sobral}, {Rettura},
  {Scoville}, {Faisst}  \& {Capak}}{{Darvish} et~al.}{2016}]{Darvish2016}
{Darvish} B.,  {Mobasher} B.,  {Sobral} D.,  {Rettura} A.,  {Scoville} N.,
  {Faisst} A.,   {Capak} P.,  2016, \mn@doi [\apj]
  {10.3847/0004-637X/825/2/113}, \href
  {https://ui.adsabs.harvard.edu/\#abs/2016ApJ...825..113D} {825, 113}

\bibitem[\protect\citeauthoryear{{Dav{\'e}}, {Thompson}  \&
  {Hopkins}}{{Dav{\'e}} et~al.}{2016}]{Dave2016}
{Dav{\'e}} R.,  {Thompson} R.,   {Hopkins} P.~F.,  2016, \mn@doi [\mnras]
  {10.1093/mnras/stw1862}, \href
  {https://ui.adsabs.harvard.edu/\#abs/2016MNRAS.462.3265D} {462, 3265}

\bibitem[\protect\citeauthoryear{{Davidzon} et~al.,}{{Davidzon}
  et~al.}{2017}]{Davidzon2017}
{Davidzon} I.,  et~al., 2017, \mn@doi [\aap] {10.1051/0004-6361/201730419},
  \href {http://adsabs.harvard.edu/abs/2017A%26A...605A..70D} {605, A70}

\bibitem[\protect\citeauthoryear{{Davidzon}, {Ilbert}, {Faisst}, {Sparre}  \&
  {Capak}}{{Davidzon} et~al.}{2018}]{Davidzon2018}
{Davidzon} I.,  {Ilbert} O.,  {Faisst} A.~L.,  {Sparre} M.,   {Capak} P.~L.,
  2018, \mn@doi [\apj] {10.3847/1538-4357/aaa19e}, \href
  {https://ui.adsabs.harvard.edu/\#abs/2018ApJ...852..107D} {852, 107}

\bibitem[\protect\citeauthoryear{{Davies} et~al.,}{{Davies}
  et~al.}{2016}]{Davies2016}
{Davies} L.~J.~M.,  et~al., 2016, \mn@doi [\mnras] {10.1093/mnras/stw1342},
  \href {https://ui.adsabs.harvard.edu/\#abs/2016MNRAS.461..458D} {461, 458}

\bibitem[\protect\citeauthoryear{{Davies} et~al.,}{{Davies}
  et~al.}{2017}]{Davies2017}
{Davies} L.~J.~M.,  et~al., 2017, \mn@doi [\mnras] {10.1093/mnras/stw3080},
  \href {https://ui.adsabs.harvard.edu/abs/2017MNRAS.466.2312D} {466, 2312}

\bibitem[\protect\citeauthoryear{{Davies} et~al.,}{{Davies}
  et~al.}{2019}]{Davies2019}
{Davies} L.~J.~M.,  et~al., 2019, \mn@doi [\mnras] {10.1093/mnras/sty2957},
  \href {https://ui.adsabs.harvard.edu/abs/2019MNRAS.483.1881D} {483, 1881}

\bibitem[\protect\citeauthoryear{{Dekel} \& {Birnboim}}{{Dekel} \&
  {Birnboim}}{2006}]{Dekel2006}
{Dekel} A.,  {Birnboim} Y.,  2006, \mn@doi [\mnras]
  {10.1111/j.1365-2966.2006.10145.x}, \href
  {https://ui.adsabs.harvard.edu/abs/2006MNRAS.368....2D} {368, 2}

\bibitem[\protect\citeauthoryear{{Dekel} \& {Birnboim}}{{Dekel} \&
  {Birnboim}}{2008}]{Dekel2008}
{Dekel} A.,  {Birnboim} Y.,  2008, \mn@doi [\mnras]
  {10.1111/j.1365-2966.2007.12569.x}, \href
  {https://ui.adsabs.harvard.edu/abs/2008MNRAS.383..119D} {383, 119}

\bibitem[\protect\citeauthoryear{{Delhaize} et~al.,}{{Delhaize}
  et~al.}{2017}]{Delhaize2017}
{Delhaize} J.,  et~al., 2017, \mn@doi [\aap] {10.1051/0004-6361/201629430},
  \href {http://adsabs.harvard.edu/abs/2017A%26A...602A...4D} {602, A4}

\bibitem[\protect\citeauthoryear{{Delvecchio} et~al.,}{{Delvecchio}
  et~al.}{2017}]{Delvecchio2017}
{Delvecchio} I.,  et~al., 2017, \mn@doi [\aap] {10.1051/0004-6361/201629367},
  \href {https://ui.adsabs.harvard.edu/abs/2017A&A...602A...3D} {602, A3}

\bibitem[\protect\citeauthoryear{{Donley} et~al.,}{{Donley}
  et~al.}{2012}]{Donley2012}
{Donley} J.~L.,  et~al., 2012, \mn@doi [\apj] {10.1088/0004-637X/748/2/142},
  \href {https://ui.adsabs.harvard.edu/abs/2012ApJ...748..142D} {748, 142}

\bibitem[\protect\citeauthoryear{{Donnari} et~al.,}{{Donnari}
  et~al.}{2018}]{Donnari2019}
{Donnari} M.,  et~al., 2018, arXiv e-prints, \href
  {https://ui.adsabs.harvard.edu/\#abs/2018arXiv181207584D} {p.
  arXiv:1812.07584}

\bibitem[\protect\citeauthoryear{{Dressler}}{{Dressler}}{1980}]{Dressler1980}
{Dressler} A.,  1980, \mn@doi [\apj] {10.1086/157753}, \href
  {https://ui.adsabs.harvard.edu/abs/1980ApJ...236..351D} {236, 351}

\bibitem[\protect\citeauthoryear{{Driver} et~al.,}{{Driver}
  et~al.}{2018}]{Driver2018}
{Driver} S.~P.,  et~al., 2018, \mn@doi [\mnras] {10.1093/mnras/stx2728}, \href
  {https://ui.adsabs.harvard.edu/abs/2018MNRAS.475.2891D} {475, 2891}

\bibitem[\protect\citeauthoryear{{Duivenvoorden} et~al.,}{{Duivenvoorden}
  et~al.}{2016}]{Duivenvoorden2016}
{Duivenvoorden} S.,  et~al., 2016, \mn@doi [\mnras] {10.1093/mnras/stw1466},
  \href {https://ui.adsabs.harvard.edu/\#abs/2016MNRAS.462..277D} {462, 277}

\bibitem[\protect\citeauthoryear{{Eales} et~al.,}{{Eales}
  et~al.}{2018a}]{Eales2018a}
{Eales} S.,  et~al., 2018a, \mn@doi [\mnras] {10.1093/mnras/stx2548}, \href
  {http://adsabs.harvard.edu/abs/2018MNRAS.473.3507E} {473, 3507}

\bibitem[\protect\citeauthoryear{{Eales} et~al.,}{{Eales}
  et~al.}{2018b}]{Eales2018}
{Eales} S.~A.,  et~al., 2018b, \mn@doi [\mnras] {10.1093/mnras/sty2220}, \href
  {https://ui.adsabs.harvard.edu/abs/2018MNRAS.481.1183E} {481, 1183}

\bibitem[\protect\citeauthoryear{{Elbaz} et~al.,}{{Elbaz}
  et~al.}{2007}]{Elbaz2007}
{Elbaz} D.,  et~al., 2007, \mn@doi [\aap] {10.1051/0004-6361:20077525}, \href
  {http://adsabs.harvard.edu/abs/2007A%26A...468...33E} {468, 33}

\bibitem[\protect\citeauthoryear{{Elitzur}, {Ho}  \& {Trump}}{{Elitzur}
  et~al.}{2014}]{Elitzur2014}
{Elitzur} M.,  {Ho} L.~C.,   {Trump} J.~R.,  2014, \mn@doi [\mnras]
  {10.1093/mnras/stt2445}, \href
  {https://ui.adsabs.harvard.edu/abs/2014MNRAS.438.3340E} {438, 3340}

\bibitem[\protect\citeauthoryear{{Erfanianfar} et~al.,}{{Erfanianfar}
  et~al.}{2016}]{Erfanianfar2016}
{Erfanianfar} G.,  et~al., 2016, \mn@doi [\mnras] {10.1093/mnras/stv2485},
  \href {https://ui.adsabs.harvard.edu/\#abs/2016MNRAS.455.2839E} {455, 2839}

\bibitem[\protect\citeauthoryear{{Fabbiano}, {Gioia}  \&
  {Trinchieri}}{{Fabbiano} et~al.}{1989}]{Fabbiano1989}
{Fabbiano} G.,  {Gioia} I.~M.,   {Trinchieri} G.,  1989, \mn@doi [\apj]
  {10.1086/168103}, \href
  {https://ui.adsabs.harvard.edu/abs/1989ApJ...347..127F} {347, 127}

\bibitem[\protect\citeauthoryear{{Faber} et~al.,}{{Faber}
  et~al.}{2007}]{Faber2007}
{Faber} S.~M.,  et~al., 2007, \mn@doi [\apj] {10.1086/519294}, \href
  {https://ui.adsabs.harvard.edu/\#abs/2007ApJ...665..265F} {665, 265}

\bibitem[\protect\citeauthoryear{{Faucher-Gigu{\`e}re}, {Kere{\v{s}}}  \&
  {Ma}}{{Faucher-Gigu{\`e}re} et~al.}{2011}]{FaucherGiguere2011}
{Faucher-Gigu{\`e}re} C.-A.,  {Kere{\v{s}}} D.,   {Ma} C.-P.,  2011, \mn@doi
  [\mnras] {10.1111/j.1365-2966.2011.19457.x}, \href
  {https://ui.adsabs.harvard.edu/abs/2011MNRAS.417.2982F} {417, 2982}

\bibitem[\protect\citeauthoryear{{Filho}, {Tabatabaei}, {S{\'a}nchez Almeida},
  {Mu{\~n}oz-Tu{\~n}{\'o}n}  \& {Elmegreen}}{{Filho} et~al.}{2019}]{Filho2019}
{Filho} M.~E.,  {Tabatabaei} F.~S.,  {S{\'a}nchez Almeida} J.,
  {Mu{\~n}oz-Tu{\~n}{\'o}n} C.,   {Elmegreen} B.~G.,  2019, \mn@doi [\mnras]
  {10.1093/mnras/sty3199}, \href
  {https://ui.adsabs.harvard.edu/abs/2019MNRAS.484..543F} {484, 543}

\bibitem[\protect\citeauthoryear{{Finoguenov} et~al.,}{{Finoguenov}
  et~al.}{2007}]{Finoguenov2007}
{Finoguenov} A.,  et~al., 2007, \mn@doi [The Astrophysical Journal Supplement
  Series] {10.1086/516577}, \href
  {https://ui.adsabs.harvard.edu/\#abs/2007ApJS..172..182F} {172, 182}

\bibitem[\protect\citeauthoryear{{Flores} et~al.,}{{Flores}
  et~al.}{1999}]{Flores1999}
{Flores} H.,  et~al., 1999, \mn@doi [\apj] {10.1086/307172}, \href
  {https://ui.adsabs.harvard.edu/abs/1999ApJ...517..148F} {517, 148}

\bibitem[\protect\citeauthoryear{Foreman-Mackey}{Foreman-Mackey}{2016}]{ForemanMackey2016}
Foreman-Mackey D.,  2016, \mn@doi [The Journal of Open Source Software]
  {10.21105/joss.00024}, 24

\bibitem[\protect\citeauthoryear{{Fragkoudi}, {Athanassoula}  \&
  {Bosma}}{{Fragkoudi} et~al.}{2016}]{Fragkoudi2016}
{Fragkoudi} F.,  {Athanassoula} E.,   {Bosma} A.,  2016, \mn@doi [\mnras]
  {10.1093/mnrasl/slw120}, \href
  {https://ui.adsabs.harvard.edu/abs/2016MNRAS.462L..41F} {462, L41}

\bibitem[\protect\citeauthoryear{{Franx}, {van Dokkum}, {F{\"o}rster
  Schreiber}, {Wuyts}, {Labb{\'e}}  \& {Toft}}{{Franx}
  et~al.}{2008}]{Franx2008}
{Franx} M.,  {van Dokkum} P.~G.,  {F{\"o}rster Schreiber} N.~M.,  {Wuyts} S.,
  {Labb{\'e}} I.,   {Toft} S.,  2008, \mn@doi [\apj] {10.1086/592431}, \href
  {https://ui.adsabs.harvard.edu/abs/2008ApJ...688..770F} {688, 770}

\bibitem[\protect\citeauthoryear{{Fudamoto} et~al.,}{{Fudamoto}
  et~al.}{2017}]{Fudamoto2017}
{Fudamoto} Y.,  et~al., 2017, \mn@doi [\mnras] {10.1093/mnras/stx1948}, \href
  {https://ui.adsabs.harvard.edu/abs/2017MNRAS.472..483F} {472, 483}

\bibitem[\protect\citeauthoryear{{Gallego}, {Zamorano}, {Aragon-Salamanca}  \&
  {Rego}}{{Gallego} et~al.}{1995}]{Gallego1995}
{Gallego} J.,  {Zamorano} J.,  {Aragon-Salamanca} A.,   {Rego} M.,  1995,
  \mn@doi [\apjl] {10.1086/309804}, \href
  {https://ui.adsabs.harvard.edu/abs/1995ApJ...455L...1G} {455, L1}

\bibitem[\protect\citeauthoryear{{Garrett}}{{Garrett}}{2002}]{Garrett2002}
{Garrett} M.~A.,  2002, \mn@doi [\aap] {10.1051/0004-6361:20020169}, \href
  {https://ui.adsabs.harvard.edu/\#abs/2002A&A...384L..19G} {384, L19}

\bibitem[\protect\citeauthoryear{{Gavazzi} et~al.,}{{Gavazzi}
  et~al.}{2015}]{Gavazzi2015}
{Gavazzi} G.,  et~al., 2015, \mn@doi [\aap] {10.1051/0004-6361/201425351},
  \href {https://ui.adsabs.harvard.edu/\#abs/2015A&A...580A.116G} {580, A116}

\bibitem[\protect\citeauthoryear{{Gensior}, {Kruijssen}  \& {Keller}}{{Gensior}
  et~al.}{2020}]{Gensior2020}
{Gensior} J.,  {Kruijssen} J.~M.~D.,   {Keller} B.~W.,  2020, arXiv e-prints,
  \href {https://ui.adsabs.harvard.edu/abs/2020arXiv200201484G} {p.
  arXiv:2002.01484}

\bibitem[\protect\citeauthoryear{{George} et~al.,}{{George}
  et~al.}{2011}]{George2011}
{George} M.~R.,  et~al., 2011, \mn@doi [\apj] {10.1088/0004-637X/742/2/125},
  \href {https://ui.adsabs.harvard.edu/\#abs/2011ApJ...742..125G} {742, 125}

\bibitem[\protect\citeauthoryear{{G{\'o}mez} et~al.,}{{G{\'o}mez}
  et~al.}{2003}]{Gomez2003}
{G{\'o}mez} P.~L.,  et~al., 2003, \mn@doi [\apj] {10.1086/345593}, \href
  {https://ui.adsabs.harvard.edu/abs/2003ApJ...584..210G} {584, 210}

\bibitem[\protect\citeauthoryear{{Gozaliasl} et~al.,}{{Gozaliasl}
  et~al.}{2019}]{Gozaliasl2019}
{Gozaliasl} G.,  et~al., 2019, \mn@doi [\mnras] {10.1093/mnras/sty3203}, \href
  {https://ui.adsabs.harvard.edu/\#abs/2019MNRAS.483.3545G} {483, 3545}

\bibitem[\protect\citeauthoryear{{Grazian} et~al.,}{{Grazian}
  et~al.}{2015}]{Grazian2015}
{Grazian} A.,  et~al., 2015, \mn@doi [\aap] {10.1051/0004-6361/201424750},
  \href {https://ui.adsabs.harvard.edu/abs/2015A&A...575A..96G} {575, A96}

\bibitem[\protect\citeauthoryear{{Grossi}, {Fernandes}, {Sobral}, {Afonso},
  {Telles}, {Bizzocchi}, {Paulino-Afonso}  \& {Matute}}{{Grossi}
  et~al.}{2018}]{Grossi2018}
{Grossi} M.,  {Fernandes} C. A.~C.,  {Sobral} D.,  {Afonso} J.,  {Telles} E.,
  {Bizzocchi} L.,  {Paulino-Afonso} A.,   {Matute} I.,  2018, \mn@doi [\mnras]
  {10.1093/mnras/stx3165}, \href
  {https://ui.adsabs.harvard.edu/abs/2018MNRAS.475..735G} {475, 735}

\bibitem[\protect\citeauthoryear{{Gu}, {Fang}, {Yuan}, {Cai}  \& {Wang}}{{Gu}
  et~al.}{2018}]{Gu2018}
{Gu} Y.,  {Fang} G.,  {Yuan} Q.,  {Cai} Z.,   {Wang} T.,  2018, \mn@doi [\apj]
  {10.3847/1538-4357/aaad0b}, \href
  {https://ui.adsabs.harvard.edu/\#abs/2018ApJ...855...10G} {855, 10}

\bibitem[\protect\citeauthoryear{{Guo}, {Zheng}  \& {Fu}}{{Guo}
  et~al.}{2013}]{Guo2013}
{Guo} K.,  {Zheng} X.~Z.,   {Fu} H.,  2013, \mn@doi [\apj]
  {10.1088/0004-637X/778/1/23}, \href
  {https://ui.adsabs.harvard.edu/\#abs/2013ApJ...778...23G} {778, 23}

\bibitem[\protect\citeauthoryear{{G{\"u}rkan} et~al.,}{{G{\"u}rkan}
  et~al.}{2018}]{Gurkan2018}
{G{\"u}rkan} G.,  et~al., 2018, \mn@doi [\mnras] {10.1093/mnras/sty016}, \href
  {https://ui.adsabs.harvard.edu/\#abs/2018MNRAS.475.3010G} {475, 3010}

\bibitem[\protect\citeauthoryear{{Haarsma}, {Partridge}, {Windhorst}  \&
  {Richards}}{{Haarsma} et~al.}{2000}]{Haarsma2000}
{Haarsma} D.~B.,  {Partridge} R.~B.,  {Windhorst} R.~A.,   {Richards} E.~A.,
  2000, \mn@doi [\apj] {10.1086/317225}, \href
  {https://ui.adsabs.harvard.edu/abs/2000ApJ...544..641H} {544, 641}

\bibitem[\protect\citeauthoryear{{Haggard}, {Green}, {Anderson}, {Constantin},
  {Aldcroft}, {Kim}  \& {Barkhouse}}{{Haggard} et~al.}{2010}]{Haggard2010}
{Haggard} D.,  {Green} P.~J.,  {Anderson} S.~F.,  {Constantin} A.,  {Aldcroft}
  T.~L.,  {Kim} D.-W.,   {Barkhouse} W.~A.,  2010, \mn@doi [\apj]
  {10.1088/0004-637X/723/2/1447}, \href
  {https://ui.adsabs.harvard.edu/abs/2010ApJ...723.1447H} {723, 1447}

\bibitem[\protect\citeauthoryear{{Hahn} et~al.,}{{Hahn}
  et~al.}{2019}]{Hahn2019}
{Hahn} C.,  et~al., 2019, \mn@doi [\apj] {10.3847/1538-4357/aafedd}, \href
  {https://ui.adsabs.harvard.edu/abs/2019ApJ...872..160H} {872, 160}

\bibitem[\protect\citeauthoryear{{Haines} et~al.,}{{Haines}
  et~al.}{2013}]{Haines2013}
{Haines} C.~P.,  et~al., 2013, \mn@doi [\apj] {10.1088/0004-637X/775/2/126},
  \href {https://ui.adsabs.harvard.edu/\#abs/2013ApJ...775..126H} {775, 126}

\bibitem[\protect\citeauthoryear{{Hales}, {Murphy}, {Curran}, {Middelberg},
  {Gaensler}  \& {Norris}}{{Hales} et~al.}{2012}]{Hales2012}
{Hales} C.~A.,  {Murphy} T.,  {Curran} J.~R.,  {Middelberg} E.,  {Gaensler}
  B.~M.,   {Norris} R.~P.,  2012, \mn@doi [\mnras]
  {10.1111/j.1365-2966.2012.21373.x}, \href
  {http://adsabs.harvard.edu/abs/2012MNRAS.425..979H} {425, 979}

\bibitem[\protect\citeauthoryear{{Hashimoto}, {Oemler}, {Lin}  \&
  {Tucker}}{{Hashimoto} et~al.}{1998}]{Hashimoto1998}
{Hashimoto} Y.,  {Oemler} Augustus J.,  {Lin} H.,   {Tucker} D.~L.,  1998,
  \mn@doi [\apj] {10.1086/305657}, \href
  {https://ui.adsabs.harvard.edu/abs/1998ApJ...499..589H} {499, 589}

\bibitem[\protect\citeauthoryear{{Heckman} \& {Best}}{{Heckman} \&
  {Best}}{2014}]{Heckman2014}
{Heckman} T.~M.,  {Best} P.~N.,  2014, \mn@doi [\araa]
  {10.1146/annurev-astro-081913-035722}, \href
  {https://ui.adsabs.harvard.edu/abs/2014ARA&A..52..589H} {52, 589}

\bibitem[\protect\citeauthoryear{{Helou}, {Soifer}  \&
  {Rowan-Robinson}}{{Helou} et~al.}{1985}]{Helou1985}
{Helou} G.,  {Soifer} B.~T.,   {Rowan-Robinson} M.,  1985, \mn@doi [\apjl]
  {10.1086/184556}, \href {http://adsabs.harvard.edu/abs/1985ApJ...298L...7H}
  {298, L7}

\bibitem[\protect\citeauthoryear{{Helou}, {Lu}, {Werner}, {Malhotra}  \&
  {Silbermann}}{{Helou} et~al.}{2000}]{Helou2000}
{Helou} G.,  {Lu} N.~Y.,  {Werner} M.~W.,  {Malhotra} S.,   {Silbermann} N.,
  2000, \mn@doi [\apjl] {10.1086/312549}, \href
  {https://ui.adsabs.harvard.edu/abs/2000ApJ...532L..21H} {532, L21}

\bibitem[\protect\citeauthoryear{{Herrera Ruiz} et~al.,}{{Herrera Ruiz}
  et~al.}{2018}]{HerreraRuiz2018}
{Herrera Ruiz} N.,  et~al., 2018, \mn@doi [\aap] {10.1051/0004-6361/201832969},
  \href {http://adsabs.harvard.edu/abs/2018A%26A...616A.128H} {616, A128}

\bibitem[\protect\citeauthoryear{{Hindson} et~al.,}{{Hindson}
  et~al.}{2018}]{Hindson2018}
{Hindson} L.,  et~al., 2018, \mn@doi [\apjs] {10.3847/1538-4365/aaa42c}, \href
  {https://ui.adsabs.harvard.edu/abs/2018ApJS..234...29H} {234, 29}

\bibitem[\protect\citeauthoryear{{Hinshaw} et~al.,}{{Hinshaw}
  et~al.}{2013}]{Hinshaw2013}
{Hinshaw} G.,  et~al., 2013, \mn@doi [The Astrophysical Journal Supplement
  Series] {10.1088/0067-0049/208/2/19}, \href
  {https://ui.adsabs.harvard.edu/abs/2013ApJS..208...19H} {208, 19}

\bibitem[\protect\citeauthoryear{{Hogg} et~al.,}{{Hogg}
  et~al.}{2004}]{Hogg2004}
{Hogg} D.~W.,  et~al., 2004, \mn@doi [\apjl] {10.1086/381749}, \href
  {https://ui.adsabs.harvard.edu/abs/2004ApJ...601L..29H} {601, L29}

\bibitem[\protect\citeauthoryear{{Hopkins} et~al.,}{{Hopkins}
  et~al.}{2003}]{Hopkins2003}
{Hopkins} A.~M.,  et~al., 2003, \mn@doi [\apj] {10.1086/379608}, \href
  {https://ui.adsabs.harvard.edu/abs/2003ApJ...599..971H} {599, 971}

\bibitem[\protect\citeauthoryear{Hunter}{Hunter}{2007}]{Hunter:2007}
Hunter J.~D.,  2007, Computing In Science \& Engineering, 9, 90

\bibitem[\protect\citeauthoryear{Hurvich \& Tsai}{Hurvich \&
  Tsai}{1989}]{Hurvich1989}
Hurvich C.,  Tsai C.-L.,  1989, \mn@doi [Biometrika] {10.1093/biomet/76.2.297},
  76, 297

\bibitem[\protect\citeauthoryear{{Ilbert} et~al.,}{{Ilbert}
  et~al.}{2009}]{Ilbert2009}
{Ilbert} O.,  et~al., 2009, \mn@doi [\apj] {10.1088/0004-637X/690/2/1236},
  \href {http://adsabs.harvard.edu/abs/2009ApJ...690.1236I} {690, 1236}

\bibitem[\protect\citeauthoryear{{Ilbert} et~al.,}{{Ilbert}
  et~al.}{2010}]{Ilbert2010}
{Ilbert} O.,  et~al., 2010, \mn@doi [\apj] {10.1088/0004-637X/709/2/644}, \href
  {http://adsabs.harvard.edu/abs/2010ApJ...709..644I} {709, 644}

\bibitem[\protect\citeauthoryear{{Ilbert} et~al.,}{{Ilbert}
  et~al.}{2013}]{Ilbert2013}
{Ilbert} O.,  et~al., 2013, \mn@doi [\aap] {10.1051/0004-6361/201321100}, \href
  {http://adsabs.harvard.edu/abs/2013A%26A...556A..55I} {556, A55}

\bibitem[\protect\citeauthoryear{{Ilbert} et~al.,}{{Ilbert}
  et~al.}{2015}]{Ilbert2015}
{Ilbert} O.,  et~al., 2015, \mn@doi [\aap] {10.1051/0004-6361/201425176}, \href
  {http://adsabs.harvard.edu/abs/2015A%26A...579A...2I} {579, A2}

\bibitem[\protect\citeauthoryear{{Ilbert} et~al.,}{{Ilbert}
  et~al.}{2017}]{Ilbert2017}
{Ilbert} O.,  et~al., 2017, VizieR Online Data Catalog, \href
  {https://ui.adsabs.harvard.edu/\#abs/2017yCat..16901236I} {p. J/ApJ/690/1236}

\bibitem[\protect\citeauthoryear{{Ivison} et~al.,}{{Ivison}
  et~al.}{2010a}]{Ivison2010a}
{Ivison} R.~J.,  et~al., 2010a, \mn@doi [\mnras]
  {10.1111/j.1365-2966.2009.15918.x}, \href
  {https://ui.adsabs.harvard.edu/\#abs/2010MNRAS.402..245I} {402, 245}

\bibitem[\protect\citeauthoryear{{Ivison} et~al.,}{{Ivison}
  et~al.}{2010b}]{Ivison2010b}
{Ivison} R.~J.,  et~al., 2010b, \mn@doi [\aap] {10.1051/0004-6361/201014552},
  \href {https://ui.adsabs.harvard.edu/\#abs/2010A&A...518L..31I} {518, L31}

\bibitem[\protect\citeauthoryear{{Iyer} et~al.,}{{Iyer}
  et~al.}{2018}]{Iyer2018}
{Iyer} K.,  et~al., 2018, \mn@doi [\apj] {10.3847/1538-4357/aae0fa}, \href
  {https://ui.adsabs.harvard.edu/#abs/2018ApJ...866..120I} {866, 120}

\bibitem[\protect\citeauthoryear{{Jarrett} et~al.,}{{Jarrett}
  et~al.}{2011}]{Jarrett2011}
{Jarrett} T.~H.,  et~al., 2011, \mn@doi [\apj] {10.1088/0004-637X/735/2/112},
  \href {https://ui.adsabs.harvard.edu/abs/2011ApJ...735..112J} {735, 112}

\bibitem[\protect\citeauthoryear{{Jim{\'e}nez-Andrade}
  et~al.,}{{Jim{\'e}nez-Andrade} et~al.}{2019}]{JimenezAndrade2019}
{Jim{\'e}nez-Andrade} E.~F.,  et~al., 2019, \mn@doi [\aap]
  {10.1051/0004-6361/201935178}, \href
  {https://ui.adsabs.harvard.edu/abs/2019A&A...625A.114J} {625, A114}

\bibitem[\protect\citeauthoryear{{Johnston}, {Vaccari}, {Jarvis}, {Smith},
  {Giovannoli}, {H{\"a}u{\ss}ler}  \& {Prescott}}{{Johnston}
  et~al.}{2015}]{Johnston2015}
{Johnston} R.,  {Vaccari} M.,  {Jarvis} M.,  {Smith} M.,  {Giovannoli} E.,
  {H{\"a}u{\ss}ler} B.,   {Prescott} M.,  2015, \mn@doi [\mnras]
  {10.1093/mnras/stv1715}, \href
  {http://adsabs.harvard.edu/abs/2015MNRAS.453.2540J} {453, 2540}

\bibitem[\protect\citeauthoryear{{Karim} et~al.,}{{Karim}
  et~al.}{2011}]{Karim2011}
{Karim} A.,  et~al., 2011, \mn@doi [\apj] {10.1088/0004-637X/730/2/61}, \href
  {http://adsabs.harvard.edu/abs/2011ApJ...730...61K} {730, 61}

\bibitem[\protect\citeauthoryear{{Kartaltepe} et~al.,}{{Kartaltepe}
  et~al.}{2010a}]{Kartaltepe2010a}
{Kartaltepe} J.~S.,  et~al., 2010a, \mn@doi [\apj]
  {10.1088/0004-637X/709/2/572}, \href
  {https://ui.adsabs.harvard.edu/abs/2010ApJ...709..572K} {709, 572}

\bibitem[\protect\citeauthoryear{{Kartaltepe} et~al.,}{{Kartaltepe}
  et~al.}{2010b}]{Kartaltepe2010b}
{Kartaltepe} J.~S.,  et~al., 2010b, \mn@doi [\apj]
  {10.1088/0004-637X/721/1/98}, \href
  {https://ui.adsabs.harvard.edu/abs/2010ApJ...721...98K} {721, 98}

\bibitem[\protect\citeauthoryear{{Katsianis} et~al.,}{{Katsianis}
  et~al.}{2020}]{Katsianis2020}
{Katsianis} A.,  et~al., 2020, \mn@doi [\mnras] {10.1093/mnras/staa157}, \href
  {https://ui.adsabs.harvard.edu/abs/2020MNRAS.492.5592K} {492, 5592}

\bibitem[\protect\citeauthoryear{{Kauffmann} et~al.,}{{Kauffmann}
  et~al.}{2003a}]{Kauffmann2003}
{Kauffmann} G.,  et~al., 2003a, \mn@doi [\mnras]
  {10.1046/j.1365-8711.2003.06291.x}, \href
  {https://ui.adsabs.harvard.edu/abs/2003MNRAS.341...33K} {341, 33}

\bibitem[\protect\citeauthoryear{{Kauffmann} et~al.,}{{Kauffmann}
  et~al.}{2003b}]{Kauffmann2003b}
{Kauffmann} G.,  et~al., 2003b, \mn@doi [\mnras]
  {10.1046/j.1365-8711.2003.06292.x}, \href
  {https://ui.adsabs.harvard.edu/abs/2003MNRAS.341...54K} {341, 54}

\bibitem[\protect\citeauthoryear{{Kennicutt} \& {Evans}}{{Kennicutt} \&
  {Evans}}{2012}]{Kennicutt2012}
{Kennicutt} R.~C.,  {Evans} N.~J.,  2012, \mn@doi [\araa]
  {10.1146/annurev-astro-081811-125610}, \href
  {http://adsabs.harvard.edu/abs/2012ARA%26A..50..531K} {50, 531}

\bibitem[\protect\citeauthoryear{{Koekemoer} et~al.,}{{Koekemoer}
  et~al.}{2007}]{Koekemoer2007}
{Koekemoer} A.~M.,  et~al., 2007, \mn@doi [\apjs] {10.1086/520086}, \href
  {http://adsabs.harvard.edu/abs/2007ApJS..172..196K} {172, 196}

\bibitem[\protect\citeauthoryear{{Koyama} et~al.,}{{Koyama}
  et~al.}{2013}]{Koyama2013}
{Koyama} Y.,  et~al., 2013, \mn@doi [\mnras] {10.1093/mnras/stt1035}, \href
  {https://ui.adsabs.harvard.edu/abs/2013MNRAS.434..423K} {434, 423}

\bibitem[\protect\citeauthoryear{{Koyama}, {Kodama}, {Tadaki}, {Hayashi},
  {Tanaka}  \& {Shimakawa}}{{Koyama} et~al.}{2014}]{Koyama2014}
{Koyama} Y.,  {Kodama} T.,  {Tadaki} K.-i.,  {Hayashi} M.,  {Tanaka} I.,
  {Shimakawa} R.,  2014, \mn@doi [\apj] {10.1088/0004-637X/789/1/18}, \href
  {https://ui.adsabs.harvard.edu/\#abs/2014ApJ...789...18K} {789, 18}

\bibitem[\protect\citeauthoryear{{Kroupa}}{{Kroupa}}{2001}]{Kroupa2001}
{Kroupa} P.,  2001, \mn@doi [\mnras] {10.1046/j.1365-8711.2001.04022.x}, \href
  {https://ui.adsabs.harvard.edu/abs/2001MNRAS.322..231K} {322, 231}

\bibitem[\protect\citeauthoryear{{Lacki}, {Thompson}  \& {Quataert}}{{Lacki}
  et~al.}{2010}]{Lacki2010}
{Lacki} B.~C.,  {Thompson} T.~A.,   {Quataert} E.,  2010, \mn@doi [\apj]
  {10.1088/0004-637X/717/1/1}, \href
  {https://ui.adsabs.harvard.edu/\#abs/2010ApJ...717....1L} {717, 1}

\bibitem[\protect\citeauthoryear{{Laigle} et~al.,}{{Laigle}
  et~al.}{2016}]{Laigle2016}
{Laigle} C.,  et~al., 2016, \mn@doi [\apjs] {10.3847/0067-0049/224/2/24}, \href
  {http://adsabs.harvard.edu/abs/2016ApJS..224...24L} {224, 24}

\bibitem[\protect\citeauthoryear{{Lang} et~al.,}{{Lang}
  et~al.}{2014}]{Lang2014}
{Lang} P.,  et~al., 2014, \mn@doi [\apj] {10.1088/0004-637X/788/1/11}, \href
  {https://ui.adsabs.harvard.edu/\#abs/2014ApJ...788...11L} {788, 11}

\bibitem[\protect\citeauthoryear{{Larson} et~al.,}{{Larson}
  et~al.}{2011}]{Larson2011}
{Larson} D.,  et~al., 2011, \mn@doi [The Astrophysical Journal Supplement
  Series] {10.1088/0067-0049/192/2/16}, \href
  {https://ui.adsabs.harvard.edu/abs/2011ApJS..192...16L} {192, 16}

\bibitem[\protect\citeauthoryear{{Leauthaud} et~al.,}{{Leauthaud}
  et~al.}{2010}]{Leauthaud2010}
{Leauthaud} A.,  et~al., 2010, \mn@doi [\apj] {10.1088/0004-637X/709/1/97},
  \href {https://ui.adsabs.harvard.edu/\#abs/2010ApJ...709...97L} {709, 97}

\bibitem[\protect\citeauthoryear{{Leauthaud} et~al.,}{{Leauthaud}
  et~al.}{2012}]{Leauthaud2012}
{Leauthaud} A.,  et~al., 2012, \mn@doi [\apj] {10.1088/0004-637X/744/2/159},
  \href {https://ui.adsabs.harvard.edu/abs/2012ApJ...744..159L} {744, 159}

\bibitem[\protect\citeauthoryear{{Lee} et~al.,}{{Lee} et~al.}{2015}]{Lee2015}
{Lee} N.,  et~al., 2015, \mn@doi [\apj] {10.1088/0004-637X/801/2/80}, \href
  {http://adsabs.harvard.edu/abs/2015ApJ...801...80L} {801, 80}

\bibitem[\protect\citeauthoryear{{Lee} et~al.,}{{Lee} et~al.}{2018}]{Lee2018}
{Lee} B.,  et~al., 2018, \mn@doi [\apj] {10.3847/1538-4357/aaa40f}, \href
  {https://ui.adsabs.harvard.edu/\#abs/2018ApJ...853..131L} {853, 131}

\bibitem[\protect\citeauthoryear{{Legrand} et~al.,}{{Legrand}
  et~al.}{2018}]{Legrand2018}
{Legrand} L.,  et~al., 2018, arXiv e-prints, \href
  {https://ui.adsabs.harvard.edu/\#abs/2018arXiv181010557L} {p.
  arXiv:1810.10557}

\bibitem[\protect\citeauthoryear{{Leroy}, {Bolatto}, {Simon}  \&
  {Blitz}}{{Leroy} et~al.}{2005}]{Leroy2005}
{Leroy} A.,  {Bolatto} A.~D.,  {Simon} J.~D.,   {Blitz} L.,  2005, \mn@doi
  [\apj] {10.1086/429578}, \href
  {https://ui.adsabs.harvard.edu/abs/2005ApJ...625..763L} {625, 763}

\bibitem[\protect\citeauthoryear{{Lilly}, {Le Fevre}, {Hammer}  \&
  {Crampton}}{{Lilly} et~al.}{1996}]{Lilly1996}
{Lilly} S.~J.,  {Le Fevre} O.,  {Hammer} F.,   {Crampton} D.,  1996, \mn@doi
  [\apjl] {10.1086/309975}, \href
  {https://ui.adsabs.harvard.edu/abs/1996ApJ...460L...1L} {460, L1}

\bibitem[\protect\citeauthoryear{{Liu} et~al.,}{{Liu} et~al.}{2018}]{Liu2018}
{Liu} D.,  et~al., 2018, \mn@doi [\apj] {10.3847/1538-4357/aaa600}, \href
  {https://ui.adsabs.harvard.edu/\#abs/2018ApJ...853..172L} {853, 172}

\bibitem[\protect\citeauthoryear{{Liu} et~al.,}{{Liu} et~al.}{2019a}]{Liu2019}
{Liu} D.,  et~al., 2019a, \mn@doi [\apjs] {10.3847/1538-4365/ab42da}, \href
  {https://ui.adsabs.harvard.edu/abs/2019ApJS..244...40L} {244, 40}

\bibitem[\protect\citeauthoryear{{Liu} et~al.,}{{Liu} et~al.}{2019b}]{Liu2019b}
{Liu} D.,  et~al., 2019b, \mn@doi [\apj] {10.3847/1538-4357/ab578d}, \href
  {https://ui.adsabs.harvard.edu/abs/2019ApJ...887..235L} {887, 235}

\bibitem[\protect\citeauthoryear{{Livermore}, {Finkelstein}  \&
  {Lotz}}{{Livermore} et~al.}{2017}]{Livermore2017}
{Livermore} R.~C.,  {Finkelstein} S.~L.,   {Lotz} J.~M.,  2017, \mn@doi [\apj]
  {10.3847/1538-4357/835/2/113}, \href
  {https://ui.adsabs.harvard.edu/abs/2017ApJ...835..113L} {835, 113}

\bibitem[\protect\citeauthoryear{{Madau} \& {Dickinson}}{{Madau} \&
  {Dickinson}}{2014}]{Madau2014}
{Madau} P.,  {Dickinson} M.,  2014, \mn@doi [\araa]
  {10.1146/annurev-astro-081811-125615}, \href
  {http://adsabs.harvard.edu/abs/2014ARA%26A..52..415M} {52, 415}

\bibitem[\protect\citeauthoryear{{Madau}, {Pozzetti}  \& {Dickinson}}{{Madau}
  et~al.}{1998}]{Madau1998}
{Madau} P.,  {Pozzetti} L.,   {Dickinson} M.,  1998, \mn@doi [\apj]
  {10.1086/305523}, \href
  {https://ui.adsabs.harvard.edu/abs/1998ApJ...498..106M} {498, 106}

\bibitem[\protect\citeauthoryear{{Madgwick}}{{Madgwick}}{2003}]{Madgwick2003}
{Madgwick} D.~S.,  2003, \mn@doi [\mnras] {10.1046/j.1365-8711.2003.06033.x},
  \href {https://ui.adsabs.harvard.edu/abs/2003MNRAS.338..197M} {338, 197}

\bibitem[\protect\citeauthoryear{{Magdis}, {Rigopoulou}, {Huang}  \&
  {Fazio}}{{Magdis} et~al.}{2010}]{Magdis2010}
{Magdis} G.~E.,  {Rigopoulou} D.,  {Huang} J.-S.,   {Fazio} G.~G.,  2010,
  \mn@doi [\mnras] {10.1111/j.1365-2966.2009.15779.x}, \href
  {http://adsabs.harvard.edu/abs/2010MNRAS.401.1521M} {401, 1521}

\bibitem[\protect\citeauthoryear{{Magnelli} et~al.,}{{Magnelli}
  et~al.}{2014}]{Magnelli2014}
{Magnelli} B.,  et~al., 2014, \mn@doi [\aap] {10.1051/0004-6361/201322217},
  \href {https://ui.adsabs.harvard.edu/abs/2014A&A...561A..86M} {561, A86}

\bibitem[\protect\citeauthoryear{{Magnelli} et~al.,}{{Magnelli}
  et~al.}{2015}]{Magnelli2015}
{Magnelli} B.,  et~al., 2015, \mn@doi [\aap] {10.1051/0004-6361/201424937},
  \href {http://adsabs.harvard.edu/abs/2015A%26A...573A..45M} {573, A45}

\bibitem[\protect\citeauthoryear{{Maier}, {Ziegler}, {Haines}  \&
  {Smith}}{{Maier} et~al.}{2019}]{Maier2019}
{Maier} C.,  {Ziegler} B.~L.,  {Haines} C.~P.,   {Smith} G.~P.,  2019, \mn@doi
  [\aap] {10.1051/0004-6361/201834290}, \href
  {https://ui.adsabs.harvard.edu/abs/2019A&A...621A.131M} {621, A131}

\bibitem[\protect\citeauthoryear{{Mao}, {Huynh}, {Norris}, {Dickinson},
  {Frayer}, {Helou}  \& {Monkiewicz}}{{Mao} et~al.}{2011}]{Mao2011}
{Mao} M.~Y.,  {Huynh} M.~T.,  {Norris} R.~P.,  {Dickinson} M.,  {Frayer} D.,
  {Helou} G.,   {Monkiewicz} J.~A.,  2011, \mn@doi [\apj]
  {10.1088/0004-637X/731/2/79}, \href
  {https://ui.adsabs.harvard.edu/\#abs/2011ApJ...731...79M} {731, 79}

\bibitem[\protect\citeauthoryear{{Mateos} et~al.,}{{Mateos}
  et~al.}{2012}]{Mateos2012}
{Mateos} S.,  et~al., 2012, \mn@doi [\mnras]
  {10.1111/j.1365-2966.2012.21843.x}, \href
  {https://ui.adsabs.harvard.edu/abs/2012MNRAS.426.3271M} {426, 3271}

\bibitem[\protect\citeauthoryear{{Matthee} \& {Schaye}}{{Matthee} \&
  {Schaye}}{2019}]{Matthee2019}
{Matthee} J.,  {Schaye} J.,  2019, \mn@doi [\mnras] {10.1093/mnras/stz030},
  \href {https://ui.adsabs.harvard.edu/abs/2019MNRAS.484..915M} {484, 915}

\bibitem[\protect\citeauthoryear{{McCracken} et~al.,}{{McCracken}
  et~al.}{2010}]{McCracken2010}
{McCracken} H.~J.,  et~al., 2010, \mn@doi [\apj] {10.1088/0004-637X/708/1/202},
  \href {http://adsabs.harvard.edu/abs/2010ApJ...708..202M} {708, 202}

\bibitem[\protect\citeauthoryear{{McCracken} et~al.,}{{McCracken}
  et~al.}{2012}]{McCracken2012}
{McCracken} H.~J.,  et~al., 2012, \mn@doi [\aap] {10.1051/0004-6361/201219507},
  \href {http://adsabs.harvard.edu/abs/2012A%26A...544A.156M} {544, A156}

\bibitem[\protect\citeauthoryear{{McGee}, {Balogh}, {Wilman}, {Bower},
  {Mulchaey}, {Parker}  \& {Oemler}}{{McGee} et~al.}{2011}]{McGee2011}
{McGee} S.~L.,  {Balogh} M.~L.,  {Wilman} D.~J.,  {Bower} R.~G.,  {Mulchaey}
  J.~S.,  {Parker} L.~C.,   {Oemler} A.,  2011, \mn@doi [\mnras]
  {10.1111/j.1365-2966.2010.18189.x}, \href
  {https://ui.adsabs.harvard.edu/\#abs/2011MNRAS.413..996M} {413, 996}

\bibitem[\protect\citeauthoryear{{McPartland}, {Sanders}, {Kewley}  \&
  {Leslie}}{{McPartland} et~al.}{2019}]{McPartland2019}
{McPartland} C.,  {Sanders} D.~B.,  {Kewley} L.~J.,   {Leslie} S.~K.,  2019,
  \mn@doi [\mnras] {10.1093/mnrasl/sly202}, \href
  {https://ui.adsabs.harvard.edu/abs/2019MNRAS.482L.129M} {482, L129}

\bibitem[\protect\citeauthoryear{{Meidt} et~al.,}{{Meidt}
  et~al.}{2018}]{Meidt2018}
{Meidt} S.~E.,  et~al., 2018, \mn@doi [\apj] {10.3847/1538-4357/aaa290}, \href
  {https://ui.adsabs.harvard.edu/\#abs/2018ApJ...854..100M} {854, 100}

\bibitem[\protect\citeauthoryear{{Mishra} \& {Dai}}{{Mishra} \&
  {Dai}}{2020}]{Mishra2020}
{Mishra} H.~D.,  {Dai} X.,  2020, \mn@doi [\aj] {10.3847/1538-3881/ab6225},
  \href {https://ui.adsabs.harvard.edu/abs/2020AJ....159...69M} {159, 69}

\bibitem[\protect\citeauthoryear{{Moln{\'a}r} et~al.,}{{Moln{\'a}r}
  et~al.}{2018}]{Molnar2018}
{Moln{\'a}r} D.~C.,  et~al., 2018, \mn@doi [\mnras] {10.1093/mnras/stx3234},
  \href {http://adsabs.harvard.edu/abs/2018MNRAS.475..827M} {475, 827}

\bibitem[\protect\citeauthoryear{{Moln{\'a}r}, {Sargent}  \&
  {Leslie}}{{Moln{\'a}r} et~al.}{2020}]{Molnar2020}
{Moln{\'a}r} D.~C.,  {Sargent} M.~T.,   {Leslie} S.,  2020, \mn@doi [\mnras,
  Submitted] {10.1093/mnras/stx3234}, 475, tbd

\bibitem[\protect\citeauthoryear{{Mori{\'c}}, {Smol{\v{c}}i{\'c}}, {Kimball},
  {Riechers}, {Ivezi{\'c}}  \& {Scoville}}{{Mori{\'c}}
  et~al.}{2010}]{Moric2010}
{Mori{\'c}} I.,  {Smol{\v{c}}i{\'c}} V.,  {Kimball} A.,  {Riechers} D.~A.,
  {Ivezi{\'c}} {\v{Z}}.,   {Scoville} N.,  2010, \mn@doi [\apj]
  {10.1088/0004-637X/724/1/779}, \href
  {https://ui.adsabs.harvard.edu/\#abs/2010ApJ...724..779M} {724, 779}

\bibitem[\protect\citeauthoryear{{Moster}, {Somerville}, {Maulbetsch}, {van den
  Bosch}, {Macci{\`o}}, {Naab}  \& {Oser}}{{Moster} et~al.}{2010}]{Moster2010}
{Moster} B.~P.,  {Somerville} R.~S.,  {Maulbetsch} C.,  {van den Bosch} F.~C.,
  {Macci{\`o}} A.~V.,  {Naab} T.,   {Oser} L.,  2010, \mn@doi [\apj]
  {10.1088/0004-637X/710/2/903}, \href
  {https://ui.adsabs.harvard.edu/abs/2010ApJ...710..903M} {710, 903}

\bibitem[\protect\citeauthoryear{{Moster}, {Somerville}, {Newman}  \&
  {Rix}}{{Moster} et~al.}{2011}]{Moster2011}
{Moster} B.~P.,  {Somerville} R.~S.,  {Newman} J.~A.,   {Rix} H.-W.,  2011,
  \mn@doi [\apj] {10.1088/0004-637X/731/2/113}, \href
  {https://ui.adsabs.harvard.edu/abs/2011ApJ...731..113M} {731, 113}

\bibitem[\protect\citeauthoryear{{Moustakas} et~al.,}{{Moustakas}
  et~al.}{2013}]{Moustakas2013}
{Moustakas} J.,  et~al., 2013, \mn@doi [\apj] {10.1088/0004-637X/767/1/50},
  \href {http://adsabs.harvard.edu/abs/2013ApJ...767...50M} {767, 50}

\bibitem[\protect\citeauthoryear{{Mowla}, {van der Wel}, {van Dokkum}  \&
  {Miller}}{{Mowla} et~al.}{2019}]{Mowla2019}
{Mowla} L.,  {van der Wel} A.,  {van Dokkum} P.,   {Miller} T.~B.,  2019,
  \mn@doi [\apjl] {10.3847/2041-8213/ab0379}, \href
  {https://ui.adsabs.harvard.edu/abs/2019ApJ...872L..13M} {872, L13}

\bibitem[\protect\citeauthoryear{{Murphy} et~al.,}{{Murphy}
  et~al.}{2011}]{Murphy2011}
{Murphy} E.~J.,  et~al., 2011, \mn@doi [\apj] {10.1088/0004-637X/737/2/67},
  \href {http://adsabs.harvard.edu/abs/2011ApJ...737...67M} {737, 67}

\bibitem[\protect\citeauthoryear{{Muzzin} et~al.,}{{Muzzin}
  et~al.}{2013}]{Muzzin2013}
{Muzzin} A.,  et~al., 2013, \mn@doi [\apj] {10.1088/0004-637X/777/1/18}, \href
  {https://ui.adsabs.harvard.edu/abs/2013ApJ...777...18M} {777, 18}

\bibitem[\protect\citeauthoryear{{Noda} \& {Done}}{{Noda} \&
  {Done}}{2018}]{Noda2018}
{Noda} H.,  {Done} C.,  2018, \mn@doi [\mnras] {10.1093/mnras/sty2032}, \href
  {https://ui.adsabs.harvard.edu/abs/2018MNRAS.480.3898N} {480, 3898}

\bibitem[\protect\citeauthoryear{{Noeske} et~al.,}{{Noeske}
  et~al.}{2007}]{Noeske2007}
{Noeske} K.~G.,  et~al., 2007, \mn@doi [\apjl] {10.1086/517926}, \href
  {http://adsabs.harvard.edu/abs/2007ApJ...660L..43N} {660, L43}

\bibitem[\protect\citeauthoryear{{Novak} et~al.,}{{Novak}
  et~al.}{2017}]{Novak2017}
{Novak} M.,  et~al., 2017, \mn@doi [\aap] {10.1051/0004-6361/201629436}, \href
  {http://adsabs.harvard.edu/abs/2017A%26A...602A...5N} {602, A5}

\bibitem[\protect\citeauthoryear{{Novak}, {Smol{\v{c}}i{\'c}}, {Schinnerer},
  {Zamorani}, {Delvecchio}, {Bondi}  \& {Delhaize}}{{Novak}
  et~al.}{2018}]{Novak2018}
{Novak} M.,  {Smol{\v{c}}i{\'c}} V.,  {Schinnerer} E.,  {Zamorani} G.,
  {Delvecchio} I.,  {Bondi} M.,   {Delhaize} J.,  2018, \mn@doi [\aap]
  {10.1051/0004-6361/201731635}, \href
  {https://ui.adsabs.harvard.edu/\#abs/2018A&A...614A..47N} {614, A47}

\bibitem[\protect\citeauthoryear{{Ocran}, {Taylor}, {Vaccari}, {Ishwara-Chand
  ra}, {Prandoni}, {Prescott}  \& {Mancuso}}{{Ocran} et~al.}{2019}]{Ocran2019}
{Ocran} E.~F.,  {Taylor} A.~R.,  {Vaccari} M.,  {Ishwara-Chand ra} C.~H.,
  {Prandoni} I.,  {Prescott} M.,   {Mancuso} C.,  2019, \mn@doi [\mnras]
  {10.1093/mnras/stz3401}, \href
  {https://ui.adsabs.harvard.edu/abs/2019MNRAS.tmp.3050O} {p.~3050}

\bibitem[\protect\citeauthoryear{{Oesch} et~al.,}{{Oesch}
  et~al.}{2013}]{Oesch2013}
{Oesch} P.~A.,  et~al., 2013, \mn@doi [\apj] {10.1088/0004-637X/773/1/75},
  \href {http://adsabs.harvard.edu/abs/2013ApJ...773...75O} {773, 75}

\bibitem[\protect\citeauthoryear{{Oesch}, {Bouwens}, {Illingworth}, {Labb{\'e}}
   \& {Stefanon}}{{Oesch} et~al.}{2018}]{Oesch2018}
{Oesch} P.~A.,  {Bouwens} R.~J.,  {Illingworth} G.~D.,  {Labb{\'e}} I.,
  {Stefanon} M.,  2018, \mn@doi [\apj] {10.3847/1538-4357/aab03f}, \href
  {https://ui.adsabs.harvard.edu/abs/2018ApJ...855..105O} {855, 105}

\bibitem[\protect\citeauthoryear{{Ogle}, {Lanz}, {Nader}  \& {Helou}}{{Ogle}
  et~al.}{2016}]{Ogle2016}
{Ogle} P.~M.,  {Lanz} L.,  {Nader} C.,   {Helou} G.,  2016, \mn@doi [\apj]
  {10.3847/0004-637X/817/2/109}, \href
  {https://ui.adsabs.harvard.edu/abs/2016ApJ...817..109O} {817, 109}

\bibitem[\protect\citeauthoryear{{Old} et~al.,}{{Old} et~al.}{2020}]{Old2020}
{Old} L.~J.,  et~al., 2020, \mn@doi [\mnras] {10.1093/mnras/staa579}, \href
  {https://ui.adsabs.harvard.edu/abs/2020MNRAS.493.5987O} {493, 5987}

\bibitem[\protect\citeauthoryear{{Paccagnella} et~al.,}{{Paccagnella}
  et~al.}{2016}]{Paccagnella2016}
{Paccagnella} A.,  et~al., 2016, \mn@doi [\apj] {10.3847/2041-8205/816/2/L25},
  \href {https://ui.adsabs.harvard.edu/\#abs/2016ApJ...816L..25P} {816, L25}

\bibitem[\protect\citeauthoryear{{Padoan} \& {Nordlund}}{{Padoan} \&
  {Nordlund}}{2002}]{Padoan2002}
{Padoan} P.,  {Nordlund} A.,  2002, \mn@doi [\apj] {10.1086/341790}, \href
  {https://ui.adsabs.harvard.edu/abs/2002ApJ...576..870P} {576, 870}

\bibitem[\protect\citeauthoryear{{Paladino}, {Murgia}, {Helfer}, {Wong},
  {Ekers}, {Blitz}, {Gregorini}  \& {Moscadelli}}{{Paladino}
  et~al.}{2006}]{Paladino2006}
{Paladino} R.,  {Murgia} M.,  {Helfer} T.~T.,  {Wong} T.,  {Ekers} R.,  {Blitz}
  L.,  {Gregorini} L.,   {Moscadelli} L.,  2006, \mn@doi [\aap]
  {10.1051/0004-6361:20065002}, \href
  {https://ui.adsabs.harvard.edu/abs/2006A&A...456..847P} {456, 847}

\bibitem[\protect\citeauthoryear{{Pannella} et~al.,}{{Pannella}
  et~al.}{2009a}]{Pannella2009}
{Pannella} M.,  et~al., 2009a, \mn@doi [\apjl] {10.1088/0004-637X/698/2/L116},
  \href {http://adsabs.harvard.edu/abs/2009ApJ...698L.116P} {698, L116}

\bibitem[\protect\citeauthoryear{{Pannella} et~al.,}{{Pannella}
  et~al.}{2009b}]{Pannella2009b}
{Pannella} M.,  et~al., 2009b, \mn@doi [\apj] {10.1088/0004-637X/701/1/787},
  \href {https://ui.adsabs.harvard.edu/\#abs/2009ApJ...701..787P} {701, 787}

\bibitem[\protect\citeauthoryear{{Pannella} et~al.,}{{Pannella}
  et~al.}{2015}]{Pannella2015}
{Pannella} M.,  et~al., 2015, \mn@doi [\apj] {10.1088/0004-637X/807/2/141},
  \href {https://ui.adsabs.harvard.edu/\#abs/2015ApJ...807..141P} {807, 141}

\bibitem[\protect\citeauthoryear{{Parkash}, {Brown}, {Jarrett}  \&
  {Bonne}}{{Parkash} et~al.}{2018}]{Parkash2018}
{Parkash} V.,  {Brown} M. J.~I.,  {Jarrett} T.~H.,   {Bonne} N.~J.,  2018,
  \mn@doi [\apj] {10.3847/1538-4357/aad3b9}, \href
  {https://ui.adsabs.harvard.edu/abs/2018ApJ...864...40P} {864, 40}

\bibitem[\protect\citeauthoryear{{Pascarelle}, {Lanzetta}  \&
  {Fern{\'a}ndez-Soto}}{{Pascarelle} et~al.}{1998}]{Pascarelle1998}
{Pascarelle} S.~M.,  {Lanzetta} K.~M.,   {Fern{\'a}ndez-Soto} A.,  1998,
  \mn@doi [\apjl] {10.1086/311708}, \href
  {https://ui.adsabs.harvard.edu/abs/1998ApJ...508L...1P} {508, L1}

\bibitem[\protect\citeauthoryear{{Paulino-Afonso} et~al.,}{{Paulino-Afonso}
  et~al.}{2019}]{Paulino-Afonso2019}
{Paulino-Afonso} A.,  et~al., 2019, \mn@doi [\aap]
  {10.1051/0004-6361/201935137}, \href
  {https://ui.adsabs.harvard.edu/abs/2019A&A...630A..57P} {630, A57}

\bibitem[\protect\citeauthoryear{{Pearson} et~al.,}{{Pearson}
  et~al.}{2018}]{Pearson2018}
{Pearson} W.~J.,  et~al., 2018, \mn@doi [\aap] {10.1051/0004-6361/201832821},
  \href {http://adsabs.harvard.edu/abs/2018A%26A...615A.146P} {615, A146}

\bibitem[\protect\citeauthoryear{{Peng} et~al.,}{{Peng}
  et~al.}{2010}]{Peng2010}
{Peng} Y.-j.,  et~al., 2010, \mn@doi [\apj] {10.1088/0004-637X/721/1/193},
  \href {http://adsabs.harvard.edu/abs/2010ApJ...721..193P} {721, 193}

\bibitem[\protect\citeauthoryear{{Peng}, {Lilly}, {Renzini}  \&
  {Carollo}}{{Peng} et~al.}{2012}]{Peng2012}
{Peng} Y.-j.,  {Lilly} S.~J.,  {Renzini} A.,   {Carollo} M.,  2012, \mn@doi
  [\apj] {10.1088/0004-637X/757/1/4}, \href
  {https://ui.adsabs.harvard.edu/abs/2012ApJ...757....4P} {757, 4}

\bibitem[\protect\citeauthoryear{{Pharo} et~al.,}{{Pharo}
  et~al.}{2020}]{Pharo2020}
{Pharo} J.,  et~al., 2020, \mn@doi [\apj] {10.3847/1538-4357/ab5f5c}, \href
  {https://ui.adsabs.harvard.edu/abs/2020ApJ...888...79P} {888, 79}

\bibitem[\protect\citeauthoryear{{Popesso} et~al.,}{{Popesso}
  et~al.}{2015}]{Popesso2015b}
{Popesso} P.,  et~al., 2015, \mn@doi [\aap] {10.1051/0004-6361/201424715},
  \href {https://ui.adsabs.harvard.edu/\#abs/2015A&A...579A.132P} {579, A132}

\bibitem[\protect\citeauthoryear{{Popesso} et~al.,}{{Popesso}
  et~al.}{2019a}]{Popesso2019}
{Popesso} P.,  et~al., 2019a, \mn@doi [\mnras] {10.1093/mnras/sty3210}, \href
  {https://ui.adsabs.harvard.edu/abs/2019MNRAS.483.3213P} {483, 3213}

\bibitem[\protect\citeauthoryear{{Popesso} et~al.,}{{Popesso}
  et~al.}{2019b}]{Popesso2019b}
{Popesso} P.,  et~al., 2019b, \mn@doi [\mnras] {10.1093/mnras/stz2635}, \href
  {https://ui.adsabs.harvard.edu/abs/2019MNRAS.490.5285P} {490, 5285}

\bibitem[\protect\citeauthoryear{{Pozzetti} et~al.,}{{Pozzetti}
  et~al.}{2010}]{Pozzetti2010}
{Pozzetti} L.,  et~al., 2010, \mn@doi [\aap] {10.1051/0004-6361/200913020},
  \href {http://adsabs.harvard.edu/abs/2010A%26A...523A..13P} {523, A13}

\bibitem[\protect\citeauthoryear{{Renzini}}{{Renzini}}{2016}]{Renzini2016}
{Renzini} A.,  2016, \mn@doi [\mnras] {10.1093/mnrasl/slw066}, \href
  {https://ui.adsabs.harvard.edu/abs/2016MNRAS.460L..45R} {460, L45}

\bibitem[\protect\citeauthoryear{{Renzini} \& {Peng}}{{Renzini} \&
  {Peng}}{2015}]{Renzini2015}
{Renzini} A.,  {Peng} Y.-j.,  2015, \mn@doi [\apj]
  {10.1088/2041-8205/801/2/L29}, \href
  {https://ui.adsabs.harvard.edu/\#abs/2015ApJ...801L..29R} {801, L29}

\bibitem[\protect\citeauthoryear{{Ricciardelli}, {Cava}, {Varela}  \&
  {Quilis}}{{Ricciardelli} et~al.}{2014}]{Ricciardelli2014}
{Ricciardelli} E.,  {Cava} A.,  {Varela} J.,   {Quilis} V.,  2014, \mn@doi
  [\mnras] {10.1093/mnras/stu2061}, \href
  {https://ui.adsabs.harvard.edu/\#abs/2014MNRAS.445.4045R} {445, 4045}

\bibitem[\protect\citeauthoryear{{Rodighiero} et~al.,}{{Rodighiero}
  et~al.}{2010}]{Rodighiero2010}
{Rodighiero} G.,  et~al., 2010, \mn@doi [\aap] {10.1051/0004-6361/201014624},
  \href {https://ui.adsabs.harvard.edu/\#abs/2010A&A...518L..25R} {518, L25}

\bibitem[\protect\citeauthoryear{{Rodighiero} et~al.,}{{Rodighiero}
  et~al.}{2011}]{Rodighiero2011}
{Rodighiero} G.,  et~al., 2011, \mn@doi [\apj] {10.1088/2041-8205/739/2/L40},
  \href {https://ui.adsabs.harvard.edu/\#abs/2011ApJ...739L..40R} {739, L40}

\bibitem[\protect\citeauthoryear{{Rodighiero} et~al.,}{{Rodighiero}
  et~al.}{2014}]{Rodighiero2014}
{Rodighiero} G.,  et~al., 2014, \mn@doi [\mnras] {10.1093/mnras/stu1110}, \href
  {https://ui.adsabs.harvard.edu/abs/2014MNRAS.443...19R} {443, 19}

\bibitem[\protect\citeauthoryear{{Rodr{\'\i}guez-Mu{\~n}oz}
  et~al.,}{{Rodr{\'\i}guez-Mu{\~n}oz} et~al.}{2019}]{RodriguezMunoz2019}
{Rodr{\'\i}guez-Mu{\~n}oz} L.,  et~al., 2019, \mn@doi [\mnras]
  {10.1093/mnras/sty3335}, \href
  {https://ui.adsabs.harvard.edu/abs/2019MNRAS.485..586R} {485, 586}

\bibitem[\protect\citeauthoryear{{Sabater} et~al.,}{{Sabater}
  et~al.}{2019}]{Sabater2019}
{Sabater} J.,  et~al., 2019, \mn@doi [\aap] {10.1051/0004-6361/201833883},
  \href {https://ui.adsabs.harvard.edu/abs/2019A&A...622A..17S} {622, A17}

\bibitem[\protect\citeauthoryear{{Sadler}}{{Sadler}}{1982}]{Sadler1982}
{Sadler} E.~M.,  1982, Proceedings of the Astronomical Society of Australia,
  \href {https://ui.adsabs.harvard.edu/abs/1982PASAu...4..454S} {4, 454}

\bibitem[\protect\citeauthoryear{{Saintonge} et~al.,}{{Saintonge}
  et~al.}{2016}]{Saintonge2016}
{Saintonge} A.,  et~al., 2016, \mn@doi [\mnras] {10.1093/mnras/stw1715}, \href
  {https://ui.adsabs.harvard.edu/abs/2016MNRAS.462.1749S} {462, 1749}

\bibitem[\protect\citeauthoryear{{Salim} et~al.,}{{Salim}
  et~al.}{2016}]{Salim2016}
{Salim} S.,  et~al., 2016, \mn@doi [\apjs] {10.3847/0067-0049/227/1/2}, \href
  {https://ui.adsabs.harvard.edu/abs/2016ApJS..227....2S} {227, 2}

\bibitem[\protect\citeauthoryear{{Santini} et~al.,}{{Santini}
  et~al.}{2017}]{Santini2017}
{Santini} P.,  et~al., 2017, \mn@doi [\apj] {10.3847/1538-4357/aa8874}, \href
  {https://ui.adsabs.harvard.edu/\#abs/2017ApJ...847...76S} {847, 76}

\bibitem[\protect\citeauthoryear{{Sargent} et~al.,}{{Sargent}
  et~al.}{2007}]{Sargent2007}
{Sargent} M.~T.,  et~al., 2007, \mn@doi [\apjs] {10.1086/516584}, \href
  {http://adsabs.harvard.edu/abs/2007ApJS..172..434S} {172, 434}

\bibitem[\protect\citeauthoryear{{Sargent} et~al.,}{{Sargent}
  et~al.}{2010}]{Sargent2010}
{Sargent} M.~T.,  et~al., 2010, \mn@doi [\apj] {10.1088/2041-8205/714/2/L190},
  \href {https://ui.adsabs.harvard.edu/\#abs/2010ApJ...714L.190S} {714, L190}

\bibitem[\protect\citeauthoryear{{Sargent}, {B{\'e}thermin}, {Daddi}  \&
  {Elbaz}}{{Sargent} et~al.}{2012}]{Sargent2012}
{Sargent} M.~T.,  {B{\'e}thermin} M.,  {Daddi} E.,   {Elbaz} D.,  2012, \mn@doi
  [\apj] {10.1088/2041-8205/747/2/L31}, \href
  {https://ui.adsabs.harvard.edu/\#abs/2012ApJ...747L..31S} {747, L31}

\bibitem[\protect\citeauthoryear{{Sargent} et~al.,}{{Sargent}
  et~al.}{2014}]{Sargent2014}
{Sargent} M.~T.,  et~al., 2014, \mn@doi [\apj] {10.1088/0004-637X/793/1/19},
  \href {http://adsabs.harvard.edu/abs/2014ApJ...793...19S} {793, 19}

\bibitem[\protect\citeauthoryear{{Scarlata} et~al.,}{{Scarlata}
  et~al.}{2007}]{Scarlata2007}
{Scarlata} C.,  et~al., 2007, \mn@doi [\apjs] {10.1086/516582}, \href
  {http://adsabs.harvard.edu/abs/2007ApJS..172..406S} {172, 406}

\bibitem[\protect\citeauthoryear{{Schaefer} et~al.,}{{Schaefer}
  et~al.}{2019}]{Schaefer2019}
{Schaefer} A.~L.,  et~al., 2019, \mn@doi [\mnras] {10.1093/mnras/sty3258},
  \href {https://ui.adsabs.harvard.edu/\#abs/2019MNRAS.483.2851S} {483, 2851}

\bibitem[\protect\citeauthoryear{{Schawinski}, {Koss}, {Berney}  \&
  {Sartori}}{{Schawinski} et~al.}{2015}]{Schawinski2015}
{Schawinski} K.,  {Koss} M.,  {Berney} S.,   {Sartori} L.~F.,  2015, \mn@doi
  [\mnras] {10.1093/mnras/stv1136}, \href
  {https://ui.adsabs.harvard.edu/abs/2015MNRAS.451.2517S} {451, 2517}

\bibitem[\protect\citeauthoryear{{Schechter}}{{Schechter}}{1976}]{Schechter1976}
{Schechter} P.,  1976, \mn@doi [\apj] {10.1086/154079}, \href
  {https://ui.adsabs.harvard.edu/abs/1976ApJ...203..297S} {203, 297}

\bibitem[\protect\citeauthoryear{{Schinnerer} et~al.,}{{Schinnerer}
  et~al.}{2004}]{Schinnerer2004}
{Schinnerer} E.,  et~al., 2004, \mn@doi [\aj] {10.1086/424860}, \href
  {https://ui.adsabs.harvard.edu/\#abs/2004AJ....128.1974S} {128, 1974}

\bibitem[\protect\citeauthoryear{{Schinnerer} et~al.,}{{Schinnerer}
  et~al.}{2007}]{Schinnerer2007}
{Schinnerer} E.,  et~al., 2007, \mn@doi [The Astrophysical Journal Supplement
  Series] {10.1086/516587}, \href
  {https://ui.adsabs.harvard.edu/\#abs/2007ApJS..172...46S} {172, 46}

\bibitem[\protect\citeauthoryear{{Schinnerer} et~al.,}{{Schinnerer}
  et~al.}{2010}]{Schinnerer2010}
{Schinnerer} E.,  et~al., 2010, \mn@doi [The Astrophysical Journal Supplement
  Series] {10.1088/0067-0049/188/2/384}, \href
  {https://ui.adsabs.harvard.edu/\#abs/2010ApJS..188..384S} {188, 384}

\bibitem[\protect\citeauthoryear{{Schreiber} et~al.,}{{Schreiber}
  et~al.}{2015}]{Schreiber2015}
{Schreiber} C.,  et~al., 2015, \mn@doi [\aap] {10.1051/0004-6361/201425017},
  \href {http://adsabs.harvard.edu/abs/2015A%26A...575A..74S} {575, A74}

\bibitem[\protect\citeauthoryear{{Schreiber}, {Elbaz}, {Pannella}, {Ciesla},
  {Wang}, {Koekemoer}, {Rafelski}  \& {Daddi}}{{Schreiber}
  et~al.}{2016}]{Schreiber2016}
{Schreiber} C.,  {Elbaz} D.,  {Pannella} M.,  {Ciesla} L.,  {Wang} T.,
  {Koekemoer} A.,  {Rafelski} M.,   {Daddi} E.,  2016, \mn@doi [\aap]
  {10.1051/0004-6361/201527200}, \href
  {https://ui.adsabs.harvard.edu/\#abs/2016A&A...589A..35S} {589, A35}

\bibitem[\protect\citeauthoryear{{Schreiber}, {Pannella}, {Leiton}, {Elbaz},
  {Wang}, {Okumura}  \& {Labb{\'e}}}{{Schreiber} et~al.}{2017}]{Schreiber2017}
{Schreiber} C.,  {Pannella} M.,  {Leiton} R.,  {Elbaz} D.,  {Wang} T.,
  {Okumura} K.,   {Labb{\'e}} I.,  2017, \mn@doi [\aap]
  {10.1051/0004-6361/201629155}, \href
  {https://ui.adsabs.harvard.edu/abs/2017A&A...599A.134S} {599, A134}

\bibitem[\protect\citeauthoryear{{Scoville} et~al.,}{{Scoville}
  et~al.}{2007}]{Scoville2007}
{Scoville} N.,  et~al., 2007, \mn@doi [\apjs] {10.1086/516580}, \href
  {http://adsabs.harvard.edu/abs/2007ApJS..172...38S} {172, 38}

\bibitem[\protect\citeauthoryear{{Scoville} et~al.,}{{Scoville}
  et~al.}{2013}]{Scoville2013}
{Scoville} N.,  et~al., 2013, \mn@doi [The Astrophysical Journal Supplement
  Series] {10.1088/0067-0049/206/1/3}, \href
  {https://ui.adsabs.harvard.edu/\#abs/2013ApJS..206....3S} {206, 3}

\bibitem[\protect\citeauthoryear{{Seymour}, {Huynh}, {Dwelly}, {Symeonidis},
  {Hopkins}, {McHardy}, {Page}  \& {Rieke}}{{Seymour}
  et~al.}{2009}]{Seymour2009}
{Seymour} N.,  {Huynh} M.,  {Dwelly} T.,  {Symeonidis} M.,  {Hopkins} A.,
  {McHardy} I.~M.,  {Page} M.~J.,   {Rieke} G.,  2009, \mn@doi [\mnras]
  {10.1111/j.1365-2966.2009.15224.x}, \href
  {https://ui.adsabs.harvard.edu/\#abs/2009MNRAS.398.1573S} {398, 1573}

\bibitem[\protect\citeauthoryear{{Shivaei} et~al.,}{{Shivaei}
  et~al.}{2015}]{Shivaei2015}
{Shivaei} I.,  et~al., 2015, \mn@doi [\apj] {10.1088/0004-637X/815/2/98}, \href
  {https://ui.adsabs.harvard.edu/abs/2015ApJ...815...98S} {815, 98}

\bibitem[\protect\citeauthoryear{{Siudek} et~al.,}{{Siudek}
  et~al.}{2017}]{Siudek2017}
{Siudek} M.,  et~al., 2017, \mn@doi [\aap] {10.1051/0004-6361/201628951}, \href
  {https://ui.adsabs.harvard.edu/\#abs/2017A&A...597A.107S} {597, A107}

\bibitem[\protect\citeauthoryear{{Smol{\v{c}}i{\'c}}}{{Smol{\v{c}}i{\'c}}}{2009}]{Smolcic2009}
{Smol{\v{c}}i{\'c}} V.,  2009, \mn@doi [\apj] {10.1088/0004-637X/699/1/L43},
  \href {https://ui.adsabs.harvard.edu/\#abs/2009ApJ...699L..43S} {699, L43}

\bibitem[\protect\citeauthoryear{{Smol{\v{c}}i{\'c}}
  et~al.,}{{Smol{\v{c}}i{\'c}} et~al.}{2017a}]{Smolcic2017b}
{Smol{\v{c}}i{\'c}} V.,  et~al., 2017a, \mn@doi [\aap]
  {10.1051/0004-6361/201628704}, \href
  {https://ui.adsabs.harvard.edu/\#abs/2017A&A...602A...1S} {602, A1}

\bibitem[\protect\citeauthoryear{{Smol{\v{c}}i{\'c}}
  et~al.,}{{Smol{\v{c}}i{\'c}} et~al.}{2017b}]{Smolcic2017}
{Smol{\v{c}}i{\'c}} V.,  et~al., 2017b, \mn@doi [\aap]
  {10.1051/0004-6361/201630223}, \href
  {https://ui.adsabs.harvard.edu/\#abs/2017A&A...602A...2S} {602, A2}

\bibitem[\protect\citeauthoryear{{Sobral}, {Best}, {Smail}, {Mobasher}, {Stott}
   \& {Nisbet}}{{Sobral} et~al.}{2014}]{Sobral2014}
{Sobral} D.,  {Best} P.~N.,  {Smail} I.,  {Mobasher} B.,  {Stott} J.,
  {Nisbet} D.,  2014, \mn@doi [\mnras] {10.1093/mnras/stt2159}, \href
  {https://ui.adsabs.harvard.edu/\#abs/2014MNRAS.437.3516S} {437, 3516}

\bibitem[\protect\citeauthoryear{{Speagle}, {Steinhardt}, {Capak}  \&
  {Silverman}}{{Speagle} et~al.}{2014}]{Speagle2014}
{Speagle} J.~S.,  {Steinhardt} C.~L.,  {Capak} P.~L.,   {Silverman} J.~D.,
  2014, \mn@doi [\apjs] {10.1088/0067-0049/214/2/15}, \href
  {http://adsabs.harvard.edu/abs/2014ApJS..214...15S} {214, 15}

\bibitem[\protect\citeauthoryear{{Steidel}, {Adelberger}, {Giavalisco},
  {Dickinson}  \& {Pettini}}{{Steidel} et~al.}{1999}]{Steidel1999}
{Steidel} C.~C.,  {Adelberger} K.~L.,  {Giavalisco} M.,  {Dickinson} M.,
  {Pettini} M.,  1999, \mn@doi [\apj] {10.1086/307363}, \href
  {https://ui.adsabs.harvard.edu/abs/1999ApJ...519....1S} {519, 1}

\bibitem[\protect\citeauthoryear{{Strateva} et~al.,}{{Strateva}
  et~al.}{2001}]{Strateva2001}
{Strateva} I.,  et~al., 2001, \mn@doi [\aj] {10.1086/323301}, \href
  {https://ui.adsabs.harvard.edu/abs/2001AJ....122.1861S} {122, 1861}

\bibitem[\protect\citeauthoryear{{Su}, {Hopkins}, {Hayward}, {Ma},
  {Faucher-Gigu{\`e}re}, {Kere{\v{s}}}, {Orr}  \& {Robles}}{{Su}
  et~al.}{2018}]{Su2018}
{Su} K.-Y.,  {Hopkins} P.~F.,  {Hayward} C.~C.,  {Ma} X.,
  {Faucher-Gigu{\`e}re} C.-A.,  {Kere{\v{s}}} D.,  {Orr} M.~E.,   {Robles}
  V.~H.,  2018, arXiv e-prints, \href
  {https://ui.adsabs.harvard.edu/abs/2018arXiv180909120S} {p. arXiv:1809.09120}

\bibitem[\protect\citeauthoryear{{Tabatabaei} et~al.,}{{Tabatabaei}
  et~al.}{2017}]{Tabatabaei2017}
{Tabatabaei} F.~S.,  et~al., 2017, \mn@doi [\apj]
  {10.3847/1538-4357/836/2/185}, \href
  {http://adsabs.harvard.edu/abs/2017ApJ...836..185T} {836, 185}

\bibitem[\protect\citeauthoryear{{Tasca} et~al.,}{{Tasca}
  et~al.}{2015}]{Tasca2015}
{Tasca} L.~A.~M.,  et~al., 2015, \mn@doi [\aap] {10.1051/0004-6361/201425379},
  \href {https://ui.adsabs.harvard.edu/abs/2015A&A...581A..54T} {581, A54}

\bibitem[\protect\citeauthoryear{{Taylor}}{{Taylor}}{2005}]{2005ASPC..347...29T}
{Taylor} M.~B.,  2005, in {Shopbell} P.,  {Britton} M.,   {Ebert} R.,  eds,
  Astronomical Society of the Pacific Conference Series Vol. 347, Astronomical
  Data Analysis Software and Systems XIV. p.~29

\bibitem[\protect\citeauthoryear{{Thanjavur}, {Simard}, {Bluck}  \&
  {Mendel}}{{Thanjavur} et~al.}{2016}]{Thanjavur2016}
{Thanjavur} K.,  {Simard} L.,  {Bluck} A. F.~L.,   {Mendel} T.,  2016, \mn@doi
  [\mnras] {10.1093/mnras/stw495}, \href
  {https://ui.adsabs.harvard.edu/abs/2016MNRAS.459...44T} {459, 44}

\bibitem[\protect\citeauthoryear{{Thomas}, {Maraston}, {Schawinski}, {Sarzi}
  \& {Silk}}{{Thomas} et~al.}{2010}]{Thomas2010}
{Thomas} D.,  {Maraston} C.,  {Schawinski} K.,  {Sarzi} M.,   {Silk} J.,  2010,
  \mn@doi [\mnras] {10.1111/j.1365-2966.2010.16427.x}, \href
  {https://ui.adsabs.harvard.edu/\#abs/2010MNRAS.404.1775T} {404, 1775}

\bibitem[\protect\citeauthoryear{{Tinsley}}{{Tinsley}}{1968}]{Tinsley1968}
{Tinsley} B.~M.,  1968, \mn@doi [\apj] {10.1086/149455}, \href
  {https://ui.adsabs.harvard.edu/abs/1968ApJ...151..547T} {151, 547}

\bibitem[\protect\citeauthoryear{{Tinsley} \& {Danly}}{{Tinsley} \&
  {Danly}}{1980}]{Tinsley1980}
{Tinsley} B.~M.,  {Danly} L.,  1980, \mn@doi [\apj] {10.1086/158477}, \href
  {https://ui.adsabs.harvard.edu/abs/1980ApJ...242..435T} {242, 435}

\bibitem[\protect\citeauthoryear{{Tisani{\'c}} et~al.,}{{Tisani{\'c}}
  et~al.}{2018}]{Tisanic2019}
{Tisani{\'c}} K.,  et~al., 2018, arXiv e-prints, \href
  {https://ui.adsabs.harvard.edu/\#abs/2018arXiv181203392T} {p.
  arXiv:1812.03392}

\bibitem[\protect\citeauthoryear{{Tomczak} et~al.,}{{Tomczak}
  et~al.}{2016}]{Tomczak2016}
{Tomczak} A.~R.,  et~al., 2016, \mn@doi [\apj] {10.3847/0004-637X/817/2/118},
  \href {http://adsabs.harvard.edu/abs/2016ApJ...817..118T} {817, 118}

\bibitem[\protect\citeauthoryear{{Tomczak} et~al.,}{{Tomczak}
  et~al.}{2019}]{Tomczak2019}
{Tomczak} A.~R.,  et~al., 2019, \mn@doi [\mnras] {10.1093/mnras/stz342}, \href
  {https://ui.adsabs.harvard.edu/abs/2019MNRAS.484.4695T} {484, 4695}

\bibitem[\protect\citeauthoryear{{Torrey} et~al.,}{{Torrey}
  et~al.}{2018}]{Torrey2018}
{Torrey} P.,  et~al., 2018, \mn@doi [\mnras] {10.1093/mnrasl/sly031}, \href
  {https://ui.adsabs.harvard.edu/\#abs/2018MNRAS.477L..16T} {477, L16}

\bibitem[\protect\citeauthoryear{{Tresse} \& {Maddox}}{{Tresse} \&
  {Maddox}}{1998}]{Tresse1998}
{Tresse} L.,  {Maddox} S.~J.,  1998, \mn@doi [\apj] {10.1086/305331}, \href
  {https://ui.adsabs.harvard.edu/abs/1998ApJ...495..691T} {495, 691}

\bibitem[\protect\citeauthoryear{{Tyler}, {Rieke}  \& {Bai}}{{Tyler}
  et~al.}{2013}]{Tyler2013}
{Tyler} K.~D.,  {Rieke} G.~H.,   {Bai} L.,  2013, \mn@doi [\apj]
  {10.1088/0004-637X/773/2/86}, \href
  {https://ui.adsabs.harvard.edu/\#abs/2013ApJ...773...86T} {773, 86}

\bibitem[\protect\citeauthoryear{{Tyler}, {Bai}  \& {Rieke}}{{Tyler}
  et~al.}{2014}]{Tyler2014}
{Tyler} K.~D.,  {Bai} L.,   {Rieke} G.~H.,  2014, \mn@doi [\apj]
  {10.1088/0004-637X/794/1/31}, \href
  {https://ui.adsabs.harvard.edu/\#abs/2014ApJ...794...31T} {794, 31}

\bibitem[\protect\citeauthoryear{{Upjohn}, {Brown}, {Hopkins}  \&
  {Bonne}}{{Upjohn} et~al.}{2019}]{Upjohn2019}
{Upjohn} J.~E.,  {Brown} M. J.~I.,  {Hopkins} A.~M.,   {Bonne} N.~J.,  2019,
  \mn@doi [\pasa] {10.1017/pasa.2019.6}, \href
  {https://ui.adsabs.harvard.edu/abs/2019PASA...36...12U} {36, e012}

\bibitem[\protect\citeauthoryear{Van Der~Walt, Colbert  \& Varoquaux}{Van
  Der~Walt et~al.}{2011}]{van2011numpy}
Van Der~Walt S.,  Colbert S.~C.,   Varoquaux G.,  2011, Computing in Science \&
  Engineering, 13, 22

\bibitem[\protect\citeauthoryear{{Vardoulaki} et~al.,}{{Vardoulaki}
  et~al.}{2019}]{Vardoulaki2019}
{Vardoulaki} E.,  et~al., 2019, arXiv e-prints, \href
  {https://ui.adsabs.harvard.edu/abs/2019arXiv190110168V} {p. arXiv:1901.10168}

\bibitem[\protect\citeauthoryear{{Viero} et~al.,}{{Viero}
  et~al.}{2013}]{Viero2013}
{Viero} M.~P.,  et~al., 2013, \mn@doi [\apj] {10.1088/0004-637X/772/1/77},
  \href {https://ui.adsabs.harvard.edu/abs/2013ApJ...772...77V} {772, 77}

\bibitem[\protect\citeauthoryear{{Vito}, {Gilli}, {Vignali}, {Comastri},
  {Brusa}, {Cappelluti}  \& {Iwasawa}}{{Vito} et~al.}{2014}]{Vito2014}
{Vito} F.,  {Gilli} R.,  {Vignali} C.,  {Comastri} A.,  {Brusa} M.,
  {Cappelluti} N.,   {Iwasawa} K.,  2014, \mn@doi [\mnras]
  {10.1093/mnras/stu2004}, \href
  {https://ui.adsabs.harvard.edu/abs/2014MNRAS.445.3557V} {445, 3557}

\bibitem[\protect\citeauthoryear{{Whitaker} et~al.,}{{Whitaker}
  et~al.}{2011}]{Whitaker2011}
{Whitaker} K.~E.,  et~al., 2011, \mn@doi [\apj] {10.1088/0004-637X/735/2/86},
  \href {https://ui.adsabs.harvard.edu/\#abs/2011ApJ...735...86W} {735, 86}

\bibitem[\protect\citeauthoryear{{Whitaker}, {van Dokkum}, {Brammer}  \&
  {Franx}}{{Whitaker} et~al.}{2012}]{Whitaker2012}
{Whitaker} K.~E.,  {van Dokkum} P.~G.,  {Brammer} G.,   {Franx} M.,  2012,
  \mn@doi [\apj] {10.1088/2041-8205/754/2/L29}, \href
  {https://ui.adsabs.harvard.edu/\#abs/2012ApJ...754L..29W} {754, L29}

\bibitem[\protect\citeauthoryear{{Whitaker} et~al.,}{{Whitaker}
  et~al.}{2014}]{Whitaker2014}
{Whitaker} K.~E.,  et~al., 2014, \mn@doi [\apj] {10.1088/0004-637X/795/2/104},
  \href {http://adsabs.harvard.edu/abs/2014ApJ...795..104W} {795, 104}

\bibitem[\protect\citeauthoryear{{Whitaker} et~al.,}{{Whitaker}
  et~al.}{2015}]{Whitaker2015}
{Whitaker} K.~E.,  et~al., 2015, \mn@doi [\apjl] {10.1088/2041-8205/811/1/L12},
  \href {https://ui.adsabs.harvard.edu/abs/2015ApJ...811L..12W} {811, L12}

\bibitem[\protect\citeauthoryear{{Whitaker}, {Pope}, {Cybulski}, {Casey},
  {Popping}  \& {Yun}}{{Whitaker} et~al.}{2017}]{Whitaker2017}
{Whitaker} K.~E.,  {Pope} A.,  {Cybulski} R.,  {Casey} C.~M.,  {Popping} G.,
  {Yun} M.~S.,  2017, \mn@doi [\apj] {10.3847/1538-4357/aa94ce}, \href
  {https://ui.adsabs.harvard.edu/abs/2017ApJ...850..208W} {850, 208}

\bibitem[\protect\citeauthoryear{{White}, {Helfand}, {Becker}, {Glikman}  \&
  {de Vries}}{{White} et~al.}{2007}]{White2007}
{White} R.~L.,  {Helfand} D.~J.,  {Becker} R.~H.,  {Glikman} E.,   {de Vries}
  W.,  2007, \mn@doi [\apj] {10.1086/507700}, \href
  {https://ui.adsabs.harvard.edu/\#abs/2007ApJ...654...99W} {654, 99}

\bibitem[\protect\citeauthoryear{{Williams} et~al.,}{{Williams}
  et~al.}{2018}]{Williams2018}
{Williams} W.~L.,  et~al., 2018, \mn@doi [\mnras] {10.1093/mnras/sty026}, \href
  {https://ui.adsabs.harvard.edu/abs/2018MNRAS.475.3429W} {475, 3429}

\bibitem[\protect\citeauthoryear{{Wong} et~al.,}{{Wong}
  et~al.}{2016}]{Wong2016}
{Wong} O.~I.,  et~al., 2016, \mn@doi [\mnras] {10.1093/mnras/stw957}, \href
  {http://adsabs.harvard.edu/abs/2016MNRAS.460.1588W} {460, 1588}

\bibitem[\protect\citeauthoryear{{Wuyts} et~al.,}{{Wuyts}
  et~al.}{2011}]{Wuyts2011}
{Wuyts} S.,  et~al., 2011, \mn@doi [\apj] {10.1088/0004-637X/742/2/96}, \href
  {http://adsabs.harvard.edu/abs/2011ApJ...742...96W} {742, 96}

\bibitem[\protect\citeauthoryear{{Yun}, {Reddy}  \& {Condon}}{{Yun}
  et~al.}{2001}]{Yun2001}
{Yun} M.~S.,  {Reddy} N.~A.,   {Condon} J.~J.,  2001, \mn@doi [\apj]
  {10.1086/323145}, \href
  {https://ui.adsabs.harvard.edu/abs/2001ApJ...554..803Y} {554, 803}

\bibitem[\protect\citeauthoryear{{Zwart}, {Jarvis}, {Deane}, {Bonfield},
  {Knowles}, {Madhanpall}, {Rahmani}  \& {Smith}}{{Zwart}
  et~al.}{2014}]{Zwart2014}
{Zwart} J. T.~L.,  {Jarvis} M.~J.,  {Deane} R.~P.,  {Bonfield} D.~G.,
  {Knowles} K.,  {Madhanpall} N.,  {Rahmani} H.,   {Smith} D. J.~B.,  2014,
  \mn@doi [\mnras] {10.1093/mnras/stu053}, \href
  {https://ui.adsabs.harvard.edu/abs/2014MNRAS.439.1459Z} {439, 1459}

\bibitem[\protect\citeauthoryear{{van der Wel} et~al.,}{{van der Wel}
  et~al.}{2014}]{vanderWel2014}
{van der Wel} A.,  et~al., 2014, \mn@doi [\apj] {10.1088/0004-637X/788/1/28},
  \href {https://ui.adsabs.harvard.edu/abs/2014ApJ...788...28V} {788, 28}

\bibitem[\protect\citeauthoryear{{von der Linden}, {Wild}, {Kauffmann}, {White}
   \& {Weinmann}}{{von der Linden} et~al.}{2010}]{vonderLinden2010}
{von der Linden} A.,  {Wild} V.,  {Kauffmann} G.,  {White} S. D.~M.,
  {Weinmann} S.,  2010, \mn@doi [\mnras] {10.1111/j.1365-2966.2010.16375.x},
  \href {https://ui.adsabs.harvard.edu/\#abs/2010MNRAS.404.1231V} {404, 1231}

\makeatother
\end{thebibliography}

\clearpage


\startlongtable
\label{tab:resultsf}
\begin{deluxetable*}{ccccccccc}
\tablewidth{0pt}
\tablecaption{3GHz stacking results for star-forming mass-selected galaxies where our sample is mass-complete. $S_p$ is peak flux and $S_t$ is integrated flux from our mean stacked images and their errors are from bootstrapping. In some of the low-mass stack, a good fit for the integrated flux could not be obtained, so these values were removed. }
\tablehead{\colhead{$\Delta z_p$} & \colhead{$\langle z_p \rangle$} &\colhead{$\Delta \log(M_*)$} &\colhead{$ \langle\log(M_*)\rangle$} & \colhead{$N_{\mathrm{obj}}$}&  \colhead{$S_p$} & \colhead{$S_t$} & \colhead{rms} & \colhead{log(SFR)} \\
\colhead{ } & \colhead{ } & \colhead{log(M$_\odot$) }  &  \colhead{log(M$_\odot$)}  & & \colhead{$\mu$Jy/beam} & \colhead{$\mu$Jy} & \colhead{$\mu$Jy/beam} & \colhead{log(M$_\odot$/yr)}}
\startdata
0.3-0.5&0.40&8.6-9.0&8.79&3484&0.26$^{+0.05}_{-0.07}$ &2.23$^{+4.32}_{-1.66}$  &0.050$^{+0.001}_{-0.001}$ &0.08$^{+0.40}_{-0.50}$ \\
&0.39&9.0-9.4&9.18&2566&0.51$^{+0.07}_{-0.08}$ &2.70$^{+1128.23}_{-0.80}$ &0.058$^{+0.000}_{-0.000}$ &0.13$^{+2.22}_{-0.13}$ \\
&0.38&9.4-9.7&9.53&1322&1.24$^{+0.09}_{-0.04}$ &5.89$^{+1.05}_{-0.93}$ &0.082$^{+0.002}_{-0.001}$ &0.41$^{+0.06}_{-0.06}$ \\
&0.38&9.7-9.9&9.79&653&1.93$^{+0.17}_{-0.17}$ &7.98$^{+1.38}_{-0.84}$ &0.118$^{+0.002}_{-0.000}$ &0.50$^{+0.06}_{-0.04}$ \\
&0.39&9.9-10.1&10.00&512&2.58$^{+0.29}_{-0.24}$ &10.19$^{+1.00}_{-1.16}$ &0.129$^{+0.002}_{-0.005}$ &0.62$^{+0.03}_{-0.04}$ \\
&0.38&10.1-10.3&10.20&430&3.99$^{+0.28}_{-0.22}$ &15.54$^{+3.77}_{-1.85}$ &0.143$^{+0.002}_{-0.002}$ &0.76$^{+0.08}_{-0.05}$ \\
&0.38&10.3-10.5&10.40&340&4.37$^{+0.43}_{-0.55}$ &18.15$^{+2.09}_{-2.20}$ &0.164$^{+0.001}_{-0.002}$ &0.81$^{+0.04}_{-0.05}$ \\
&0.39&10.5-10.7&10.60&227&5.58$^{+0.62}_{-0.90}$ &21.13$^{+2.22}_{-1.72}$ &0.197$^{+0.002}_{-0.002}$ &0.89$^{+0.04}_{-0.03}$ \\
&0.38&10.7-10.9&10.79&139&5.12$^{+1.68}_{-0.84}$ &25.88$^{+3.05}_{-3.52}$ &0.250$^{+0.005}_{-0.002}$ &0.94$^{+0.04}_{-0.05}$ \\
&0.39&10.9-11.6&11.00&80&4.40$^{+1.31}_{-1.13}$ &25.31$^{+6.39}_{-4.39}$ &0.322$^{+0.004}_{-0.008}$ &0.95$^{+0.08}_{-0.07}$ \\
\hline
0.5-0.8&0.68&9.0-9.4&9.18&6623&0.46$^{+0.05}_{-0.03}$ &1.80$^{+1.42}_{-0.70}$ &0.036$^{+0.001}_{-0.000}$ &0.45$^{+0.21}_{-0.18}$ \\
 &0.67&9.4-9.7&9.54&3371&0.89$^{+0.04}_{-0.04}$ &3.29$^{+0.71}_{-0.50}$ &0.050$^{+0.000}_{-0.000}$ &0.66$^{+0.07}_{-0.06}$ \\
 &0.67&9.7-9.9&9.80&1633&1.56$^{+0.10}_{-0.13}$ &5.55$^{+0.77}_{-0.66}$ &0.073$^{+0.002}_{-0.002}$ &0.85$^{+0.05}_{-0.05}$ \\
 &0.67&9.9-10.1&9.99&1277&2.24$^{+0.10}_{-0.10}$ &7.44$^{+0.62}_{-0.63}$ &0.083$^{+0.003}_{-0.001}$ &0.96$^{+0.03}_{-0.03}$ \\
 &0.67&10.1-10.3&10.20&1031&3.32$^{+0.12}_{-0.17}$ &10.09$^{+0.83}_{-0.50}$ &0.093$^{+0.002}_{-0.003}$ &1.07$^{+0.03}_{-0.02}$ \\
 &0.67&10.3-10.5&10.40&782&3.64$^{+0.18}_{-0.18}$ &11.92$^{+0.83}_{-0.77}$ &0.106$^{+0.002}_{-0.002}$ &1.14$^{+0.02}_{-0.02}$ \\
 &0.67&10.5-10.7&10.59&602&4.50$^{+0.14}_{-0.21}$ &15.81$^{+1.17}_{-1.23}$ &0.118$^{+0.001}_{-0.001}$ &1.24$^{+0.03}_{-0.03}$ \\
 &0.67&10.7-10.9&10.80&368&5.11$^{+0.27}_{-0.24}$ &17.24$^{+1.52}_{-1.40}$ &0.161$^{+0.006}_{-0.008}$ &1.27$^{+0.03}_{-0.03}$ \\
 &0.68&10.9-11.6&11.03&233&4.91$^{+0.29}_{-0.40}$ &18.81$^{+2.02}_{-1.71}$ &0.187$^{+0.002}_{-0.001}$ &1.31$^{+0.04}_{-0.04}$ \\
 \hline
0.8-1.1&0.95&9.0-9.4&9.19&8935&0.38$^{+0.05}_{-0.03}$ & -- &0.032$^{+0.001}_{-0.001}$ & --  \\
 &0.95&9.4-9.7&9.54&4916&0.79$^{+0.05}_{-0.05}$ &2.97$^{+48.38}_{-1.19}$ &0.042$^{+0.001}_{-0.001}$ &0.92$^{+1.05}_{-0.19}$ \\
 &0.94&9.7-9.9&9.80&2495&1.31$^{+0.04}_{-0.06}$ &3.30$^{+0.55}_{-0.37}$ &0.060$^{+0.001}_{-0.002}$ &0.95$^{+0.06}_{-0.04}$ \\
 &0.94&9.9-10.1&9.99&1978&1.97$^{+0.07}_{-0.07}$ &5.06$^{+0.57}_{-0.54}$ &0.065$^{+0.001}_{-0.001}$ &1.11$^{+0.04}_{-0.04}$ \\
 &0.93&10.1-10.3&10.20&1526&2.62$^{+0.11}_{-0.11}$ &7.50$^{+0.54}_{-0.45}$ &0.075$^{+0.001}_{-0.002}$ &1.25$^{+0.03}_{-0.02}$ \\
 &0.93&10.3-10.5&10.40&1176&3.65$^{+0.14}_{-0.13}$ &10.19$^{+0.67}_{-0.59}$ &0.086$^{+0.002}_{-0.002}$ &1.36$^{+0.02}_{-0.02}$ \\
 &0.93&10.5-10.7&10.60&878&4.28$^{+0.23}_{-0.13}$ &11.85$^{+0.68}_{-0.63}$ &0.100$^{+0.001}_{-0.002}$ &1.42$^{+0.02}_{-0.02}$ \\
 &0.92&10.7-10.9&10.79&509&4.84$^{+0.50}_{-0.37}$ &14.89$^{+1.31}_{-1.17}$ &0.132$^{+0.002}_{-0.004}$ &1.49$^{+0.03}_{-0.03}$ \\
 &0.93&10.9-11.6&11.01&331&5.99$^{+0.47}_{-0.33}$ &18.76$^{+1.65}_{-1.25}$ &0.161$^{+0.001}_{-0.001}$ &1.58$^{+0.03}_{-0.03}$ \\
 \hline
1.1-1.5&1.29&9.4-9.7&9.54&6850&0.62$^{+0.04}_{-0.04}$ &1.25$^{+0.24}_{-0.17}$ &0.036$^{+0.000}_{-0.000}$ &0.86$^{+0.06}_{-0.05}$ \\
 &1.29&9.7-9.9&9.79&3480&1.02$^{+0.03}_{-0.05}$ &2.20$^{+0.30}_{-0.25}$ &0.050$^{+0.000}_{-0.000}$ &1.07$^{+0.05}_{-0.04}$ \\
 &1.28&9.9-10.1&9.99&2748&1.56$^{+0.06}_{-0.06}$ &3.83$^{+1.12}_{-0.53}$ &0.057$^{+0.000}_{-0.001}$ &1.27$^{+0.09}_{-0.05}$ \\
 &1.31&10.1-10.3&10.19&2218&2.02$^{+0.08}_{-0.07}$ &5.21$^{+0.39}_{-0.33}$ &0.062$^{+0.000}_{-0.000}$ &1.40$^{+0.03}_{-0.02}$ \\
 &1.32&10.3-10.5&10.39&1641&2.66$^{+0.10}_{-0.07}$ &7.02$^{+0.52}_{-0.49}$ &0.074$^{+0.001}_{-0.001}$ &1.52$^{+0.03}_{-0.03}$ \\
 &1.30&10.5-10.7&10.59&1284&3.35$^{+0.16}_{-0.15}$ &9.09$^{+0.53}_{-0.57}$ &0.083$^{+0.002}_{-0.000}$ &1.60$^{+0.02}_{-0.02}$ \\
 &1.31&10.7-10.9&10.80&813&4.55$^{+0.24}_{-0.27}$ &11.72$^{+0.71}_{-0.64}$ &0.102$^{+0.003}_{-0.003}$ &1.70$^{+0.02}_{-0.02}$ \\
 &1.32&10.9-11.6&11.03&611&4.74$^{+0.40}_{-0.39}$ &15.88$^{+3.42}_{-1.52}$ &0.118$^{+0.003}_{-0.002}$ &1.82$^{+0.07}_{-0.04}$ \\
1.5-2.0&1.75&9.9-10.1&9.99&3050&1.17$^{+0.07}_{-0.05}$ &2.56$^{+0.26}_{-0.30}$ &0.053$^{+0.000}_{-0.000}$ &1.39$^{+0.03}_{-0.05}$ \\
 &1.76&10.1-10.3&10.19&2449&1.65$^{+0.08}_{-0.08}$ &4.19$^{+0.60}_{-0.51}$ &0.060$^{+0.001}_{-0.001}$ &1.57$^{+0.05}_{-0.05}$ \\
 &1.75&10.3-10.5&10.39&1940&2.31$^{+0.07}_{-0.10}$ &6.00$^{+0.57}_{-0.48}$ &0.067$^{+0.001}_{-0.000}$ &1.70$^{+0.03}_{-0.03}$ \\
 &1.76&10.5-10.7&10.59&1398&3.06$^{+0.10}_{-0.13}$ &7.28$^{+0.41}_{-0.43}$ &0.079$^{+0.003}_{-0.001}$ &1.77$^{+0.02}_{-0.02}$ \\
 &1.77&10.7-10.9&10.79&913&3.71$^{+0.19}_{-0.15}$ &9.74$^{+0.83}_{-0.66}$ &0.097$^{+0.002}_{-0.002}$ &1.88$^{+0.03}_{-0.03}$ \\
 &1.78&10.9-11.6&11.03&677&4.56$^{+0.18}_{-0.35}$ &13.41$^{+95.65}_{-1.11}$ &0.112$^{+0.002}_{-0.007}$ &2.01$^{+0.77}_{-0.03}$ \\
 \hline
2.0-2.5&2.19&10.1-10.3&10.19&1310&1.39$^{+0.06}_{-0.06}$ &2.72$^{+0.40}_{-0.42}$ &0.081$^{+0.001}_{-0.002}$ &1.59$^{+0.05}_{-0.06}$ \\
 &2.18&10.3-10.5&10.40&969&1.93$^{+0.15}_{-0.20}$ &4.05$^{+0.55}_{-0.51}$ &0.096$^{+0.004}_{-0.000}$ &1.74$^{+0.05}_{-0.05}$ \\
 &2.18&10.5-10.7&10.59&824&2.83$^{+0.15}_{-0.19}$ &6.97$^{+0.59}_{-0.62}$ &0.105$^{+0.003}_{-0.003}$ &1.94$^{+0.03}_{-0.03}$ \\
 &2.20&10.7-10.9&10.79&590&3.30$^{+0.26}_{-0.23}$ &7.40$^{+0.62}_{-0.61}$ &0.122$^{+0.002}_{-0.002}$ &1.96$^{+0.03}_{-0.03}$ \\
 &2.19&10.9-11.6&11.02&478&5.94$^{+0.41}_{-0.25}$ &14.17$^{+1.36}_{-1.16}$ &0.135$^{+0.004}_{-0.003}$ &2.20$^{+0.03}_{-0.03}$ \\
 \hline
2.5-3.0&2.66&10.1-10.3&10.19&1332&1.27$^{+0.07}_{-0.09}$ &3.80$^{+1.05}_{-0.87}$ &0.080$^{+0.002}_{-0.001}$ &1.88$^{+0.09}_{-0.10}$ \\
 &2.66&10.3-10.5&10.40&925&1.62$^{+0.07}_{-0.10}$ &6.41$^{+2.02}_{-1.76}$ &0.098$^{+0.001}_{-0.001}$ &2.07$^{+0.10}_{-0.12}$ \\
 &2.67&10.5-10.7&10.60&646&2.33$^{+0.19}_{-0.20}$ &6.79$^{+1.01}_{-0.81}$ &0.115$^{+0.002}_{-0.002}$ &2.09$^{+0.05}_{-0.05}$ \\
 &2.68&10.7-10.9&10.80&363&3.80$^{+0.30}_{-0.38}$ &9.09$^{+5.77}_{-1.11}$ &0.158$^{+0.002}_{-0.002}$ &2.20$^{+0.18}_{-0.05}$ \\
 &2.67&10.9-11.6&11.05&280&6.92$^{+0.35}_{-0.29}$ &14.08$^{+1.19}_{-1.31}$ &0.174$^{+0.002}_{-0.002}$ &2.36$^{+0.03}_{-0.04}$ \\
 \hline
3.0-4.0&3.37&10.3-10.5&10.39&659&1.49$^{+0.13}_{-0.10}$ &3.30$^{+0.69}_{-0.57}$ &0.113$^{+0.001}_{-0.002}$ &2.01$^{+0.07}_{-0.07}$ \\
 &3.36&10.5-10.7&10.58&399&1.80$^{+0.34}_{-0.14}$ &5.20$^{+0.91}_{-0.95}$ &0.148$^{+0.003}_{-0.005}$ &2.18$^{+0.06}_{-0.07}$ \\
 &3.31&10.7-10.9&10.78&161&2.58$^{+0.35}_{-0.50}$ &7.53$^{+1.42}_{-1.46}$ &0.234$^{+0.002}_{-0.003}$ &2.30$^{+0.06}_{-0.08}$ \\
 &3.28&10.9-11.6&11.05&140&4.79$^{+0.52}_{-0.54}$ &9.45$^{+2.29}_{-1.31}$ &0.252$^{+0.003}_{-0.003}$ &2.38$^{+0.08}_{-0.05}$ \\
 \hline
4.0-6.0&4.52&10.3-10.5&10.39&126&0.96$^{+0.42}_{-0.13}$ &3.26$^{+7.63}_{-2.14}$ &0.258$^{+0.003}_{-0.000}$ &2.24$^{+0.44}_{-0.39}$ \\
 &4.56&10.5-10.7&10.57&65&0.83$^{+0.39}_{-0.80}$ &7.52$^{+8.47}_{-3.73}$ &0.357$^{+0.001}_{-0.002}$ &2.55$^{+0.28}_{-0.25}$ \\
 &4.59&10.7-10.9&10.80&38&2.54$^{+0.79}_{-0.40}$ &3.93$^{+1.83}_{-2.15}$ &0.466$^{+0.007}_{-0.005}$ &2.32$^{+0.14}_{-0.29}$ \\
 &4.57&10.9-11.6&11.23&60&4.15$^{+1.37}_{-0.92}$ &9.25$^{+2.30}_{-2.65}$ &0.371$^{+0.005}_{-0.004}$ &2.63$^{+0.08}_{-0.12}$\\
\enddata
\label{tab:resultsf}
\end{deluxetable*}

\startlongtable
\label{tab:resultall}
\begin{deluxetable*}{ccccccccc}
\tablewidth{0pt}
\tablecaption{3GHz stacking results for all mass-selected galaxies where our sample is mass-complete. See Table 5 for more details. }\tablehead{\colhead{$\Delta z_p$} & \colhead{$\langle z_p \rangle$} &\colhead{$\Delta \log(M_*)$} &\colhead{$ \langle\log(M_*)\rangle$} & \colhead{$N_{\mathrm{obj}}$}&  \colhead{$S_p$} & \colhead{$S_t$} & \colhead{rms} & \colhead{log(SFR)} \\
\colhead{ } & \colhead{ } & \colhead{log(M$_\odot$) }  &  \colhead{log(M$_\odot$)}  & & \colhead{$\mu$Jy/beam} & \colhead{$\mu$Jy} & \colhead{$\mu$Jy/beam} & \colhead{log(M$_\odot$/yr)}}
\startdata
0.3-0.5&0.39&8.6-9.0&8.79&3786&0.24$^{+0.05}_{-0.06}$ &2.80$^{+5.10}_{-2.11}$ &0.048$^{+0.000}_{-0.000}$ &0.15$^{+0.38}_{-0.52}$ \\
 &0.39&9.0-9.4&9.18&2794&0.47$^{+0.08}_{-0.06}$ &3.00$^{+466}_{-1.27}$ &0.055$^{+0.001}_{-0.001}$ &0.16$^{+1.85}_{-0.20}$ \\
 &0.38&9.4-9.7&9.54&1507&1.09$^{+0.05}_{-0.08}$ &5.31$^{+1.13}_{-0.84}$ &0.077$^{+0.002}_{-0.002}$ &0.36$^{+0.07}_{-0.06}$ \\
 &0.38&9.7-9.9&9.80&806&1.53$^{+0.14}_{-0.18}$ &7.88$^{+1.25}_{-1.25}$ &0.107$^{+0.000}_{-0.001}$ &0.50$^{+0.05}_{-0.06}$ \\ 
 &0.38&9.9-10.1&10.00&677&2.01$^{+0.13}_{-0.16}$ &15.36$^{+22.21}_{-7.25}$ &0.112$^{+0.000}_{-0.002}$ &0.75$^{+0.33}_{-0.23}$ \\
 &0.38&10.1-10.3&10.20&616&2.57$^{+0.29}_{-0.27}$ &13.25$^{+4.24}_{-1.51}$ &0.120$^{+0.002}_{-0.002}$ &0.70$^{+0.10}_{-0.04}$ \\
 &0.38&10.3-10.5&10.40&549&2.54$^{+0.16}_{-0.20}$ &13.47$^{+1.58}_{-1.56}$ &0.131$^{+0.002}_{-0.004}$ &0.70$^{+0.04}_{-0.05}$ \\
 &0.38&10.5-10.7&10.60&446&2.35$^{+0.15}_{-0.14}$ &12.91$^{+1.87}_{-1.42}$ &0.141$^{+0.002}_{-0.001}$ &0.69$^{+0.05}_{-0.04}$ \\
 &0.39&10.7-10.9&10.80&322&2.92$^{+0.36}_{-0.20}$ &14.97$^{+2.08}_{-2.18}$ &0.166$^{+0.001}_{-0.001}$ &0.76$^{+0.05}_{-0.06}$ \\
 &0.39&10.9-11.6&11.05&310&2.97$^{+0.20}_{-0.24}$ &12.19$^{+2.18}_{-1.72}$ &0.170$^{+0.001}_{-0.001}$ &0.69$^{+0.06}_{-0.06}$ \\
 \hline
0.5-0.8&0.68&9.0-9.4&9.18&6965&0.44$^{+0.03}_{-0.03}$ &2.03$^{+1.53}_{-0.71}$ &0.035$^{+0.001}_{-0.000}$ &0.49$^{+0.21}_{-0.16}$ \\
 &0.67&9.4-9.7&9.54&3612&0.84$^{+0.09}_{-0.06}$ &3.54$^{+0.88}_{-0.41}$ &0.049$^{+0.000}_{-0.000}$ &0.69$^{+0.08}_{-0.05}$ \\
 &0.67&9.7-9.9&9.80&1833&1.31$^{+0.07}_{-0.13}$ &5.67$^{+0.77}_{-0.74}$ &0.070$^{+0.001}_{-0.002}$ &0.86$^{+0.05}_{-0.05}$ \\
 &0.67&9.9-10.1&10.00&1545&1.85$^{+0.07}_{-0.09}$ &7.37$^{+0.92}_{-0.62}$ &0.076$^{+0.002}_{-0.000}$ &0.96$^{+0.04}_{-0.03}$ \\
 &0.67&10.1-10.3&10.20&1371&2.59$^{+0.17}_{-0.12}$ &8.59$^{+0.65}_{-0.72}$ &0.081$^{+0.001}_{-0.001}$ &1.01$^{+0.03}_{-0.03}$ \\
 &0.67&10.3-10.5&10.40&1204&2.13$^{+0.16}_{-0.11}$ &9.17$^{+0.68}_{-0.80}$ &0.086$^{+0.001}_{-0.000}$ &1.04$^{+0.03}_{-0.03}$ \\
 &0.67&10.5-10.7&10.60&1031&2.64$^{+0.16}_{-0.20}$ &11.27$^{+1.17}_{-0.94}$ &0.092$^{+0.001}_{-0.002}$ &1.12$^{+0.04}_{-0.03}$ \\
 &0.68&10.7-10.9&10.80&810&2.44$^{+0.23}_{-0.14}$ &9.99$^{+1.14}_{-1.01}$ &0.109$^{+0.003}_{-0.001}$ &1.08$^{+0.04}_{-0.04}$ \\
 &0.68&10.9-11.6&11.05&758&2.38$^{+0.18}_{-0.25}$ &10.54$^{+3.43}_{-1.54}$ &0.109$^{+0.002}_{-0.000}$ &1.10$^{+0.10}_{-0.06}$ \\
 \hline
0.8-1.1&0.94&9.0-9.4&9.19&9127&0.38$^{+0.03}_{-0.03}$ & -- &0.031$^{+0.001}_{-0.001}$ &-- \\
 &0.94&9.4-9.7&9.54&5188&0.76$^{+0.05}_{-0.05}$ &11.53$^{+112.05}_{-9.46}$ &0.041$^{+0.001}_{-0.001}$ &1.42$^{+0.87}_{-0.63}$ \\
 &0.94&9.7-9.9&9.80&2751&1.22$^{+0.08}_{-0.05}$ &3.50$^{+0.70}_{-0.44}$ &0.057$^{+0.002}_{-0.003}$ &0.97$^{+0.07}_{-0.05}$ \\
 &0.94&9.9-10.1&9.99&2335&1.65$^{+0.04}_{-0.08}$ &4.91$^{+0.63}_{-0.56}$ &0.060$^{+0.000}_{-0.001}$ &1.10$^{+0.04}_{-0.04}$ \\
 &0.93&10.1-10.3&10.20&2053&1.93$^{+0.05}_{-0.11}$ &6.73$^{+0.59}_{-0.58}$ &0.065$^{+0.001}_{-0.001}$ &1.21$^{+0.03}_{-0.03}$ \\
 &0.93&10.3-10.5&10.40&1852&2.27$^{+0.13}_{-0.14}$ &7.80$^{+0.62}_{-0.40}$ &0.069$^{+0.001}_{-0.001}$ &1.26$^{+0.03}_{-0.02}$ \\
 &0.93&10.5-10.7&10.60&1623&2.21$^{+0.10}_{-0.07}$ &8.16$^{+1.17}_{-0.77}$ &0.071$^{+0.001}_{-0.002}$ &1.28$^{+0.05}_{-0.04}$ \\
 &0.92&10.7-10.9&10.80&1230&2.25$^{+0.09}_{-0.09}$ &8.83$^{+0.72}_{-0.80}$ &0.084$^{+0.001}_{-0.001}$ &1.30$^{+0.03}_{-0.03}$ \\
 &0.92&10.9-11.6&11.05&1226&2.45$^{+0.15}_{-0.14}$ &10.32$^{+1.11}_{-0.94}$ &0.084$^{+0.002}_{-0.001}$ &1.36$^{+0.04}_{-0.03}$ \\
\hline
1.1-1.5&1.29&9.4-9.7&9.54&6912&0.61$^{+0.05}_{-0.03}$ &1.45$^{+0.18}_{-0.22}$ &0.036$^{+0.000}_{-0.000}$ &0.92$^{+0.04}_{-0.06}$ \\
 &1.29&9.7-9.9&9.80&3567&1.02$^{+0.03}_{-0.05}$ &2.51$^{+0.35}_{-0.27}$ &0.050$^{+0.000}_{-0.000}$ &1.12$^{+0.05}_{-0.04}$ \\
 &1.28&9.9-10.1&9.99&2923&1.47$^{+0.10}_{-0.06}$ &3.97$^{+1.01}_{-0.56}$ &0.055$^{+0.000}_{-0.000}$ &1.28$^{+0.08}_{-0.06}$ \\
 &1.29&10.1-10.3&10.19&2527&1.76$^{+0.07}_{-0.05}$ &5.10$^{+0.51}_{-0.35}$ &0.058$^{+0.001}_{-0.001}$ &1.39$^{+0.03}_{-0.03}$ \\
 &1.30&10.3-10.5&10.40&2035&2.10$^{+0.09}_{-0.08}$ &6.57$^{+0.66}_{-0.46}$ &0.066$^{+0.000}_{-0.000}$ &1.48$^{+0.04}_{-0.03}$ \\
 &1.28&10.5-10.7&10.60&1829&2.11$^{+0.10}_{-0.06}$ &7.54$^{+1.26}_{-0.63}$ &0.069$^{+0.001}_{-0.001}$ &1.52$^{+0.06}_{-0.03}$ \\
 &1.28&10.7-10.9&10.80&1396&2.38$^{+0.13}_{-0.13}$ &8.20$^{+0.53}_{-0.61}$ &0.079$^{+0.001}_{-0.001}$ &1.55$^{+0.02}_{-0.03}$ \\
 &1.27&10.9-11.6&11.05&1252&2.91$^{+0.29}_{-0.24}$ &11.45$^{+1.19}_{-1.19}$ &0.084$^{+0.002}_{-0.002}$ &1.67$^{+0.04}_{-0.04}$ \\
 \hline
1.5-2.0&1.75&9.9-10.1&9.99&3136&1.15$^{+0.07}_{-0.05}$ &2.79$^{+0.31}_{-0.32}$ &0.052$^{+0.000}_{-0.000}$ &1.42$^{+0.04}_{-0.05}$ \\
 &1.75&10.1-10.3&10.19&2621&1.55$^{+0.04}_{-0.09}$ &4.52$^{+0.64}_{-0.49}$ &0.058$^{+0.001}_{-0.000}$ &1.59$^{+0.05}_{-0.04}$ \\
 &1.75&10.3-10.5&10.39&2142&2.13$^{+0.09}_{-0.10}$ &6.33$^{+0.60}_{-0.39}$ &0.065$^{+0.001}_{-0.001}$ &1.72$^{+0.03}_{-0.02}$ \\
 &1.74&10.5-10.7&10.59&1692&2.56$^{+0.10}_{-0.11}$ &7.11$^{+0.51}_{-0.46}$ &0.071$^{+0.001}_{-0.001}$ &1.76$^{+0.03}_{-0.02}$ \\
 &1.73&10.7-10.9&10.79&1286&2.62$^{+0.14}_{-0.12}$ &7.97$^{+0.56}_{-0.60}$ &0.083$^{+0.000}_{-0.001}$ &1.79$^{+0.02}_{-0.03}$ \\
 &1.73&10.9-11.6&11.04&1169&2.85$^{+0.19}_{-0.21}$ &10.01$^{+25.11}_{-0.70}$ &0.088$^{+0.001}_{-0.000}$ &1.88$^{+0.46}_{-0.03}$ \\
 \hline
2.0-2.5&2.19&10.1-10.3&10.19&1353&1.37$^{+0.06}_{-0.08}$ &3.08$^{+0.61}_{-0.52}$ &0.080$^{+0.001}_{-0.002}$ &1.64$^{+0.07}_{-0.07}$ \\
 &2.18&10.3-10.5&10.40&1024&1.81$^{+0.12}_{-0.12}$ &4.31$^{+0.64}_{-0.51}$ &0.093$^{+0.004}_{-0.001}$ &1.76$^{+0.05}_{-0.05}$ \\
 &2.19&10.5-10.7&10.60&902&2.65$^{+0.21}_{-0.15}$ &7.35$^{+0.84}_{-0.68}$ &0.101$^{+0.002}_{-0.001}$ &1.96$^{+0.04}_{-0.04}$ \\
 &2.21&10.7-10.9&10.79&663&2.91$^{+0.20}_{-0.19}$ &7.62$^{+0.66}_{-0.71}$ &0.116$^{+0.002}_{-0.002}$ &1.98$^{+0.03}_{-0.04}$ \\
 &2.20&10.9-11.6&11.03&582&4.68$^{+0.23}_{-0.36}$ &13.92$^{+1.52}_{-1.24}$ &0.121$^{+0.002}_{-0.001}$ &2.20$^{+0.04}_{-0.03}$ \\
 \hline
2.5-3.0&2.66&10.1-10.3&10.19&1353&1.26$^{+0.07}_{-0.08}$ &4.29$^{+1.10}_{-0.92}$ &0.080$^{+0.002}_{-0.000}$ &1.92$^{+0.08}_{-0.09}$ \\
 &2.66&10.3-10.5&10.40&963&1.59$^{+0.08}_{-0.10}$ &6.94$^{+1.93}_{-1.89}$ &0.095$^{+0.001}_{-0.001}$ &2.10$^{+0.09}_{-0.12}$ \\
 &2.67&10.5-10.7&10.60&701&2.17$^{+0.25}_{-0.22}$ &6.89$^{+0.99}_{-0.87}$ &0.111$^{+0.003}_{-0.003}$ &2.09$^{+0.05}_{-0.05}$ \\
 &2.68&10.7-10.9&10.80&409&3.53$^{+0.21}_{-0.20}$ &9.40$^{+1.65}_{-0.88}$ &0.150$^{+0.003}_{-0.001}$ &2.21$^{+0.06}_{-0.04}$ \\
 &2.67&10.9-11.6&11.05&326&5.75$^{+0.39}_{-0.38}$ &14.04$^{+1.21}_{-1.18}$ &0.163$^{+0.004}_{-0.002}$ &2.36$^{+0.03}_{-0.03}$ \\
 \hline
3.0-4.0&3.37&10.3-10.5&10.39&664&1.49$^{+0.12}_{-0.09}$ &3.84$^{+0.58}_{-0.60}$ &0.113$^{+0.001}_{-0.002}$ &2.07$^{+0.05}_{-0.06}$ \\
 &3.36&10.5-10.7&10.58&412&1.76$^{+0.37}_{-0.13}$ &5.64$^{+1.15}_{-0.95}$ &0.145$^{+0.003}_{-0.004}$ &2.21$^{+0.07}_{-0.07}$ \\
 &3.32&10.7-10.9&10.78&186&2.40$^{+0.22}_{-0.28}$ &8.43$^{+1.21}_{-1.76}$ &0.218$^{+0.003}_{-0.003}$ &2.34$^{+0.05}_{-0.09}$ \\
 &3.31&10.9-11.6&11.06&159&4.52$^{+0.52}_{-0.29}$ &10.45$^{+1.66}_{-1.88}$ &0.237$^{+0.002}_{-0.003}$ &2.42$^{+0.05}_{-0.07}$ \\
 \hline
4.0-6.0&4.51&10.3-10.5&10.39&129&0.95$^{+0.51}_{-0.17}$ &3.23$^{+13.29}_{-2.04}$ &0.254$^{+0.001}_{-0.001}$ &2.23$^{+0.60}_{-0.37}$ \\
 &4.56&10.5-10.7&10.57&69&0.76$^{+0.29}_{-0.74}$ &9.21$^{+14.79}_{-3.83}$ &0.348$^{+0.003}_{-0.002}$ &2.63$^{+0.35}_{-0.20}$ \\
 &4.62&10.7-10.9&10.80&39&2.39$^{+0.99}_{-0.72}$ &4.39$^{+3.36}_{-2.57}$ &0.465$^{+0.007}_{-0.005}$ &2.36$^{+0.21}_{-0.32}$ \\
 &4.53&10.9-11.6&11.20&71&3.11$^{+0.79}_{-1.04}$ &9.19$^{+3.34}_{-2.12}$ &0.345$^{+0.007}_{-0.008}$ &2.62$^{+0.11}_{-0.10}$ \\
 \enddata
 \label{tab:resultall}
\end{deluxetable*}

\end{document}